\documentclass[12pt, letterpaper]{article}

\usepackage[margin = 1in]{geometry}
\usepackage[utf8]{inputenc}
\usepackage[english]{babel}
\usepackage{amssymb}
\usepackage{amsmath}
\usepackage{amsthm}
\usepackage{epsfig}
\usepackage{graphicx}
\usepackage{graphics}
\usepackage{float}
\usepackage{subfigure}
\usepackage{multirow}
\usepackage{color}
\usepackage{fullpage}
\usepackage[normalem]{ulem} 
\usepackage{makeidx}
\usepackage{xspace}
\usepackage{wrapfig}
\usepackage{import}
\usepackage{pdfpages}
\usepackage[export]{adjustbox}
\usepackage{anyfontsize}
\usepackage{hyperref}

\theoremstyle{definition}

\theoremstyle{remark}

\title{Estimating SARS-CoV-2 Infections from Deaths, Confirmed Cases,
Tests, and Random Surveys}

\author{Nicholas J. Irons\footnote{Department of Statistics}\ \ and Adrian E. Raftery$^*$\footnote{Department of Sociology}\ \footnote{To whom correspondence should be addressed. E-mail: raftery@uw.edu} \\
University of Washington 
}

\begin{document}

\maketitle

\begin{center}
\textbf{Abstract}
\end{center}
\vspace{-1ex}

There are many sources of data giving information about the number of 
SARS-CoV-2 infections in the population, but all have major drawbacks,
including biases and delayed reporting.
For example, the number of confirmed cases largely
underestimates the number of infections, deaths lag infections
substantially, while test positivity rates tend to greatly overestimate
prevalence. Representative random prevalence surveys, 
the only putatively unbiased source,
are sparse in time and space, and the results come with a big delay.
Reliable estimates of population prevalence are necessary for understanding the spread of the virus and the effects of mitigation strategies. We develop a simple Bayesian framework to estimate viral prevalence by combining the main available data sources. It is based on a discrete-time SIR model with time-varying reproductive parameter. Our model includes likelihood components that incorporate data of deaths due to the virus, confirmed cases, and the number of tests administered on each day. We anchor our inference with data from random sample testing surveys in Indiana and Ohio. We use the results from these two states to calibrate the model on positive test counts and proceed to estimate the infection fatality rate and the number of new infections on each day in each state in the USA. We estimate the extent to which reported COVID cases have underestimated true infection counts, which was large, especially in the first months of the pandemic.
We explore the implications of our results for progress towards herd immunity.



\section{Introduction}
\label{sec:intro}
SARS-CoV-2 test data are fraught with biases that obscure the true rate of infection in the population. Lack of access to viral tests, which was particularly pronounced in the early days of the pandemic, in conjunction with selection bias due to asymptomatic and mild infections yield case counts that tend to underestimate the true number of infections in the population. By the same token, test positivity rates tend to overestimate viral prevalence. Hospitalization rates and emergency room visits do not estimate the overall infection rate, and are not comparable between states or counties, or over time. Reported deaths due to COVID are considered less problematic as an estimate of the true death count and provide a more accurate reflection of the course of the pandemic \cite{nas}. 

We combine several of the main sources of data relevant to the number of infections
using a simple Bayesian model that accounts for the biases and delays
in the data.
Our model relies on data on deaths due to COVID, confirmed cases, and testing reported by the COVID Tracking Project \cite{ctp}. 
We use a modified Susceptible-Infected-Removed (SIR) model, a compartmental epidemiological model widely used to simulate the spread of disease in a population. We combine this with a Poisson likelihood for death counts and a normal likelihood for estimates of viral and seroprevalence from random sample testing surveys conducted in Indiana and Ohio \cite{mmwr,ohio}. 

With these data we infer the infection fatality rate (IFR) and obtain statistically principled estimates of the number of new infections on each day since March 2020 in Indiana and Ohio. We then leverage our results from these states to build a model for confirmed cases that accounts for preferential testing as a function of the cumulative number of tests administered in each state. This allows us to pin down the IFR and infection counts for the vast majority of states that have not conducted representative testing surveys.

Our simple Bayesian model takes inspiration from Johndrow et al. \cite{johndrow}, although it differs in significant ways. Whereas Johndrow et al. model the effect of social distancing measures by allowing the SIR contact parameter to change pre- and post-lockdown, we allow it to vary in time to account for fluctuation in the tightening and loosening of restrictions, as well as in adherence to the restrictions. Furthermore, we incorporate testing data, develop a novel preferential testing model, and include the IFR as a parameter in the model to be estimated, rather than a fixed constant. Finally, to simplify model implementation we use a discrete time SIR model, rather than a continuous time model based on differential equations.


\section{Methods}
\label{sec:methods}
\subsection{SIR model}
We first define our discrete-time SIR model for infections in each state. Let $S_t$ denote the number of susceptible people in the population on day $t$, $I_t$ the number of infections, and $R_t$ the number removed. The number removed includes those who have died of the disease and those who have recovered, and are assumed immune for the rest of the period of our study.  
With $N$ denoting the state population,
these quantities evolve in time according to the equations
\begin{equation}
\begin{cases}
S_{t+1} - S_t &= -\frac{\beta_t}{N} I_t S_t, \\
I_{t+1} - I_t &= \frac{\beta_t}{N} I_t S_t -\gamma I_t, \\
R_{t+1} - R_t &= \gamma I_t.
\end{cases}
\label{eq:sir}
\end{equation}

Note that $\nu_t = S_{t-1}-S_t$ is the number of new infections on day $t$. We allow the parameters $\beta_t$, interpreted as the mean number of contacts per person on day $t$, to vary over time. This accounts for variation in exposure due to implementation or loosening of social distancing and other policy measures over time. We model $\beta_t$ as a random walk with step size $\sigma$ estimated from the data, $\beta_{t+1} \sim \text{Normal}(\beta_t,\sigma^2).$
We assume that $\gamma^{-1}$, the average length in days of the infectious period, is determined by the disease and is therefore constant over time.

\subsection{Likelihood on deaths}

Let $\tau = \{\tau_0,\tau_1,\ldots,\tau_m\}$ denote the distribution of time to death for those infected individuals who die from the disease, i.e., $\tau_s$ is the probability of death $s$ days after infection, conditional on death occurring. Similar to Johndrow et al., who calibrate $\tau$ by matching quantiles of a negative binomial distribution to case data from China \cite{zhou,lauer}, we assume that $\tau$ follows a NegativeBinomial$(\alpha,1/(\beta+1))$ distribution with parameters $\alpha = 21, \beta=1.1$, and we truncate the distribution at the 99th percentile, or $m=40$ days, to rule out extremely delayed deaths. We denote by $D_t$ the reported deaths due to COVID on day $t$, which we obtain from the COVID Tracking Project \cite{ctp}. We link the daily new infection counts $\nu = (\nu_t)_t$ to reported deaths via the likelihood
$
D_t
\stackrel{\text{ind.}}{\sim} \text{Poisson}\left(\text{IFR}\sum_{k=1}^t \nu_k\tau_{t-k}\right).$

\subsection{Representative random prevalence surveys}

To pin down the IFR, we add likelihood components incorporating the Indiana and Ohio prevalence survey data \cite{mmwr,ohio}. Active viral prevalence in Indiana in the period April 25--29, 2020 was estimated as $\hat\theta_v = 1.74\%$.
We model this quantity using a normal approximation to the binomial distribution,
$
\hat\theta_v\sim\text{Normal}\left(\theta_v,\frac{\theta_v(1-\theta_v)}{n_v} \right)$, where $\theta_v = (\sum_{t=T_1}^{T_2} I_t)/N(T_1-T_2)$ is the average viral prevalence between days $T_1=$ April 25 and $T_2=$ April 29. Here $n_v=3,605$ is the number of viral tests administered. Similarly, the estimated seroprevalence in the testing period, $\hat\theta_s = 1.09\%$, is modeled as $\hat\theta_s \sim \text{Normal}\left(\theta_s,\frac{\theta_s(1-\theta_s)}{n_s} \right)$, where $\theta_s = \sum_{t=T_1}^{T_2} R_t/N(T_1-T_2)$ and $n_s = 3518$. These results come from the first phase of the Indiana prevalence survey described in Menachemi et al. \cite{mmwr}. Due to low response rates -- less than 8\% in the second and third phases -- we do not include data from the subsequent phases of the study in our analysis. The response rates reported in Ohio and in the first phase in Indiana were significantly higher at 18.5\% and 23.4\%, respectively.

The likelihood for the prevalence survey data from Ohio is analogous. As reported in \cite{ohio}, the estimated seroprevalence in the state is $\hat\theta_s=1.3\%$ in the period July 9--28, with a sample size of $n_s=667$. Results from the PCR tests in the same study were reported in a press conference on October 1 available on YouTube \cite{ohio-youtube}. The viral prevalence in that period is estimated as $\hat\theta_v=0.9\%$ with sample size $n_v=727$. To the best of our knowledge, these numbers have not yet been published.

\subsection{Modeling preferential testing}

\begin{figure}[htbp!]
\centering
\begin{tabular}{ll}
\includegraphics[scale=0.185]{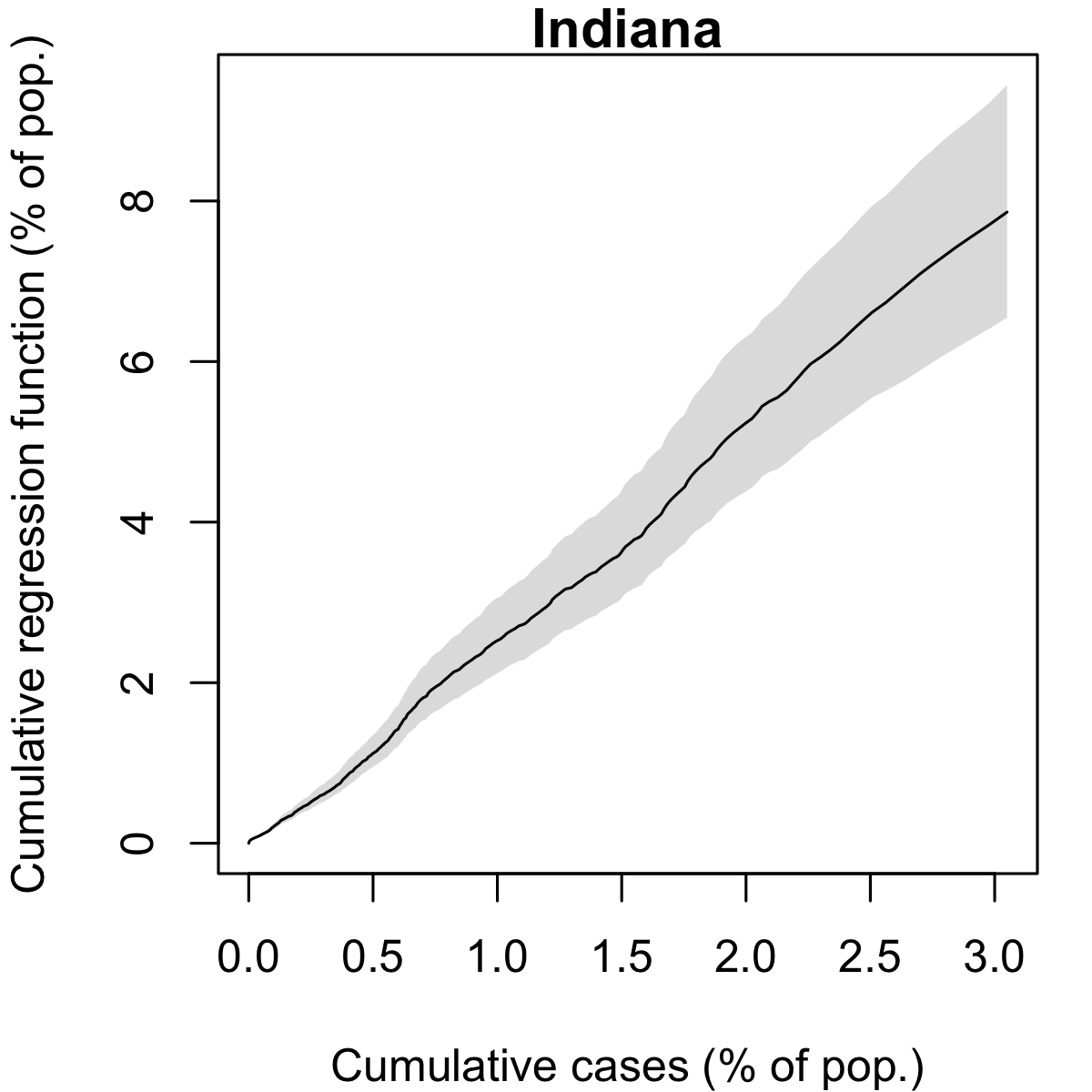}
&
\includegraphics[scale=0.185]{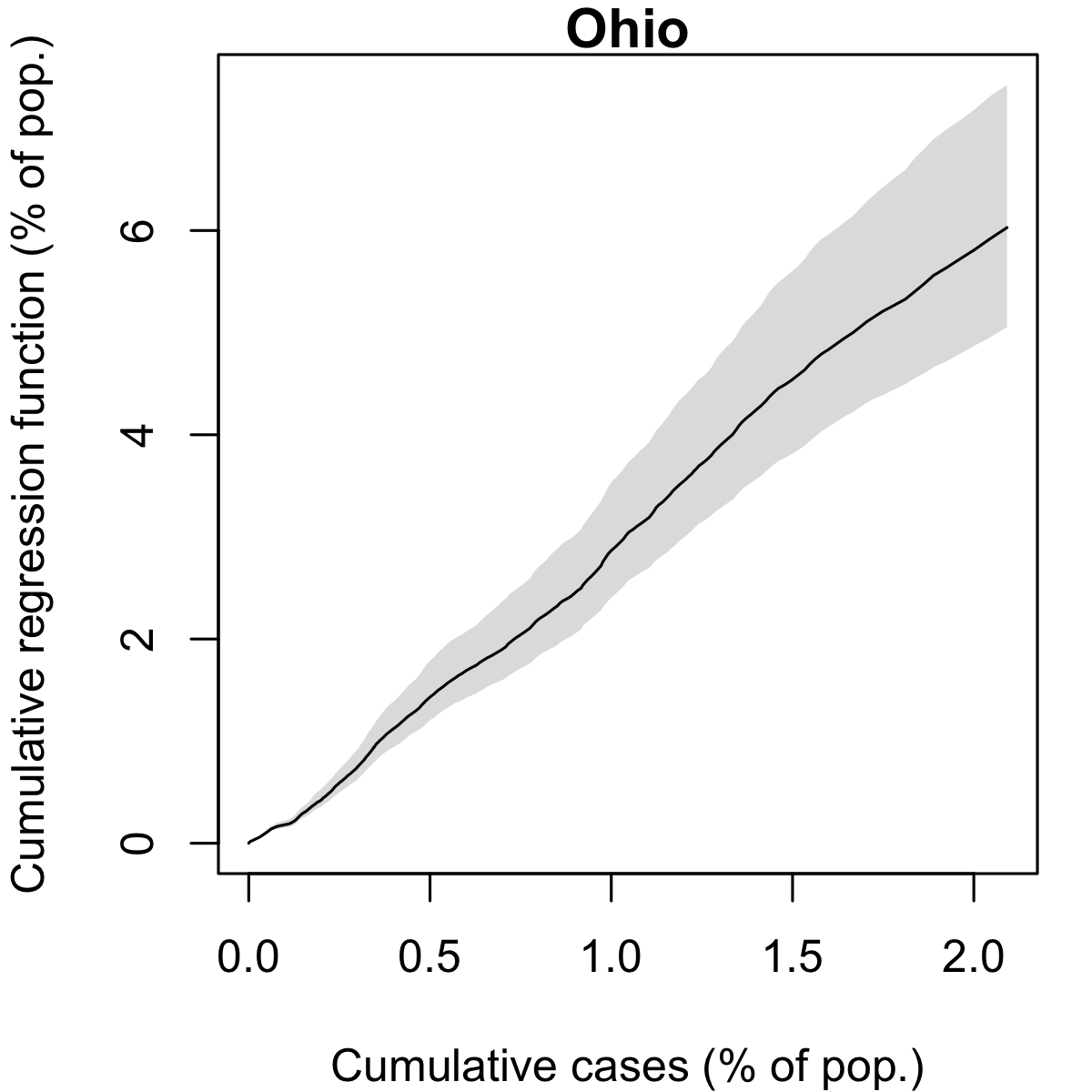}
\\
\includegraphics[scale=0.185]{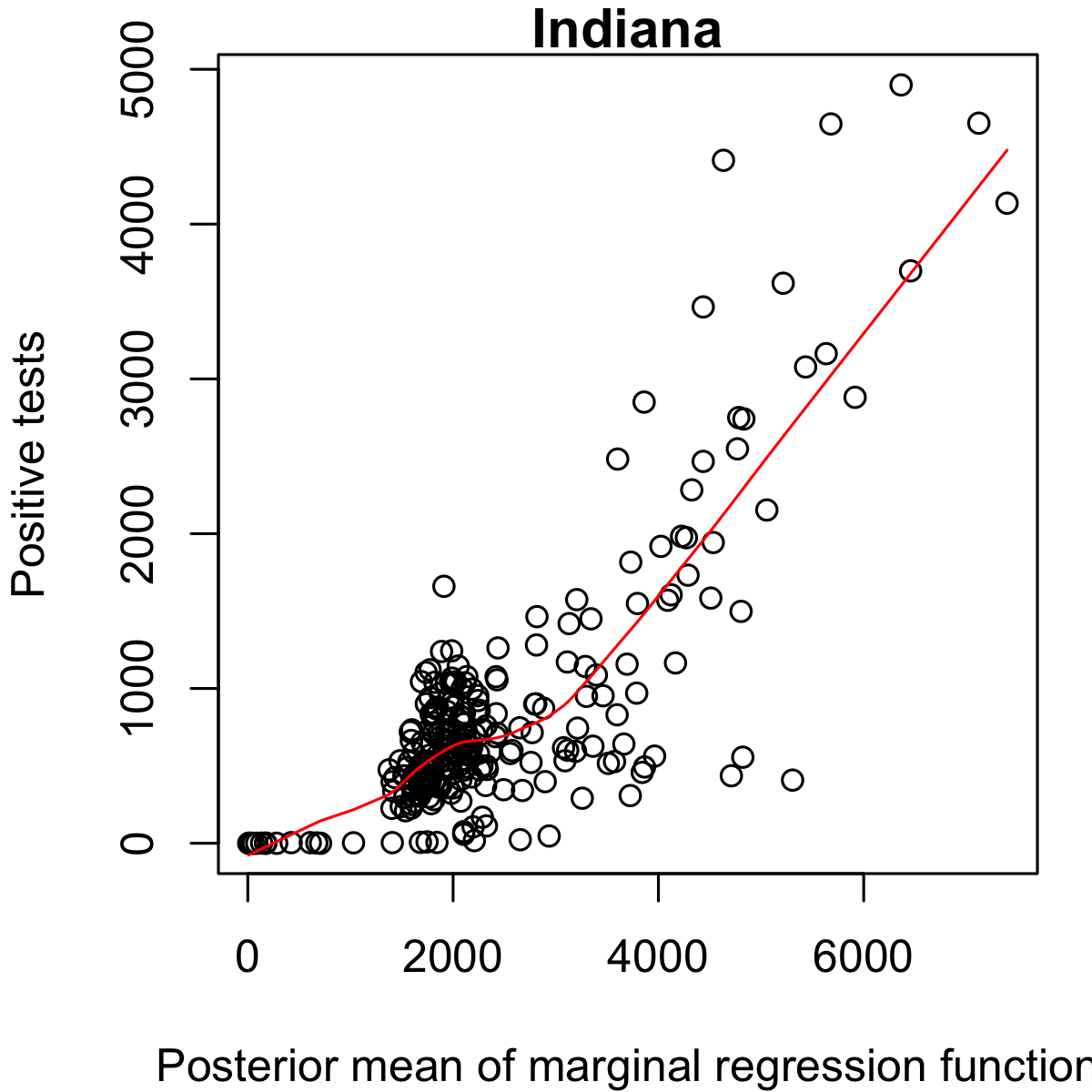}
&
\includegraphics[scale=0.185]{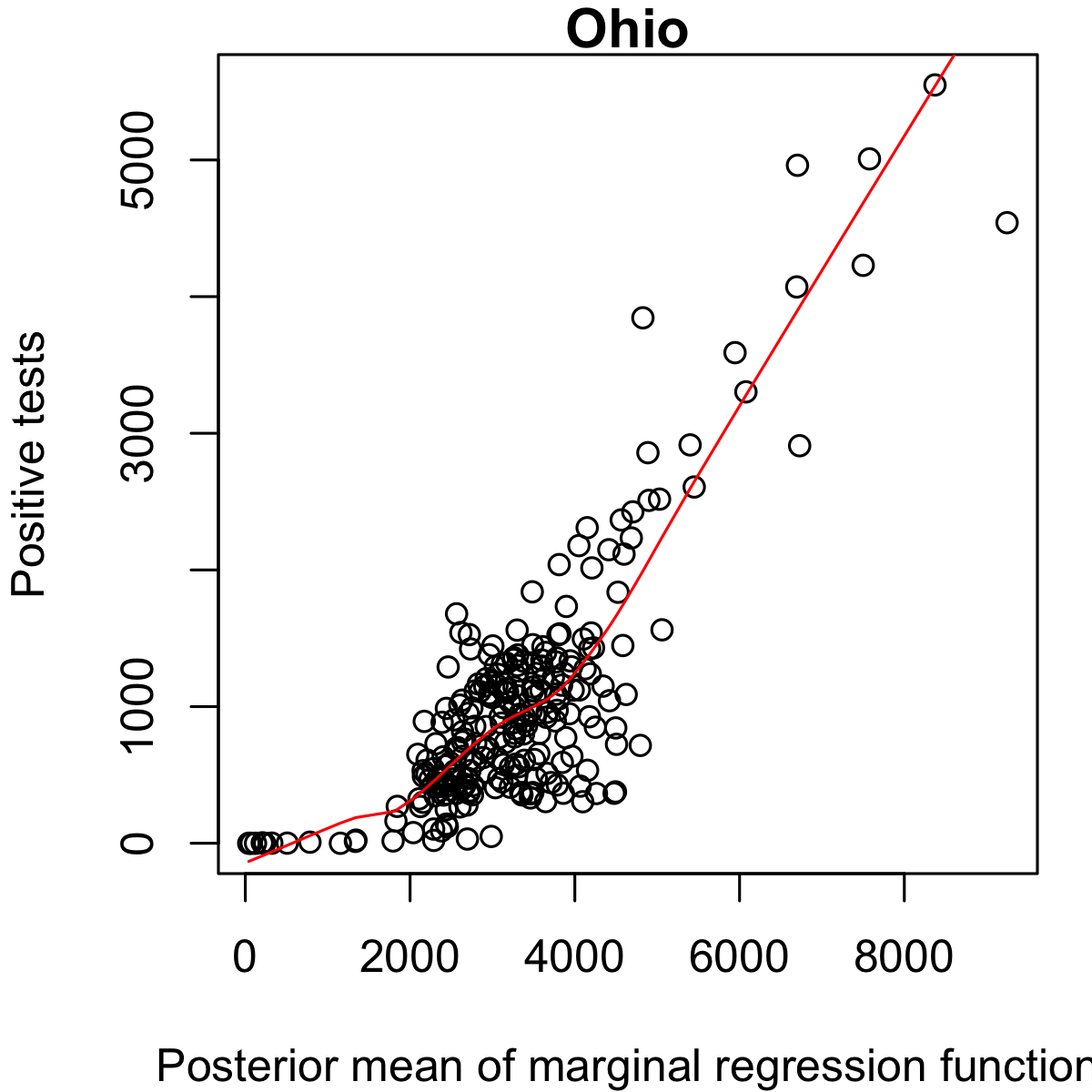}
\end{tabular}
\caption{Upper panels: Posterior median and 95\% confidence bands for the cumulative regression function in equation (\ref{eq:cum_cc}) plotted against cumulative cases in Indiana and Ohio. Lower panels: Positive tests on each day plotted against the posterior mean of the marginal regression function in equation (\ref{eq:cc_day}). LOESS curves are plotted in red.}
\label{fig:ptmodel}
\end{figure}

As shown in Figures \ref{fig:IN} and \ref{fig:OH}, the undercount curve $(I_t+R_t)/(\sum_{k\le t} C_k)$ has a common shape in Indiana and Ohio. Here, $I_t$ and $R_t$ are the SIR parameters on day $t$ and $C_t$ are the positive tests in the state on day $t$, as reported by the COVID Tracking Project \cite{ctp}.
We found that the reciprocal of the undercount is approximately linear when plotted against the square root of the cumulative number of tests administered in the state on each day, and that the slopes of these lines for the two states are similar; see Figure \ref{fig:ptmodel}.

This led to the following model for the test data:
\begin{equation}
\sum_{k=1}^t C_k \sim \text{Normal}\left(\phi_t (I_t+R_t), \eta_t^2\right).
\label{eq:cum_cc}
\end{equation}
Here the parameters $\phi_t$ and $\eta_t$ are proportional to the square root of the fraction of the population tested up to day $t$,
\begin{align*}
\phi_t = \phi\sqrt{\frac{\sum_{k=1}^t T_k}{N}}, \qquad
\eta_t^2 = \eta^2 \frac{\sum_{k=1}^t T_k}{N},
\end{align*} 
so that $\phi_t$ is the overall fraction of infections that appear in the cumulative number of positive tests. We assume that this fraction grows as the state's test capacity ramps up and that the variance in this relationship, $\eta_t^2$, grows linearly with the total number of tests administered. 

To arrive at the distribution in (\ref{eq:cum_cc}), we can model the cases on each day independently as
\begin{equation}
C_t \stackrel{ind.}{\sim} \text{Normal}\left(\phi_t(I_t+R_t)-\phi_{t-1}(I_{t-1}+R_{t-1}),\eta^2\frac{T_t}{N}\right).
\label{eq:cc_day}
\end{equation}
Noting that 
$
\nu_t = (I_t+R_t)-(I_{t-1}+R_{t-1}),
$
we can write the mean of $C_t$ as 
\[
\phi_t\cdot \nu_t + (\phi_t-\phi_{t-1})(I_{t-1}+R_{t-1}).
\]
Hence,
in expectation $C_t$ can be decomposed as a fraction of the new infections on day $t$, $\nu_t$, and a smaller fraction of the cumulative incidence on day $t-1$, $I_{t-1}+R_{t-1}$.

In fitting the model, we do not use the likelihood on each day (\ref{eq:cc_day}) due to inconsistent reporting of cases and tests, as well as weekly oscillations in these numbers due to reduced reporting on weekends. Rather, in each state we combine cases and tests into non-overlapping consecutive $L$-day periods, where $L$ is at least 7 to account for weekend effects, and model the counts in these periods independently.

We first fit the model in Indiana and Ohio without the likelihood on cases described in section 2.4. That is, initially we used only deaths data and the random sample surveys in each state. With the resulting posterior samples of cumulative incidence $I_t+R_t$ on each day, we arrived at the likelihood on cases. Figure \ref{fig:ptmodel} demonstrates the relationships defined in equations (\ref{eq:cum_cc}) and (\ref{eq:cc_day}). We refer to the normal means in (\ref{eq:cum_cc}) and (\ref{eq:cc_day}) (divided by the parameter $\phi$) as the cumulative and marginal regression functions, respectively. The lower panels of Figure \ref{fig:ptmodel} reveal a comparable slope $\phi$ for Indiana and Ohio after a brief initial period when testing and cases were very low. The widening confidence intervals in the upper panels exhibit the growth of the variance in (\ref{eq:cum_cc}) as a function of cumulative testing.

A number of other models for case and test data have been proposed. Campbell et al. introduced a binomial likelihood on cases, $C_t\sim \text{Binomial}(T_t,1-(1-I_t/N)^\alpha)$, where $I_t/N$ is the infection rate on day $t$ and $\alpha > 0$ is a parameter representing the degree of preferential testing \cite{campbell}. Assuming the infection rate is small, a binomial expansion of the test positivity rate yields the approximation $1-(1-I_t/N)^\alpha \approx \alpha I_t/N$. An application of Bayes' rule to the latter model shows that $\alpha = P(\text{tested}|\text{infected})/P(\text{tested})$. This model has some limitations in the context of our study. Firstly, the degree of preferential testing $\alpha$ is likely to decrease as testing increases, and it is not obvious how one might parametrize $\alpha=\alpha_t$ to account for this. Secondly, the model is not additive, as the test positivity relies on the active infection rate. As a result, it is not well suited to handling state-level testing data, which can be unreliable on the daily level. 

Youyang Gu \cite{gu} and Peter Ellis \cite{ellis} proposed similar models to correct case counts using test positivity rates. They take the form $\nu_t = C_t[m\cdot(C_t/T_t)^k+b]$ where $m>0,k\in[0,1],b\ge 0$ are parameters. Benatia et al. \cite{benatia} also estimate population prevalence on day $t$ by the number of positive tests on day $t$ scaled by a multiplicative factor depending on the number of tests administered on day $t$ as a fraction of the state population. These models are susceptible to the same issues as that of Campbell et al. They rely on daily test positivity rates, which are reported inconsistently across states \cite{ctp-test}. And as Youyang Gu notes, the parameters estimated at one point in time do not carry over to other time periods \cite{gu}. Furthermore, by assuming that new infections are a function only of cases and tests on that day, these models ignore the lag between infections and their confirmation via testing. They also presume that there are no new infections on days in which no positive tests are reported. Our likelihood on cases (\ref{eq:cc_day}) allows for new infections to be reflected in case counts at a later date.

\subsection{Prior specification}

Lastly, we specify prior distributions for the model parameters $ \{\text{IFR},\beta_1,\sigma,\gamma^{-1}, (S_1,I_1),\phi,\eta\}$. We use a weakly informative $\text{Uniform}(0,0.03)$ prior distribution for the IFR in each state.
For Indiana, we use a truncated normal prior for the mean infectious period, $\gamma^{-1}\sim$ \newline $\text{Normal}_{[5.5,11.5]}(8.5,1.5^2)$.
This is motivated by clinical data, which show that most infected individuals remain infectious no longer than 10 days after symptom onset \cite{cdc-duration, wolfel, arons, bullard, lu, quicke, cdc-korea, van-kampen}, and that patients can be highly infectious several days before symptom onset \cite{temporal}. 

We assume that the removal rate $\gamma$ is determined by the disease and so does not vary between states. Therefore, after fitting the model to the data for Indiana, we use the posterior distribution of $\gamma$ for Indiana  as the prior distribution of $\gamma$ for Ohio. We then use the posterior distribution from Ohio as the prior distribution for the remaining states, each of which we model independently.
The prior distributions of the remaining parameters are diffuse independent uniform priors. To estimate $\phi$, we use the same process as described for $\gamma$.

\subsection{Implementation}
We built the model in R and fit it with the RStan software package, which implements the 
No-U-Turn-Sampler for Bayesian inference \cite{r,rstan,nuts}. 
For each state, we ran 4 chains in parallel for 20,000 steps each with the first 10,000 as burn-in to obtain 40,000 samples from the posterior distribution of the model parameters.

\section{Results}
\label{sec:results}

Here we present detailed results for Indiana, Ohio,
and Connecticut -- which, to our knowledge, are the only states that have conducted representative testing surveys -- as well as New York, which has the highest number of reported deaths due to COVID. We also present aggregated estimates for the entire United States.
Table \ref{tab:states} in the appendix includes estimates of the IFR and the cumulative incidence (i.e, the percent of the state's population having been infected) and undercount factor as of January 6, 2021 for the 50 states and the District of Columbia. Results for the 50 states and DC are shown in the appendix.

We have created an online dashboard\footnote{ \url{https://rsc.stat.washington.edu/covid-dashboard}} where updated results can be found, including estimated daily infections, the IFR, and the reproductive number $r(t)$ in each state.

\subsection{Indiana}

\begin{figure}[htbp!]
\textbf{Indiana}
\centering
\begin{tabular}{ll}
\includegraphics[scale=0.185]{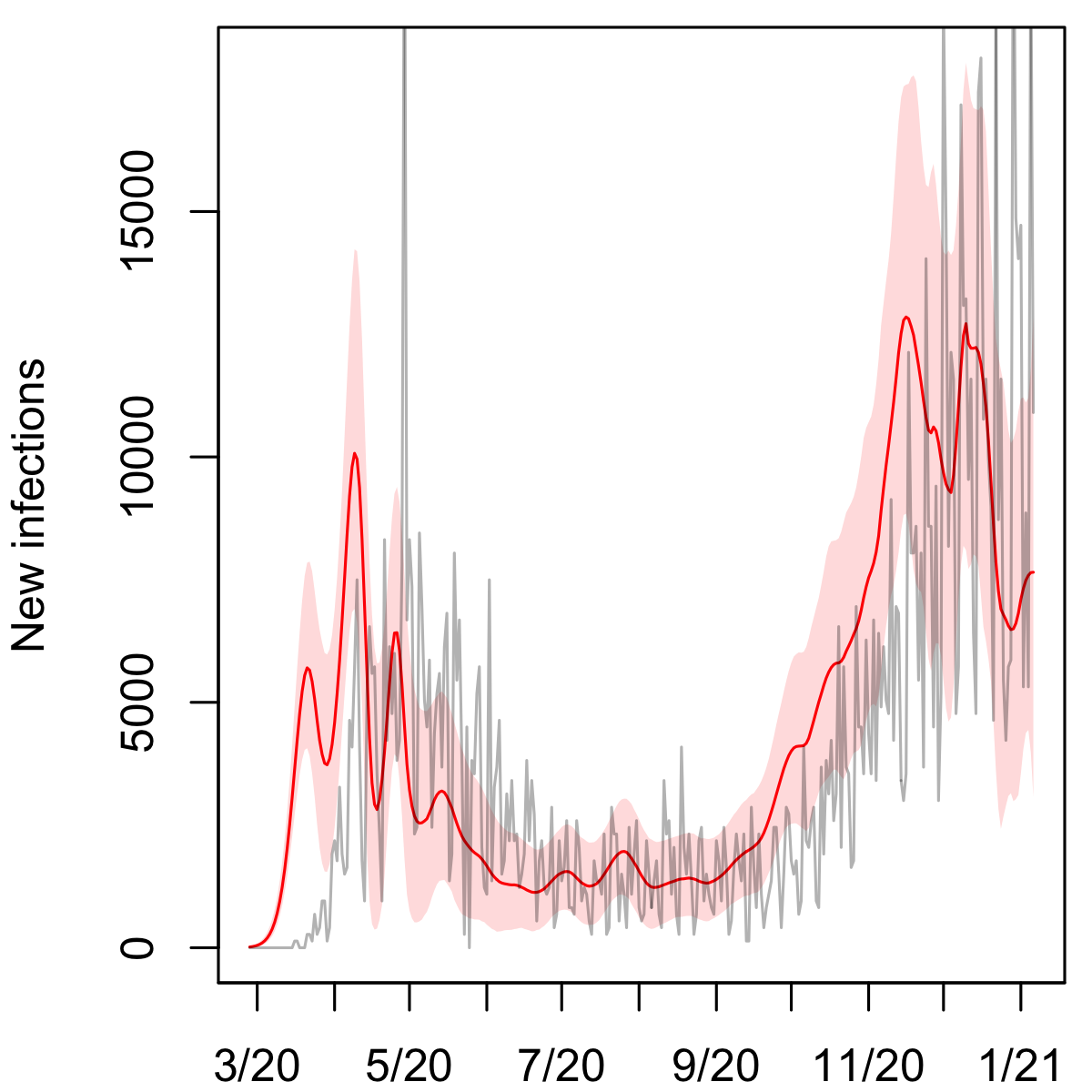}
&
\includegraphics[scale=0.185]{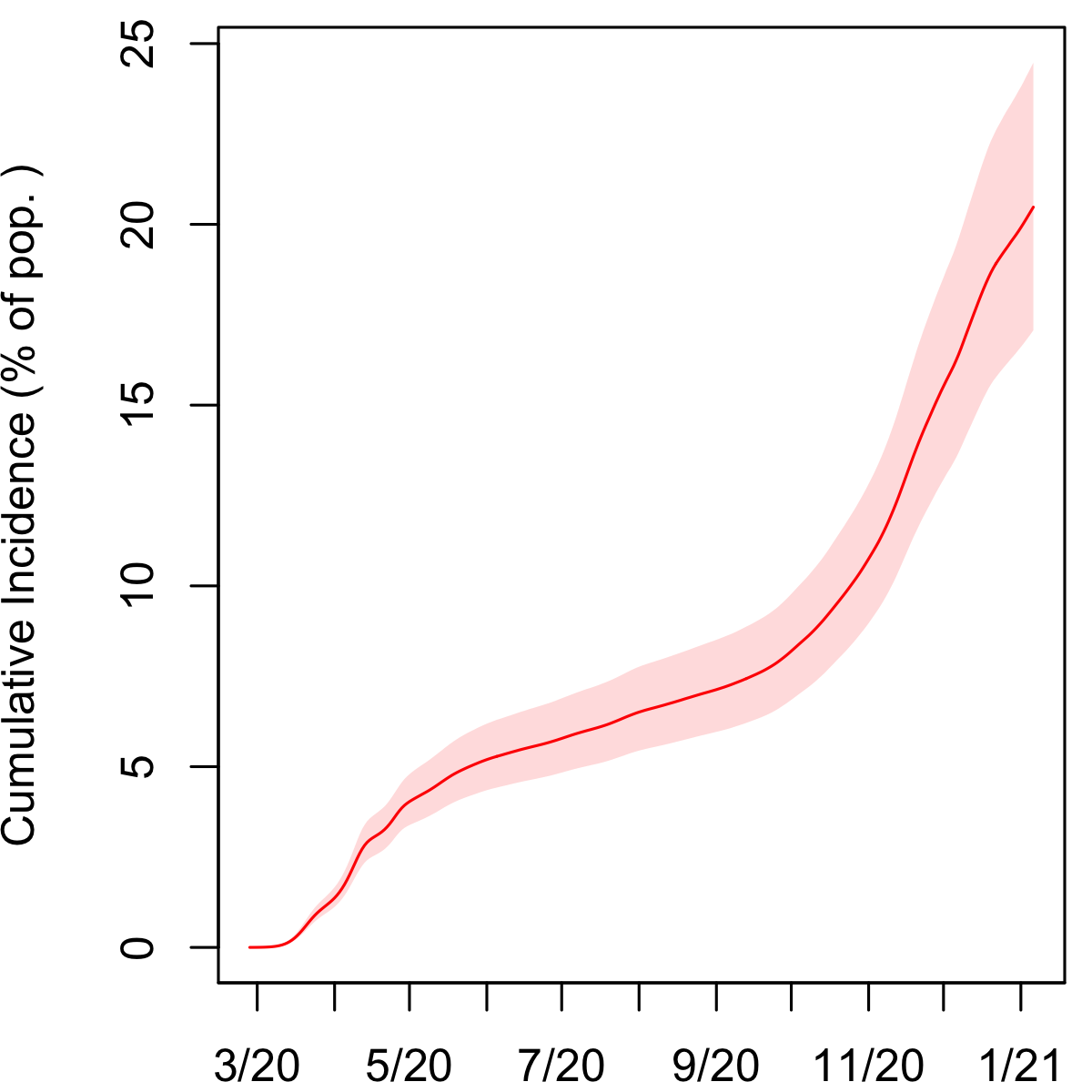} 
\\
\includegraphics[scale=0.185]{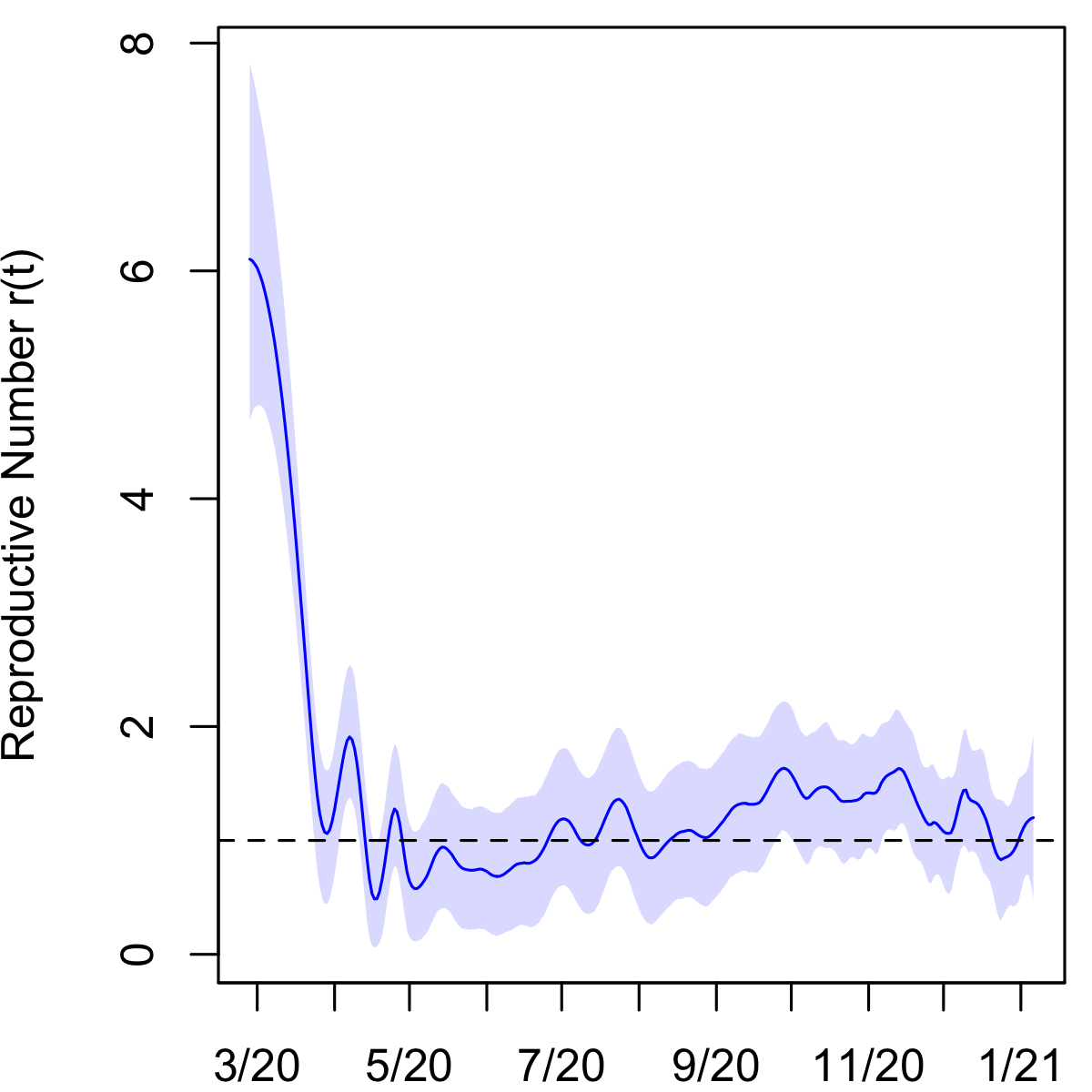}
&
\includegraphics[scale=0.185]{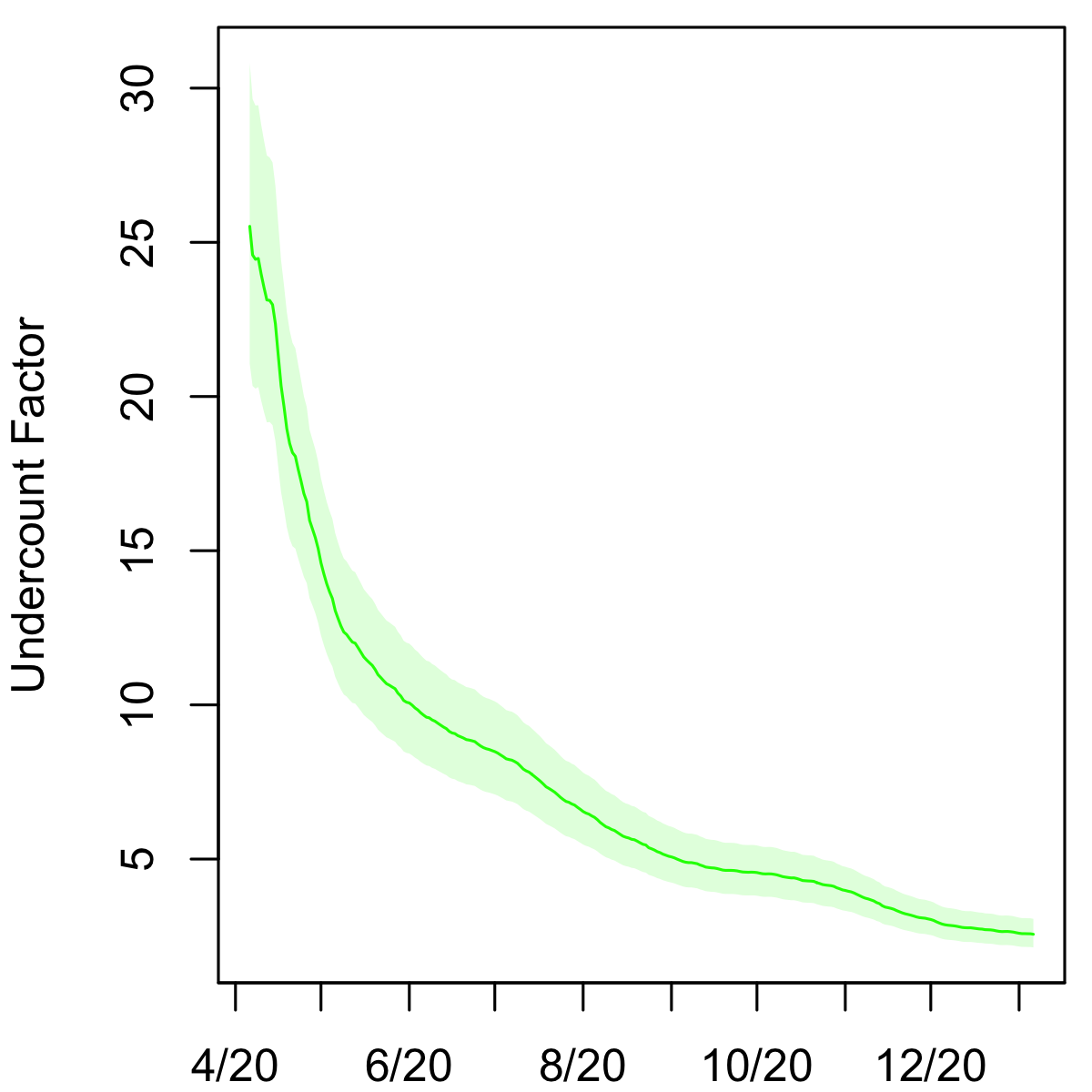} \\
\end{tabular}
\caption{Posterior median and middle 95\% intervals for daily new infections, cumulative incidence, $r(t)$, and cumulative undercount in Indiana from March 2020 to January 2021. In the top left panel, deaths divided by the posterior median IFR are plotted in grey for comparison.}
\label{fig:IN}
\end{figure}

We estimate an IFR of 0.73\% (95\% interval 0.61--0.88) and a cumulative incidence of 20.5\% (17.1--24.5) in Indiana as of January 6, 2021. There have been 2.6 (2.1--3.1) infections for every confirmed case in the state through this date. This suggests that a large majority of infections in the course of the pandemic have gone unreported, although Figure 1 shows that undercounting was most pronounced early on and has improved substantially over time.
Figure 1 exhibits posterior estimates of new infections on each day, $\nu_t$, as well as the cumulative undercount factor, which is the ratio of estimated cumulative infections to cumulative confirmed cases. Figure \ref{fig:IN} displays the viral prevalence, the cumulative incidence, and the reproductive number $r(t)=\beta_t/\gamma$ on each day. 

By the time that the first confirmed case was reported in Indiana on March 6, 2020, there had likely been more than 1,000 infections in the state (95\% interval 539--1,526). We estimate that as of May 1 there were 272,000 cumulative infections (95\% interval 228,000--323,000), compared to 18,630 confirmed cases by that date. This yields a cumulative incidence of 4.0\% (3.4--4.8) and an undercount factor of 14.6 (12.3--17.4). This estimate is comparable to others in the literature for that period \cite{johndrow,undercount,undercount2}. Between the 16th and 19th of March, the state's Governor Eric Holcomb ordered a stop to indoor dining, declared a state of emergency, and closed schools; on March 23rd he issued a stay-at-home order. According to our model, the first wave of infections reached its peak about two weeks later in early April.

\subsection{Ohio}

\begin{figure}[htbp!]
\textbf{Ohio}
\centering
\begin{tabular}{ll}
\includegraphics[scale=0.185]{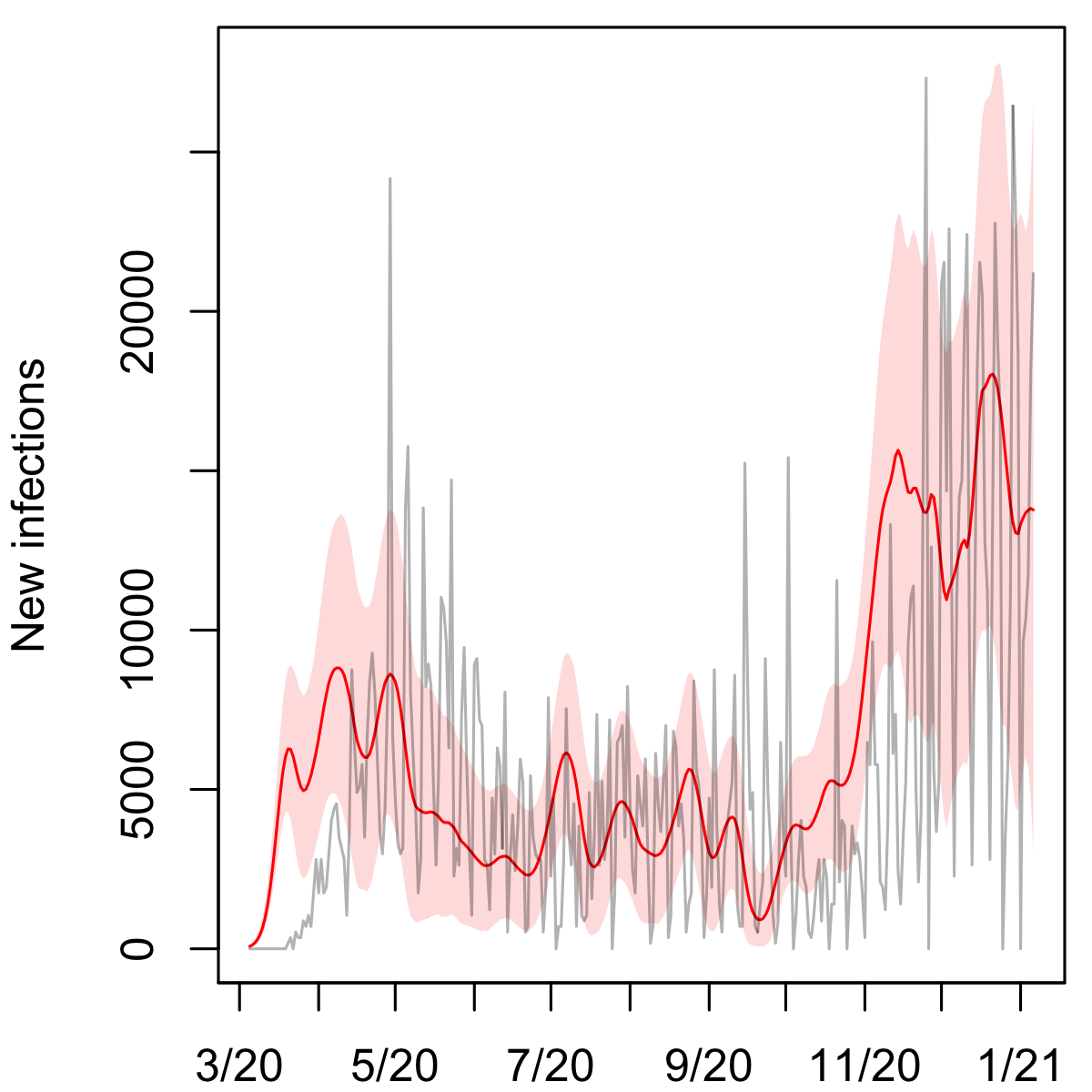}
&
\includegraphics[scale=0.185]{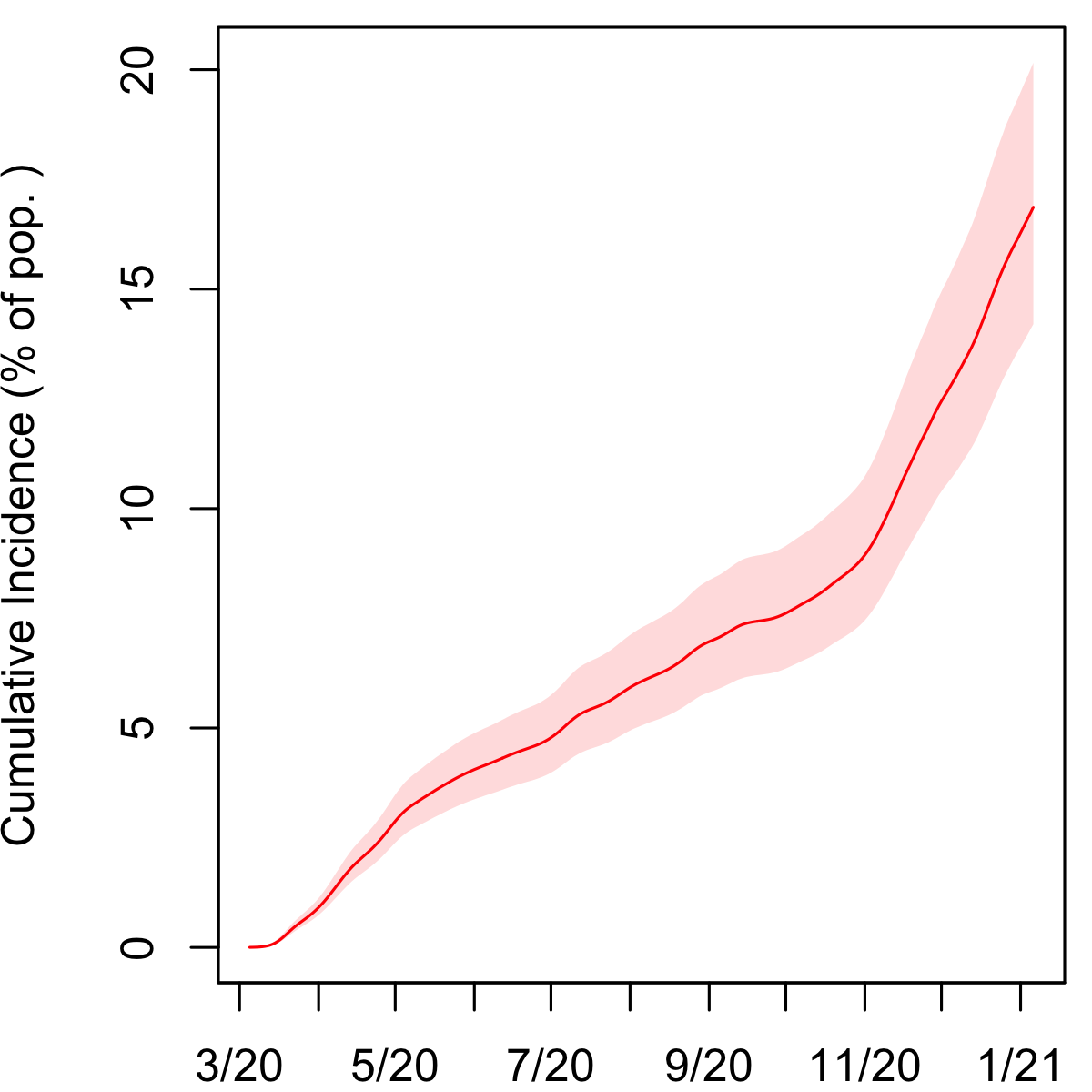} \\
\includegraphics[scale=0.185]{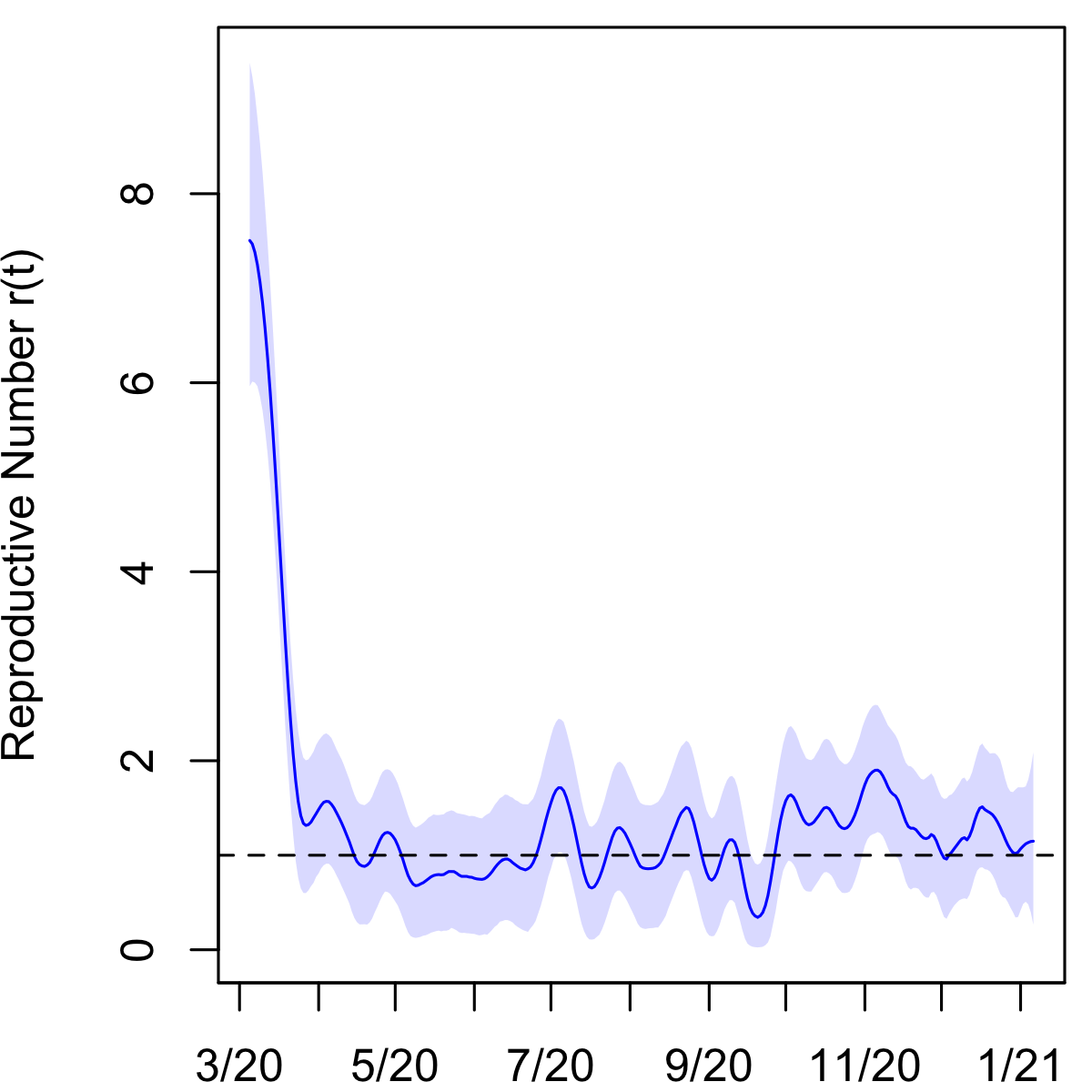}
&
\includegraphics[scale=0.185]{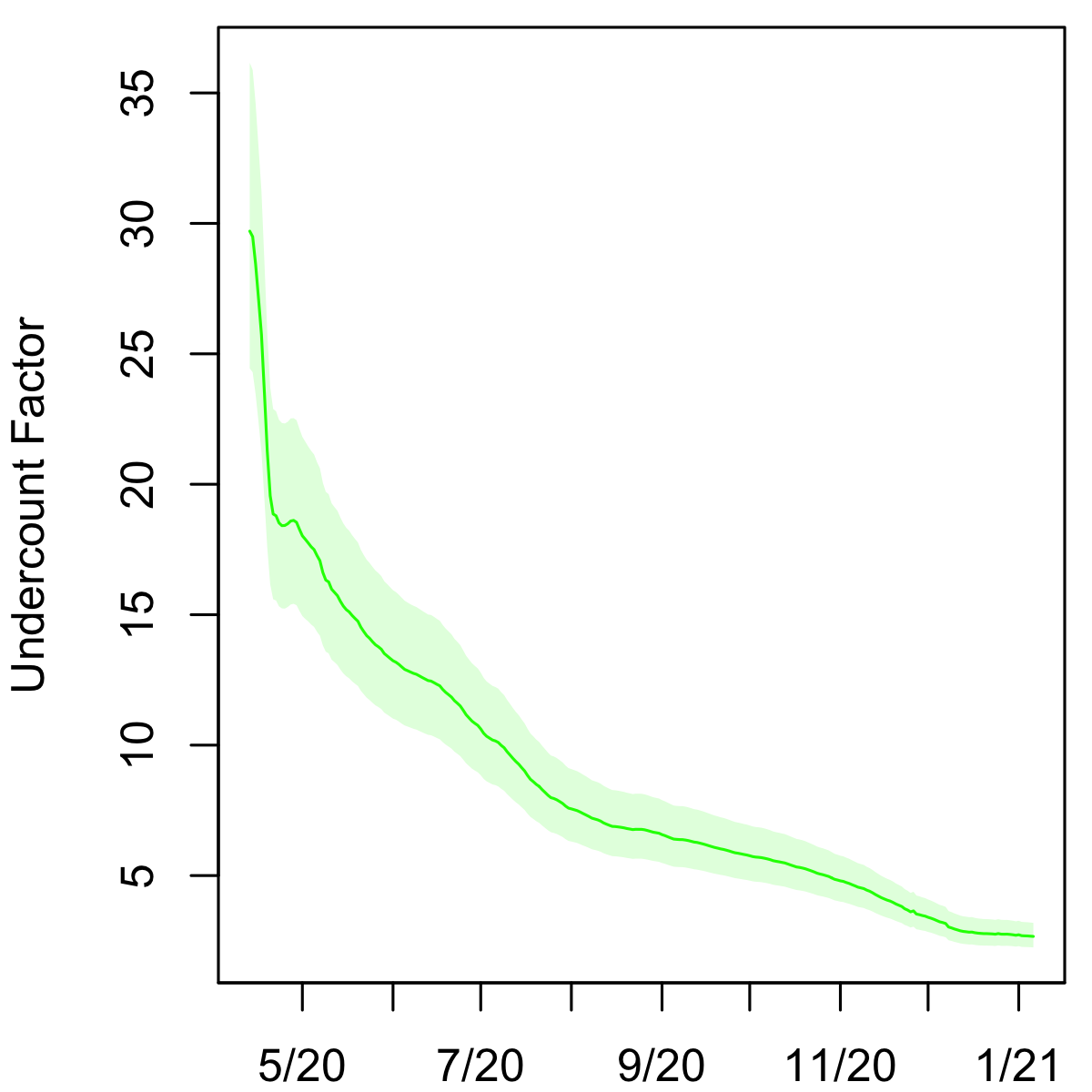} 
\end{tabular}
\caption{Posterior median and middle 95\% intervals for daily new infections, cumulative incidence, $r(t)$, and cumulative undercount in Ohio from March 2020 to January 2021. In the top left panel, deaths divided by the posterior median IFR are plotted in grey for comparison.}
\label{fig:OH}
\end{figure}

We estimate an IFR of 0.57\% (95\% interval 0.47--0.69) in Ohio. As of January 6, 2021, the cumulative incidence in the state was 16.9\% (14.2--20.2) and the cumulative undercount factor was 2.7 (2.2--3.2). 


Ohio Governor Mike Dewine declared a state of emergency on March 9th and the state's first stay-at-home order took effect on March 23rd. In mid-April, the Governor declared that businesses could begin to reopen on May 1st. Figure \ref{fig:OH} shows that the first wave of infections, which picked up in March and likely peaked by late April, did not die out but rather leveled out to a sustained spread through the summer of 2020. 
The posterior median of the reproductive number $r(t)$ in the state hovered around 1
from early April through mid-September and increased thereafter as the second wave of infections began in the fall.

\subsection{Connecticut}


We estimate an IFR of 1.54\% (95\% interval 1.22--1.94) in Connecticut. Further, as of January 6, 2021, 12.9\% (10.4--16.1) of the state's population has been infected, leading to an undercount factor of 2.3 (1.9--2.9).

According to our model as of April 30, 2020, 5.6\% (4.4--7.0) of the state's population had recovered from COVID. In comparison, Havers et al. estimated a seroprevalence of 4.9\% (95\% interval 3.6--6.5) in the state in the period April 26--May 3 \cite{undercount2}. That study relied on a convenience sample of residual blood specimens collected for clinical purposes, and so it may have been affected by selection bias, as well as imperfect sensitivity and specificity of the antibody test used. Nevertheless, their estimate agrees well with the result from our model.

By July 5, 2020, our estimate of the recovered population increased to 7.9\% (6.3--10.0). By comparison, in a random sample blood test survey, Mahajan et al. reported a seroprevalence of 4.0\% (90\% interval 2.0--6.0) for the period June 10--July 29 \cite{mahajan}, which is significantly lower. While our estimates disagree with those of Mahajan et al., we note that the survey response rate was low at 7.8\%, raising the possibility of nonresponse bias. For this reason, we did not include the Connecticut survey as a source of data in our analysis. 

\subsection{New York}


We estimate an IFR of 1.26\% (95\% interval 0.97--1.67) for New York state. As of January 6, 2021, 14.5\% (11.1--18.8) of the state has been infected, yielding an undercount factor of 2.7 (2.0--3.5) through that date.

We know of no other estimates of the IFR in New York in the literature. However, Yang et al. estimated an IFR of 1.39\% (95\% interval 1.04--1.77) for the first wave in New York City through June 6, 2020, based on available testing, mortality, and mobility data \cite{yang}. According to NYC Health Department data \cite{nyc-health}, this period accounted for more than 85\% of COVID deaths in the city and 57\% of all confirmed COVID deaths (not including probable deaths) in the state through the first week of January 2021. As such, we expect the IFR for the state as a whole to have been similar to that of NYC during the spring of 2020, and our results are consistent with those of Yang et al.

We estimate that by June 6, 10.2\% of the state's population (95\% interval 7.7--13.2), or about 2 million people, had been infected with the novel coronavirus. Multiplying that number by the fraction of confirmed COVID deaths in the state occurring in NYC during that period yields 1.5 million infections, or 18\% of the city's population. This number is compatible with that of Stadlbauer et al., who measured 20\% seroprevalence in NYC at that time based on randomly sampled residual plasma collected from patients at Mount Sinai Hospital scheduled for routine care visits unrelated to COVID-19 \cite{stadlbauer}.

\subsection{United States}

\begin{figure}[htbp!]
\textbf{USA}
\centering
\begin{tabular}{ll}
\includegraphics[scale=0.185]{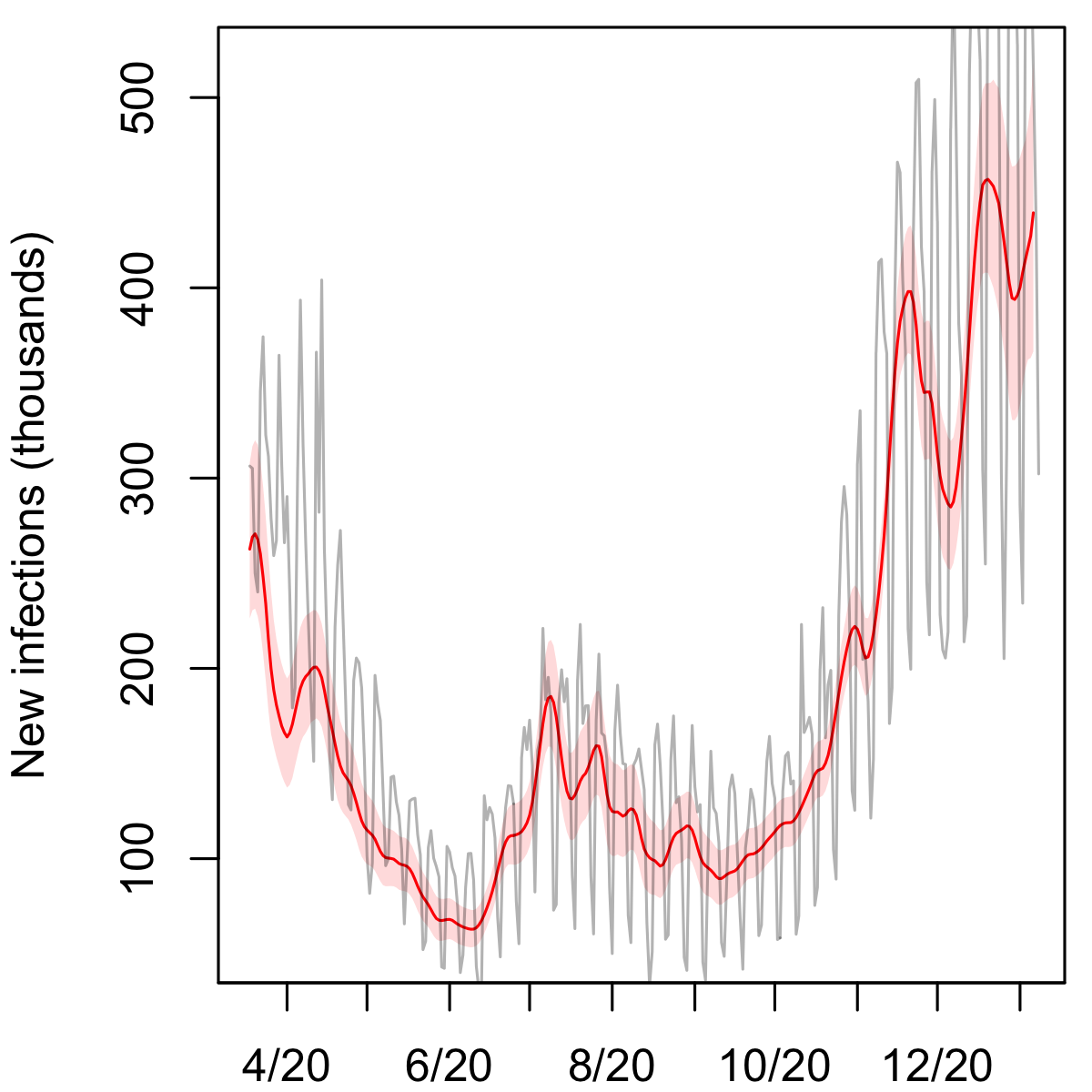}
&
\includegraphics[scale=0.185]{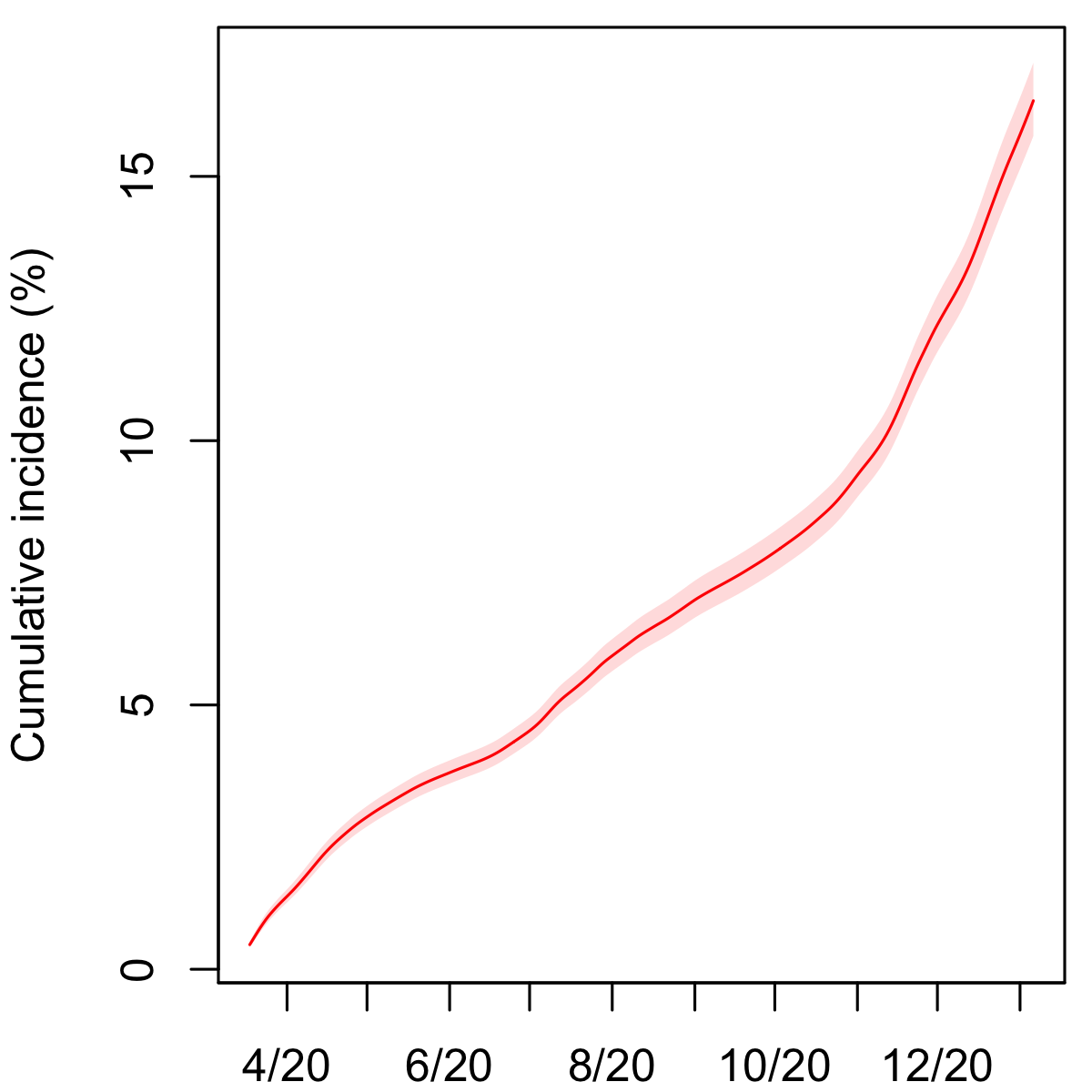} 
\\
\includegraphics[scale=0.185]{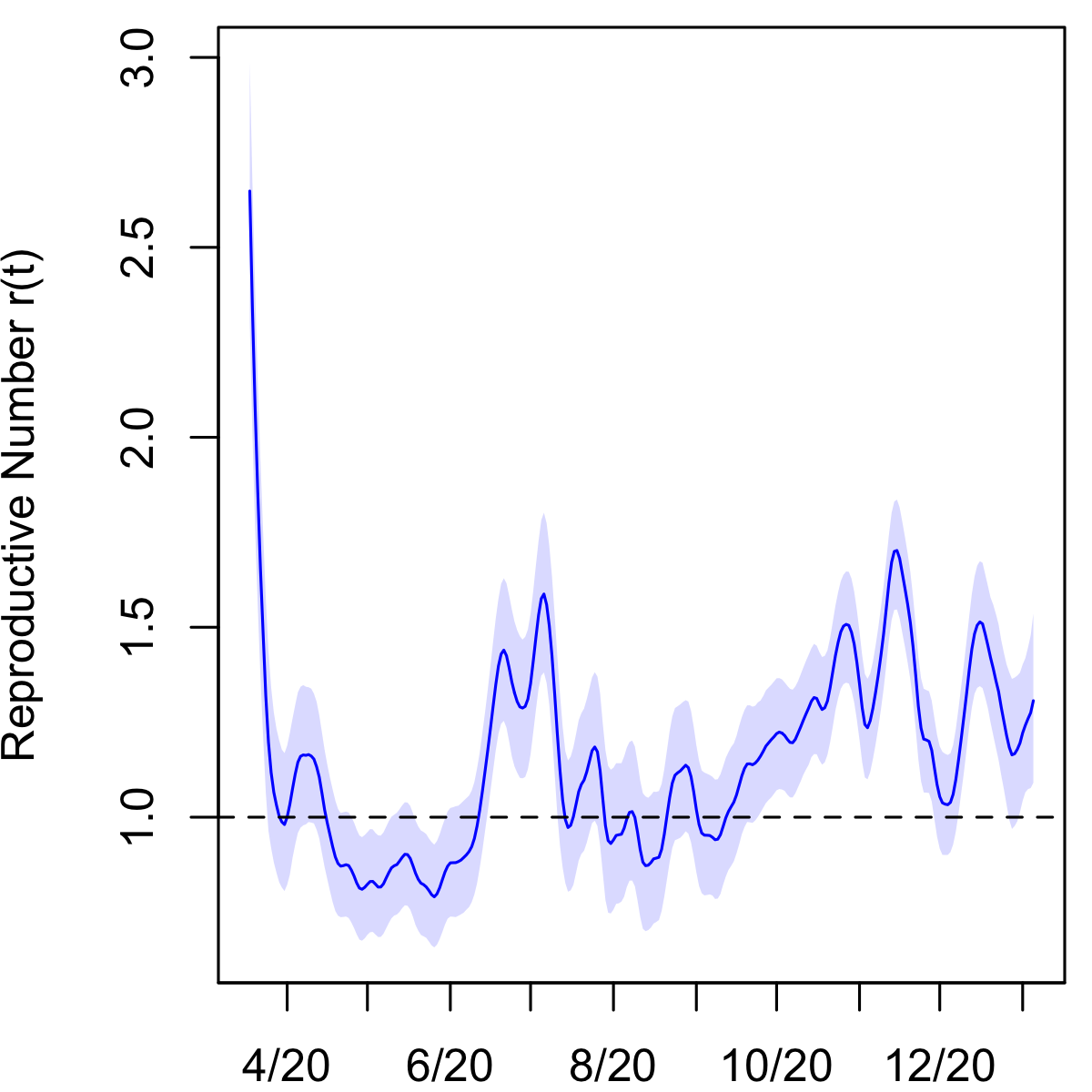} 
&
\includegraphics[scale=0.185]{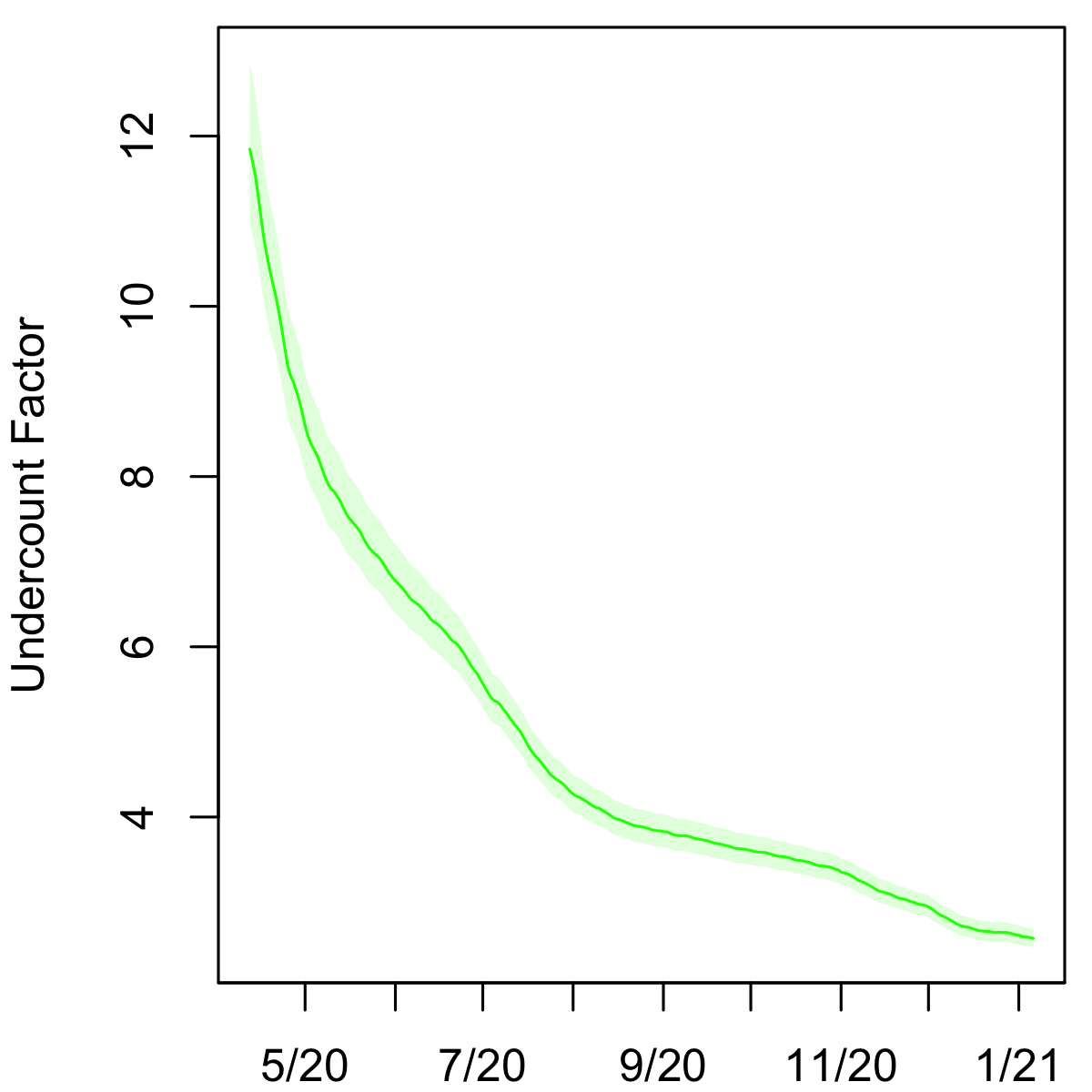}
\end{tabular}
\caption{Aggregated estimates of new infections, cumulative incidence, $r(t)$, and cumulative undercount for the United States from March 2020 to January 2021. In the top left panel, deaths (in thousands) divided by 0.0068 and shifted back 23 days are plotted in grey for comparison.
}
\label{fig:us}
\end{figure}

We summed posterior samples of the SIR trajectories from all the states to obtain estimates of viral prevalence in the United States on each day. The results are summarized in Figure \ref{fig:us}. For each sampled trajectory of the infection curve, we calculated an effective contact parameter $\beta_t$ for the entire country for each day from the SIR equations (\ref{eq:sir}). 

As of January 6, 2021, we estimate that 16.4\% (95\% interval 15.8--17.2) of the US population, or about 54 million people, had been infected with SARS-CoV-2. This suggests that the USA was far from reaching herd immunity and that it was unlikely to do so from infections alone in the short term while state and local governments continue to implement lockdowns and other mitigations. Up to that date, we estimate that one out of every 2.6 infections (2.5--2.7) in the US had been confirmed via testing. This implies that approximately 60\% of all infections in the country have gone unreported.

In the top left panel of Figure \ref{fig:us}, which exhibits estimates of new infections on each day in the US, we plot reported COVID deaths per 1000 population shifted back 23 days (which is the mean of the time-to-death distribution $\tau$). In the plot, we divide deaths per 1000 by 0.0068. This is the point estimate of IFR reported by Meyerowitz-Katz and Merone in their meta-analysis of 24 IFR estimates from a wide range of countries published between February and June 2020 \cite{meyerowitz-katz}. The two curves have a substantial overlap, suggesting that the IFR implied by our estimates of true infections in the USA is consistent with their findings.

\subsection{Implications for herd immunity}

We conducted a simulation study to assess the implications of our results for herd immunity in the US. We project the SIR model for the US forward from January 6, 2021, and incorporate vaccine administration into the dynamics. We make the following strong assumptions:
\begin{enumerate}
    \item Recovered individuals are immune to the virus, i.e., reinfection does not occur.
    
    \item Immunity is conferred after receiving the second vaccine dose. The number of individuals receiving the second dose increases linearly from 0 to 500--750 thousand per day from early January to late February, and remains at that level thereafter. This aligns with President Biden's goal of 1.0--1.5 million doses per day in the first 100 days of his administration.
    
    \item Previously infected individuals who have tested positive for the virus do not receive the vaccine. All others are equally likely to be vaccinated.
    
    \item The fraction of infections confirmed by testing (i.e., the reciprocal of the cumulative undercount) does not change after January 6. Similarly, the reproductive number $r(t)$ remains fixed after January 6.
    
\end{enumerate}

\begin{figure}[htbp!]
\textbf{USA}
\centering
\begin{tabular}{ll}
\includegraphics[scale=0.77]{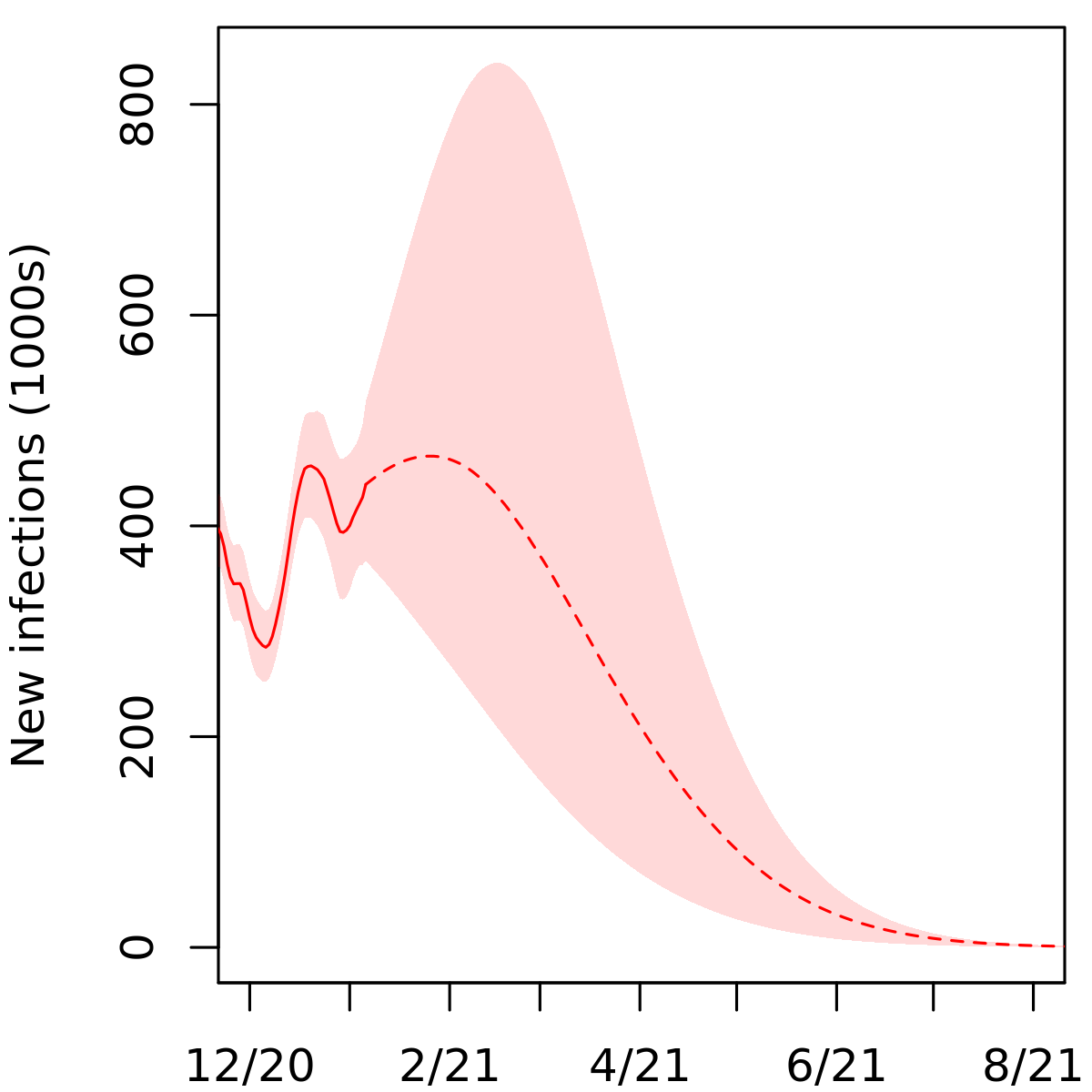}
&
\includegraphics[scale=0.77]{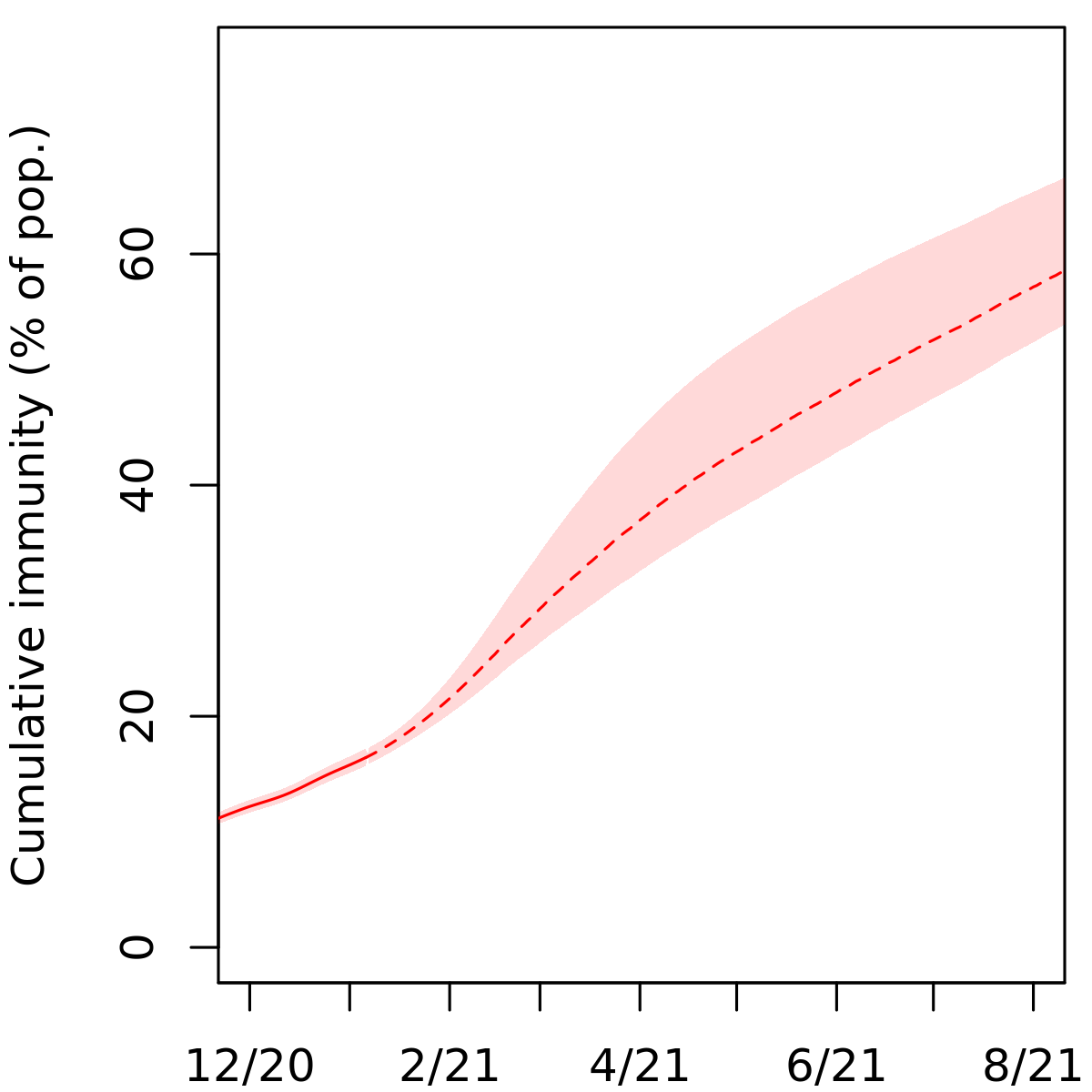}
\end{tabular}
\caption{95\% credible intervals for new infections and cumulative immunity (viral incidence and second vaccinations) in the US projected out to August 2021. 
}
\label{fig:vax}
\end{figure}

The first point merits further discussion. Our projections that follow are particularly sensitive to this assumption. It may turn out that individuals who have been vaccinated or previously infected are still susceptible to new variants of the virus that are cropping up and will continue to spread. It is also possible that the natural immunity conferred by asymptomatic and mild infections that elicited minimal immune response, which constitute a large portion of the total, will not last long enough to prevent widespread reinfection in the next few months. In either case, if Assumption 1 is violated then we may experience further waves of infection and delayed progress towards herd immunity. 

We project the 40,000 samples from the posterior distribution of the US infection trajectory forward under the modified SIR model described above. New infections and cumulative immunity (the percentage of the population previously infected or fully vaccinated) on each day are plotted in Figure \ref{fig:vax}. Based on our simulation, we find that the number of new infections per day in the country would likely fall below 5,000, about one hundredth of the second wave peak, by July 2021, if our assumptions are valid. At this point, the virus' spread through the population will have been effectively suppressed. In getting there, it is plausible that we will incur another 30--50 million new infections, beginning from January 6. These numbers are obtained as the interquartile range of the projected cumulative incidence. Note that at that point, our model suggests that cumulative immunity will be 70\% or less, although if our assumptions are violated, it could be higher.

To put this in perspective, there were about 360,000 confirmed COVID deaths and 54 million infections (by our reckoning) as of January 6, 2021. Assuming an IFR of 0.68\%, this would lead to an additional 200--350 thousand COVID deaths. (However, given that vaccine administration is prioritized for high risk groups, the effective IFR in the coming months could decline significantly, which would lead to fewer deaths.) We find that the projections given here are not very sensitive to plausible modifications of Assumptions 2--4 (e.g., that individuals with confirmed cases can receive the vaccine).

\section{Discussion}
\label{sec:discussion}
To craft and implement effective policy and mitigation strategies, policymakers need reliable assessments of the impact of previous non-pharmaceutical interventions on the transmission rate of the disease. We have developed a simple Bayesian model of the dynamics of SARS-CoV-2 transmission incorporating readily available time series data tracking the virus, as well as statewide representative point prevalence surveys conducted in Indiana and Ohio, which are the highest quality random testing surveys carried out to date. We present estimates of the infection fatality rate and the time-varying viral prevalence and reproductive number $r(t)$ in each US state on each day. Our results indicate that a large majority of COVID infections go unreported. Even so, we find that the US was still far from reaching herd immunity to the virus in early January 2021 from infections alone. This suggests that continued mitigation and an aggressive vaccination effort are necessary to surpass the herd immunity threshold without incurring many more deaths due to the disease. We hope that this work demonstrates the value of random sample COVID testing in our ongoing pandemic response.


By incorporating testing and case data aggregated over any period of time, our additive model for positive tests in equation (\ref{eq:cum_cc}) allows us to avoid using data at the daily level, which can be very unreliable. For example, the reported cumulative number of tests administered in a state may not be updated for up to two weeks at a time, or it may decrease from one day to the next as data are deduplicated upon further review. The latter scenario frequently occurs with reported cases as well. Working with data at the daily level generally requires using some kind of moving average, which washes out stochasticity in the data and leads to oversmoothing inconsistent with the high overdispersion of SARS-CoV-2 transmission \cite{endo}.

Our inference relies on daily reported deaths due to COVID in each state as opposed to excess deaths. Because of the possibility of death misclassification, excess death data represent a mix of confirmed COVID deaths and deaths from other causes. Nevertheless, relying on reported deaths is a potential source of bias, as they are affected by the accuracy of cause-of-death determinations. Their numbers can fall significantly below excess death counts and may undershoot the true number of deaths due to the disease \cite{nas}. Ascertainment of COVID deaths may vary widely between states, with the cumulative excess death count since the start of the pandemic exceeding reported COVID deaths by upwards of 50\% in some states, according to a New York Times analysis of CDC mortality data \cite{nyt-mortality}. Consequently, our results may underestimate viral incidence in those states. 

\section*{Acknowledgments}

This research was supported by NIH grant R01 HD-070936.







\section{Appendix}

\subsection{Results for all states}

\begin{table}

{\fontsize{10}{11.1}\selectfont

\begin{tabular}{|l|l|l|l|}
\hline
\textbf{State} & \textbf{IFR (\%)} & \textbf{Cumulative Incidence (\%)} & \textbf{Undercount}\\
\hline
Alabama & 0.48  (0.39--0.58) & 25.9  (21.6--31.3) & 3.3 (2.8--4.0)\\
Alaska & 0.31  (0.25--0.38) & 10.6  (9.1--12.5) & 1.6 (1.4--1.9)\\
Arizona & 0.84  (0.68--1.03) & 19.7  (16.3--24.0) & 2.5 (2.1--3.1)\\
Arkansas & 0.88  (0.72--1.07) & 18.4  (15.4--22.3) & 2.3 (1.9--2.8)\\
California & 0.62  (0.51--0.73) & 14.8  (12.5--17.7) & 2.4 (2.0--2.9)\\
Colorado & 0.64  (0.53--0.75) & 14.9  (12.7--17.9) & 2.5 (2.1--3.0)\\
Connecticut & 1.54  (1.22--1.94) & 12.9  (10.4--16.1) & 2.3 (1.9--2.9)\\
Delaware & 0.79  (0.64--0.96) & 14.1  (11.8--17.1) & 2.2 (1.9--2.7)\\
Florida & 0.76  (0.62--0.93) & 15.5  (12.8--18.9) & 2.5 (2.0--3.0)\\
Georgia & 0.82  (0.67--0.99) & 15.2  (12.7--18.5) & 2.7 (2.3--3.3)\\
Hawaii & 0.52  (0.41--0.65) & 4.5  (3.7--5.5) & 2.8 (2.3--3.4)\\
Idaho & 0.30  (0.25--0.35) & 31.4  (27.0--37.5) & 4.0 (3.4--4.7)\\
Illinois & 0.98  (0.81--1.17) & 17.0  (14.3--20.6) & 2.2 (1.8--2.6)\\
Indiana & 0.73  (0.61--0.88) & 20.5  (17.1--24.5) & 2.6 (2.1--3.1)\\
Iowa & 0.58  (0.48--0.71) & 24.2  (20.0--29.4) & 3.1 (2.6--3.8)\\
Kansas & 0.51  (0.41--0.63) & 25.5  (20.9--31.1) & 3.1 (2.6--3.8)\\
Kentucky & 0.45  (0.37--0.52) & 16.7  (14.3--19.9) & 2.6 (2.2--3.1)\\
Louisiana & 1.14  (0.92--1.38) & 16.7  (13.9--20.5) & 2.3 (1.9--2.9)\\
Maine & 0.94  (0.73--1.31) & 4.9  (3.9--6.0) & 2.4 (1.9--3.0)\\
Maryland & 0.98  (0.79--1.19) & 11.5  (9.5--14.1) & 2.4 (2.0--2.9)\\
Massachusetts & 1.89  (1.50--2.38) & 11.5  (9.2--14.4) & 2.0 (1.6--2.5)\\
Michigan & 1.27  (1.00--1.66) & 13.0  (10.2--16.3) & 2.4 (1.9--3.0)\\
Minnesota & 0.63  (0.53--0.75) & 16.8  (14.2--20.1) & 2.2 (1.9--2.7)\\
Mississippi & 0.79  (0.65--0.95) & 25.3  (21.3--30.5) & 3.3 (2.8--4.0)\\
Missouri & 0.60  (0.50--0.72) & 18.0  (15.1--21.7) & 2.7 (2.3--3.3)\\
Montana & 0.53  (0.45--0.63) & 19.3  (16.7--22.9) & 2.5 (2.2--3.0)\\
Nebraska & 0.50  (0.41--0.59) & 18.8  (16.0--22.7) & 2.1 (1.8--2.6)\\
Nevada & 0.62  (0.51--0.72) & 19.8  (17.0--23.7) & 2.6 (2.2--3.1)\\
New Hampshire & 0.84  (0.69--1.01) & 9.5  (8.1--11.4) & 2.7 (2.3--3.2)\\
New Jersey & 1.44  (1.11--1.90) & 17.3  (13.3--22.2) & 2.8 (2.1--3.6)\\
New Mexico & 0.90  (0.75--1.04) & 16.1  (13.9--19.1) & 2.2 (2.0--2.7)\\
New York & 1.26  (0.97--1.67) & 14.5  (11.1--18.8) & 2.7 (2.0--3.5)\\
North Carolina & 0.58  (0.48--0.68) & 14.1  (11.9--16.8) & 2.6 (2.2--3.1)\\
North Dakota & 0.88  (0.73--1.03) & 20.6  (17.8--24.5) & 1.7 (1.4--2.0)\\
Ohio & 0.57  (0.47--0.69) & 16.9  (14.2--20.2) & 2.7 (2.2--3.2)\\
Oklahoma & 0.40  (0.34--0.47) & 20.6  (17.8--24.3) & 2.6 (2.3--3.1)\\
Oregon & 0.56  (0.47--0.66) & 7.4  (6.4--8.8) & 2.7 (2.3--3.2)\\
Pennsylvania & 0.82  (0.67--0.98) & 21.2  (17.8--25.6) & 4.0 (3.3--4.8)\\
Rhode Island & 1.34  (1.09--1.63) & 15.0  (12.5--18.3) & 1.7 (1.4--2.0)\\
South Carolina & 0.78  (0.64--0.95) & 16.9  (14.1--20.7) & 2.6 (2.2--3.2)\\
South Dakota & 0.52  (0.43--0.62) & 35.7  (30.5--42.8) & 3.2 (2.7--3.8)\\
Tennessee & 0.74  (0.60--0.90) & 19.1  (15.9--23.2) & 2.1 (1.8--2.6)\\
Texas & 0.80  (0.62--1.11) & 16.4  (12.5--20.5) & 2.6 (2.0--3.2)\\
Utah & 0.21  (0.17--0.24) & 23.2  (20.1--27.5) & 2.6 (2.2--3.1)\\
Vermont & 1.01  (0.76--1.37) & 2.8  (2.2--3.5) & 2.2 (1.7--2.7)\\
Virginia & 0.57  (0.47--0.68) & 12.9  (10.9--15.6) & 3.0 (2.5--3.6)\\
Washington & 0.56  (0.43--0.78) & 10.1  (7.6--13.0) & 3.0 (2.3--3.9)\\
West Virginia & 0.97  (0.81--1.14) & 11.8  (10.2--14.0) & 2.2 (1.9--2.6)\\
Wisconsin & 0.52  (0.43--0.61) & 19.9  (17.0--23.9) & 2.2 (1.9--2.6)\\
Wyoming & 0.51  (0.42--0.61) & 18.0  (15.5--21.4) & 2.2 (1.9--2.6)\\
District of Columbia & 1.16  (0.91--1.48) & 10.2  (8.1--12.9) & 2.4 (1.9--3.0)\\
\hline
\end{tabular}
}
\caption{Posterior median and 95\% intervals for IFR, cumulative incidence as of January 6, 2021, and undercount factor as of January 6, 2021.}
\label{tab:states}
\end{table}
\newpage
\begin{figure}[htbp!]
\textbf{Alaska}
\centering
\begin{tabular}{ll}
\includegraphics[scale=0.77]{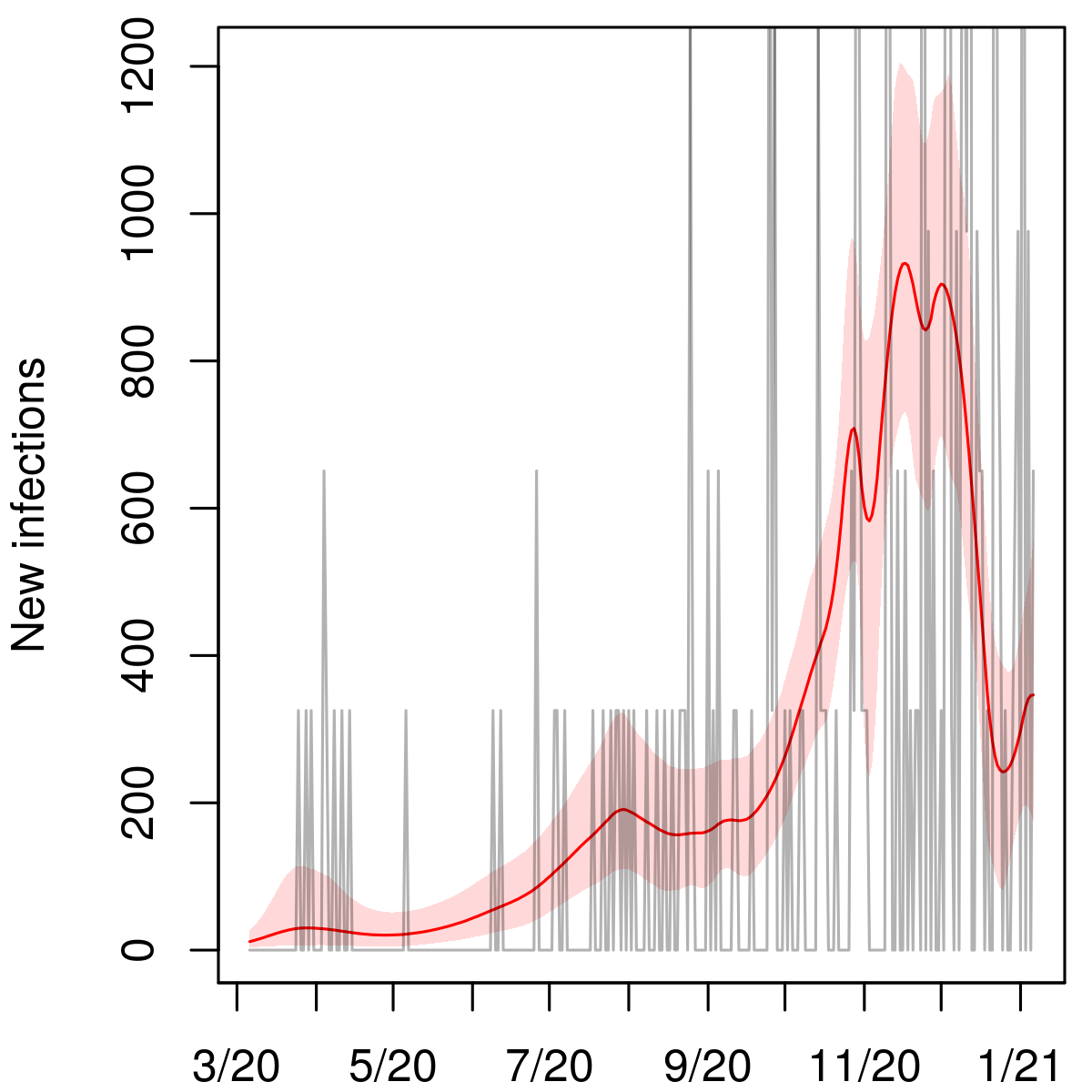}
&
\includegraphics[scale=0.77]{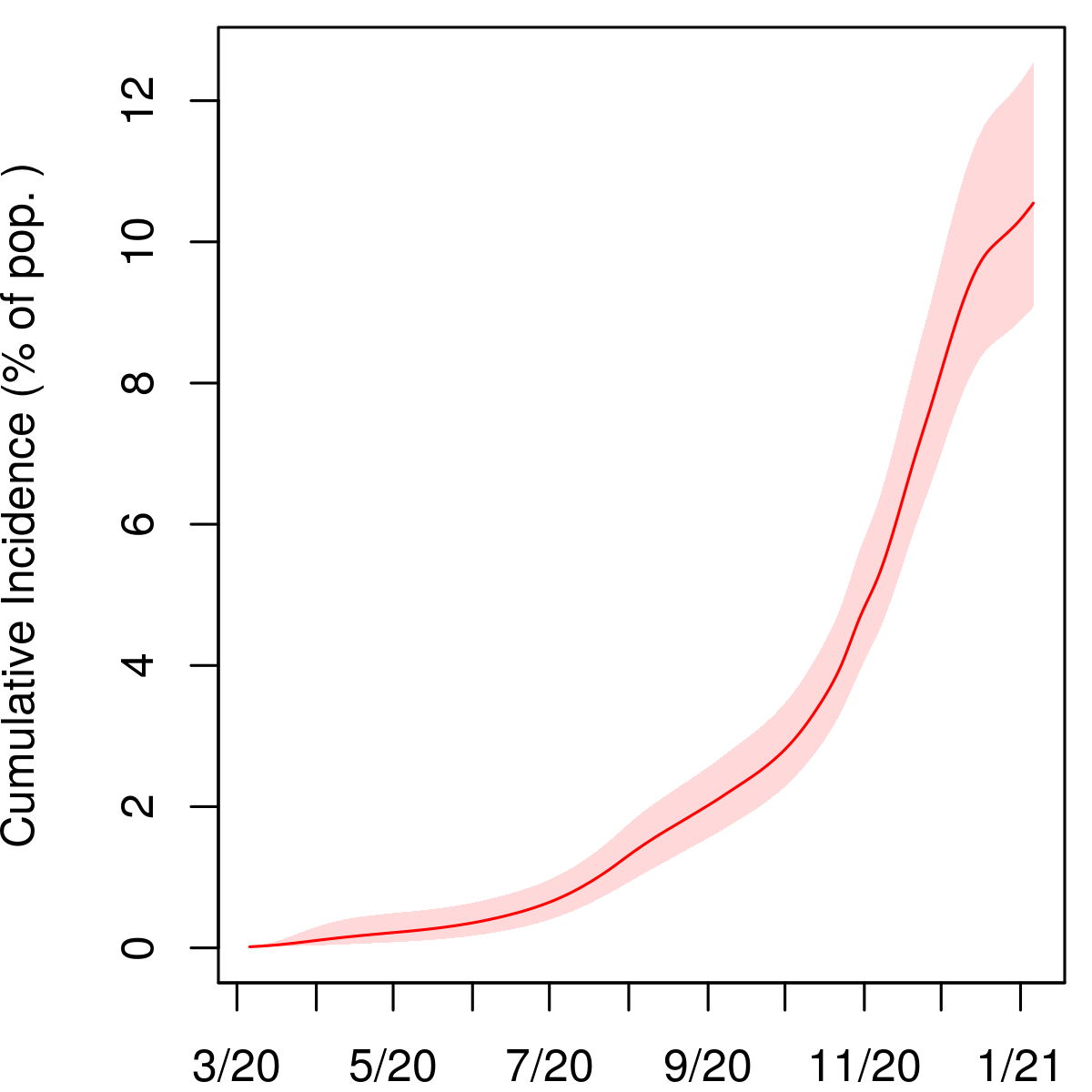} \\
\includegraphics[scale=0.77]{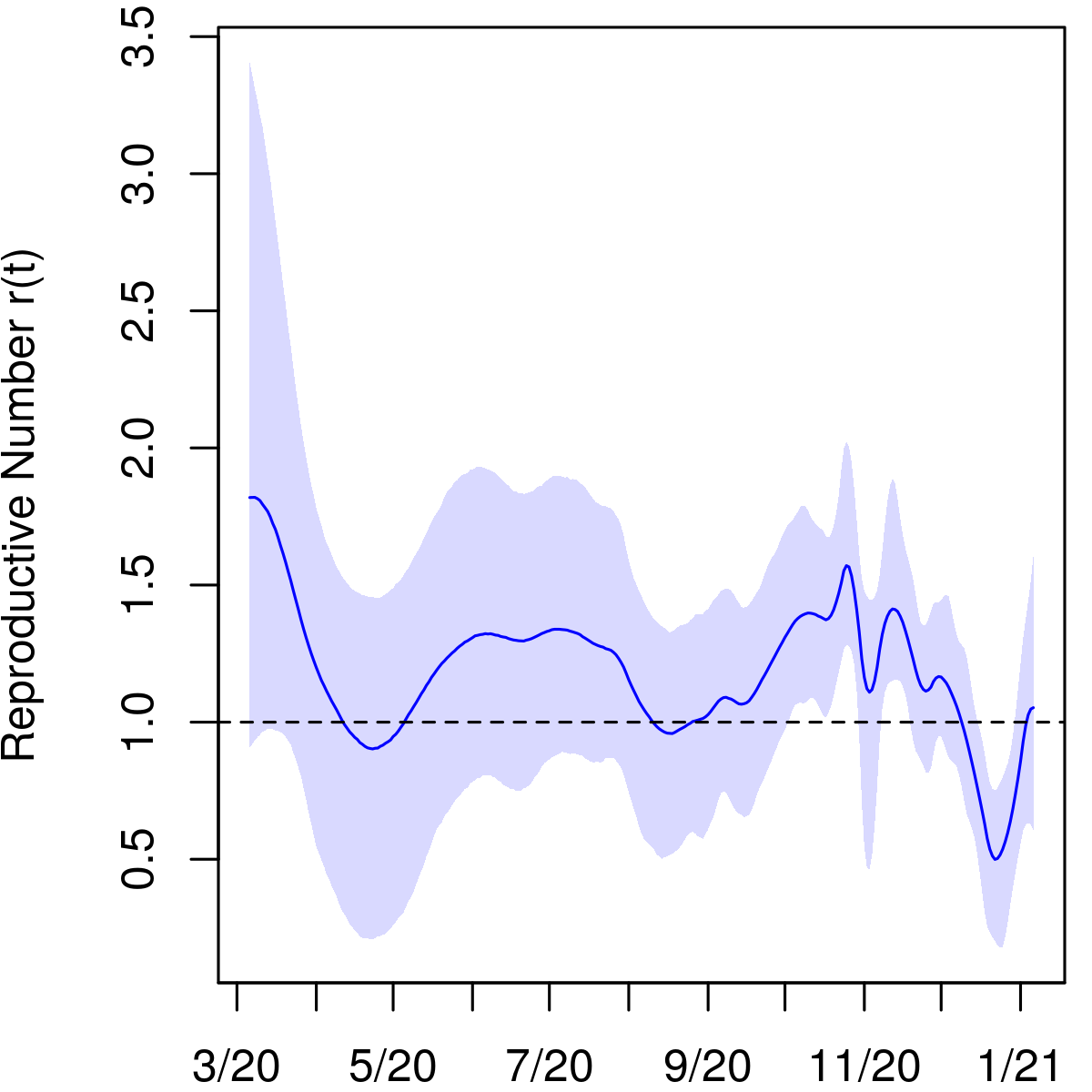}
&
\includegraphics[scale=0.77]{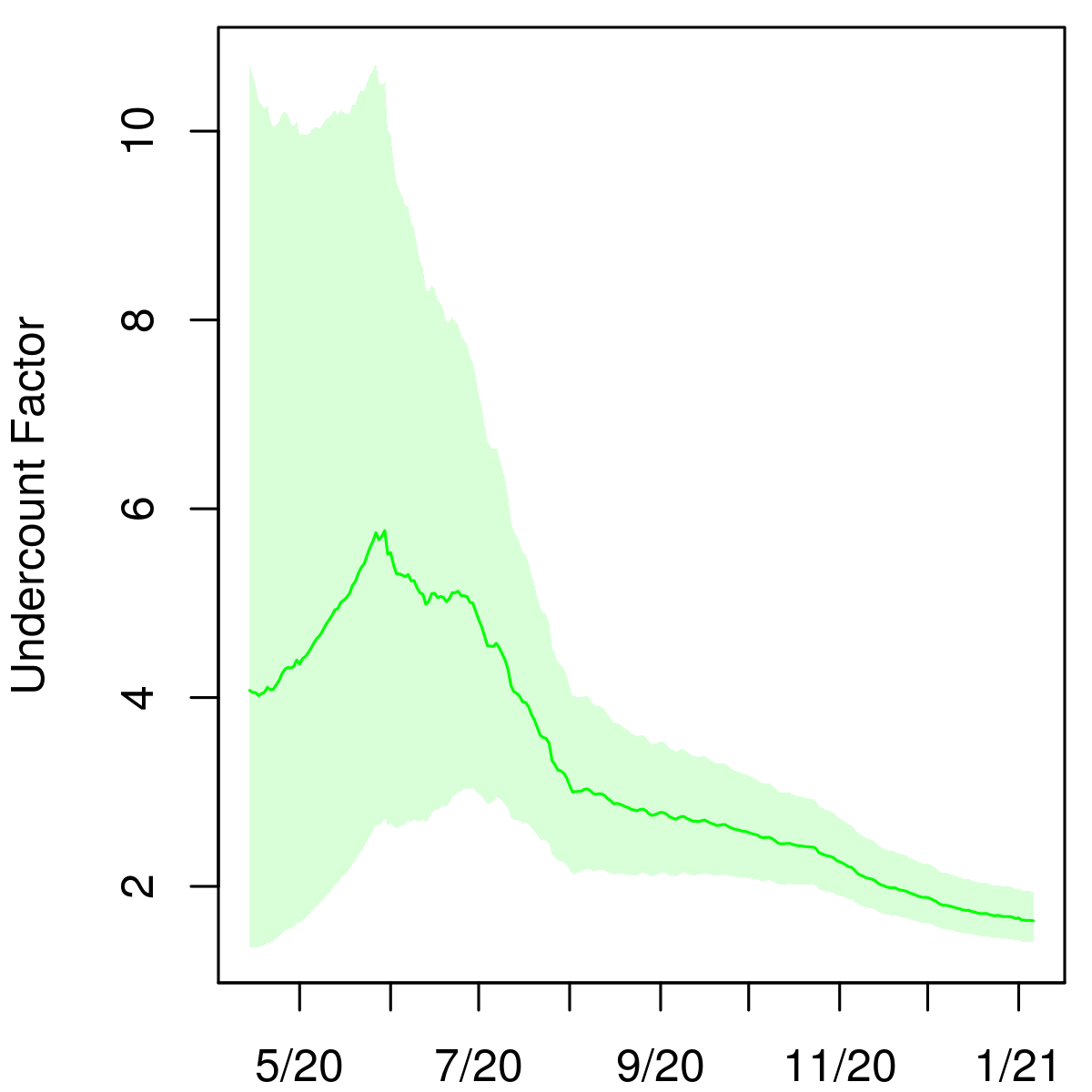} 
\end{tabular}
\caption{Posterior median and middle 95\% intervals for daily new infections, cumulative incidence, $r(t)$, and cumulative undercount from March 2020 to January 2021. In the top left panel, deaths divided by the posterior median IFR are plotted in grey for comparison.}
\end{figure}
\newpage
\begin{figure}[htbp!]
\textbf{Alabama}
\centering
\begin{tabular}{ll}
\includegraphics[scale=0.77]{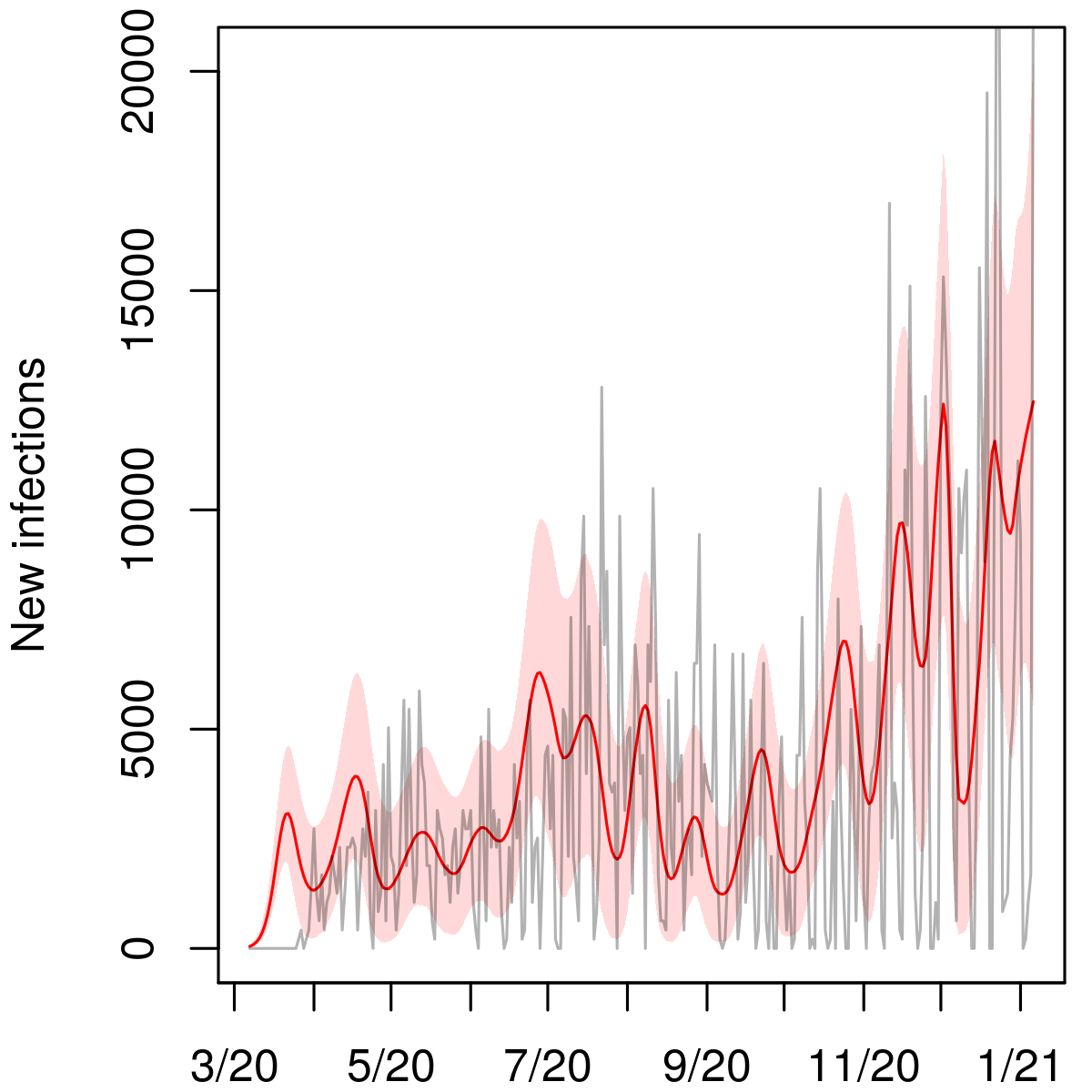}
&
\includegraphics[scale=0.77]{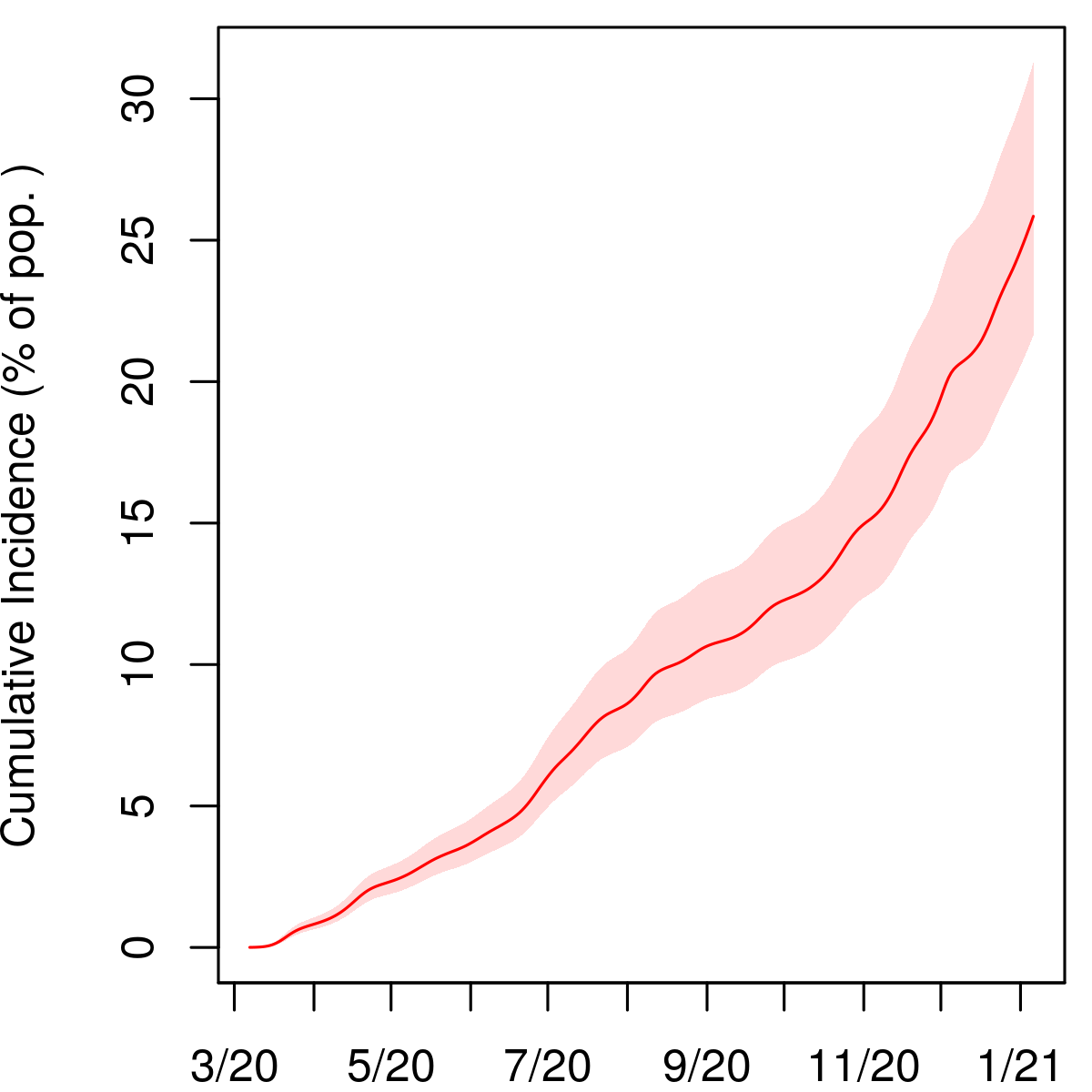} \\
\includegraphics[scale=0.77]{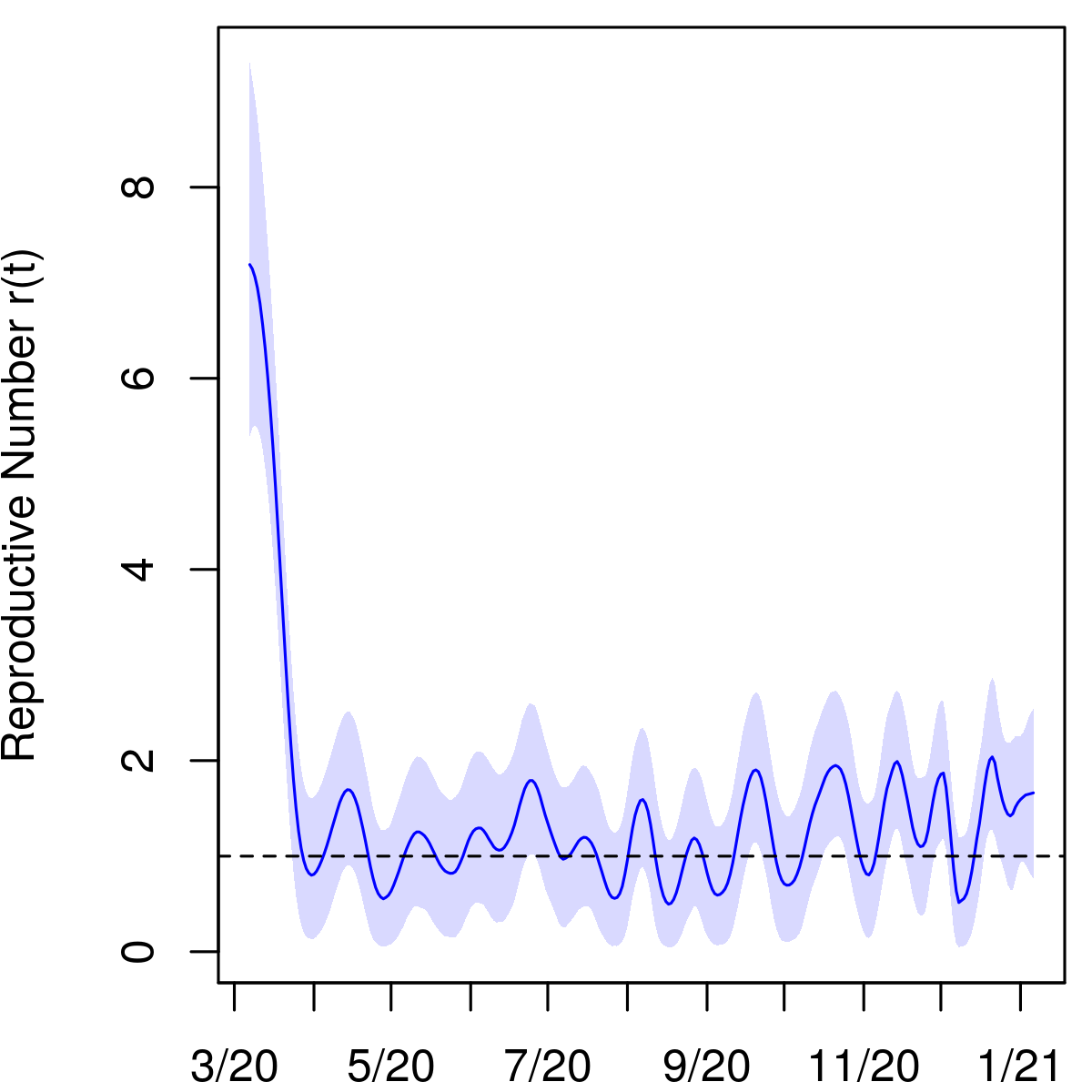}
&
\includegraphics[scale=0.77]{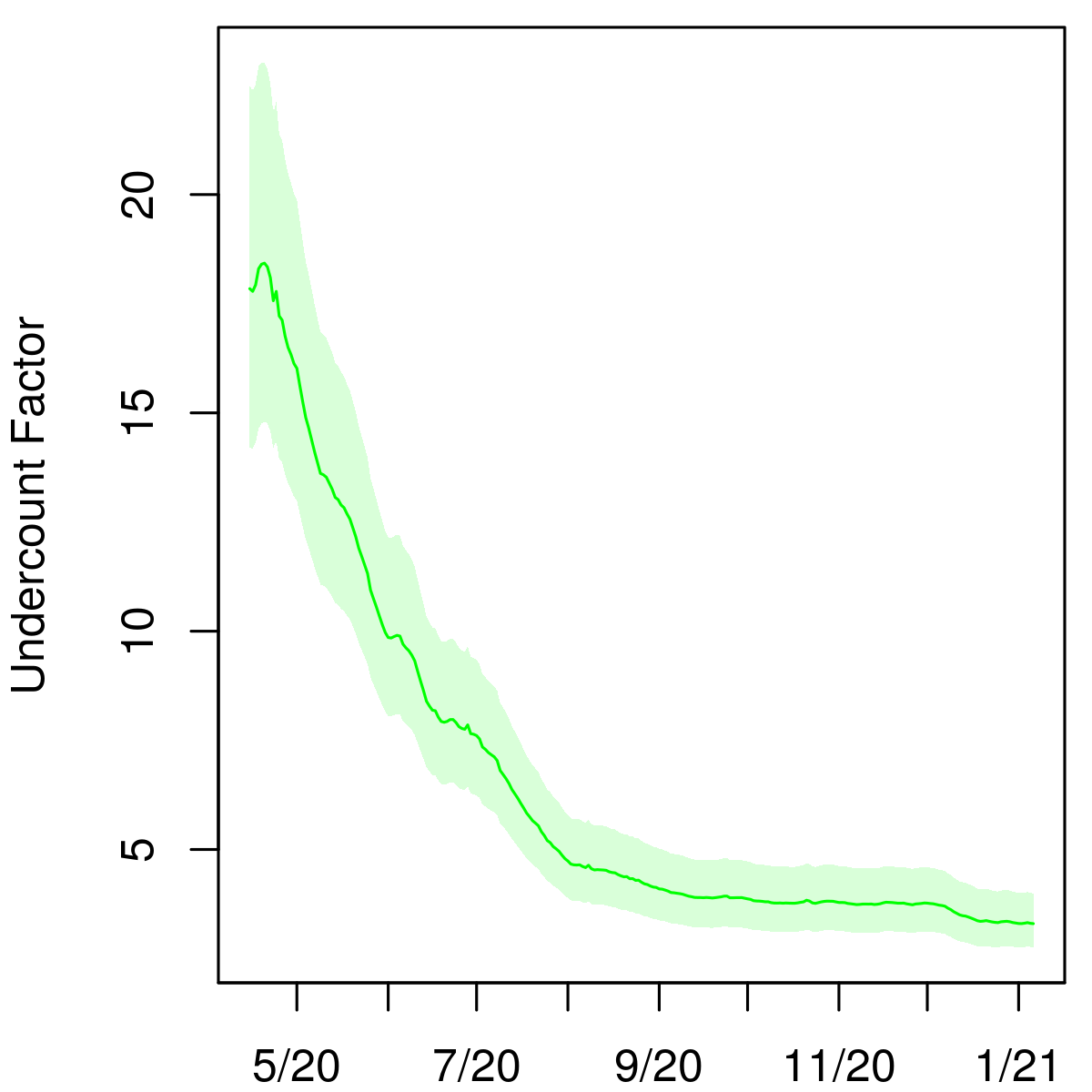} 
\end{tabular}
\caption{Posterior median and middle 95\% intervals for daily new infections, cumulative incidence, $r(t)$, and cumulative undercount from March 2020 to January 2021. In the top left panel, deaths divided by the posterior median IFR are plotted in grey for comparison.}
\end{figure}
\newpage
\begin{figure}[htbp!]
\textbf{Arkansas}
\centering
\begin{tabular}{ll}
\includegraphics[scale=0.77]{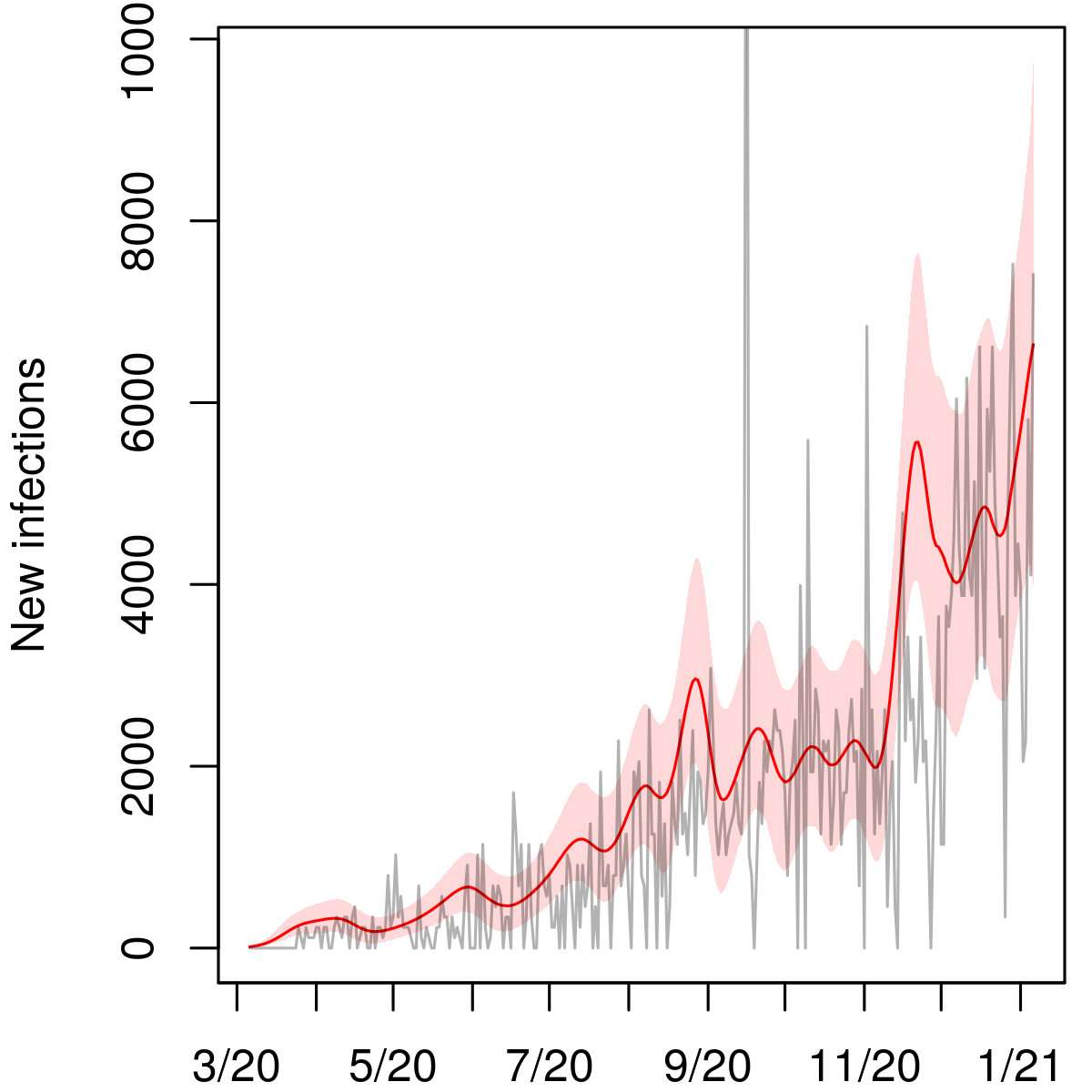}
&
\includegraphics[scale=0.77]{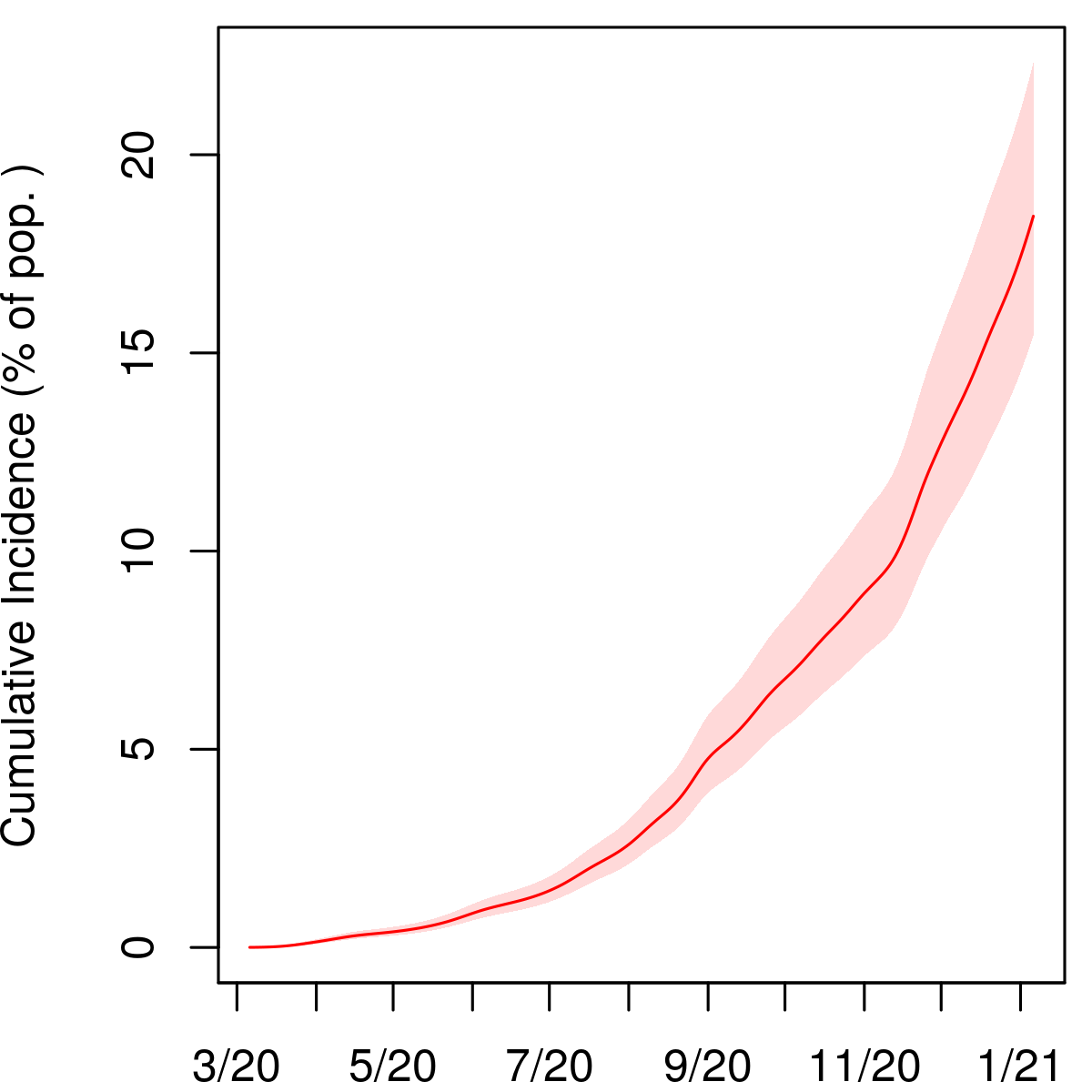} \\
\includegraphics[scale=0.77]{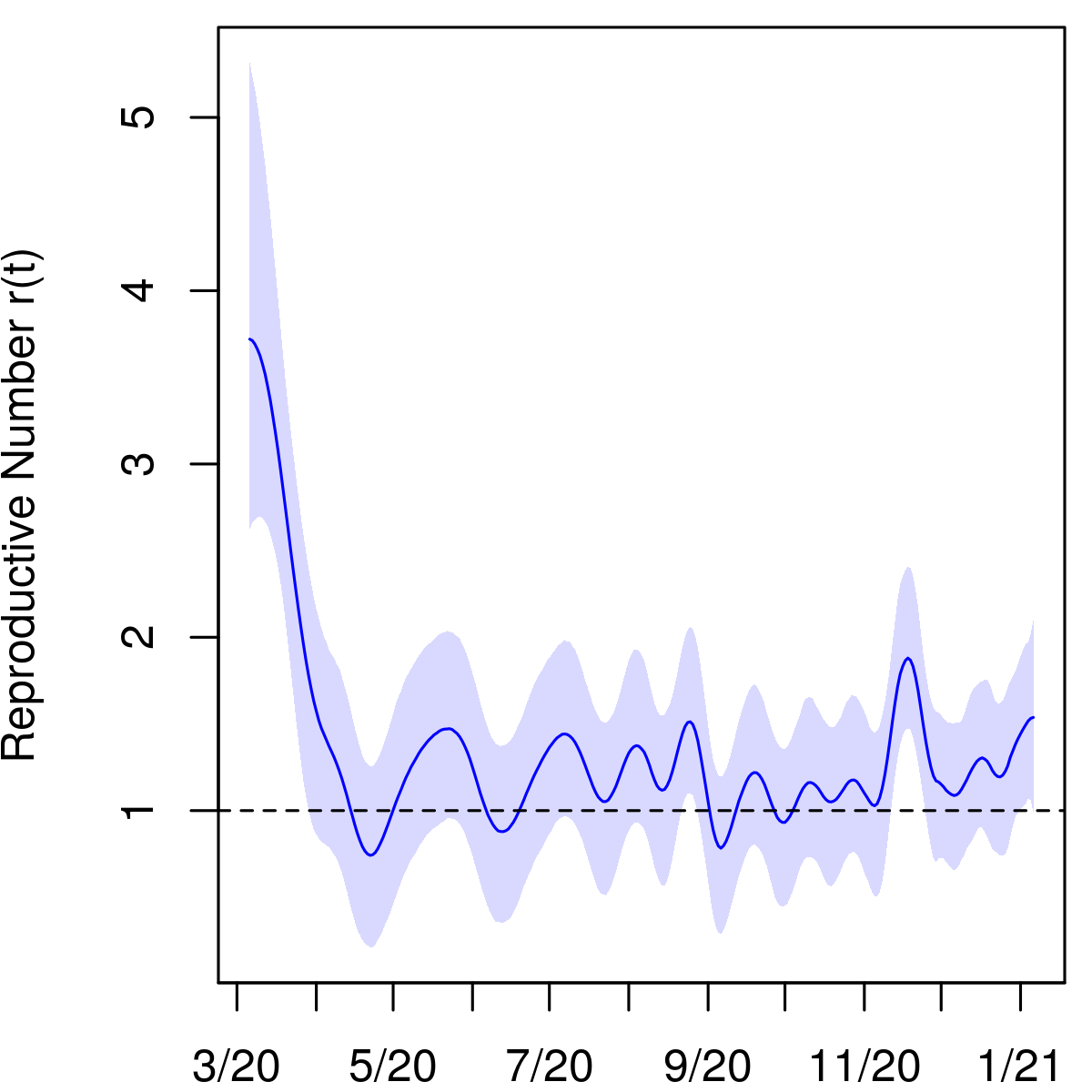}
&
\includegraphics[scale=0.77]{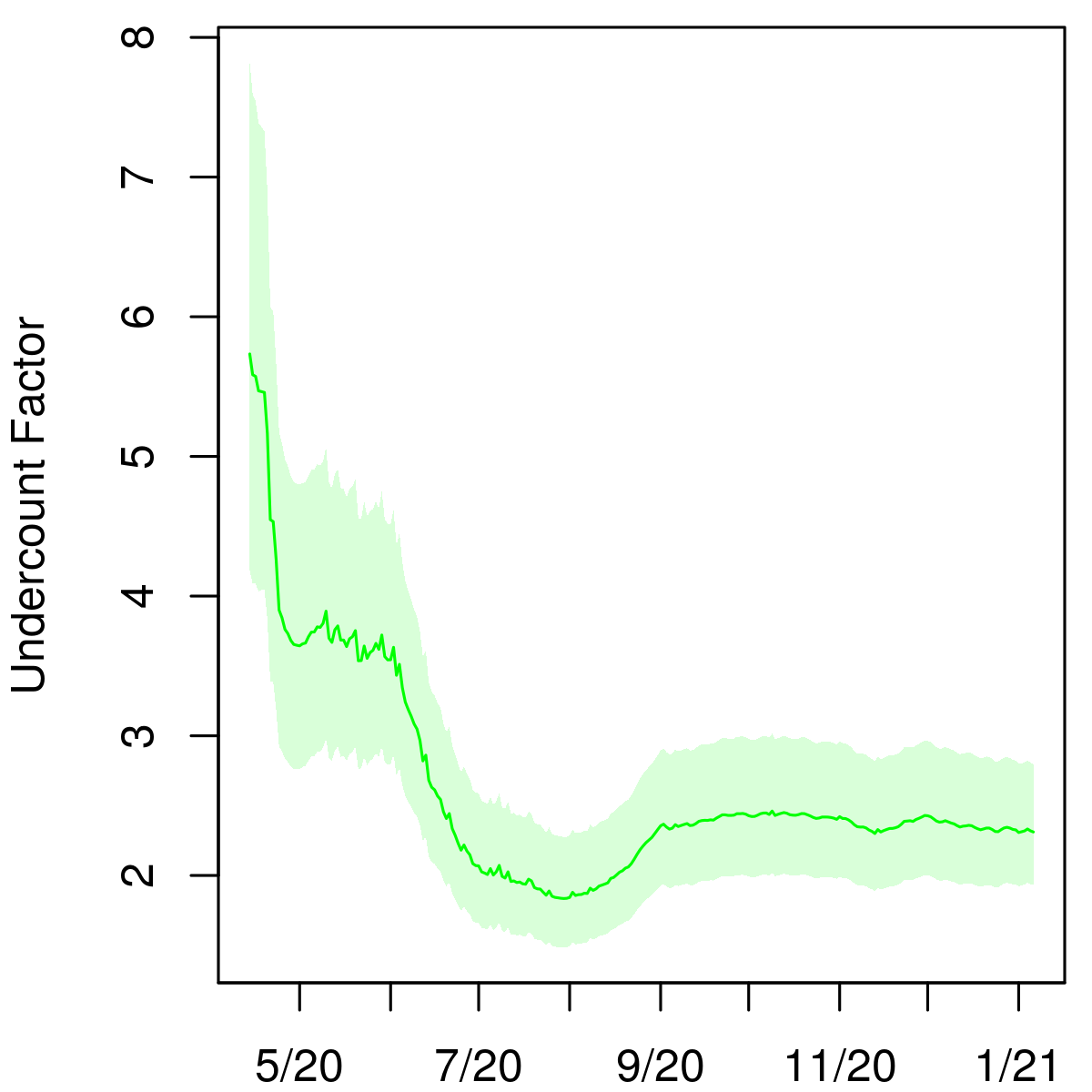} 
\end{tabular}
\caption{Posterior median and middle 95\% intervals for daily new infections, cumulative incidence, $r(t)$, and cumulative undercount from March 2020 to January 2021. In the top left panel, deaths divided by the posterior median IFR are plotted in grey for comparison.}
\end{figure}
\newpage
\begin{figure}[htbp!]
\textbf{Arizona}
\centering
\begin{tabular}{ll}
\includegraphics[scale=0.77]{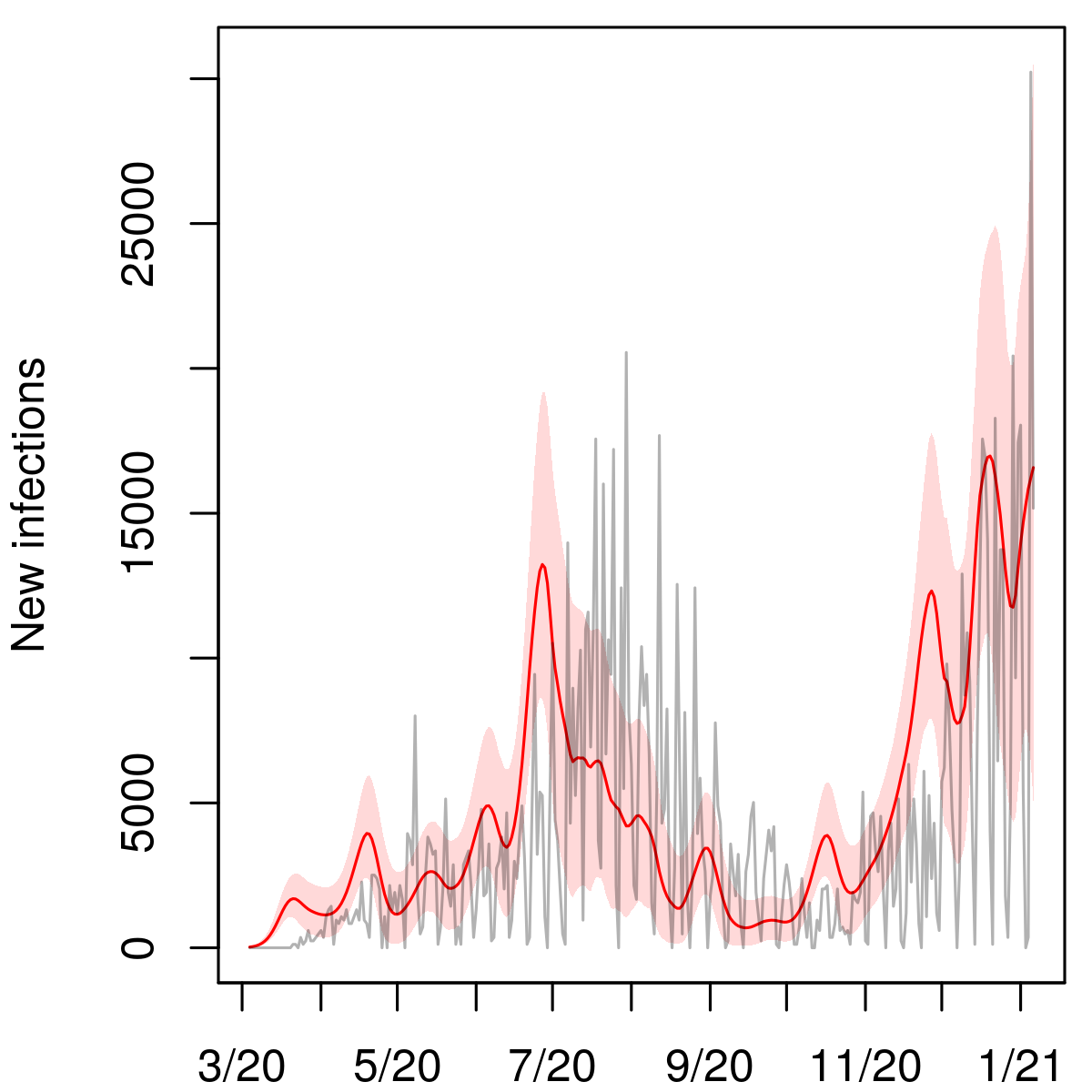}
&
\includegraphics[scale=0.77]{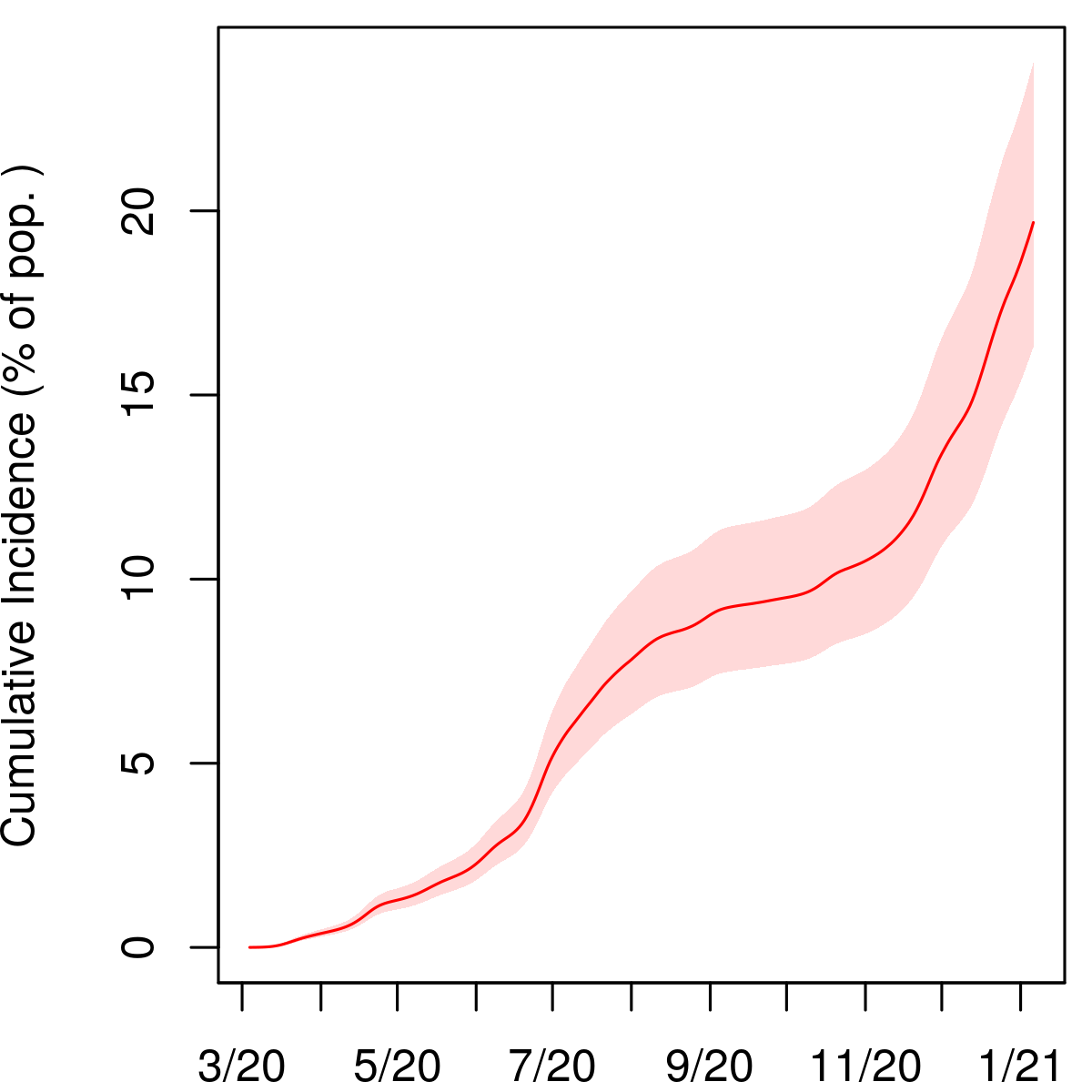} \\
\includegraphics[scale=0.77]{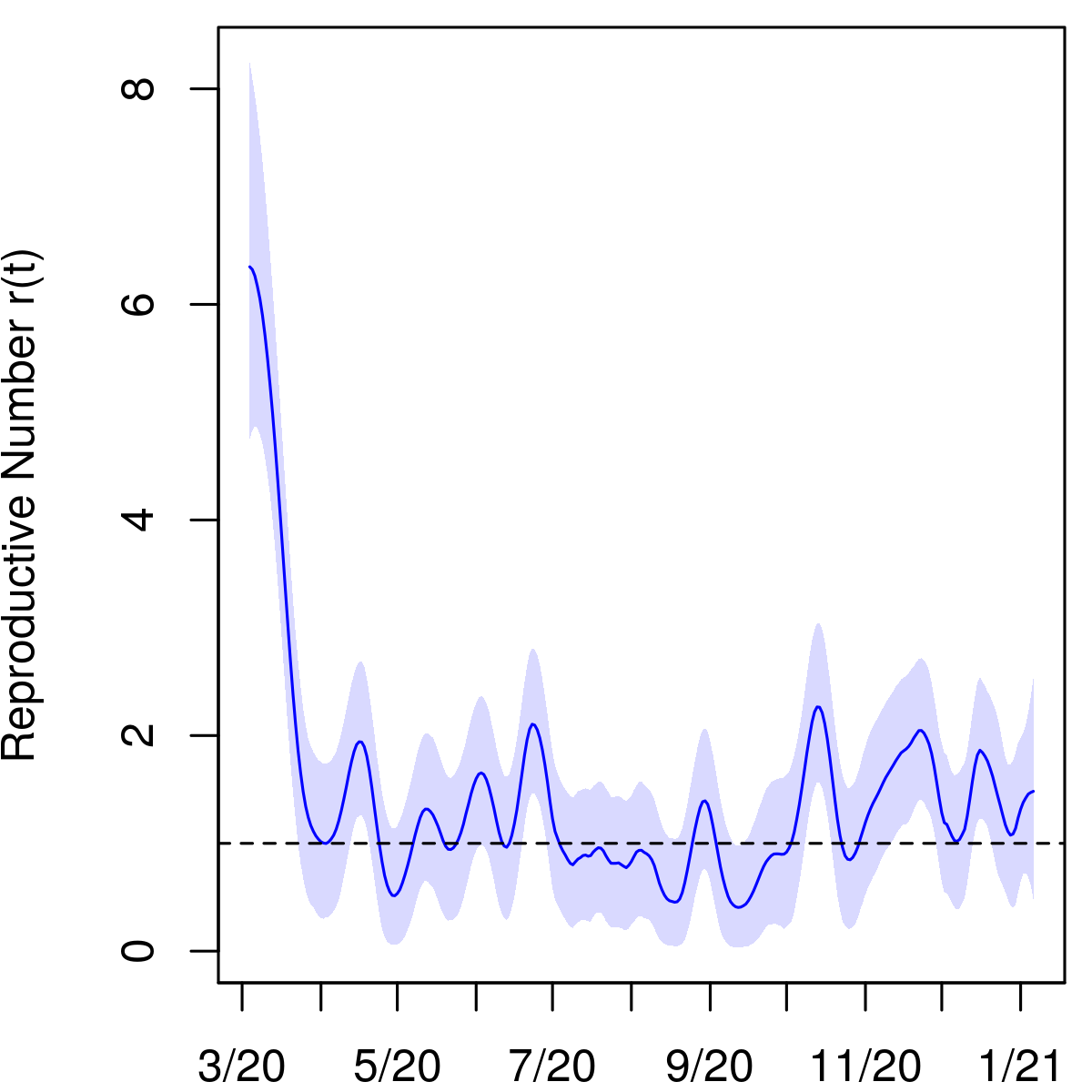}
&
\includegraphics[scale=0.77]{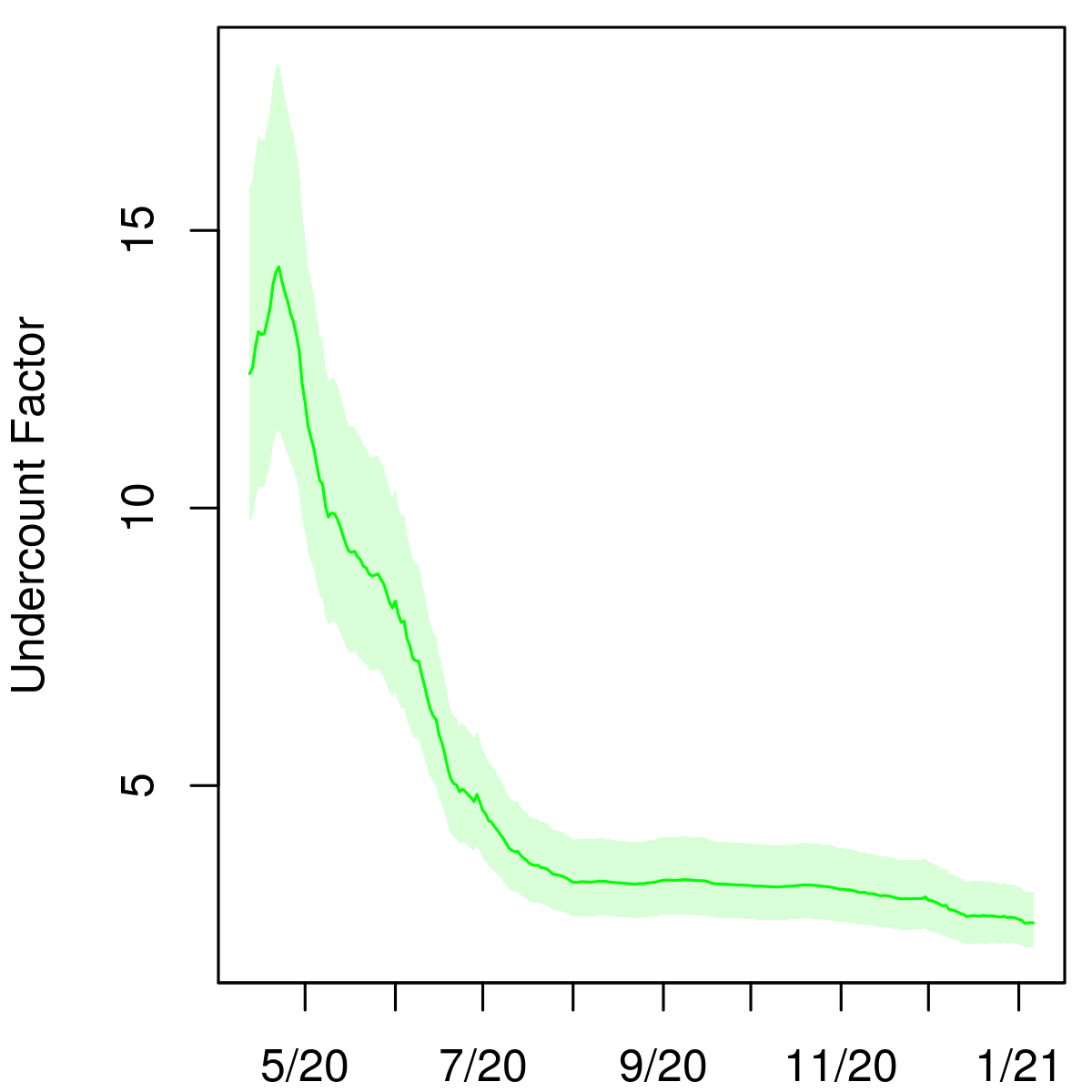} 
\end{tabular}
\caption{Posterior median and middle 95\% intervals for daily new infections, cumulative incidence, $r(t)$, and cumulative undercount from March 2020 to January 2021. In the top left panel, deaths divided by the posterior median IFR are plotted in grey for comparison.}
\end{figure}
\newpage
\begin{figure}[htbp!]
\textbf{California}
\centering
\begin{tabular}{ll}
\includegraphics[scale=0.77]{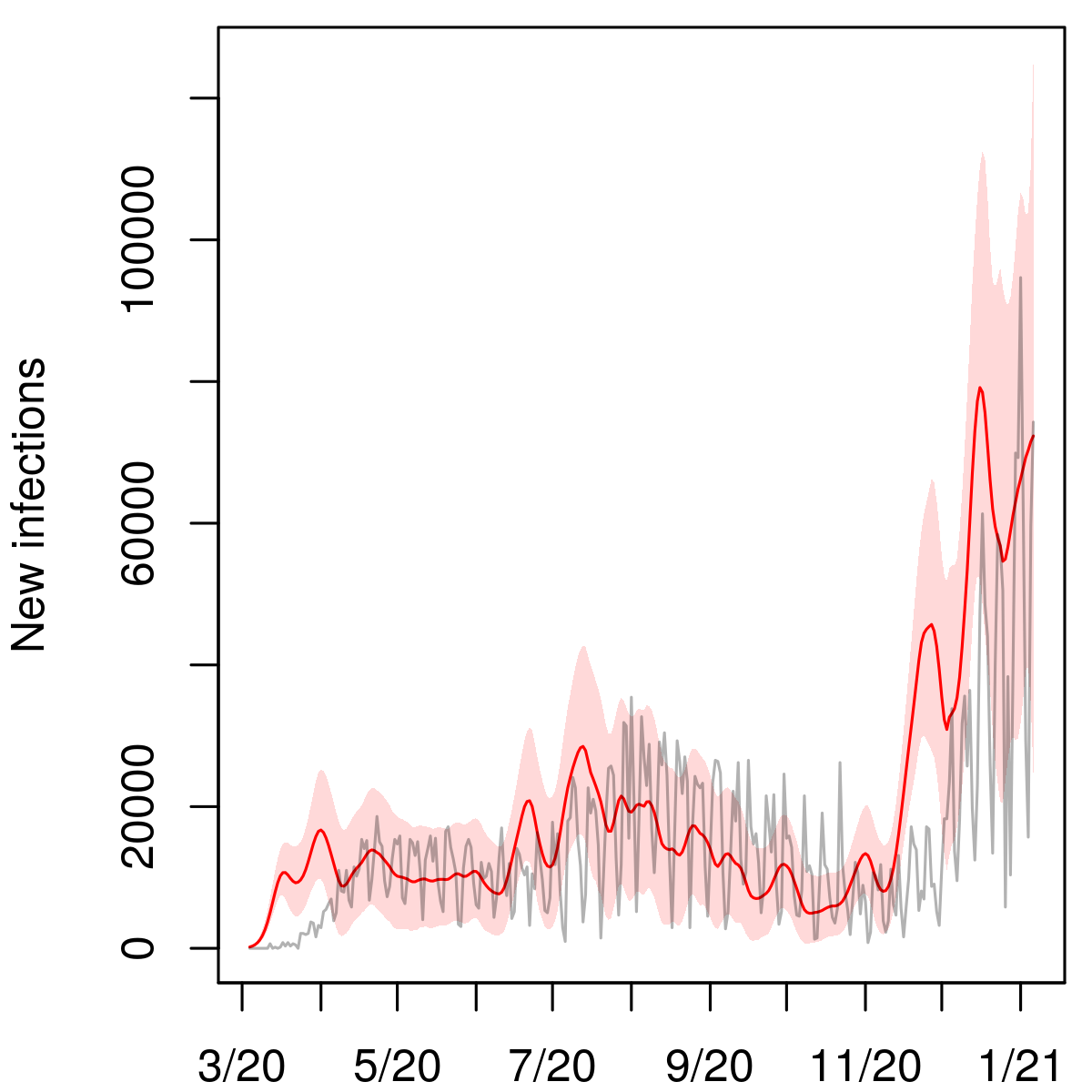}
&
\includegraphics[scale=0.77]{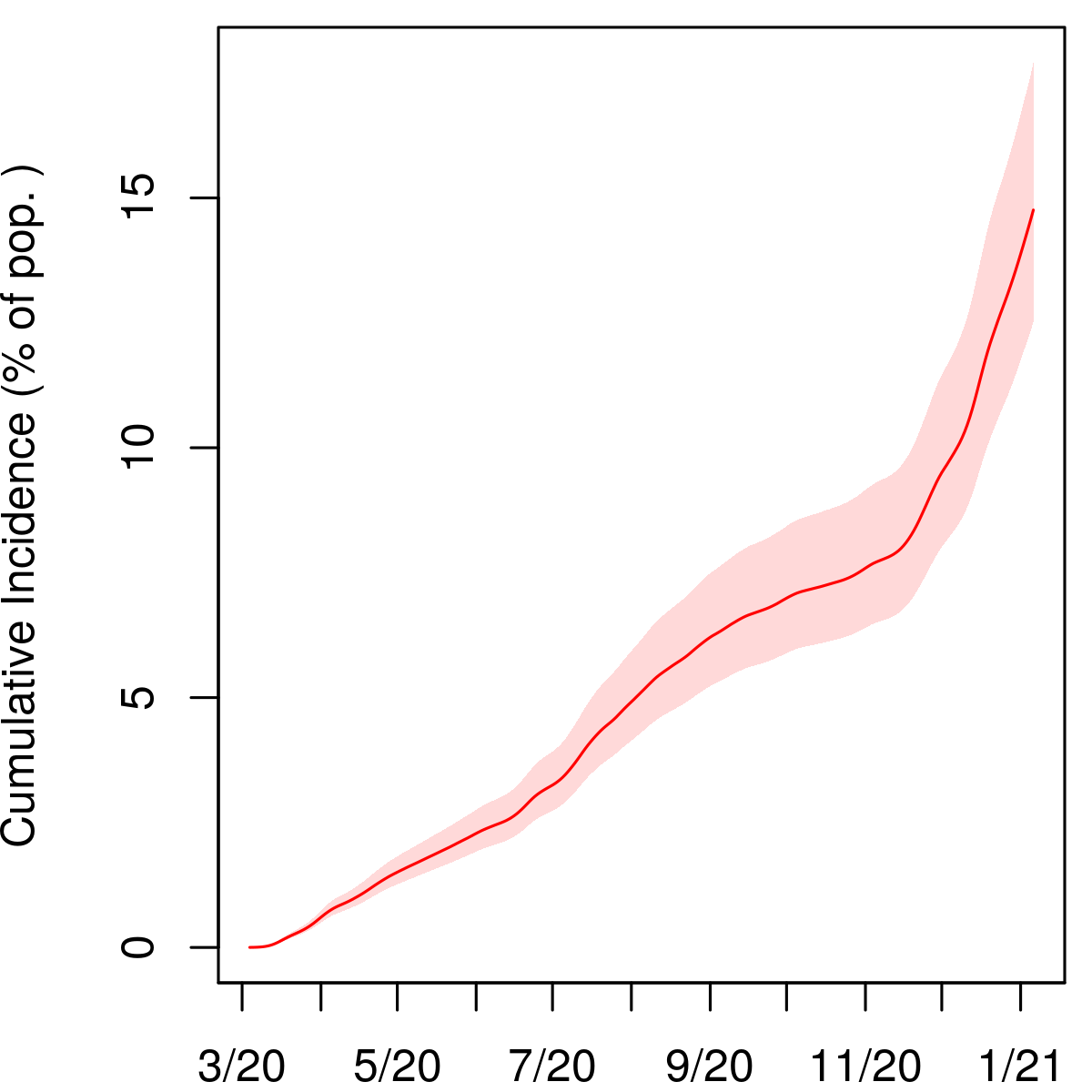} \\
\includegraphics[scale=0.77]{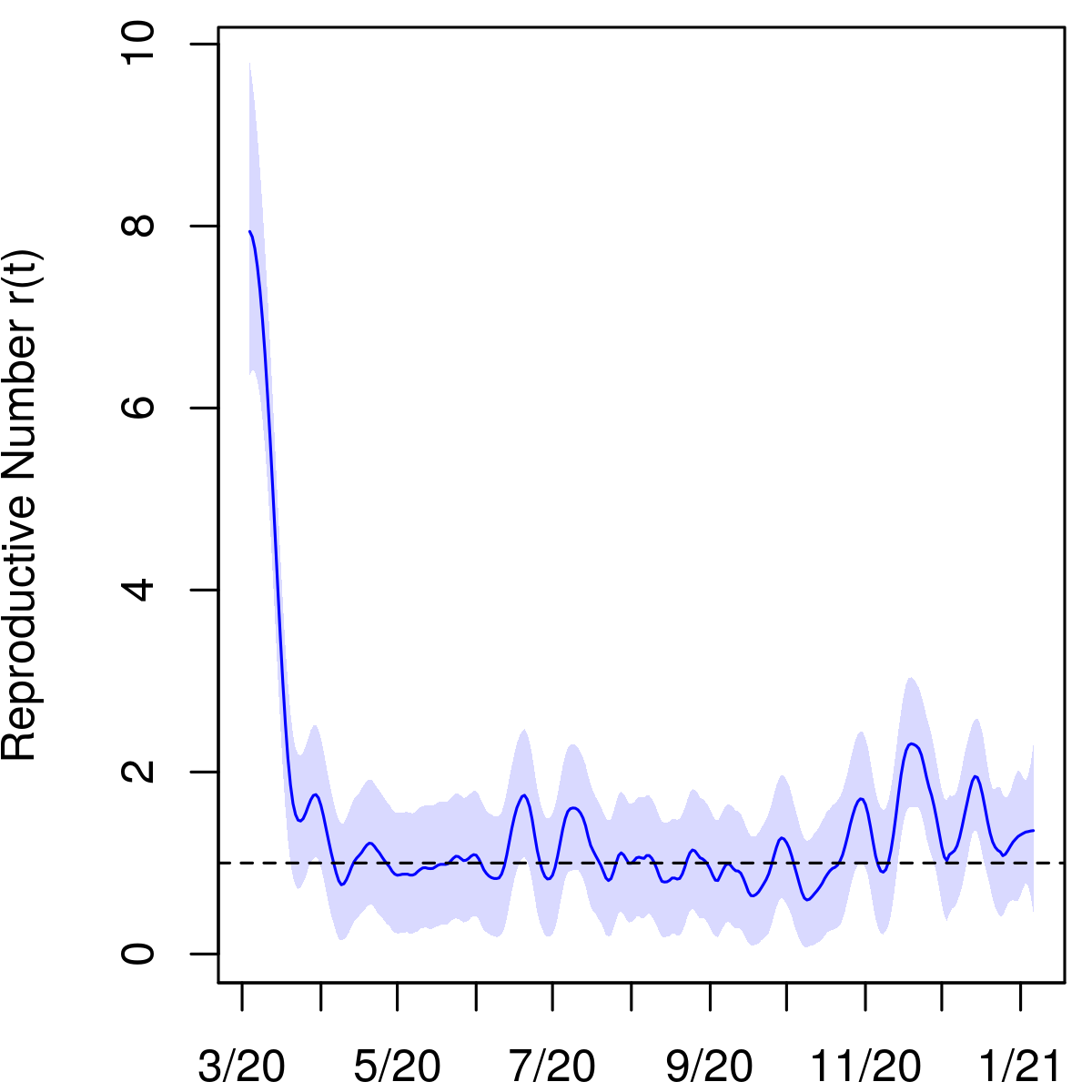}
&
\includegraphics[scale=0.77]{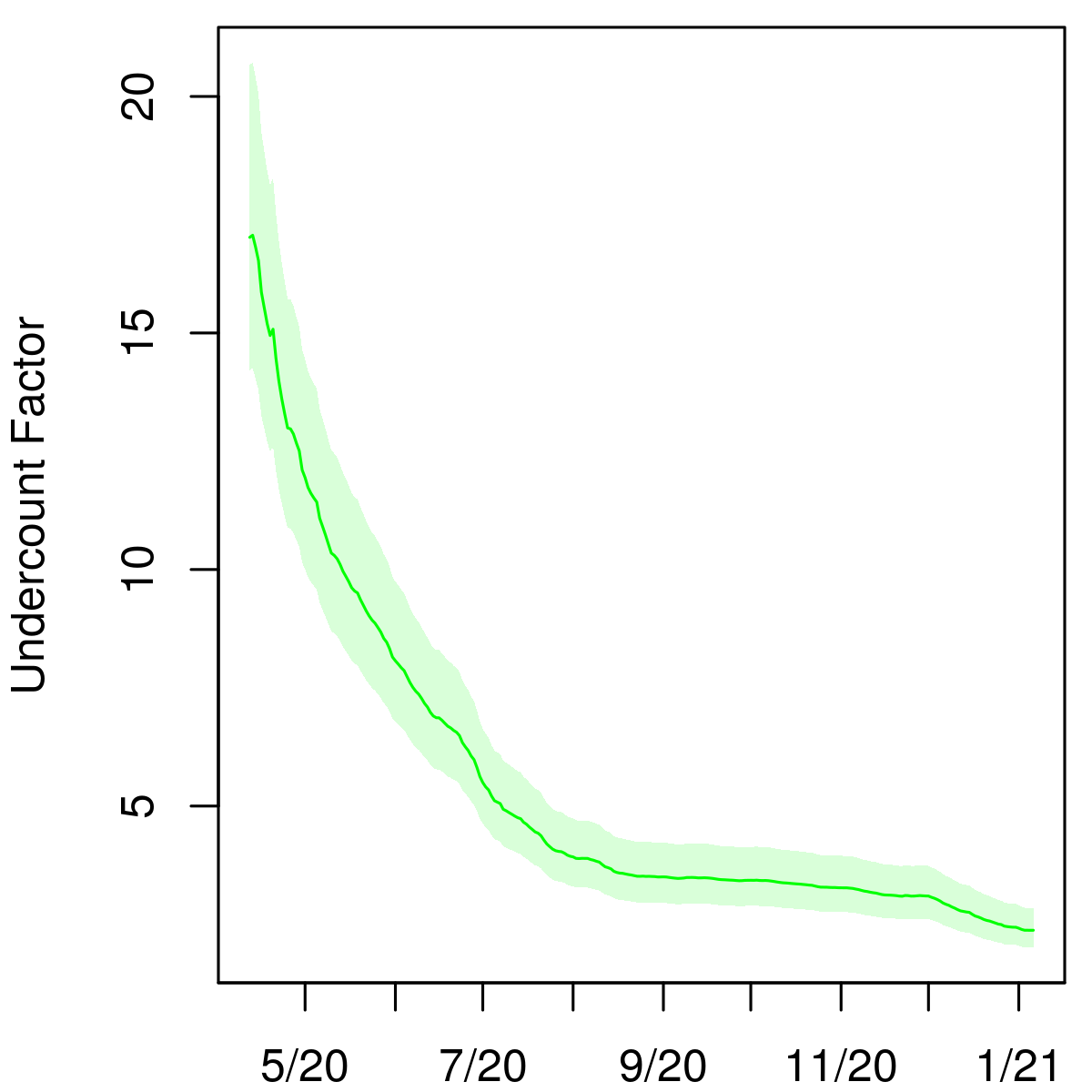} 
\end{tabular}
\caption{Posterior median and middle 95\% intervals for daily new infections, cumulative incidence, $r(t)$, and cumulative undercount from March 2020 to January 2021. In the top left panel, deaths divided by the posterior median IFR are plotted in grey for comparison.}
\end{figure}
\newpage
\begin{figure}[htbp!]
\textbf{Colorado}
\centering
\begin{tabular}{ll}
\includegraphics[scale=0.77]{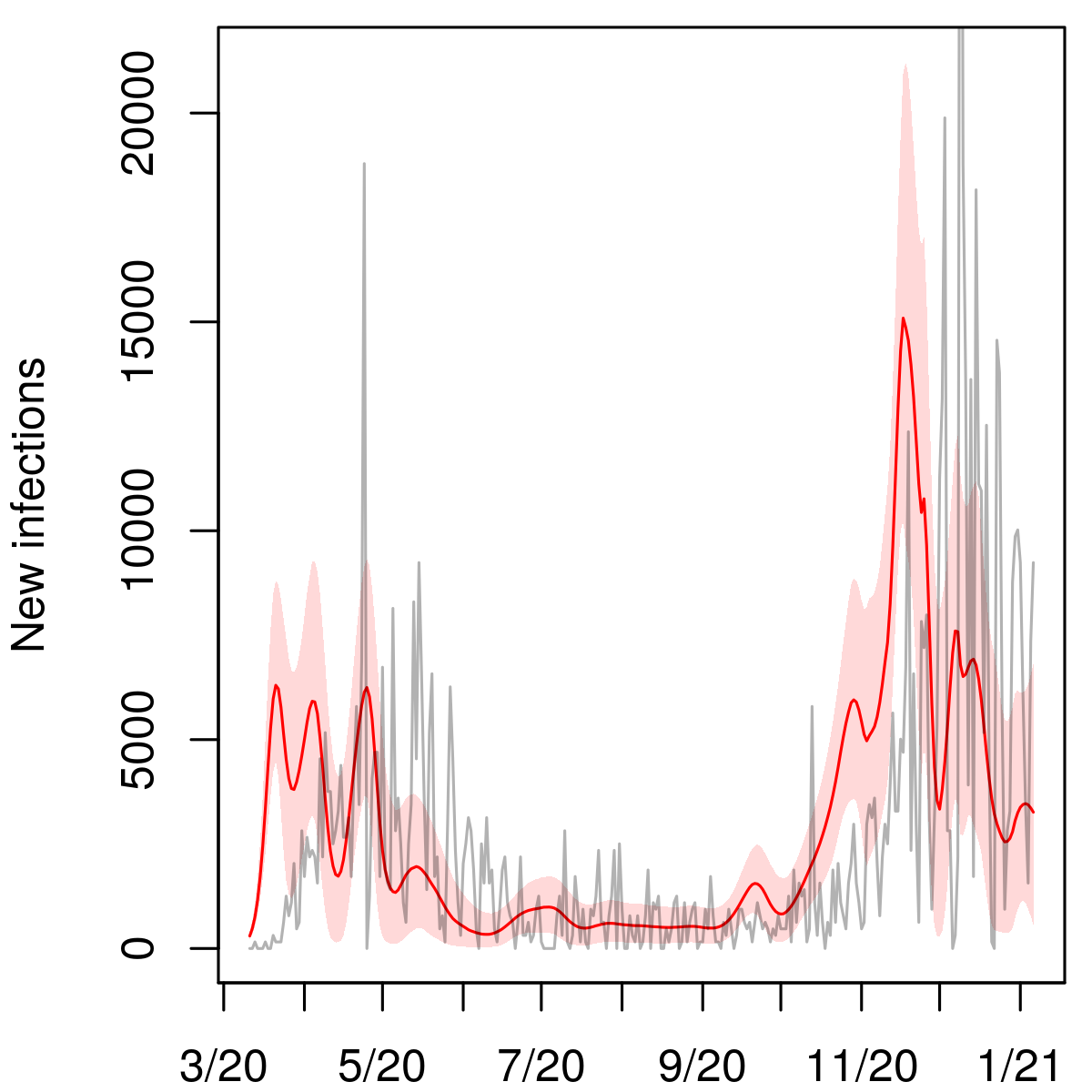}
&
\includegraphics[scale=0.77]{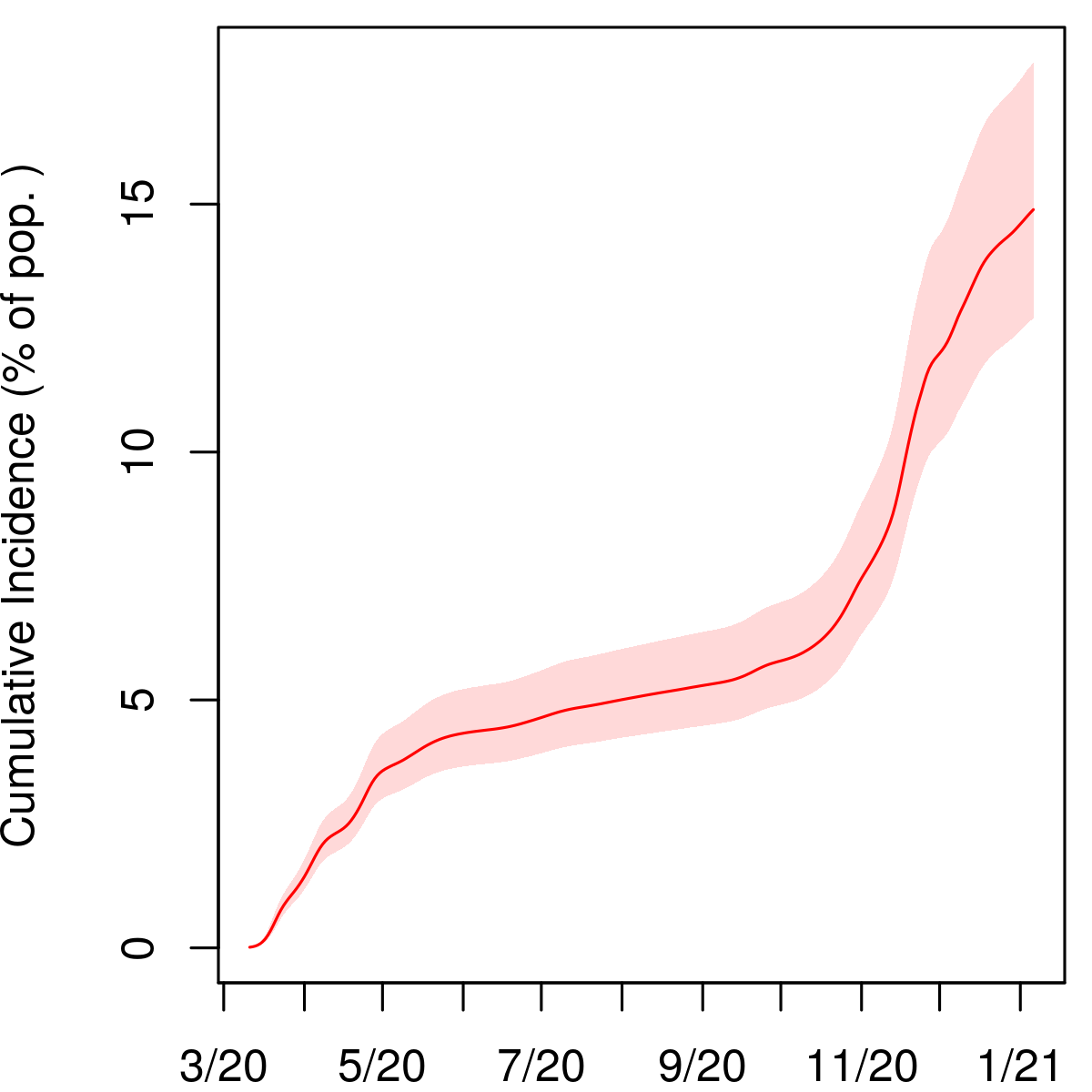} \\
\includegraphics[scale=0.77]{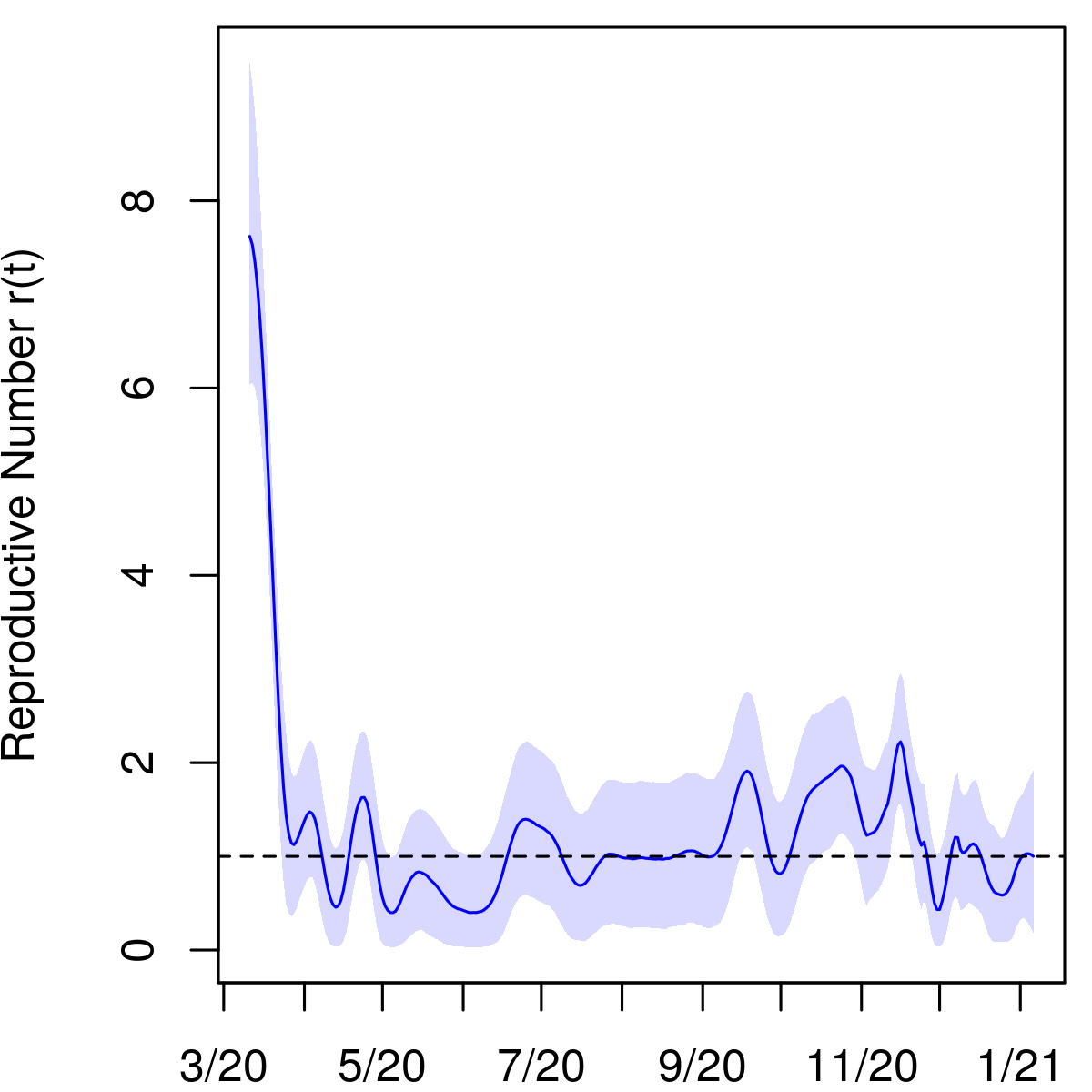}
&
\includegraphics[scale=0.77]{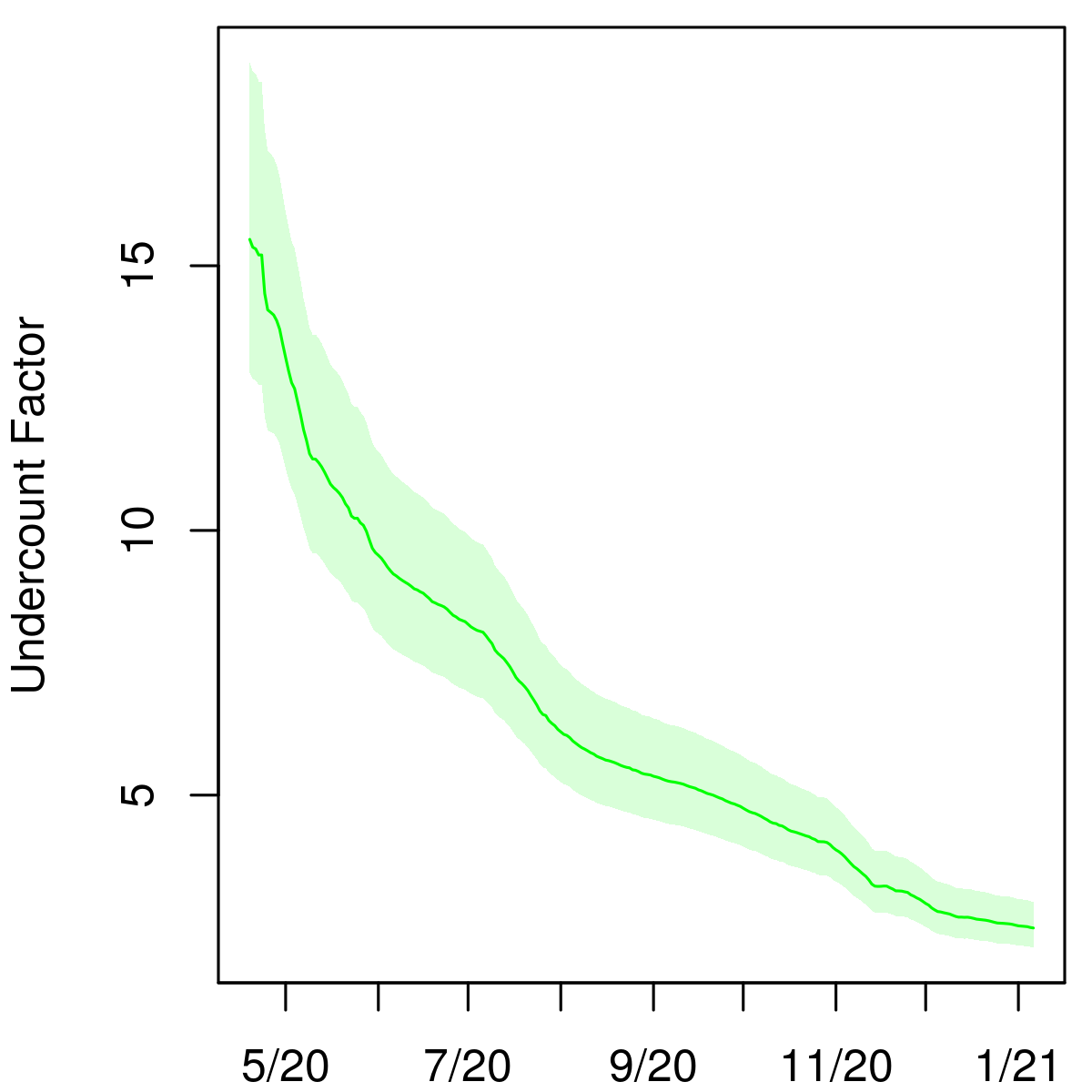} 
\end{tabular}
\caption{Posterior median and middle 95\% intervals for daily new infections, cumulative incidence, $r(t)$, and cumulative undercount from March 2020 to January 2021. In the top left panel, deaths divided by the posterior median IFR are plotted in grey for comparison.}
\end{figure}
\newpage
\begin{figure}[htbp!]
\textbf{Connecticut}
\centering
\begin{tabular}{ll}
\includegraphics[scale=0.77]{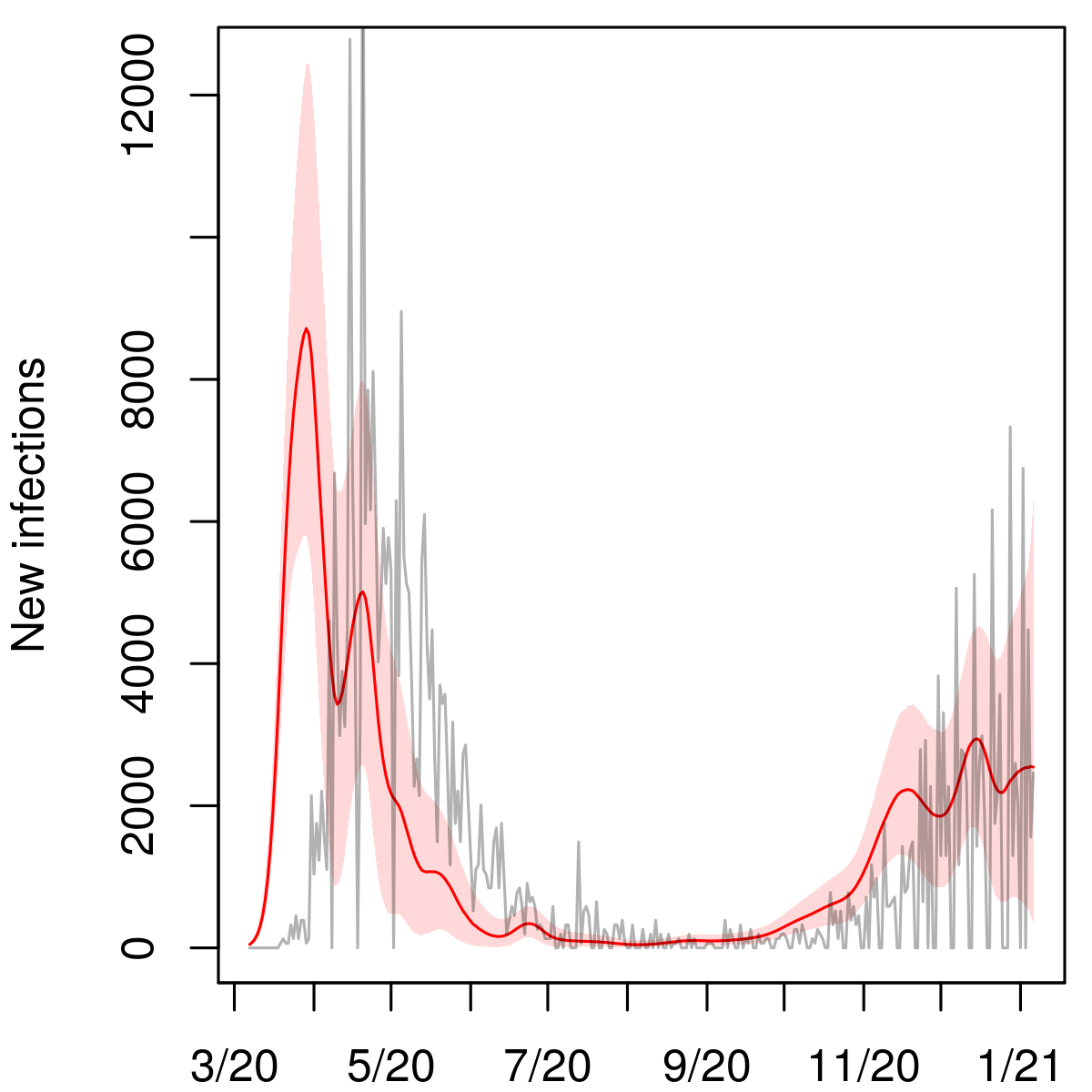}
&
\includegraphics[scale=0.77]{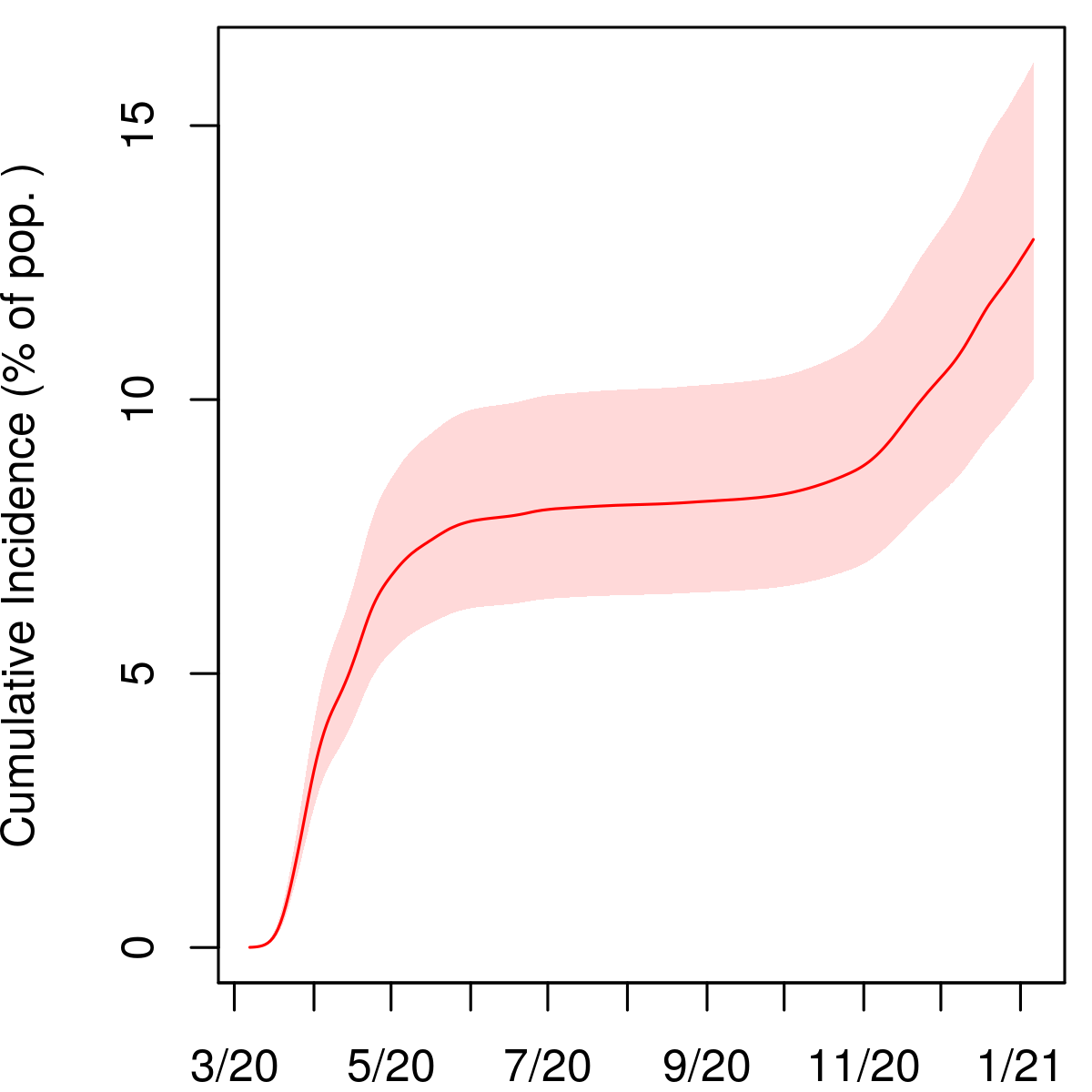} \\
\includegraphics[scale=0.77]{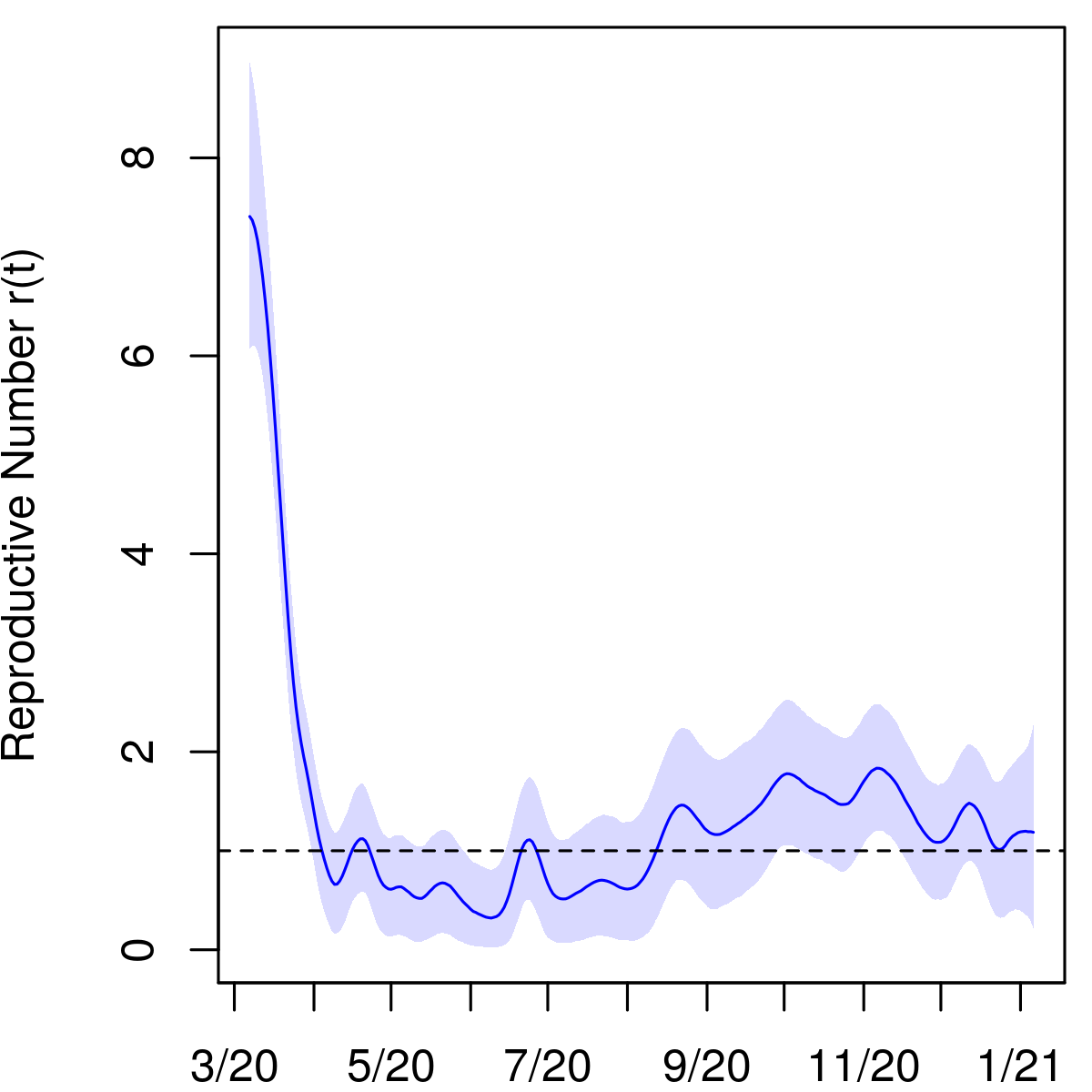}
&
\includegraphics[scale=0.77]{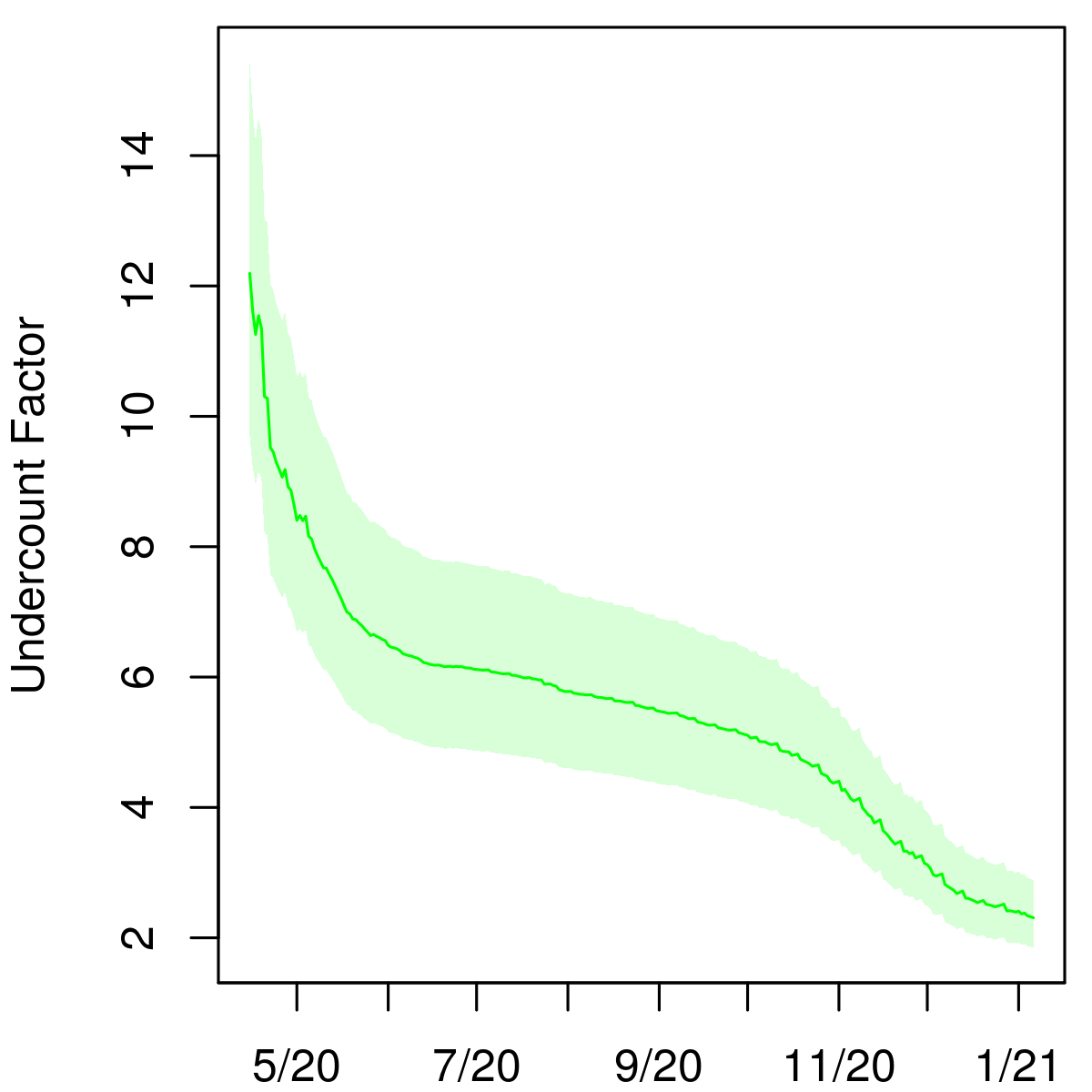} 
\end{tabular}
\caption{Posterior median and middle 95\% intervals for daily new infections, cumulative incidence, $r(t)$, and cumulative undercount from March 2020 to January 2021. In the top left panel, deaths divided by the posterior median IFR are plotted in grey for comparison.}
\end{figure}
\newpage
\begin{figure}[htbp!]
\textbf{District of Columbia}
\centering
\begin{tabular}{ll}
\includegraphics[scale=0.77]{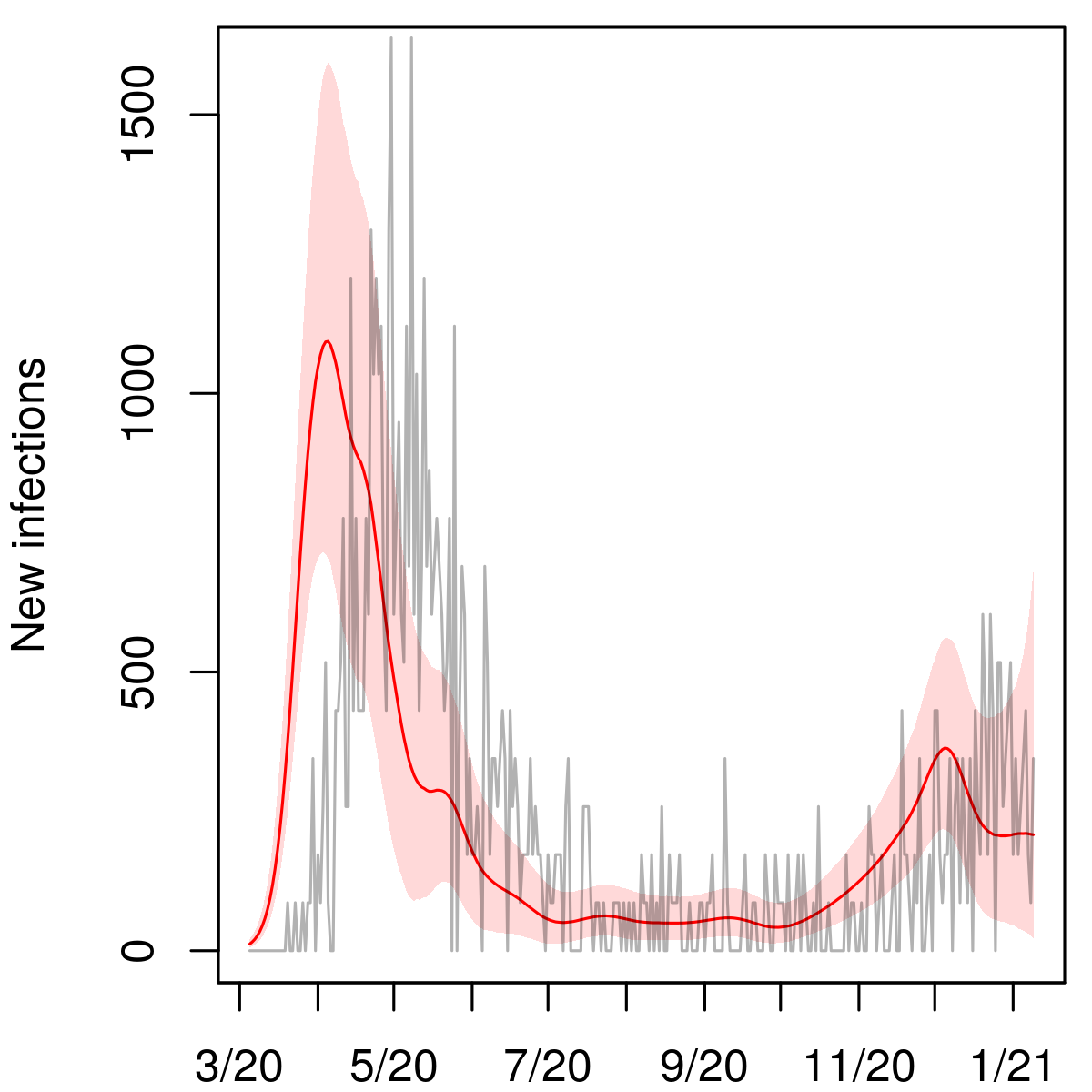}
&
\includegraphics[scale=0.77]{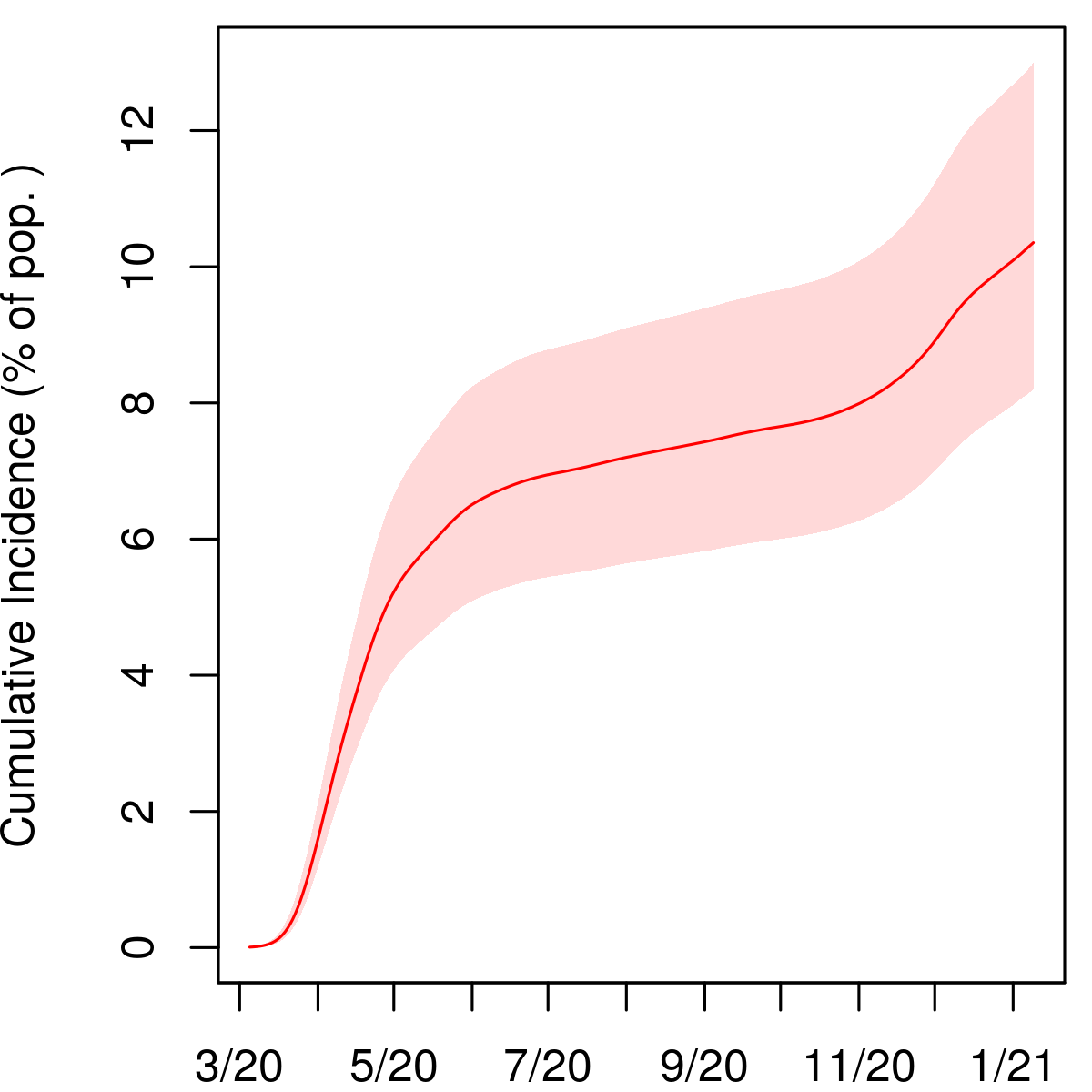} \\
\includegraphics[scale=0.77]{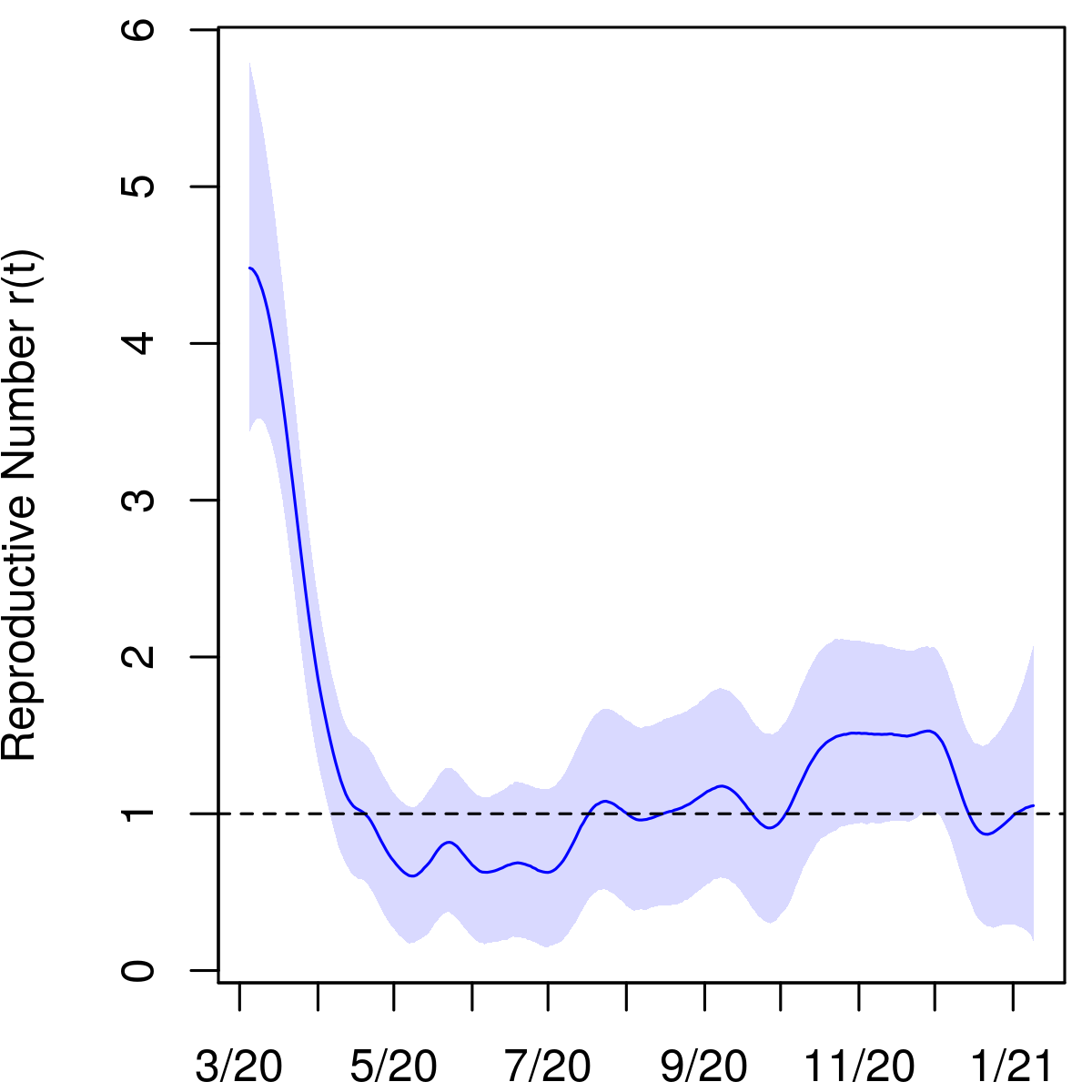}
&
\includegraphics[scale=0.77]{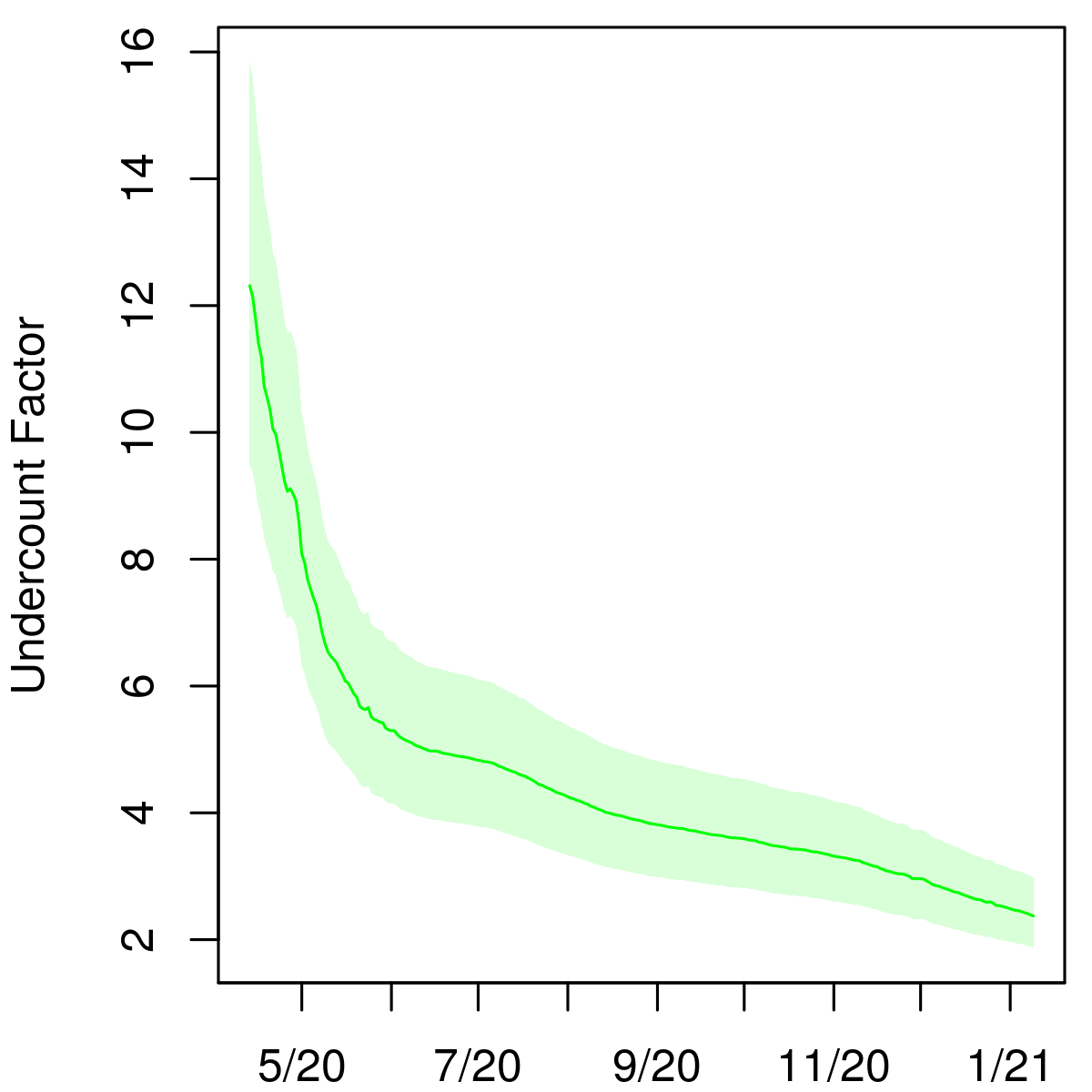} 
\end{tabular}
\caption{Posterior median and middle 95\% intervals for daily new infections, cumulative incidence, $r(t)$, and cumulative undercount from March 2020 to January 2021. In the top left panel, deaths divided by the posterior median IFR are plotted in grey for comparison.}
\end{figure}
\newpage
\begin{figure}[htbp!]
\textbf{Delaware}
\centering
\begin{tabular}{ll}
\includegraphics[scale=0.77]{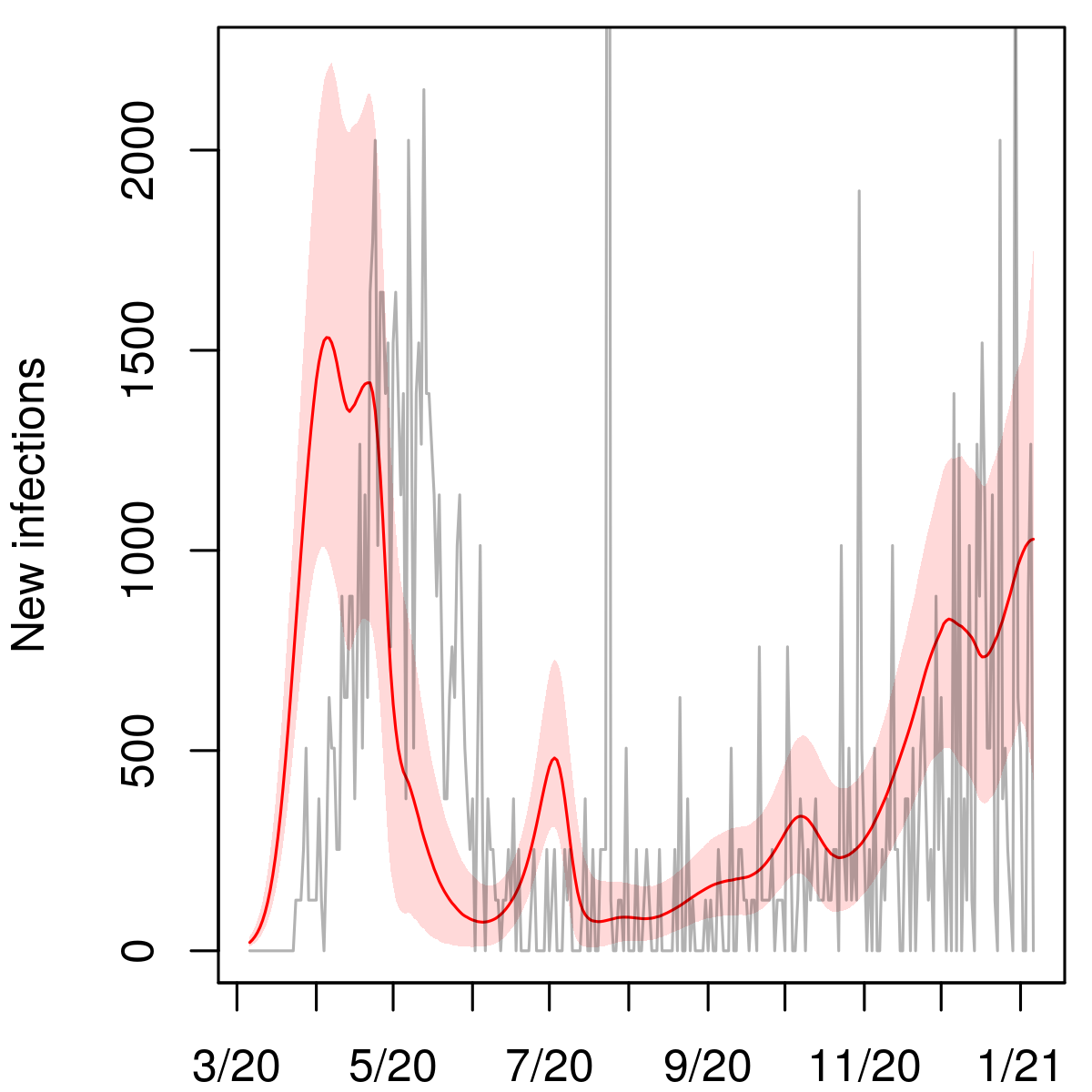}
&
\includegraphics[scale=0.77]{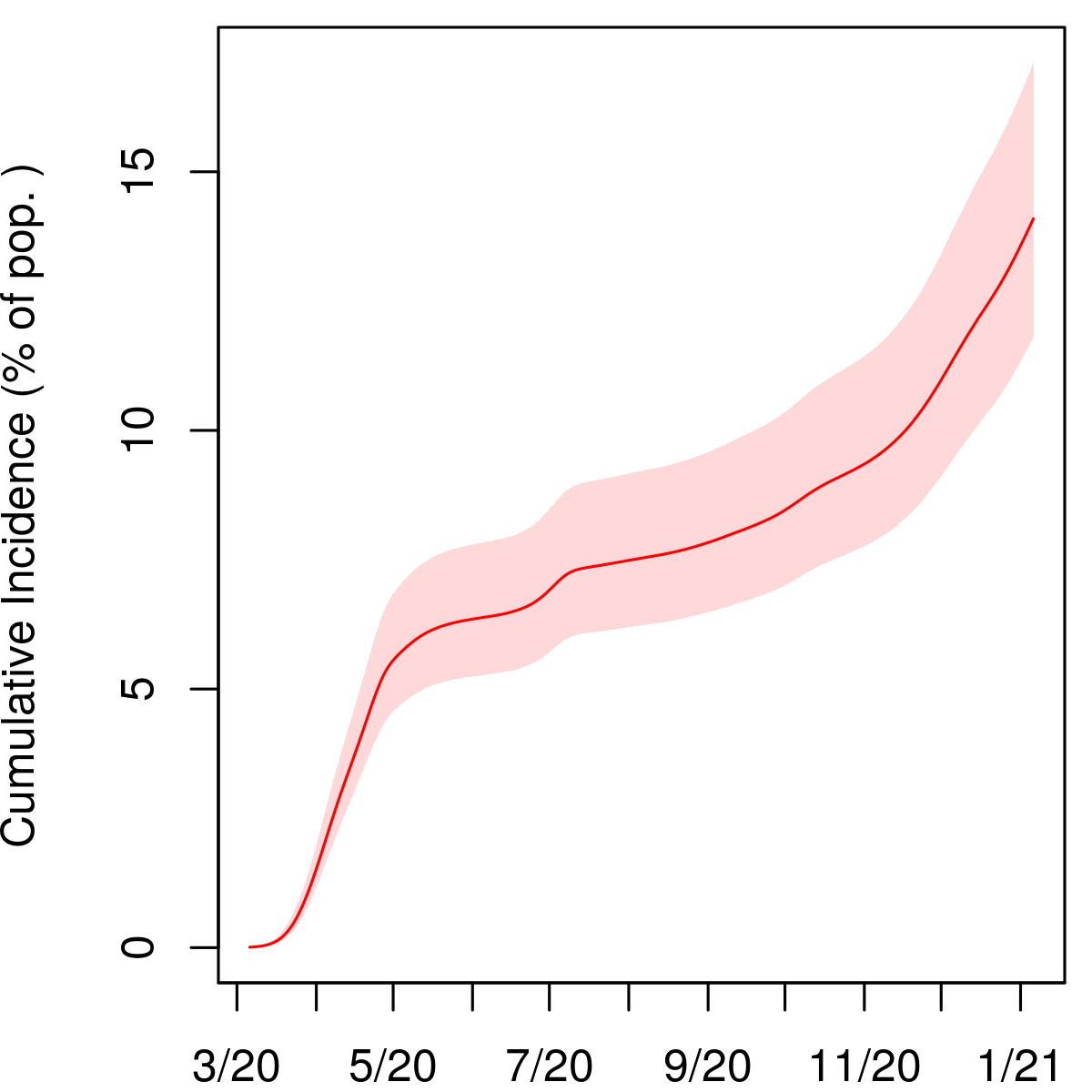} \\
\includegraphics[scale=0.77]{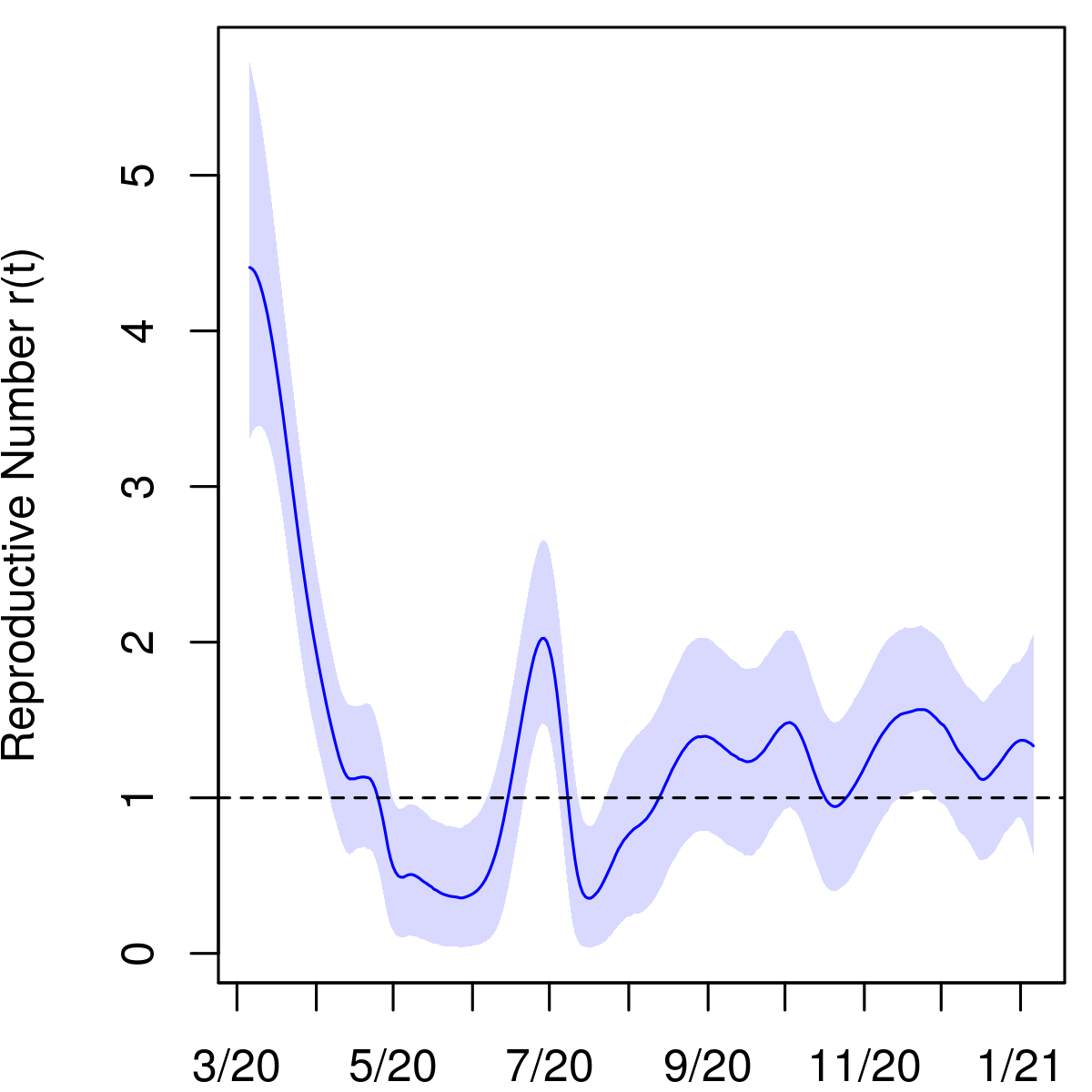}
&
\includegraphics[scale=0.77]{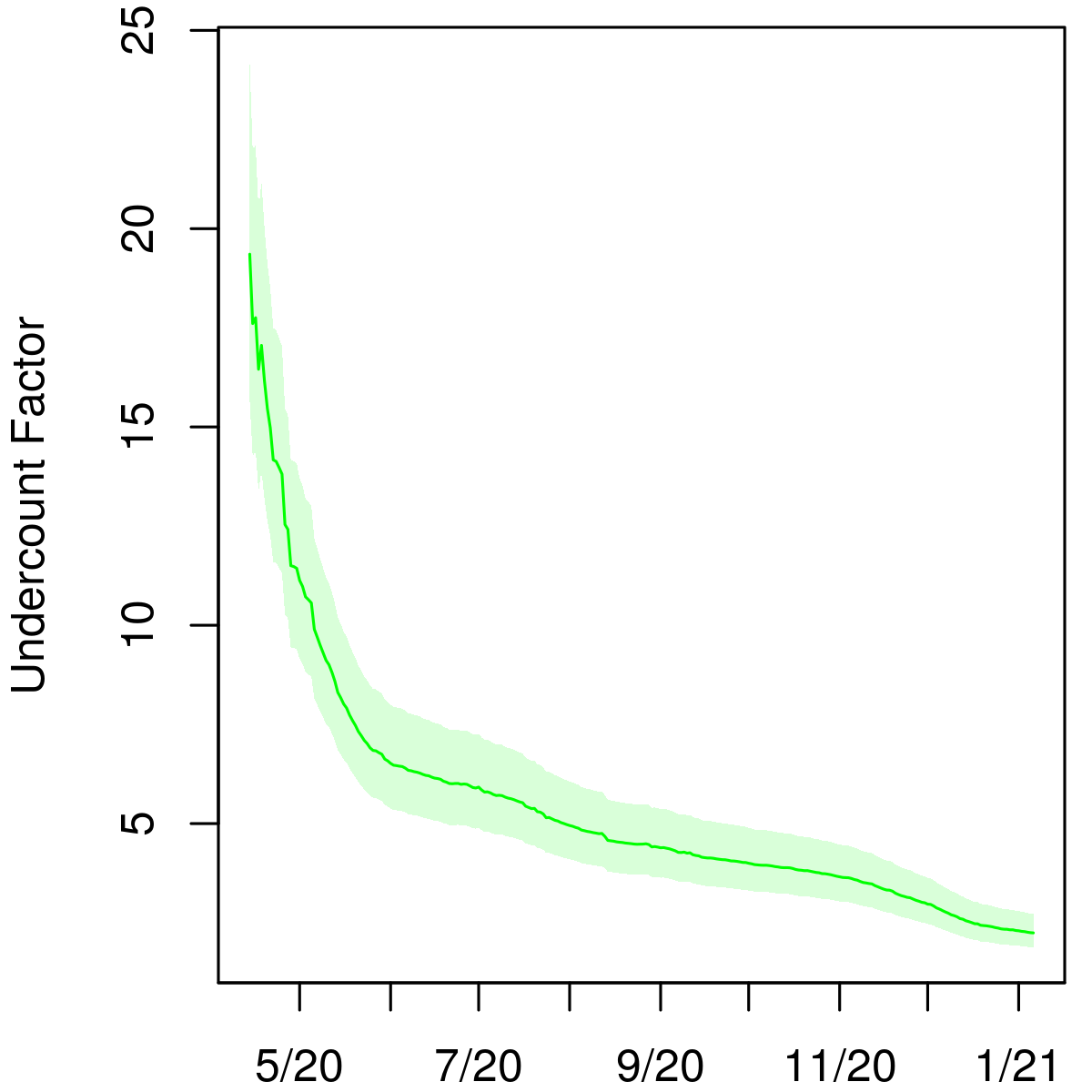} 
\end{tabular}
\caption{Posterior median and middle 95\% intervals for daily new infections, cumulative incidence, $r(t)$, and cumulative undercount from March 2020 to January 2021. In the top left panel, deaths divided by the posterior median IFR are plotted in grey for comparison.}
\end{figure}
\newpage
\begin{figure}[htbp!]
\textbf{Florida}
\centering
\begin{tabular}{ll}
\includegraphics[scale=0.77]{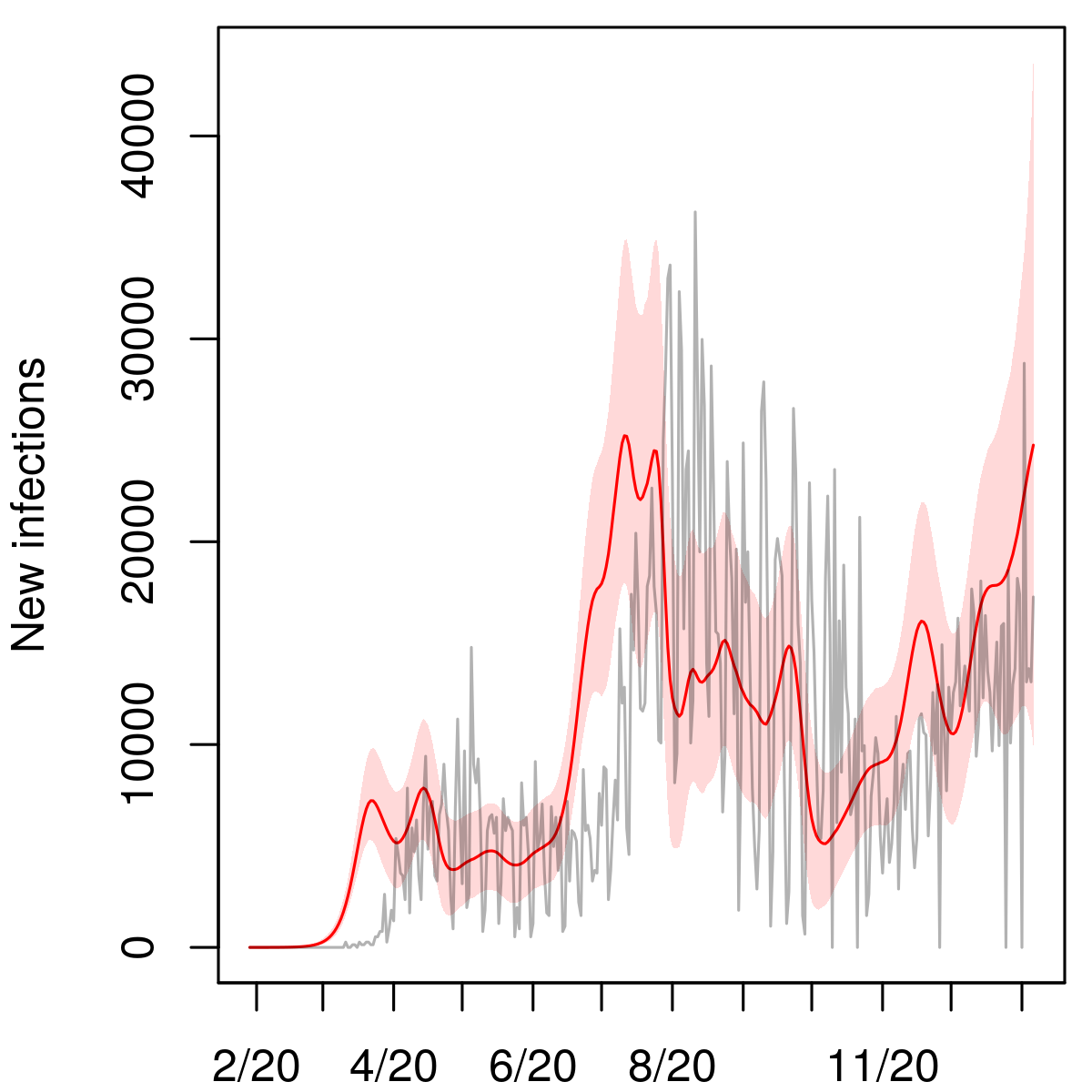}
&
\includegraphics[scale=0.77]{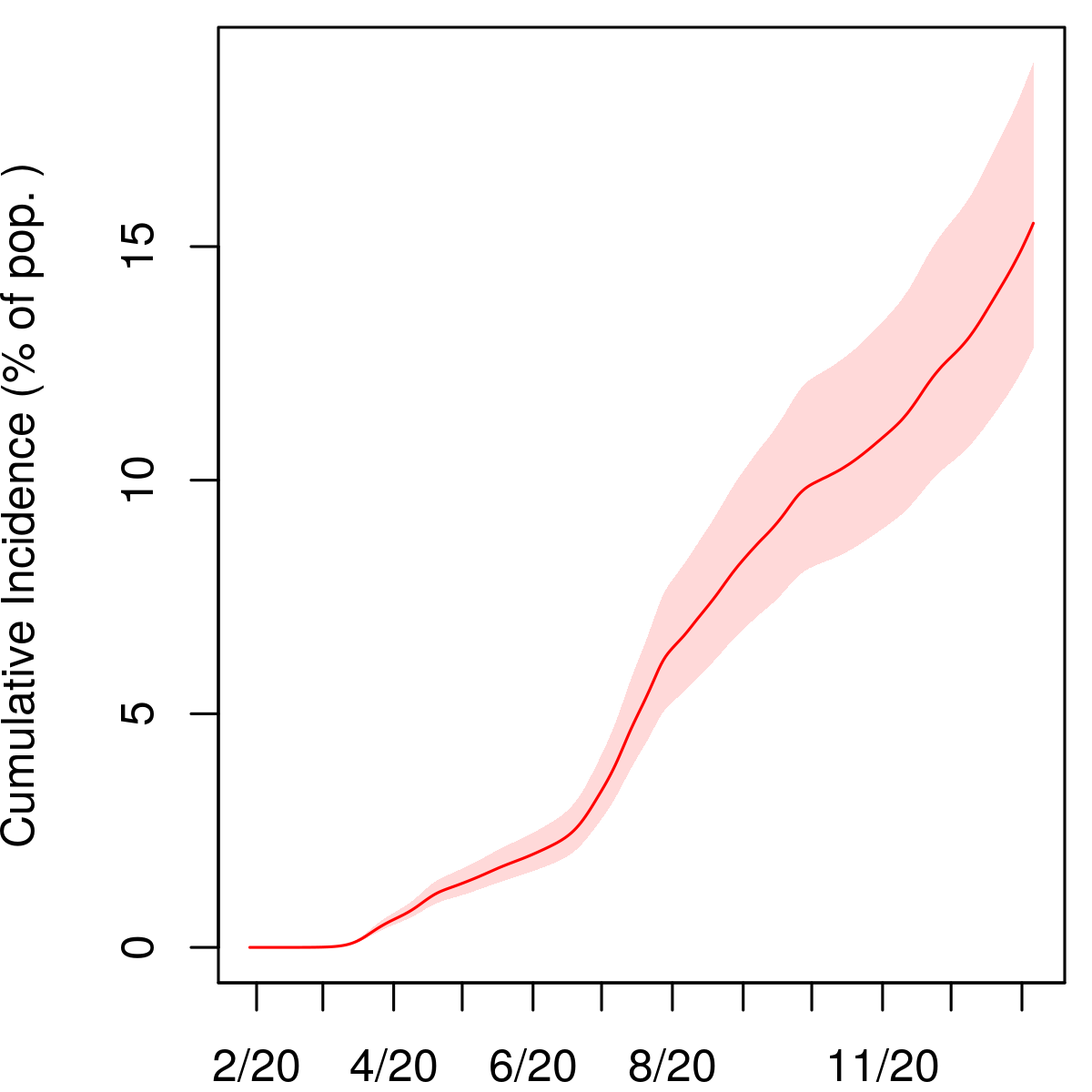} \\
\includegraphics[scale=0.77]{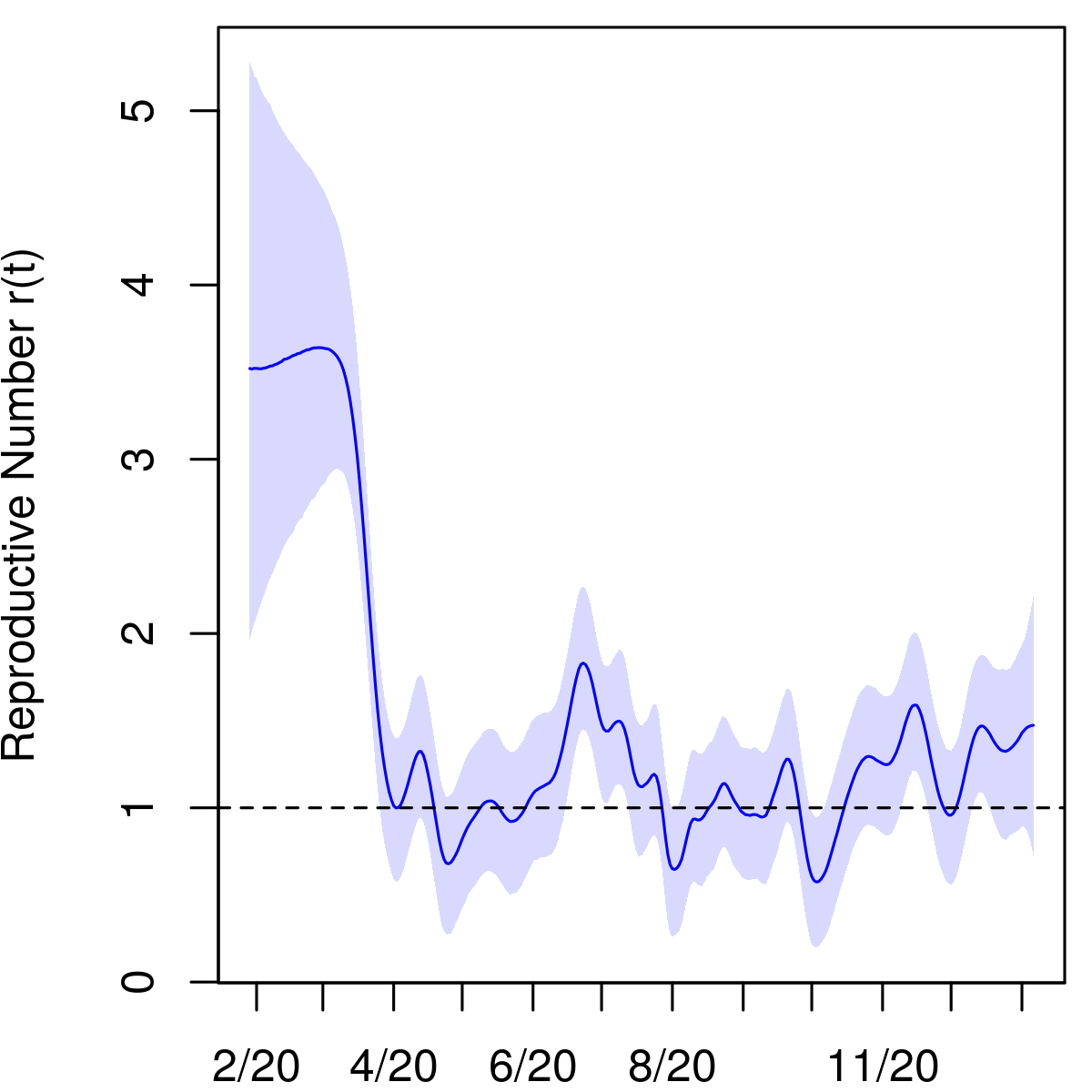}
&
\includegraphics[scale=0.77]{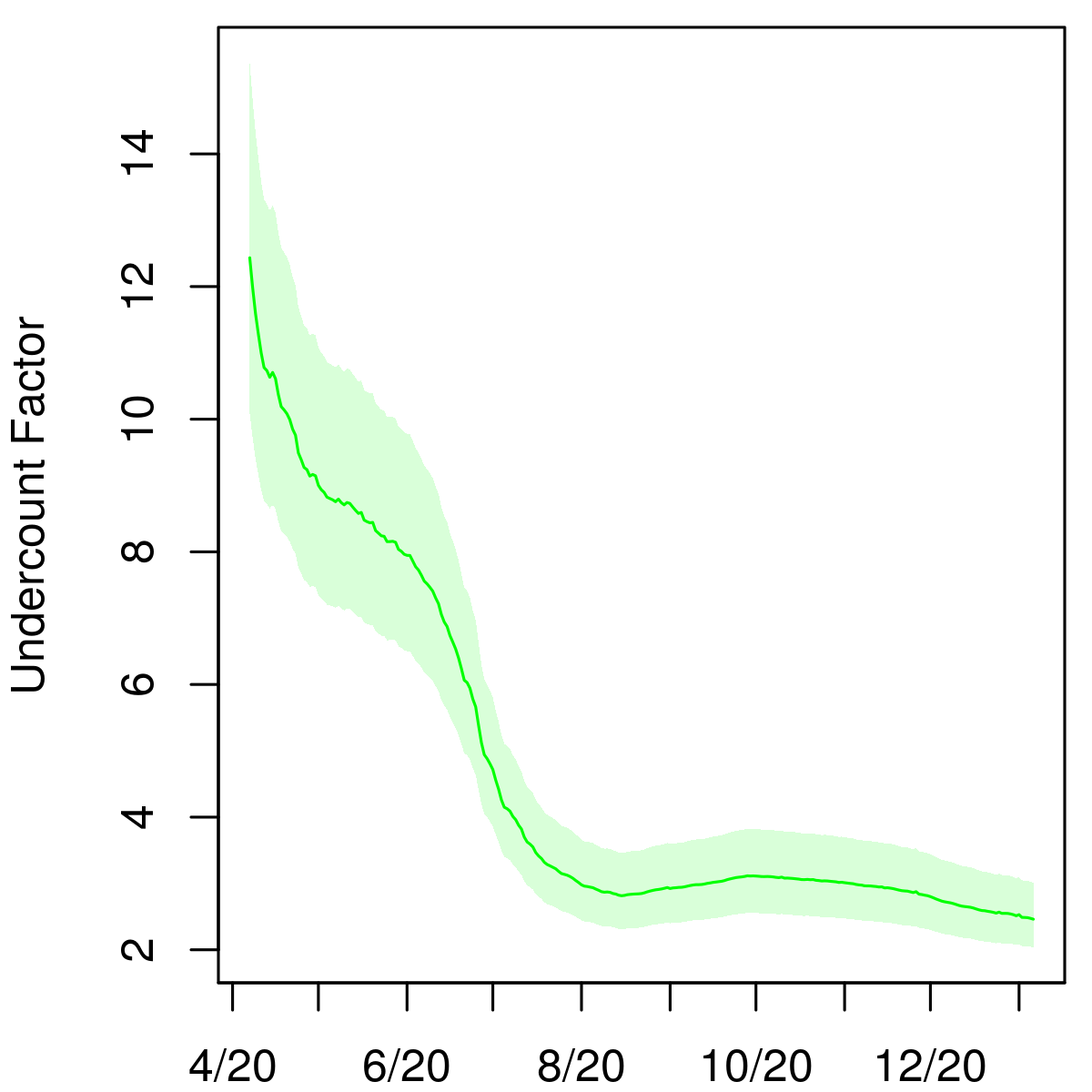} 
\end{tabular}
\caption{Posterior median and middle 95\% intervals for daily new infections, cumulative incidence, $r(t)$, and cumulative undercount from March 2020 to January 2021. In the top left panel, deaths divided by the posterior median IFR are plotted in grey for comparison.}
\end{figure}
\newpage
\begin{figure}[htbp!]
\textbf{Georgia}
\centering
\begin{tabular}{ll}
\includegraphics[scale=0.77]{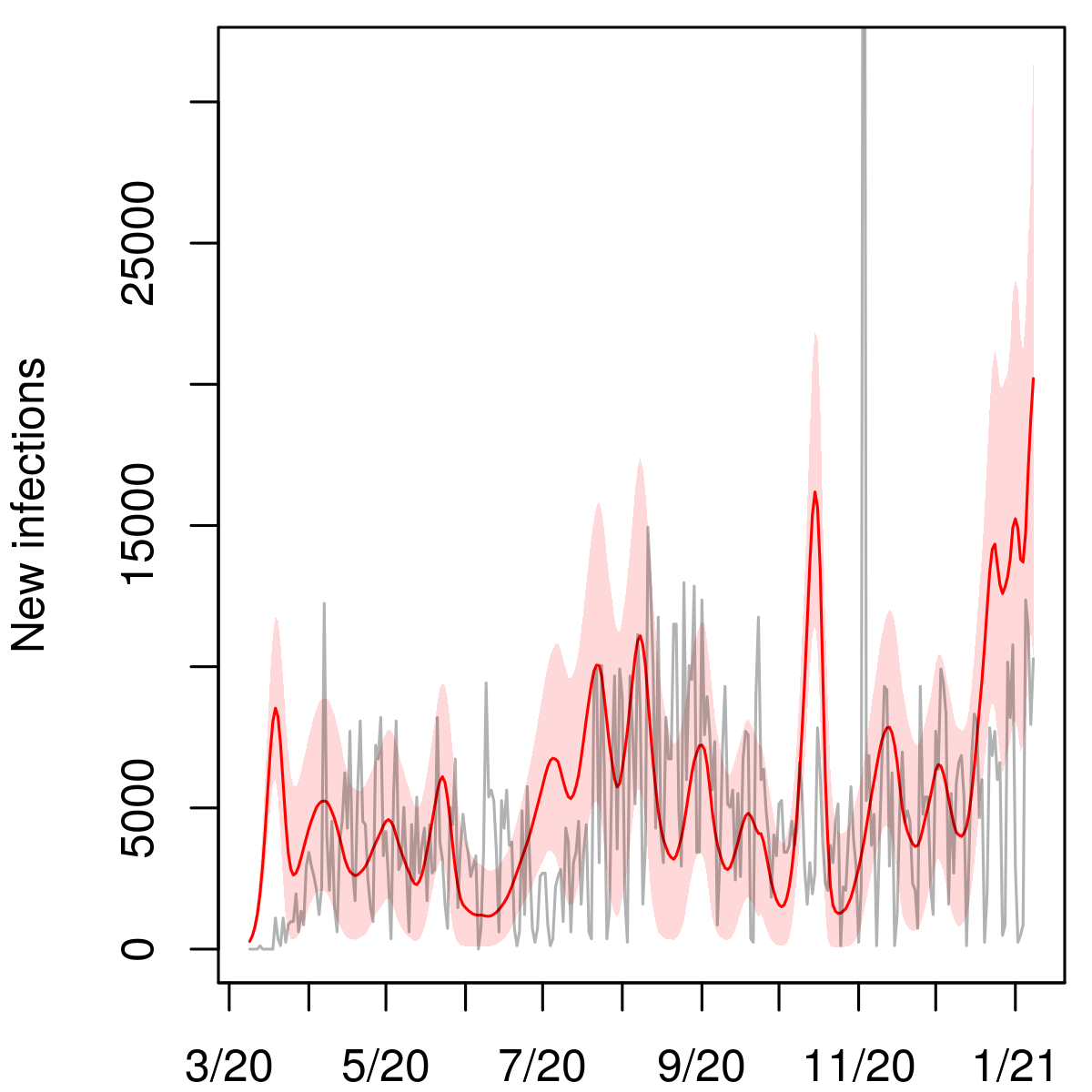}
&
\includegraphics[scale=0.77]{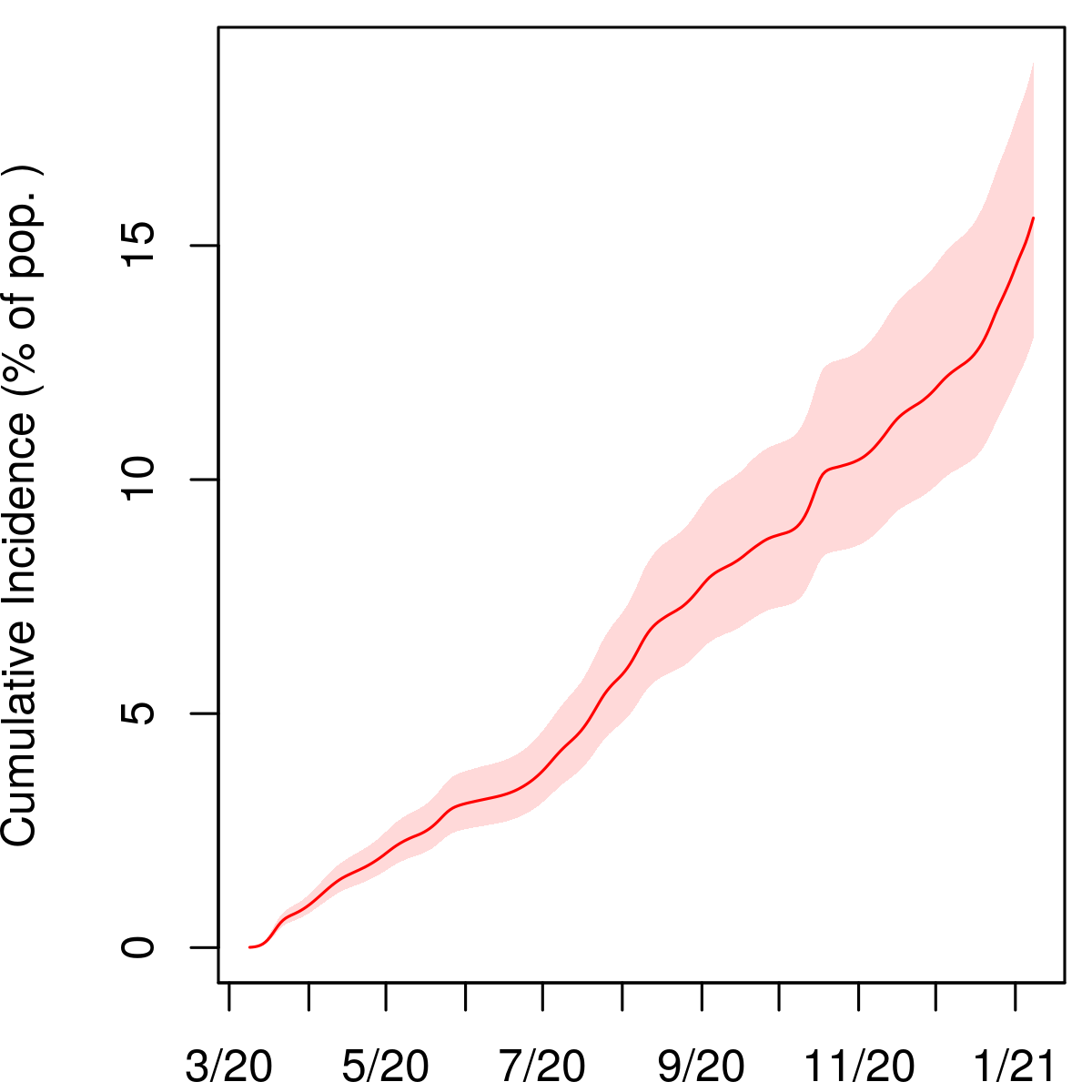} \\
\includegraphics[scale=0.77]{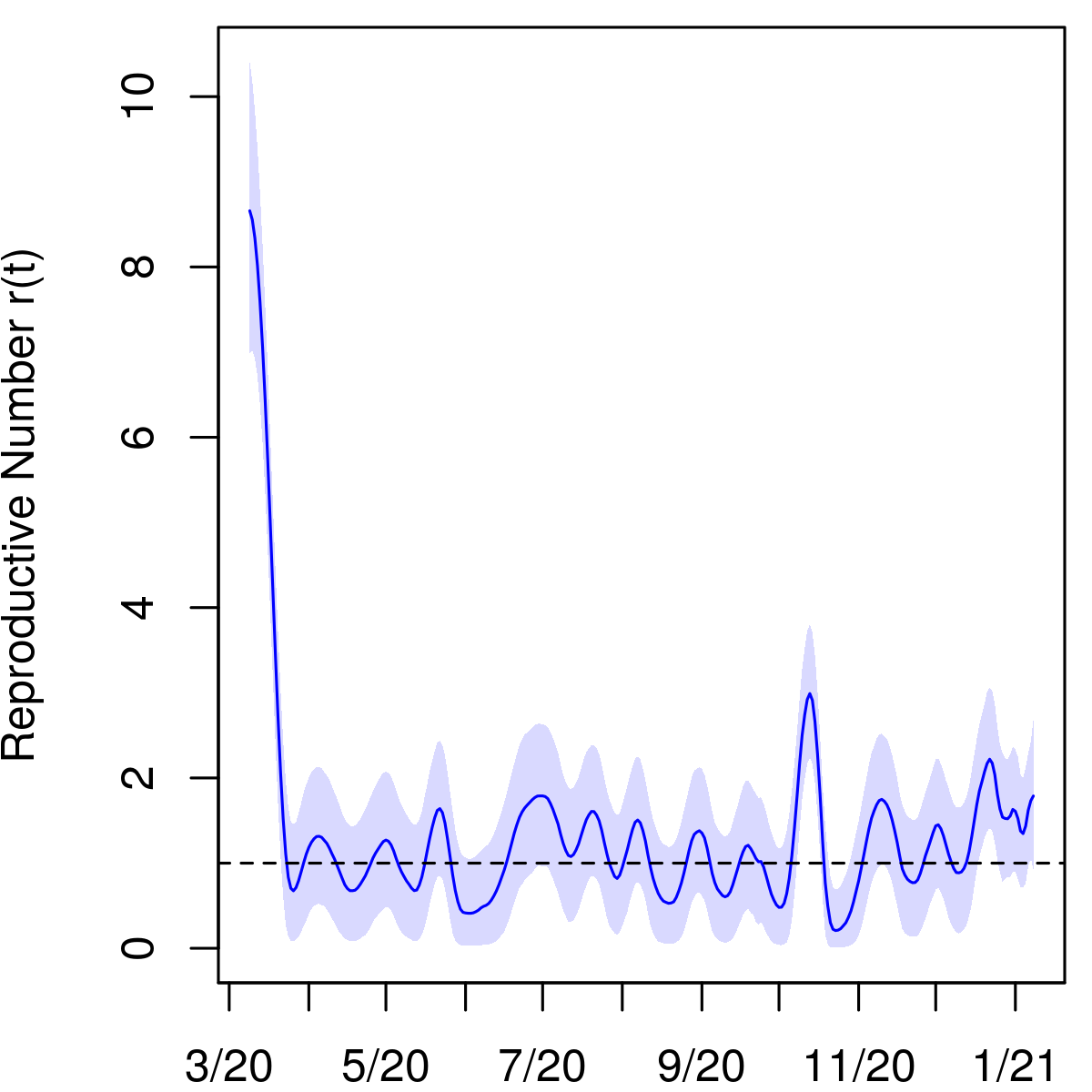}
&
\includegraphics[scale=0.77]{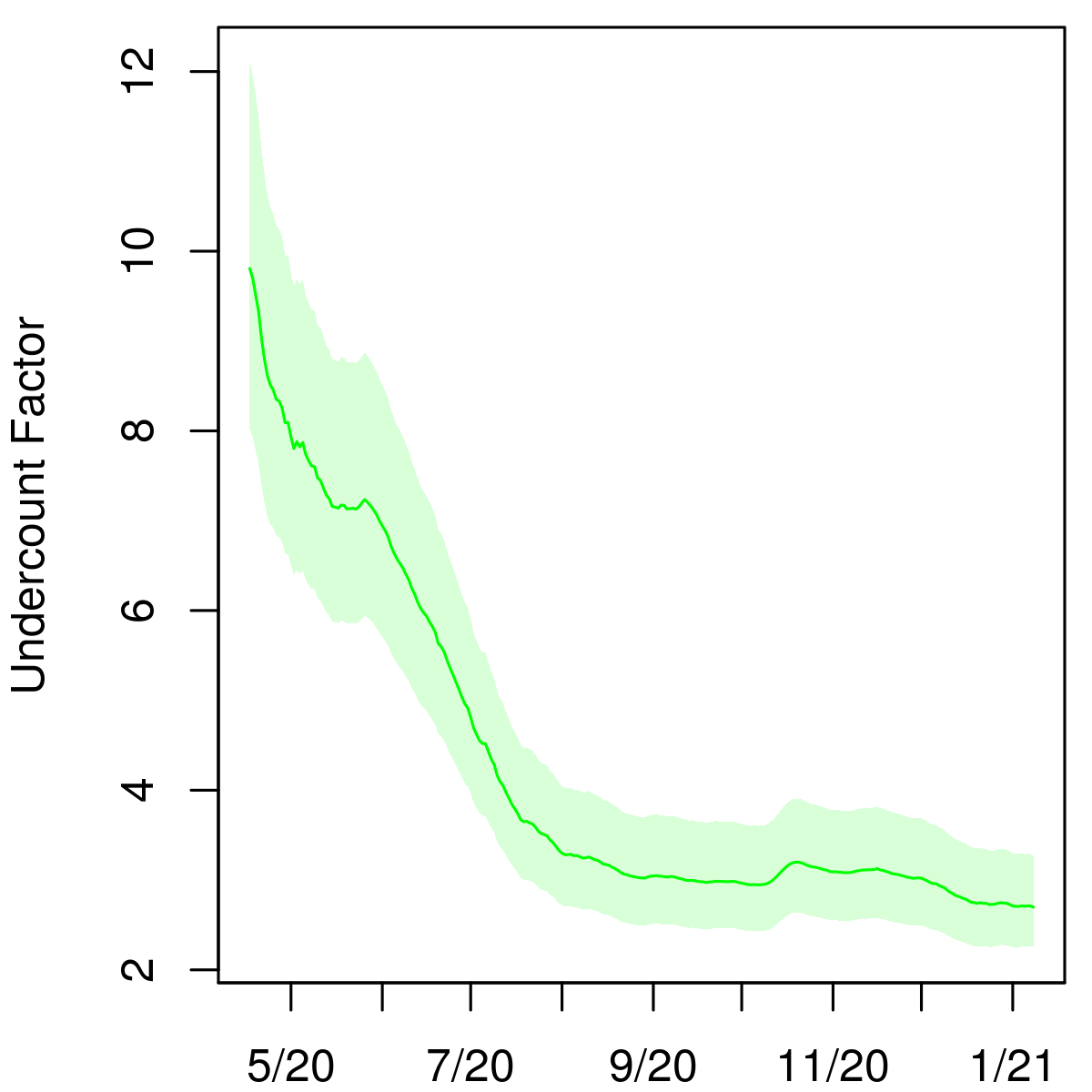} 
\end{tabular}
\caption{Posterior median and middle 95\% intervals for daily new infections, cumulative incidence, $r(t)$, and cumulative undercount from March 2020 to January 2021. In the top left panel, deaths divided by the posterior median IFR are plotted in grey for comparison.}
\end{figure}
\newpage
\begin{figure}[htbp!]
\textbf{Hawaii}
\centering
\begin{tabular}{ll}
\includegraphics[scale=0.77]{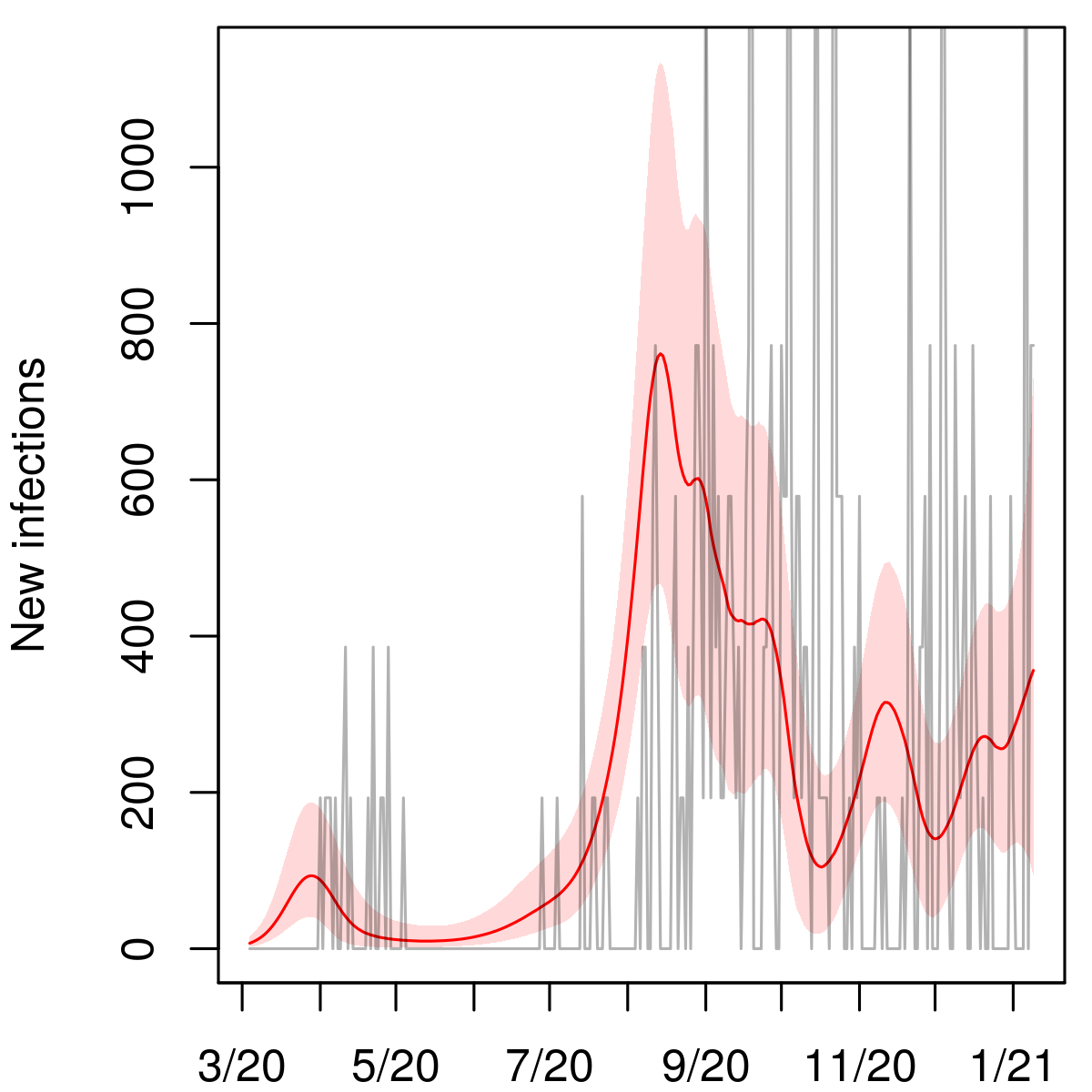}
&
\includegraphics[scale=0.77]{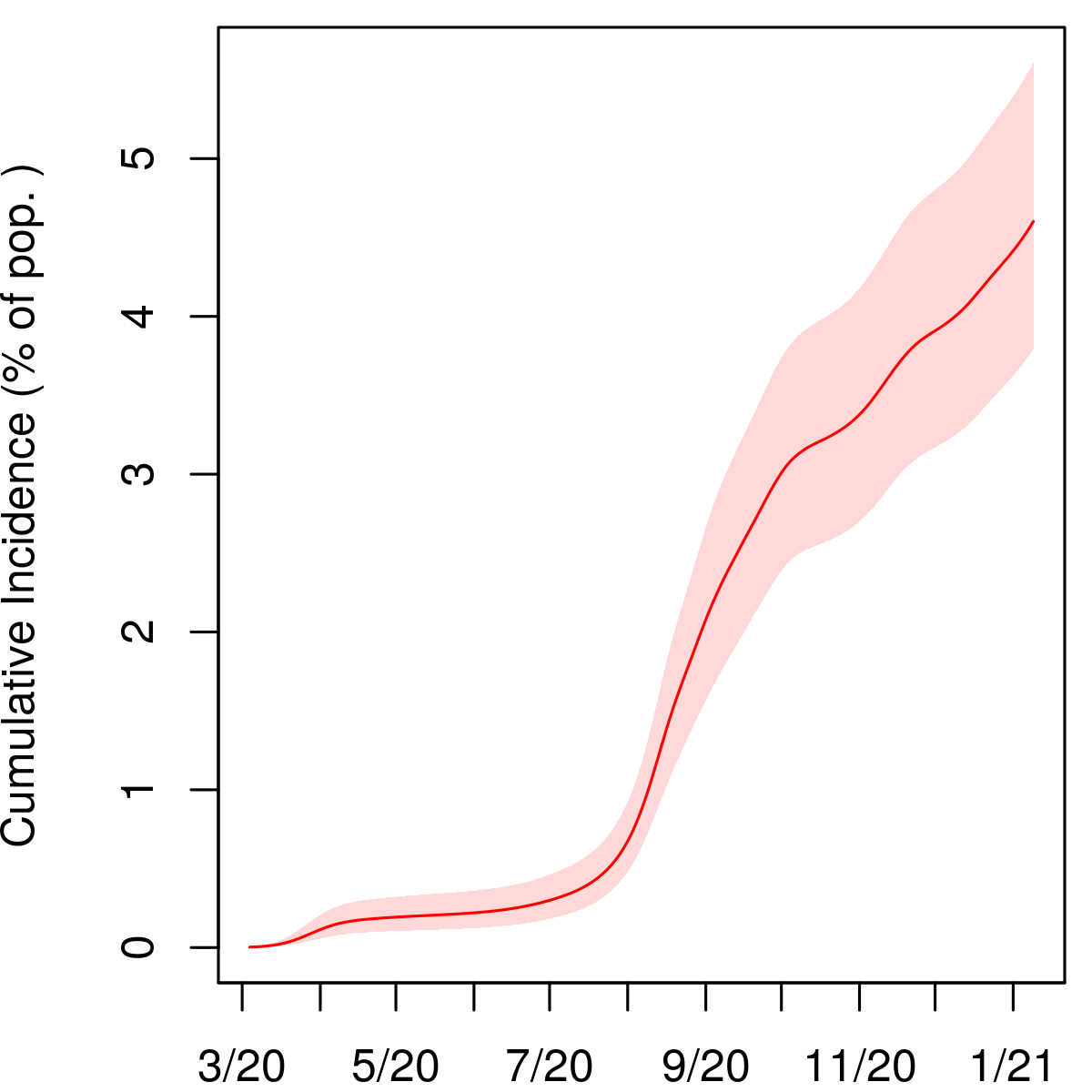} \\
\includegraphics[scale=0.77]{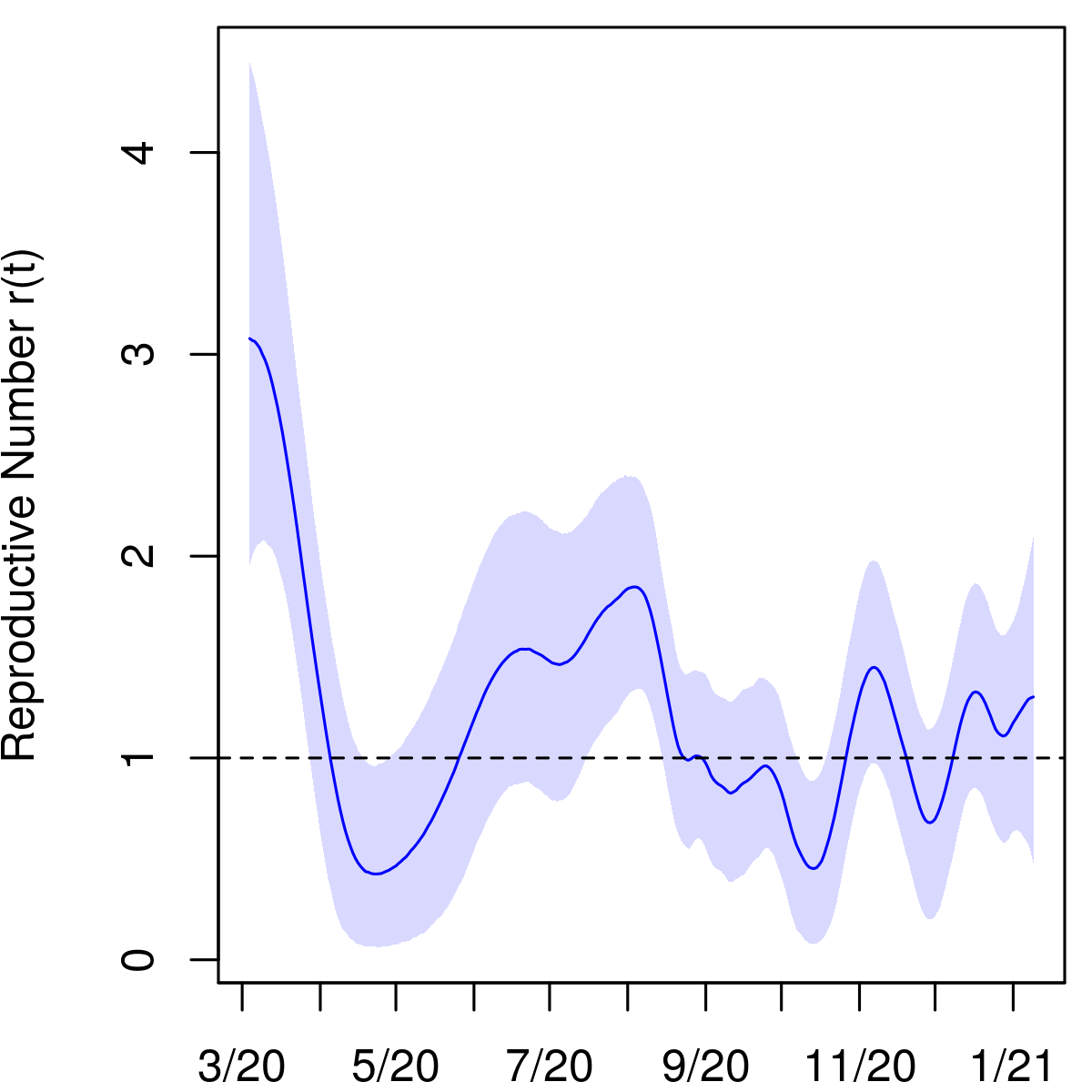}
&
\includegraphics[scale=0.77]{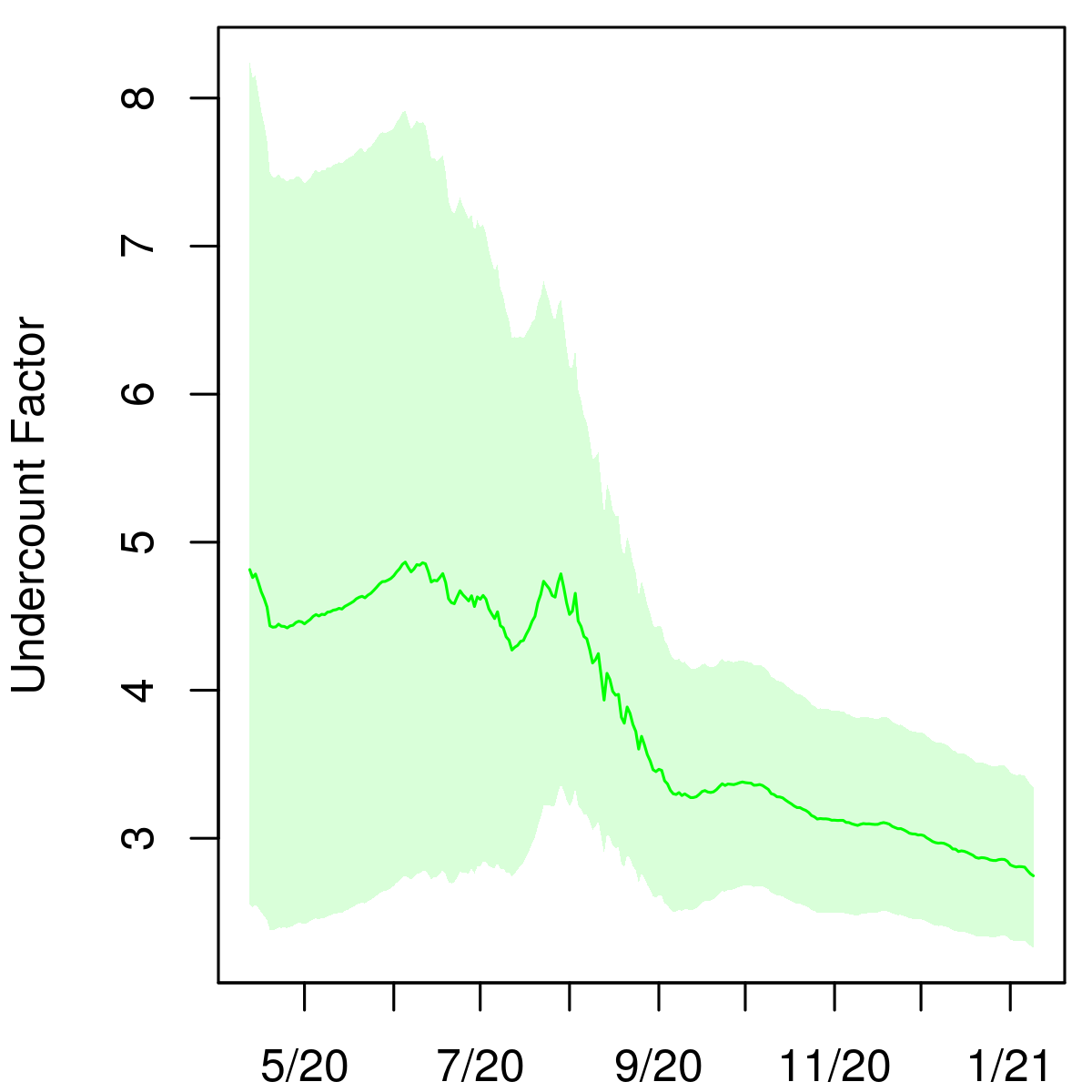} 
\end{tabular}
\caption{Posterior median and middle 95\% intervals for daily new infections, cumulative incidence, $r(t)$, and cumulative undercount from March 2020 to January 2021. In the top left panel, deaths divided by the posterior median IFR are plotted in grey for comparison.}
\end{figure}
\newpage
\begin{figure}[htbp!]
\textbf{Iowa}
\centering
\begin{tabular}{ll}
\includegraphics[scale=0.77]{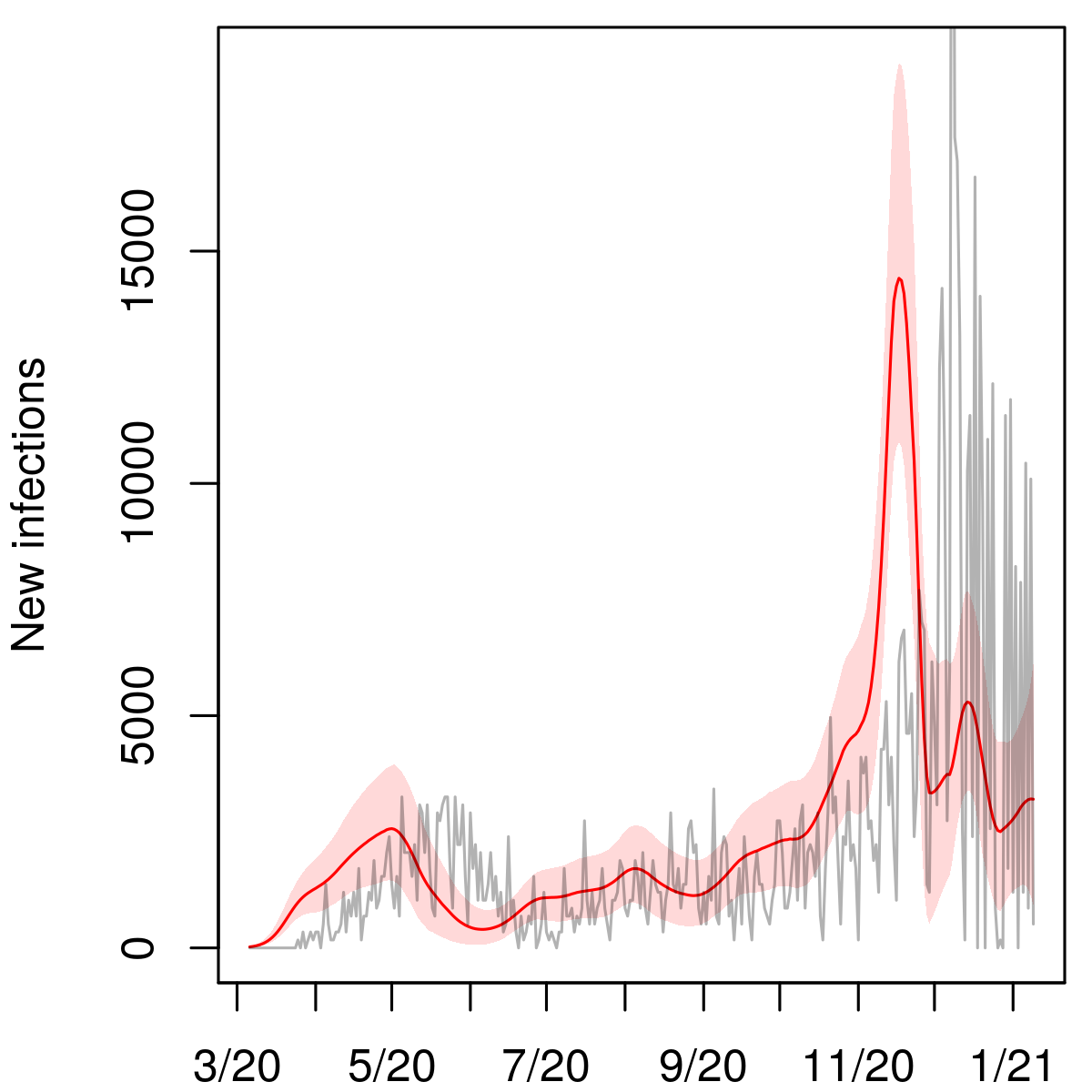}
&
\includegraphics[scale=0.77]{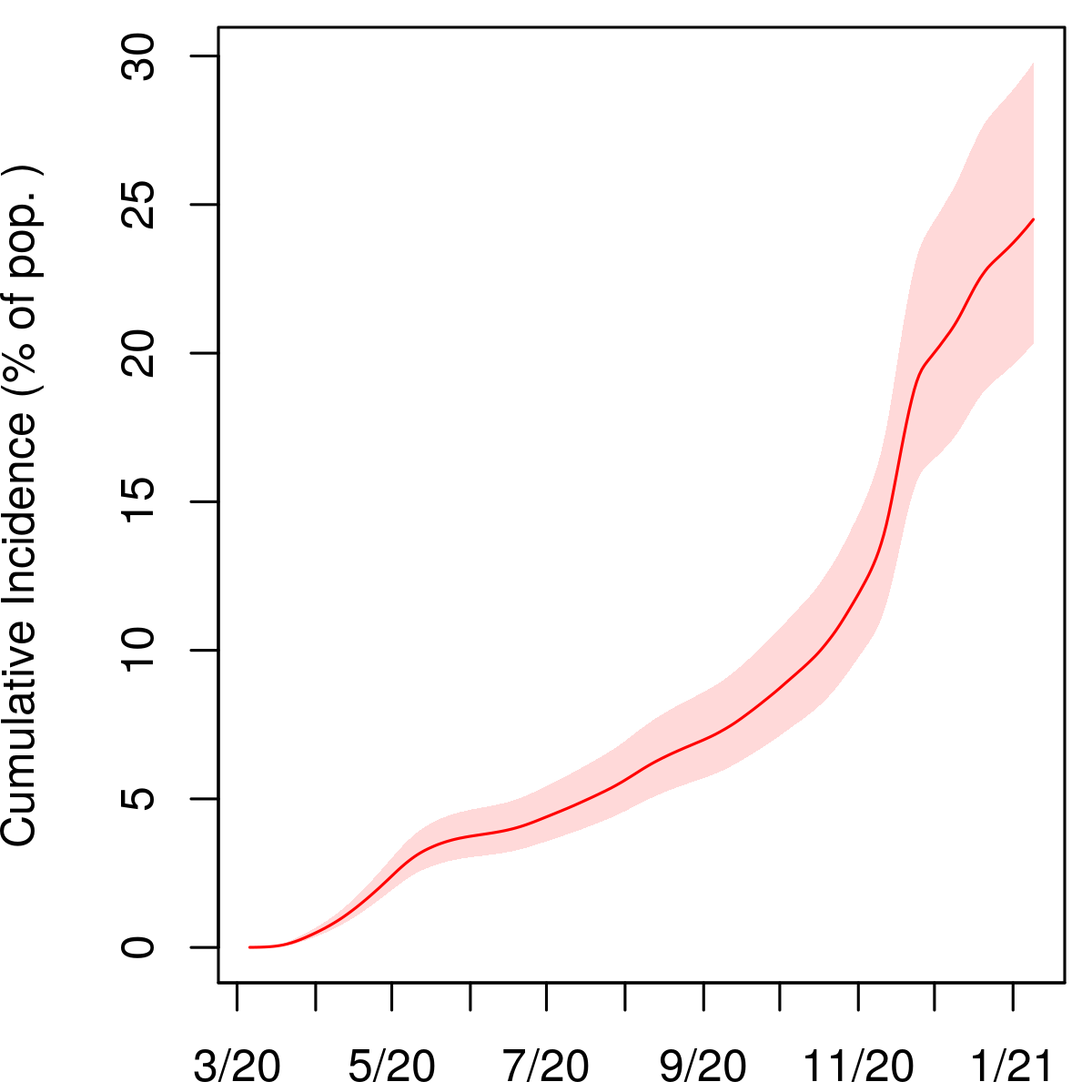} \\
\includegraphics[scale=0.77]{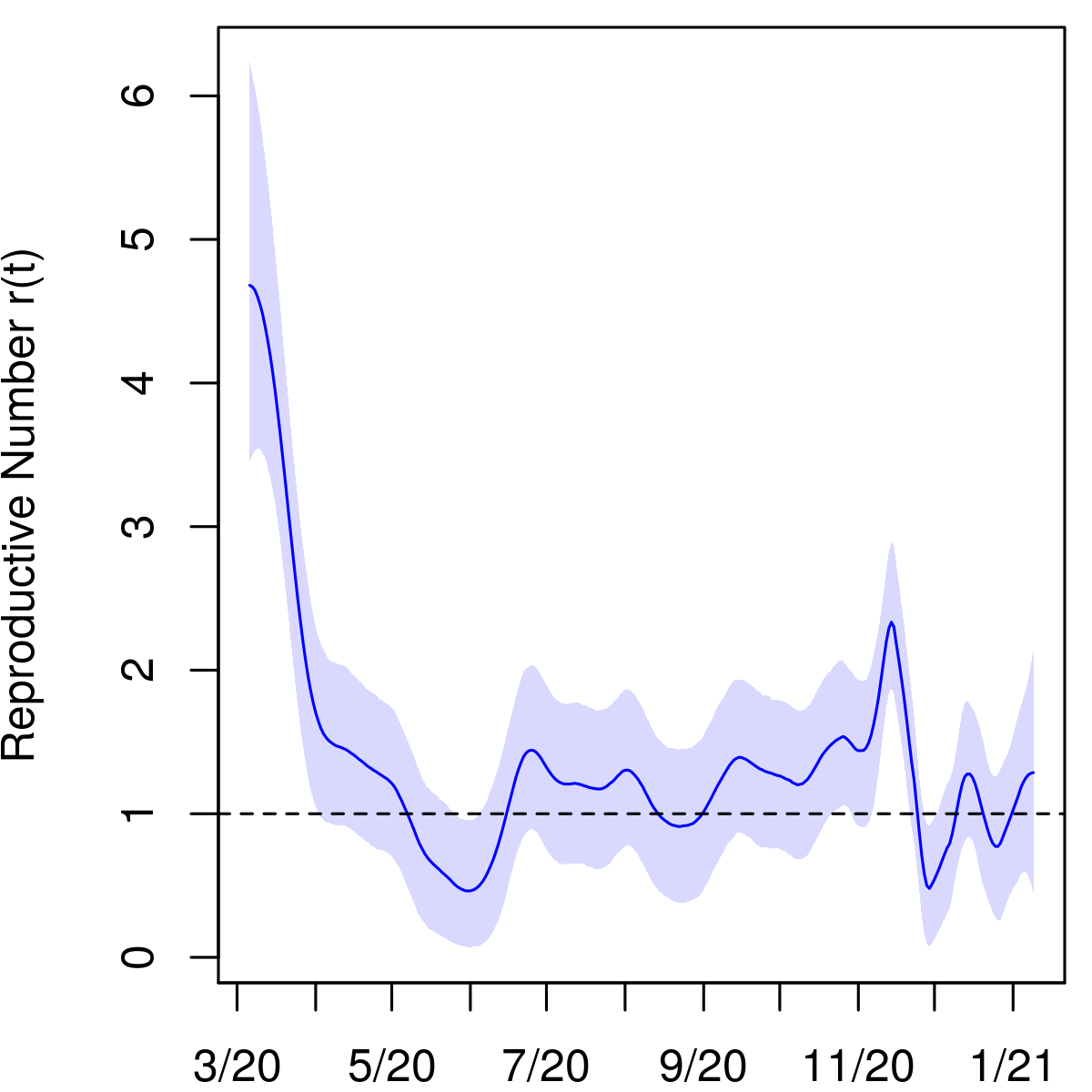}
&
\includegraphics[scale=0.77]{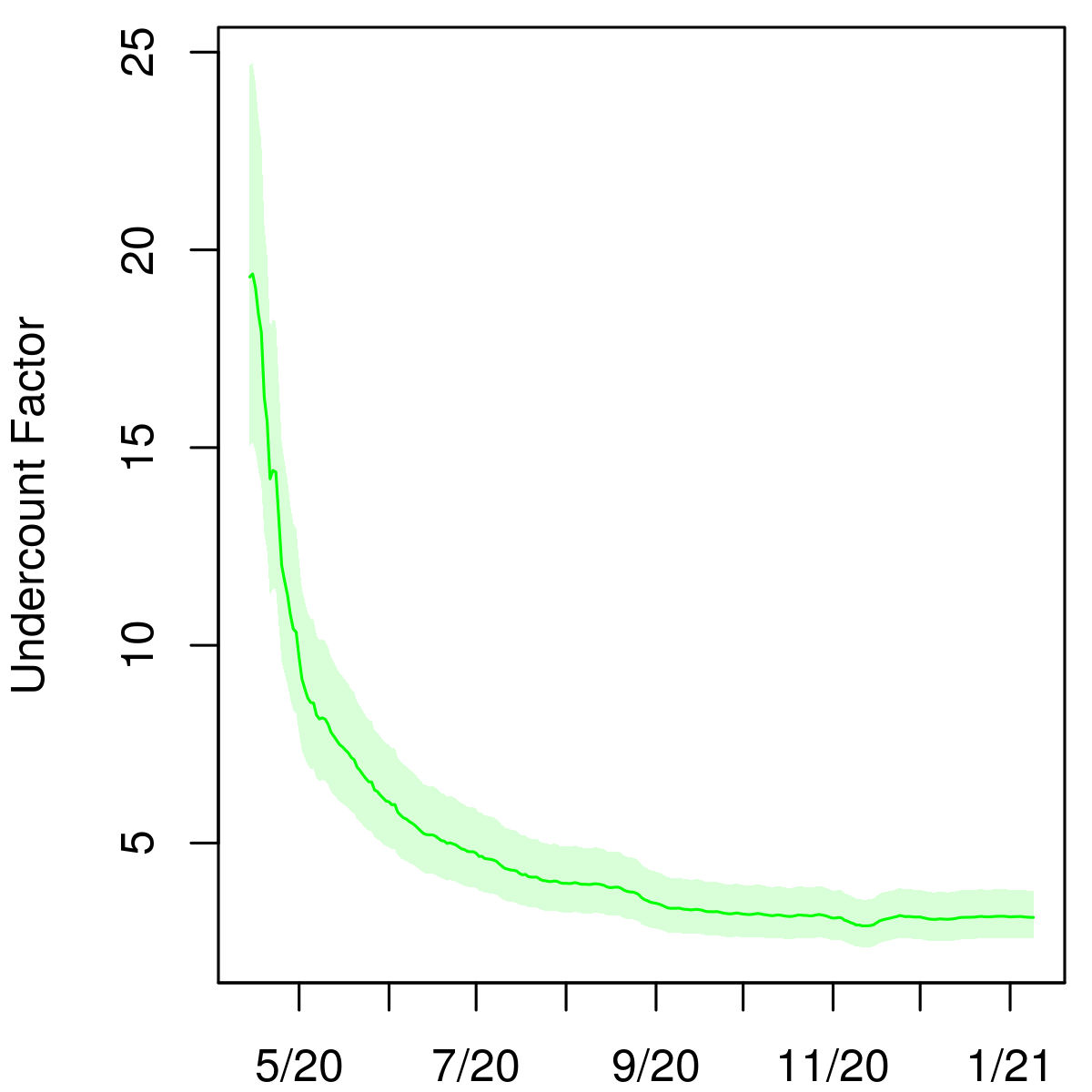} 
\end{tabular}
\caption{Posterior median and middle 95\% intervals for daily new infections, cumulative incidence, $r(t)$, and cumulative undercount from March 2020 to January 2021. In the top left panel, deaths divided by the posterior median IFR are plotted in grey for comparison.}
\end{figure}
\newpage
\begin{figure}[htbp!]
\textbf{Idaho}
\centering
\begin{tabular}{ll}
\includegraphics[scale=0.77]{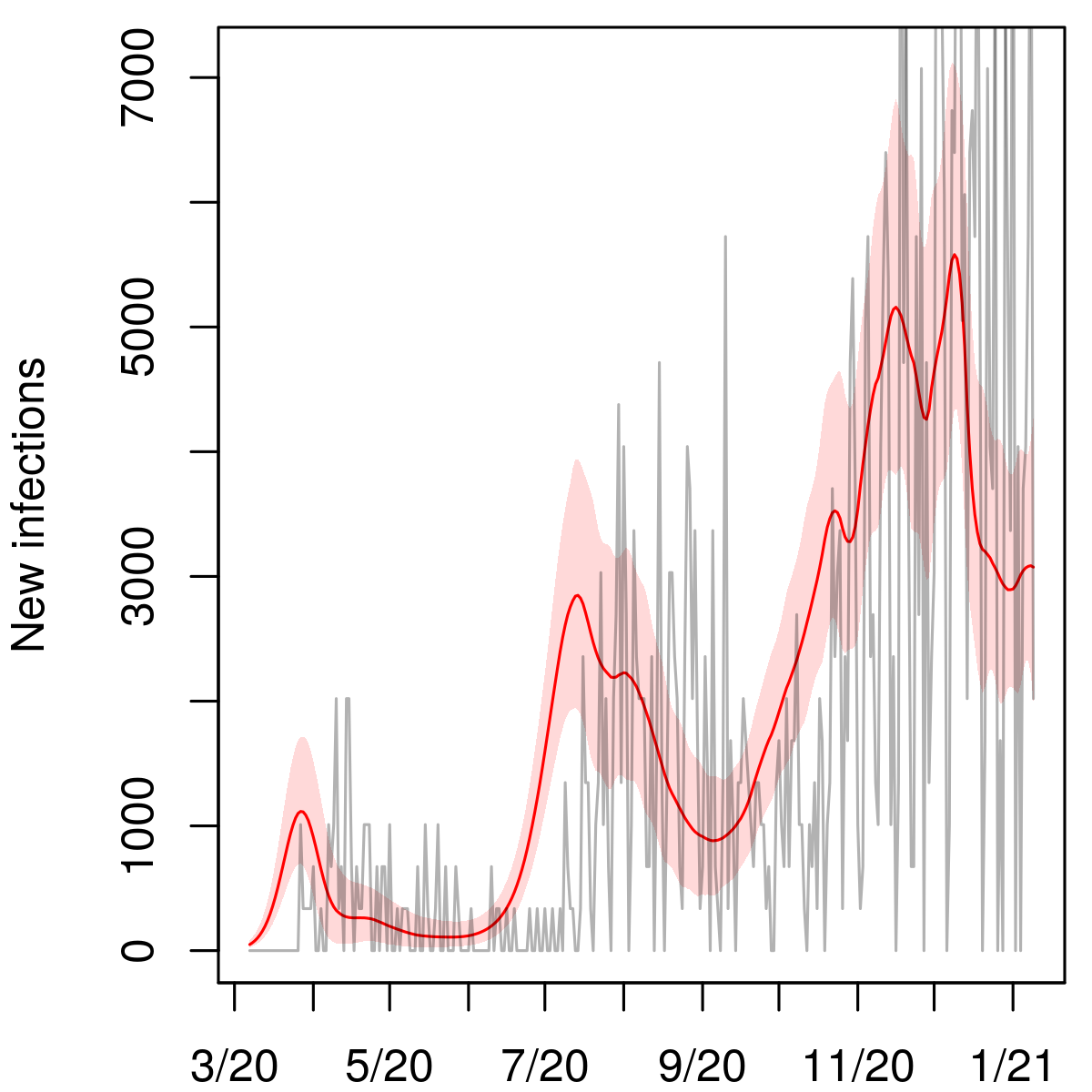}
&
\includegraphics[scale=0.77]{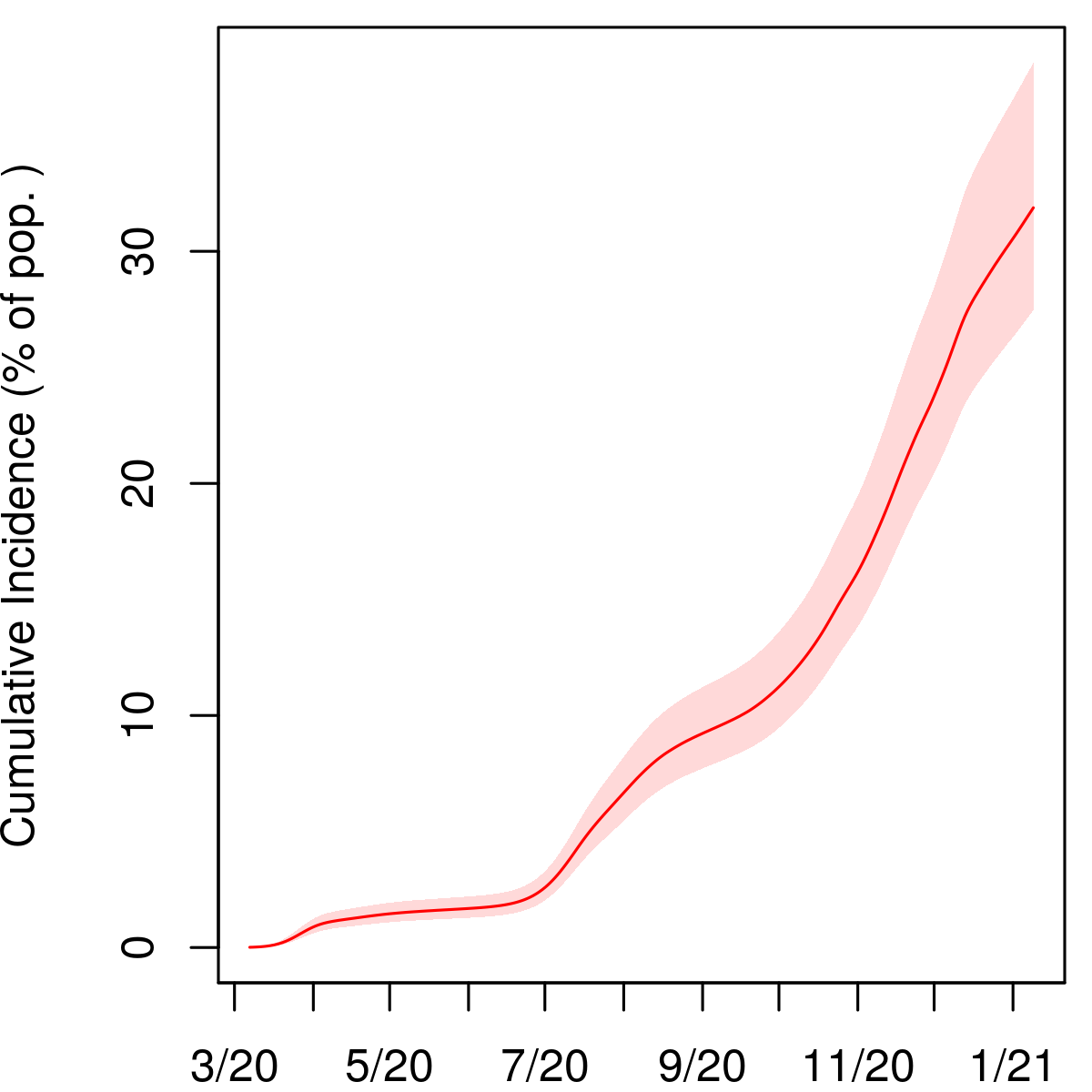} \\
\includegraphics[scale=0.77]{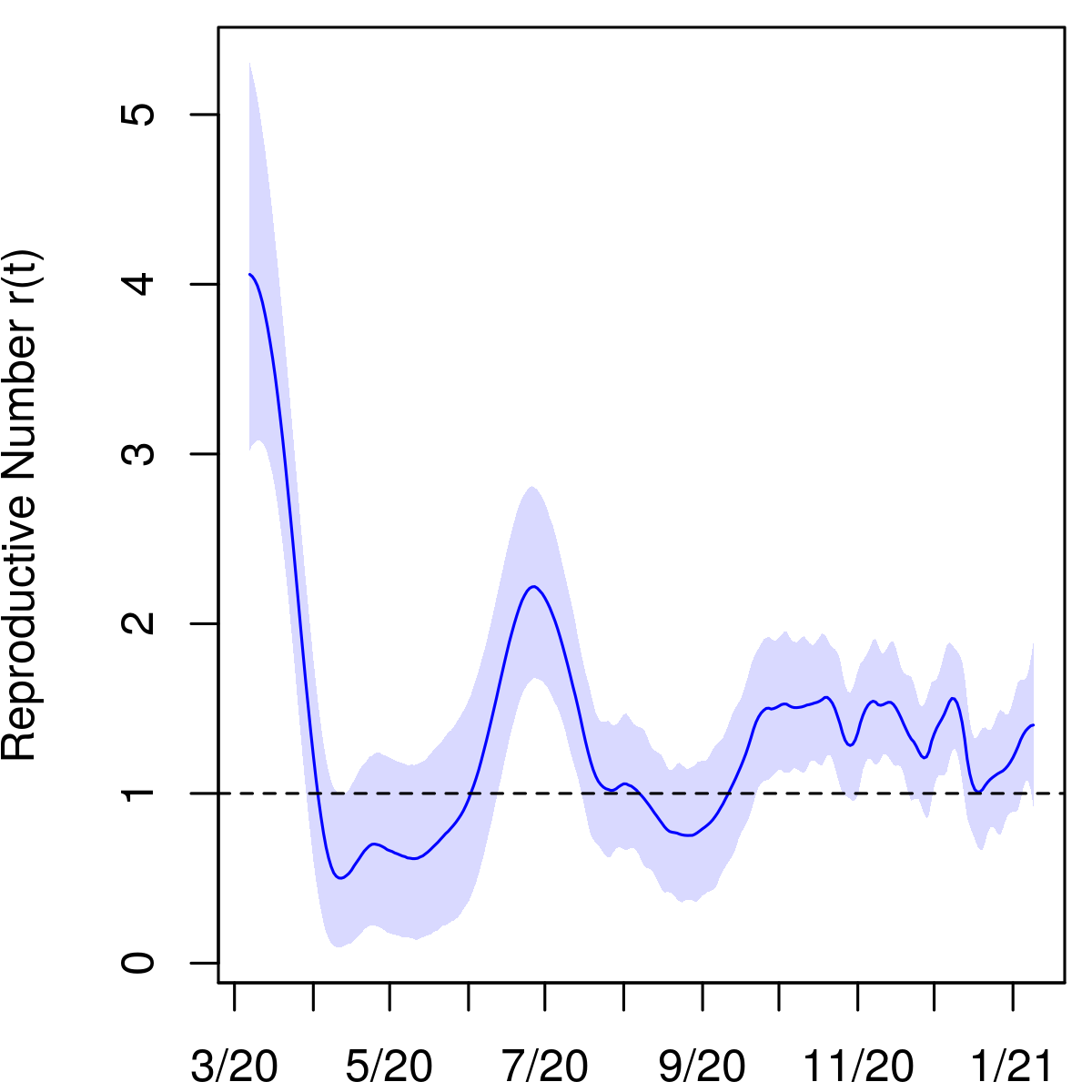}
&
\includegraphics[scale=0.77]{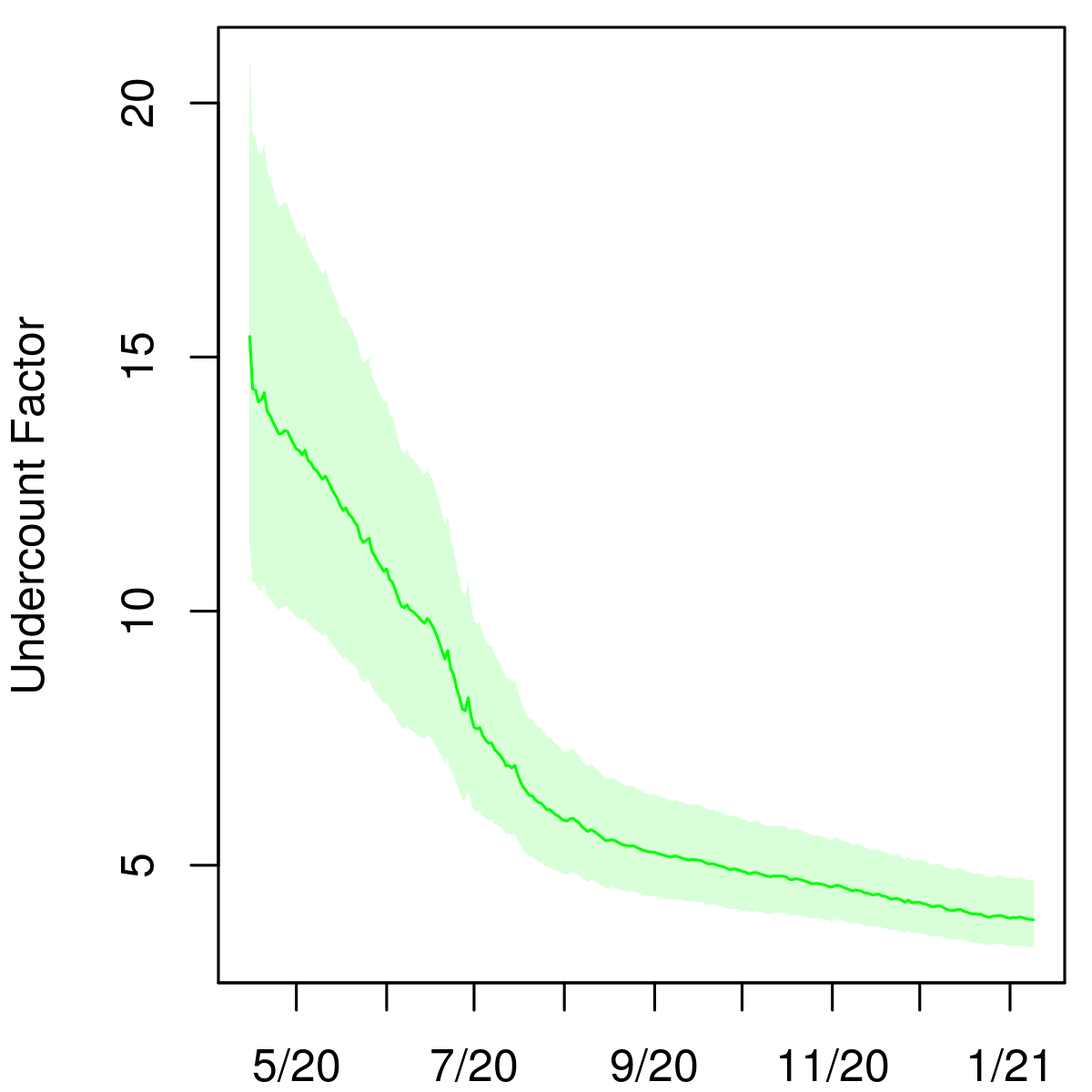} 
\end{tabular}
\caption{Posterior median and middle 95\% intervals for daily new infections, cumulative incidence, $r(t)$, and cumulative undercount from March 2020 to January 2021. In the top left panel, deaths divided by the posterior median IFR are plotted in grey for comparison.}
\end{figure}
\newpage
\begin{figure}[htbp!]
\textbf{Illinois}
\centering
\begin{tabular}{ll}
\includegraphics[scale=0.77]{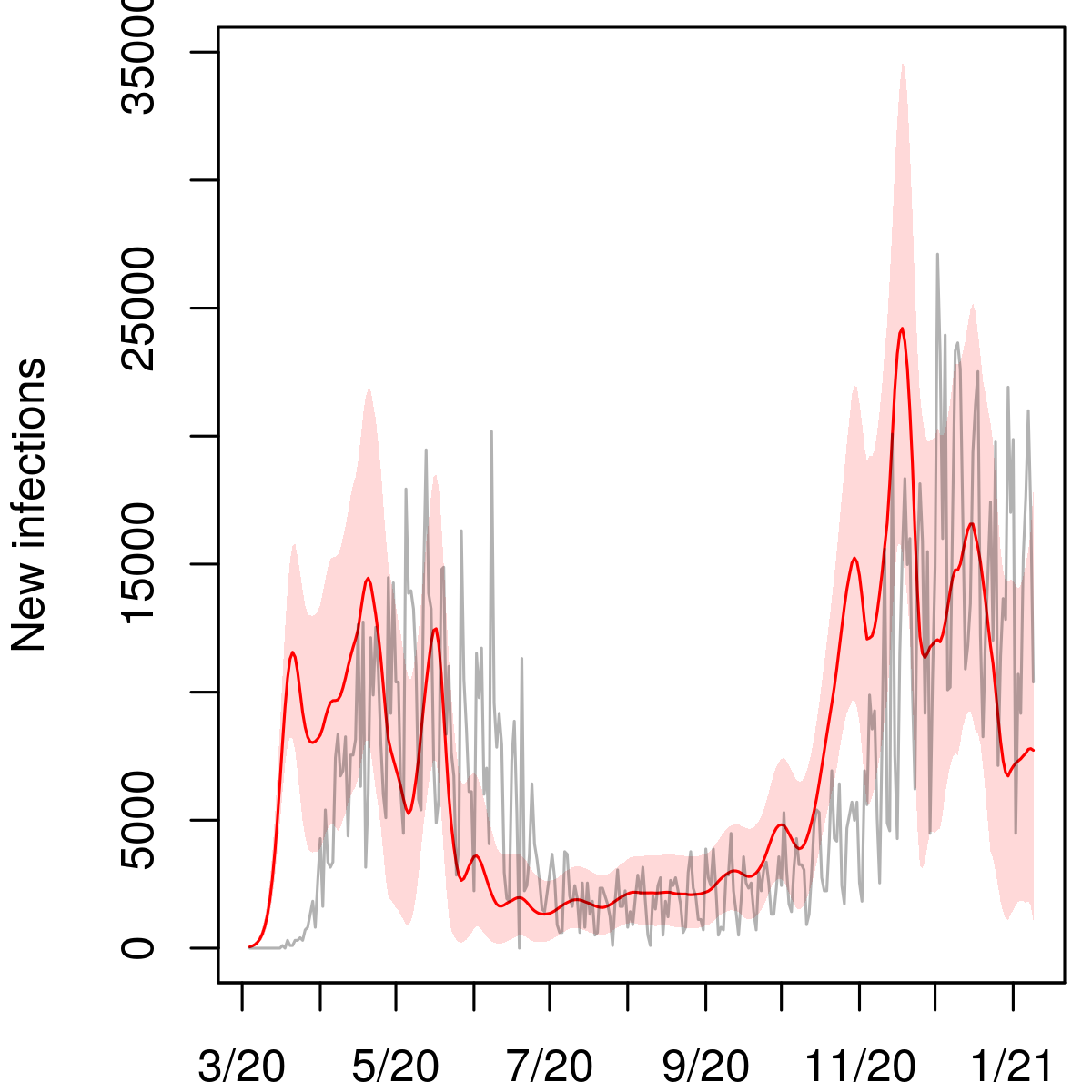}
&
\includegraphics[scale=0.77]{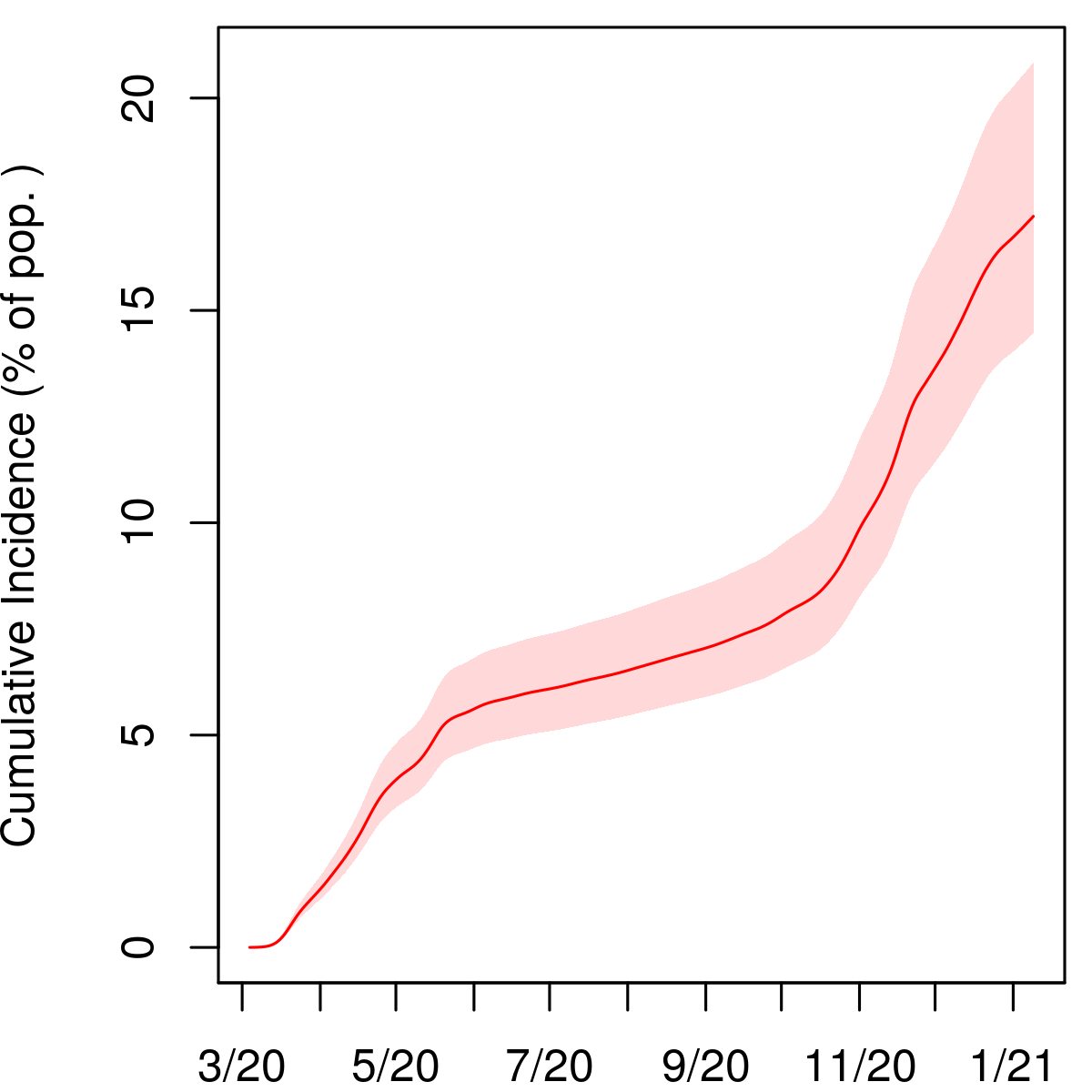} \\
\includegraphics[scale=0.77]{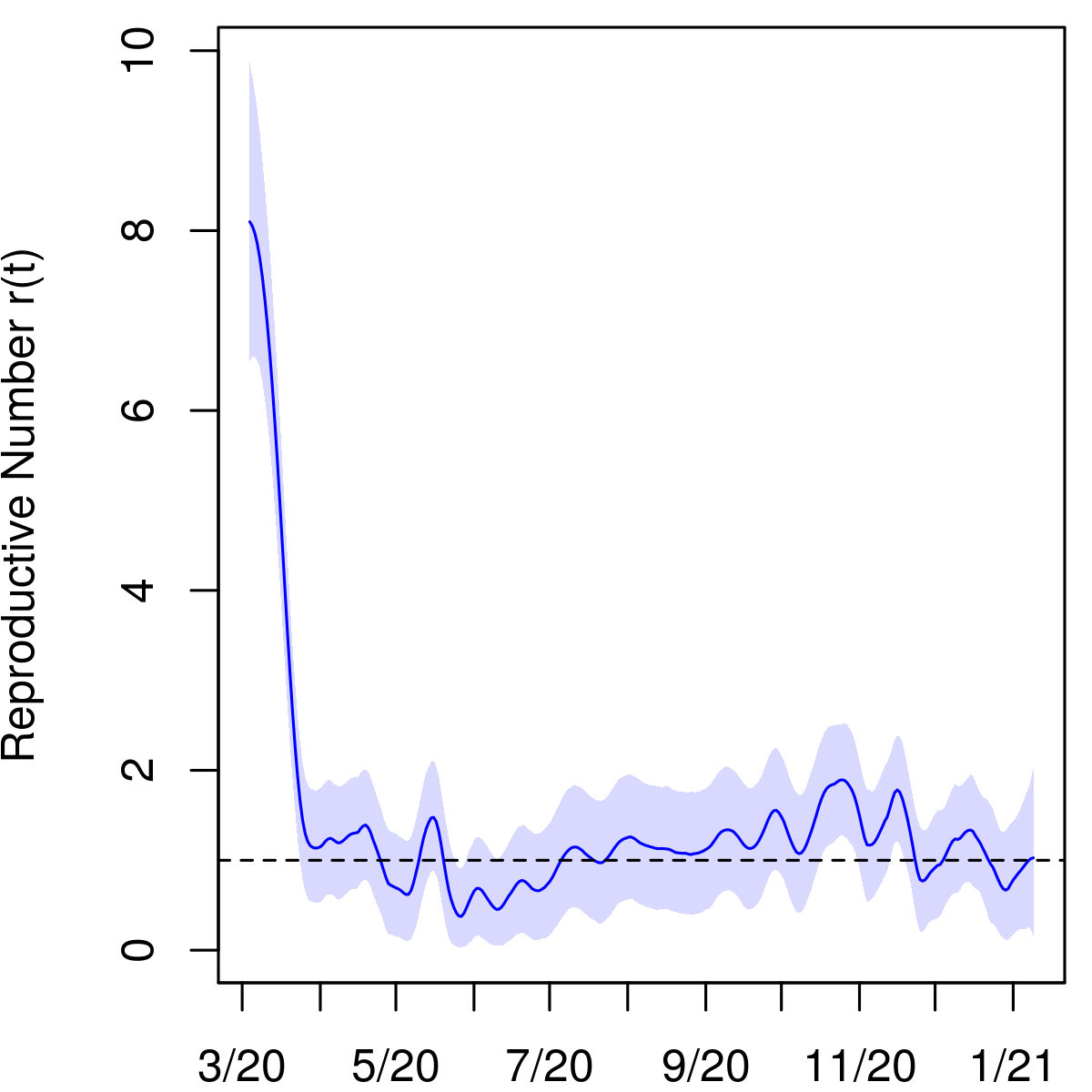}
&
\includegraphics[scale=0.77]{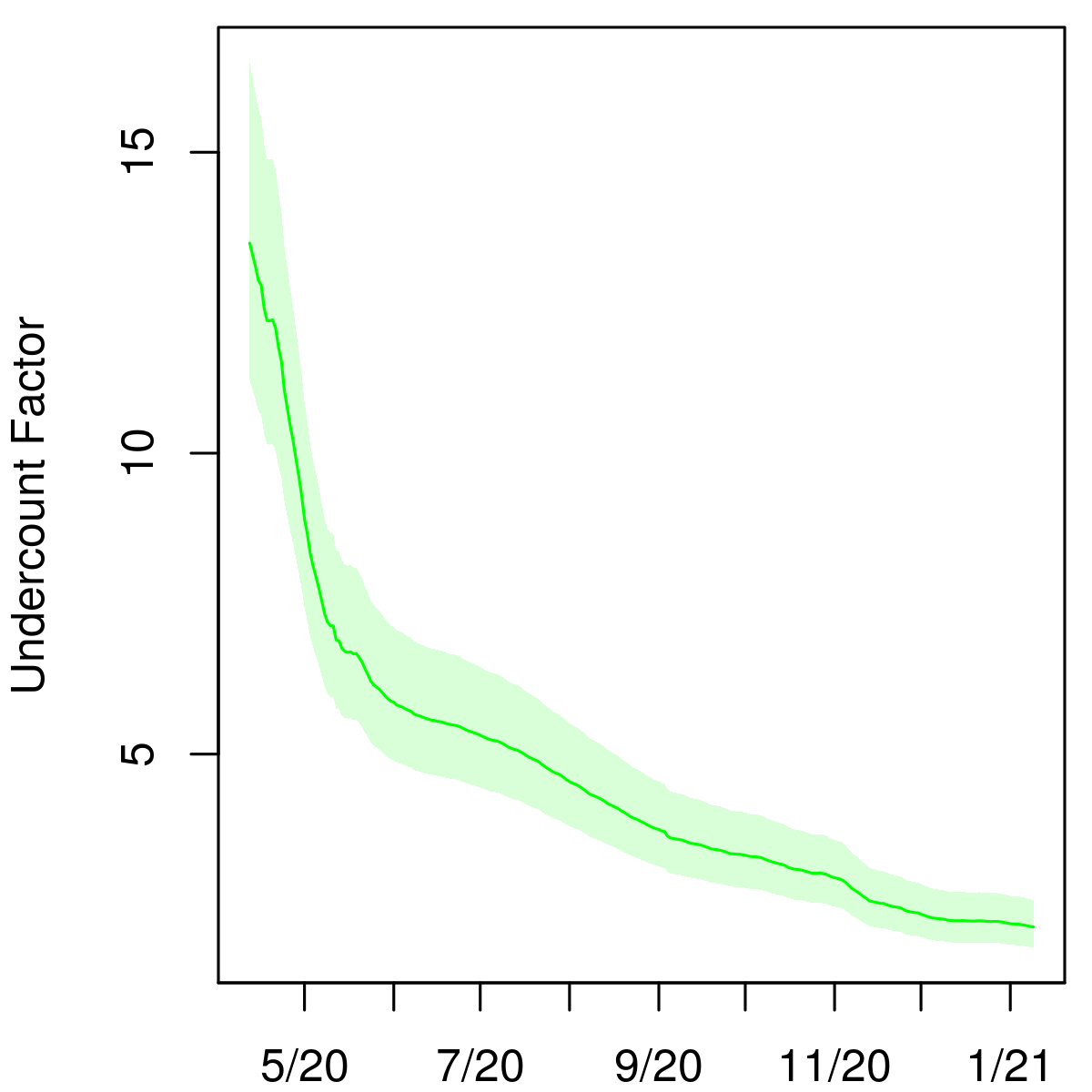} 
\end{tabular}
\caption{Posterior median and middle 95\% intervals for daily new infections, cumulative incidence, $r(t)$, and cumulative undercount from March 2020 to January 2021. In the top left panel, deaths divided by the posterior median IFR are plotted in grey for comparison.}
\end{figure}
\newpage
\begin{figure}[htbp!]
\textbf{Indiana}
\centering
\begin{tabular}{ll}
\includegraphics[scale=0.185]{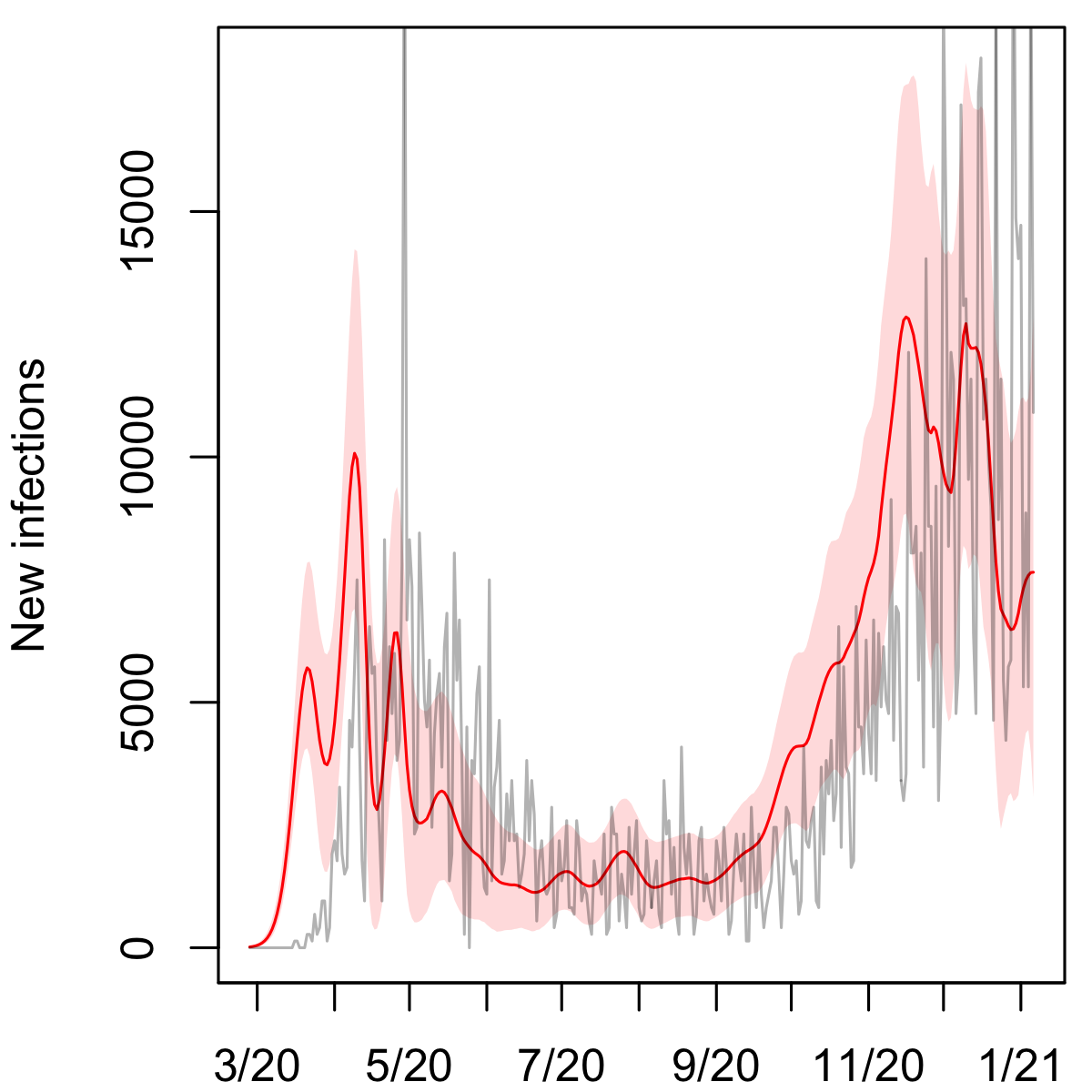}
&
\includegraphics[scale=0.185]{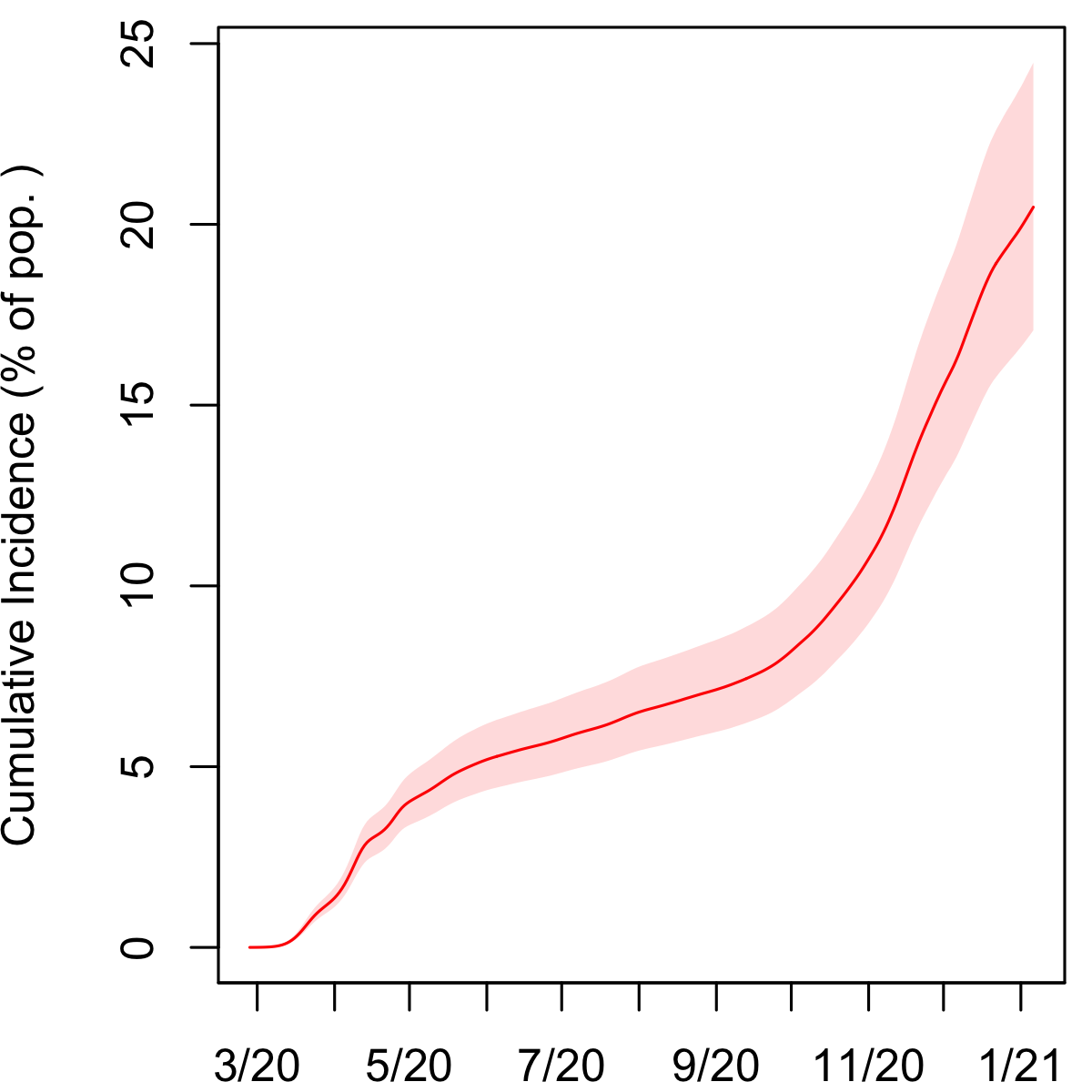} \\
\includegraphics[scale=0.185]{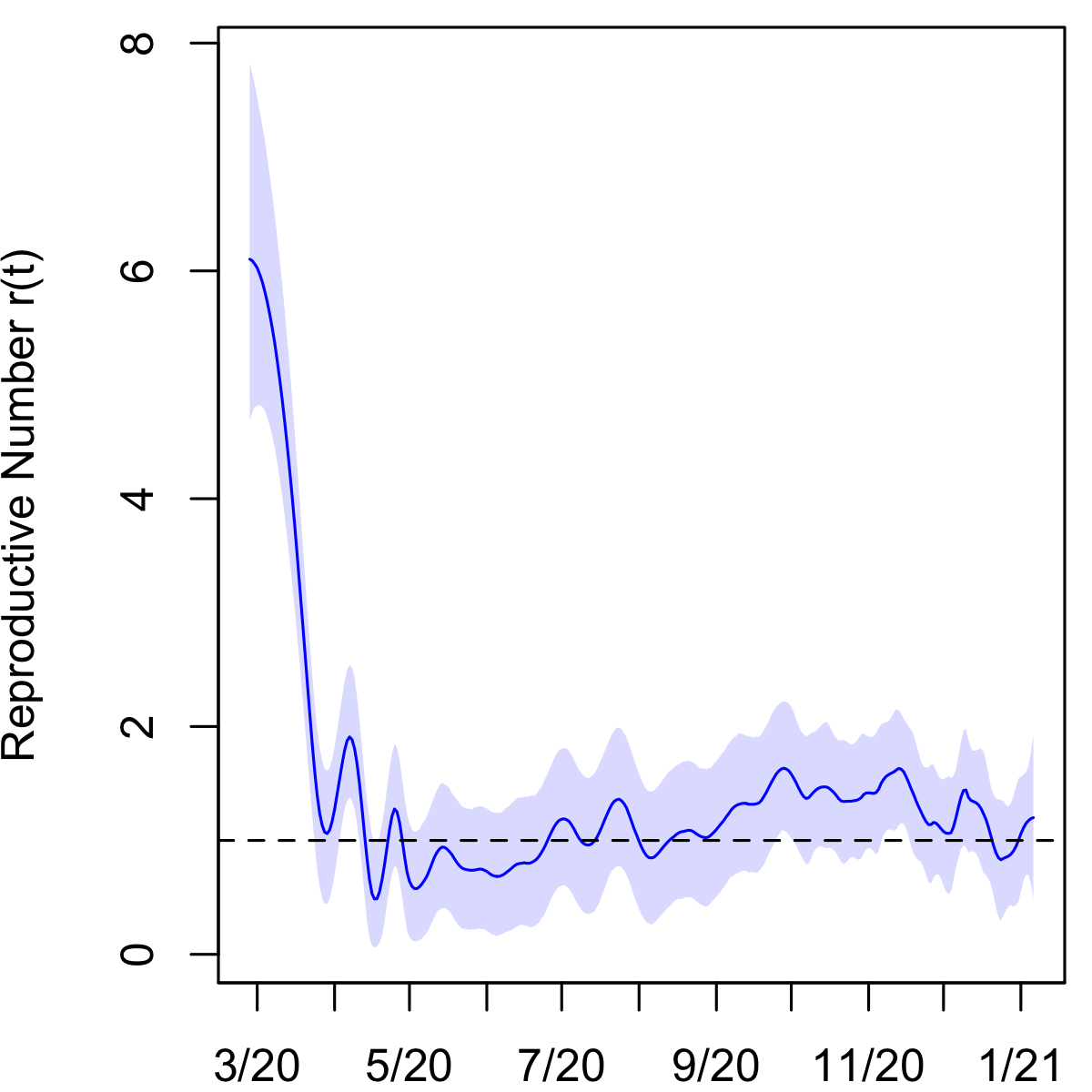}
&
\includegraphics[scale=0.185]{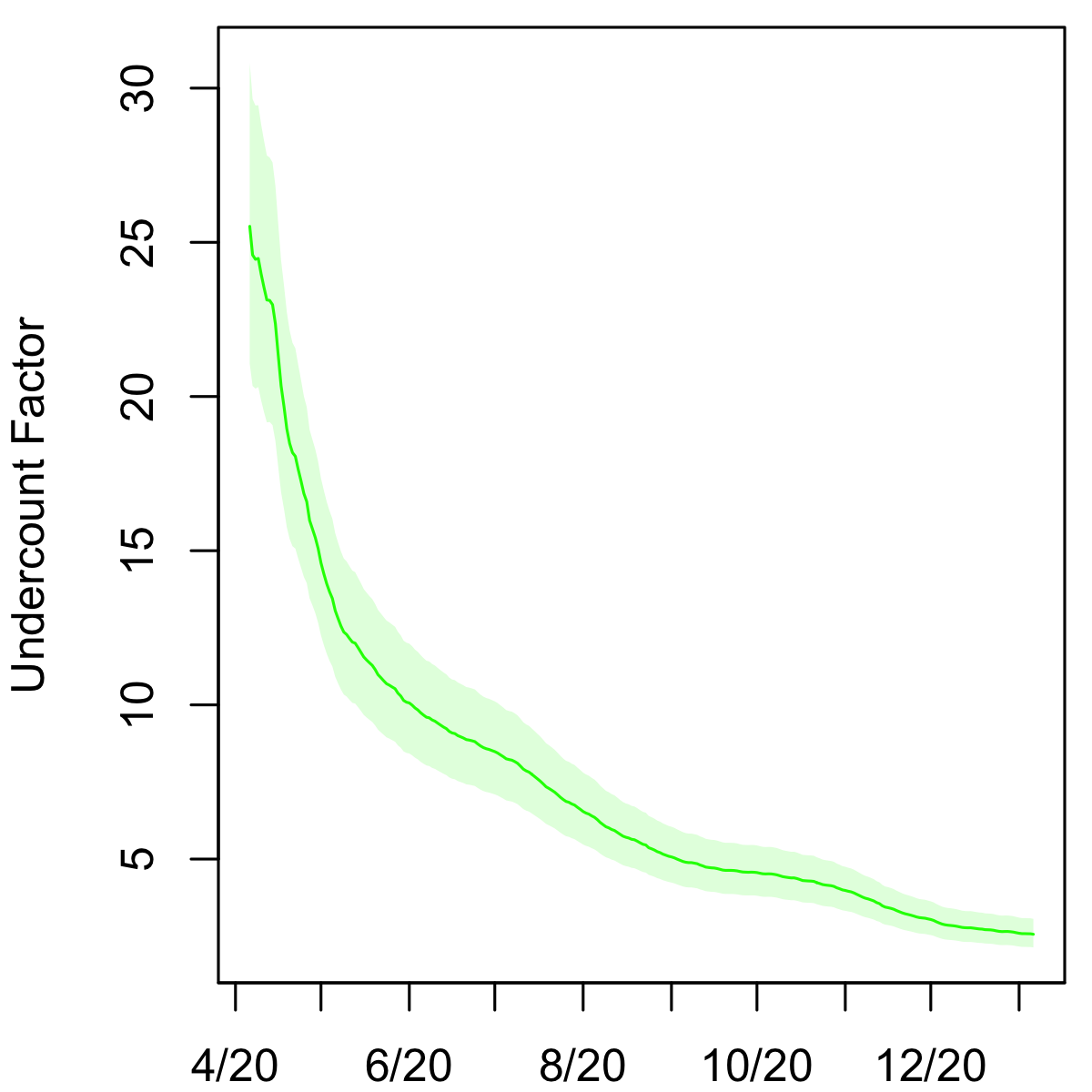} 
\end{tabular}
\caption{Posterior median and middle 95\% intervals for daily new infections, cumulative incidence, $r(t)$, and cumulative undercount from March 2020 to January 2021. In the top left panel, deaths divided by the posterior median IFR are plotted in grey for comparison.}
\end{figure}
\newpage
\begin{figure}[htbp!]
\textbf{Kansas}
\centering
\begin{tabular}{ll}
\includegraphics[scale=0.77]{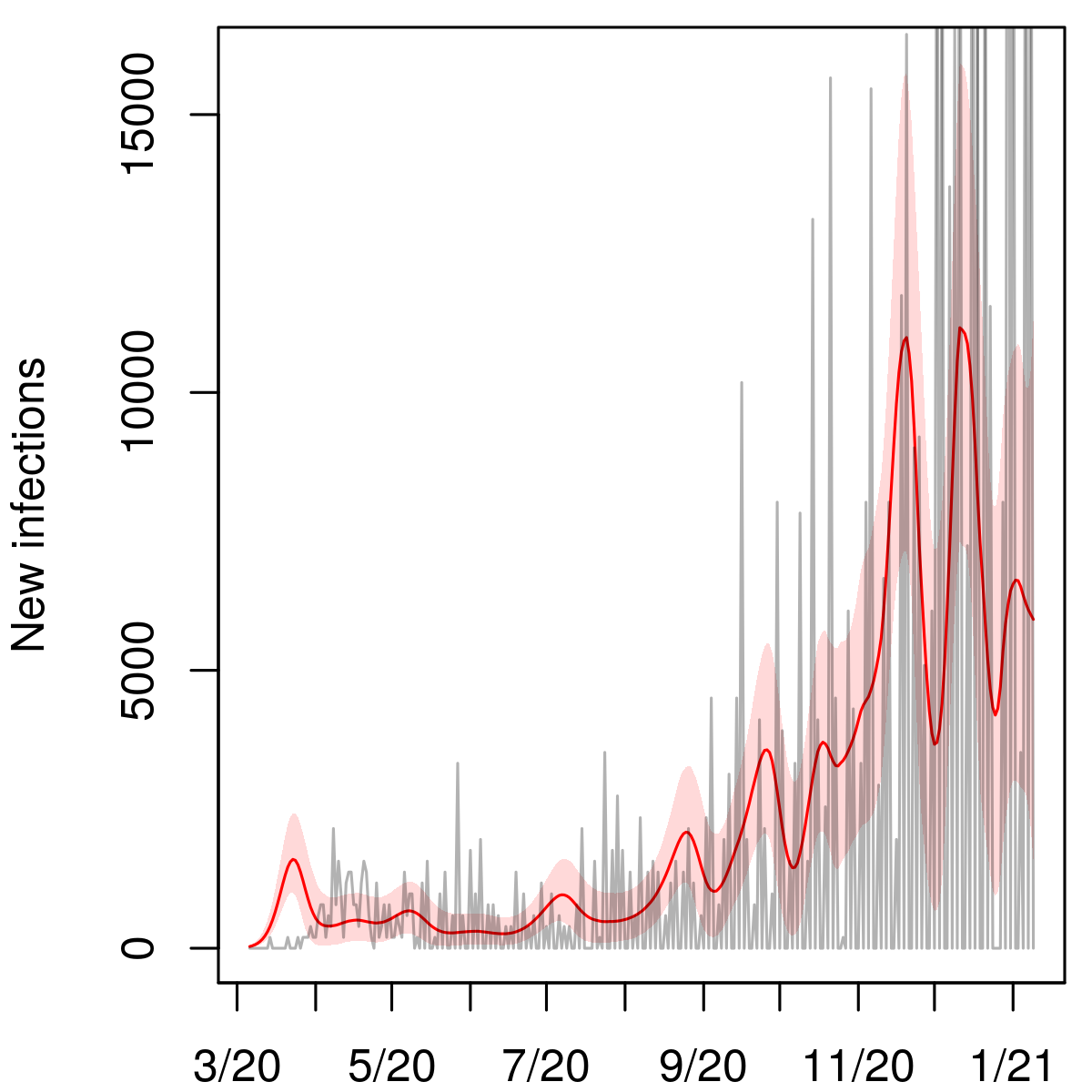}
&
\includegraphics[scale=0.77]{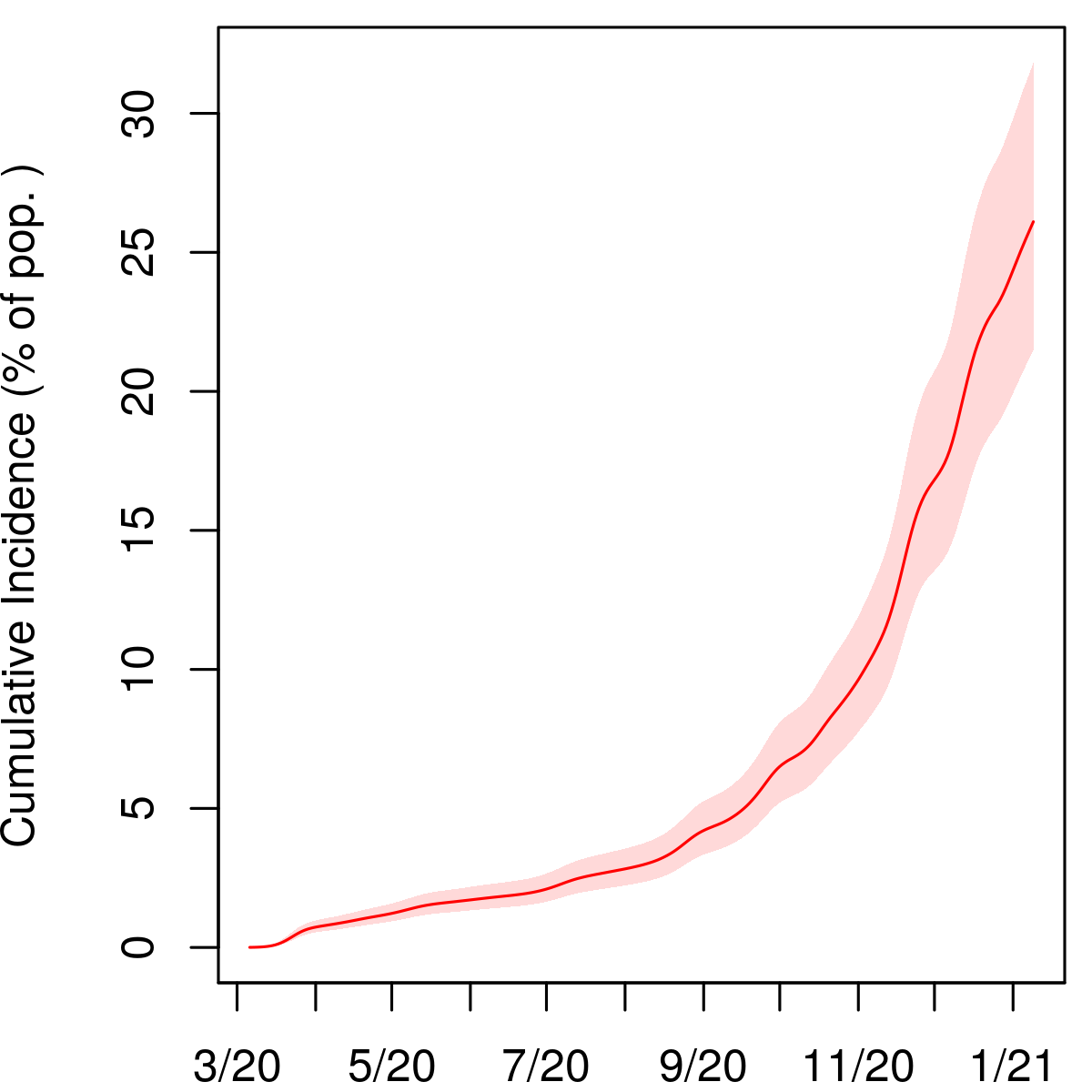} \\
\includegraphics[scale=0.77]{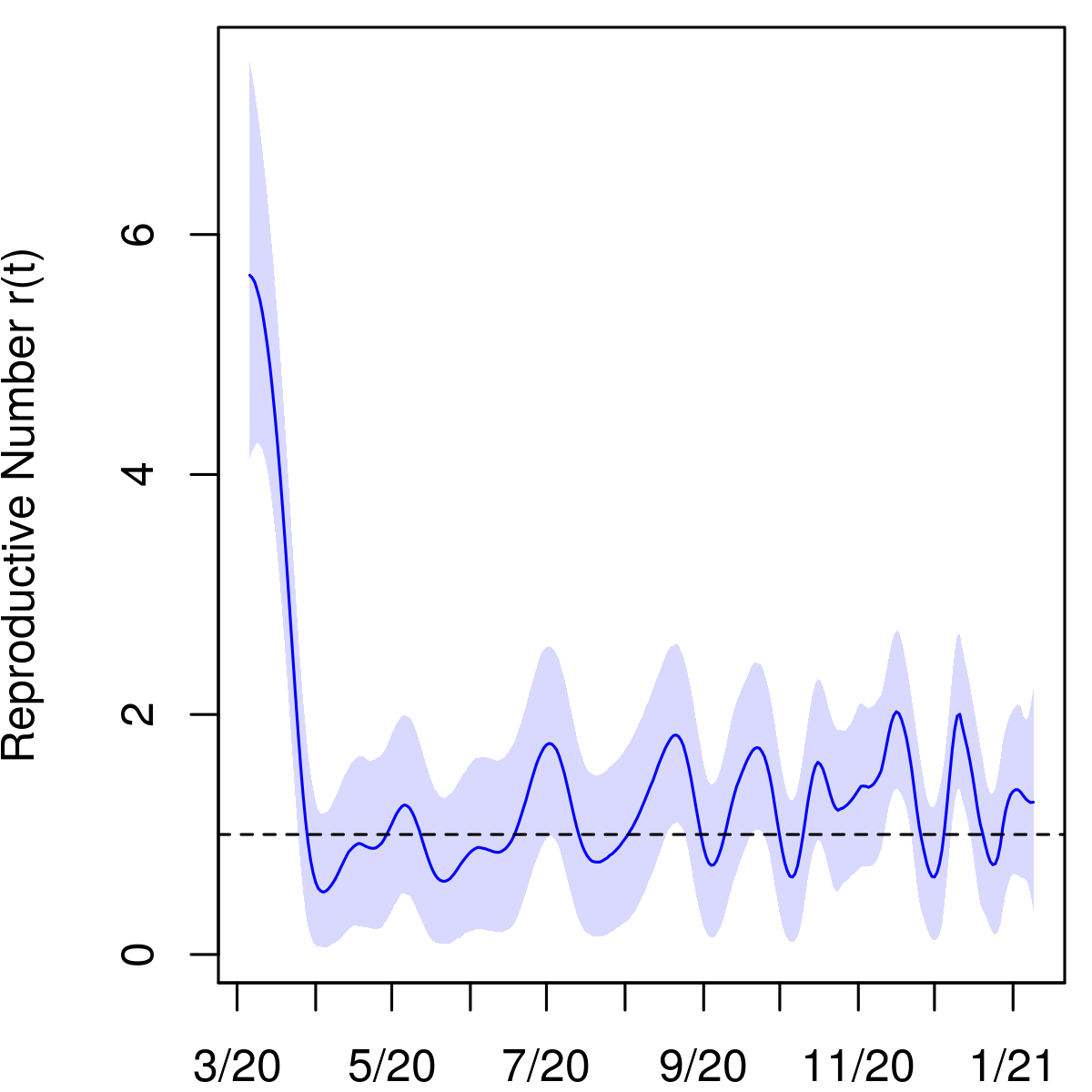}
&
\includegraphics[scale=0.77]{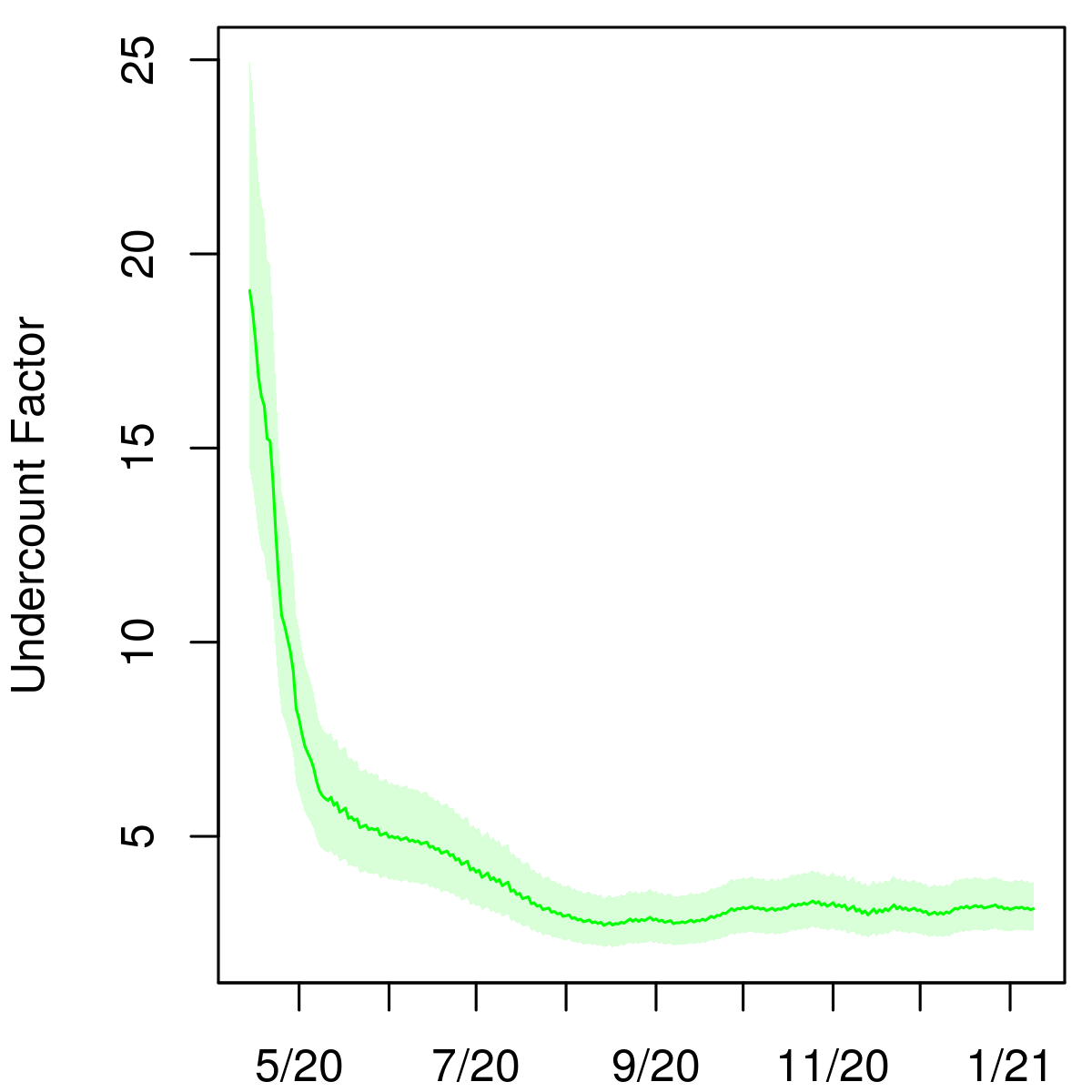} 
\end{tabular}
\caption{Posterior median and middle 95\% intervals for daily new infections, cumulative incidence, $r(t)$, and cumulative undercount from March 2020 to January 2021. In the top left panel, deaths divided by the posterior median IFR are plotted in grey for comparison.}
\end{figure}
\newpage
\begin{figure}[htbp!]
\textbf{Kentucky}
\centering
\begin{tabular}{ll}
\includegraphics[scale=0.77]{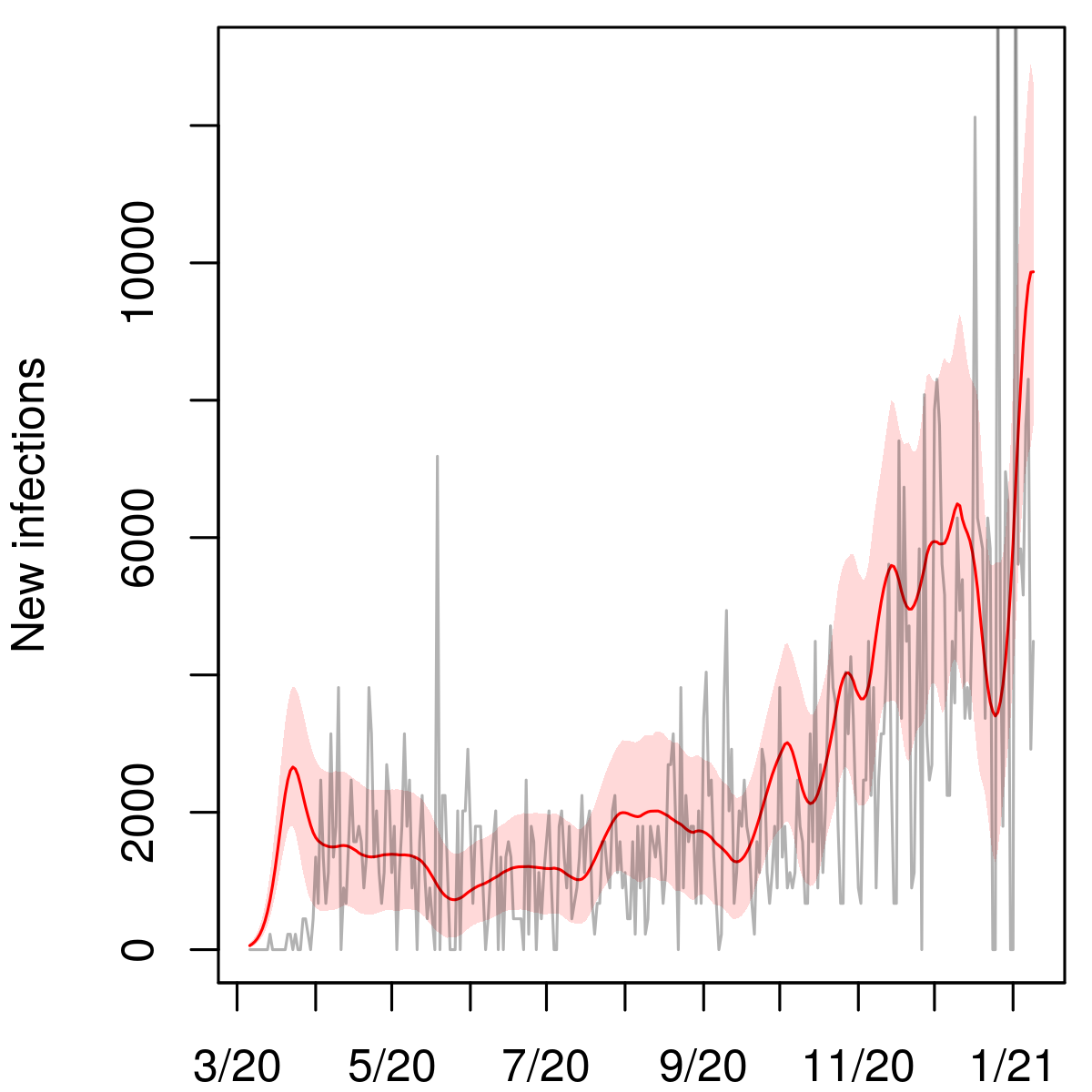}
&
\includegraphics[scale=0.77]{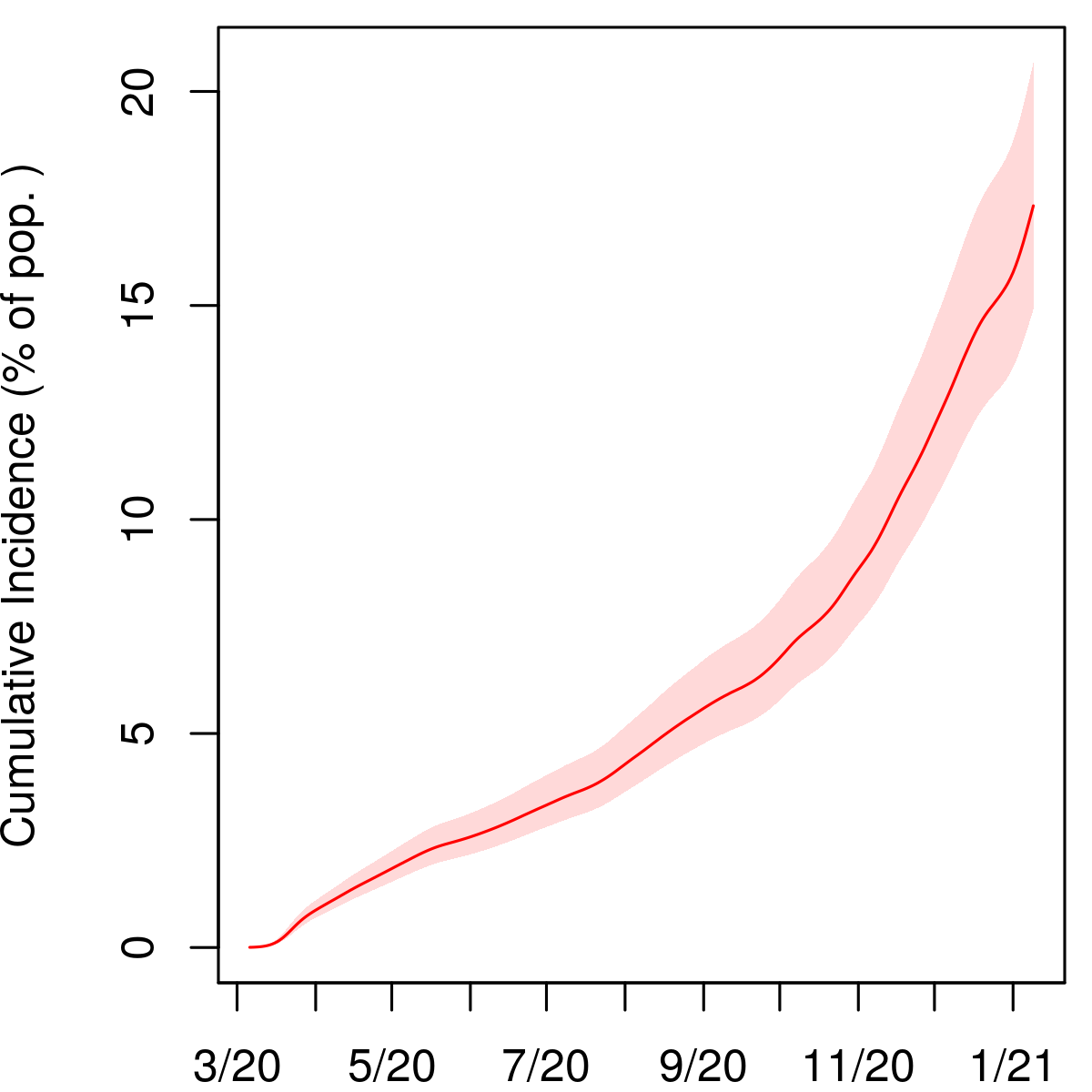} \\
\includegraphics[scale=0.77]{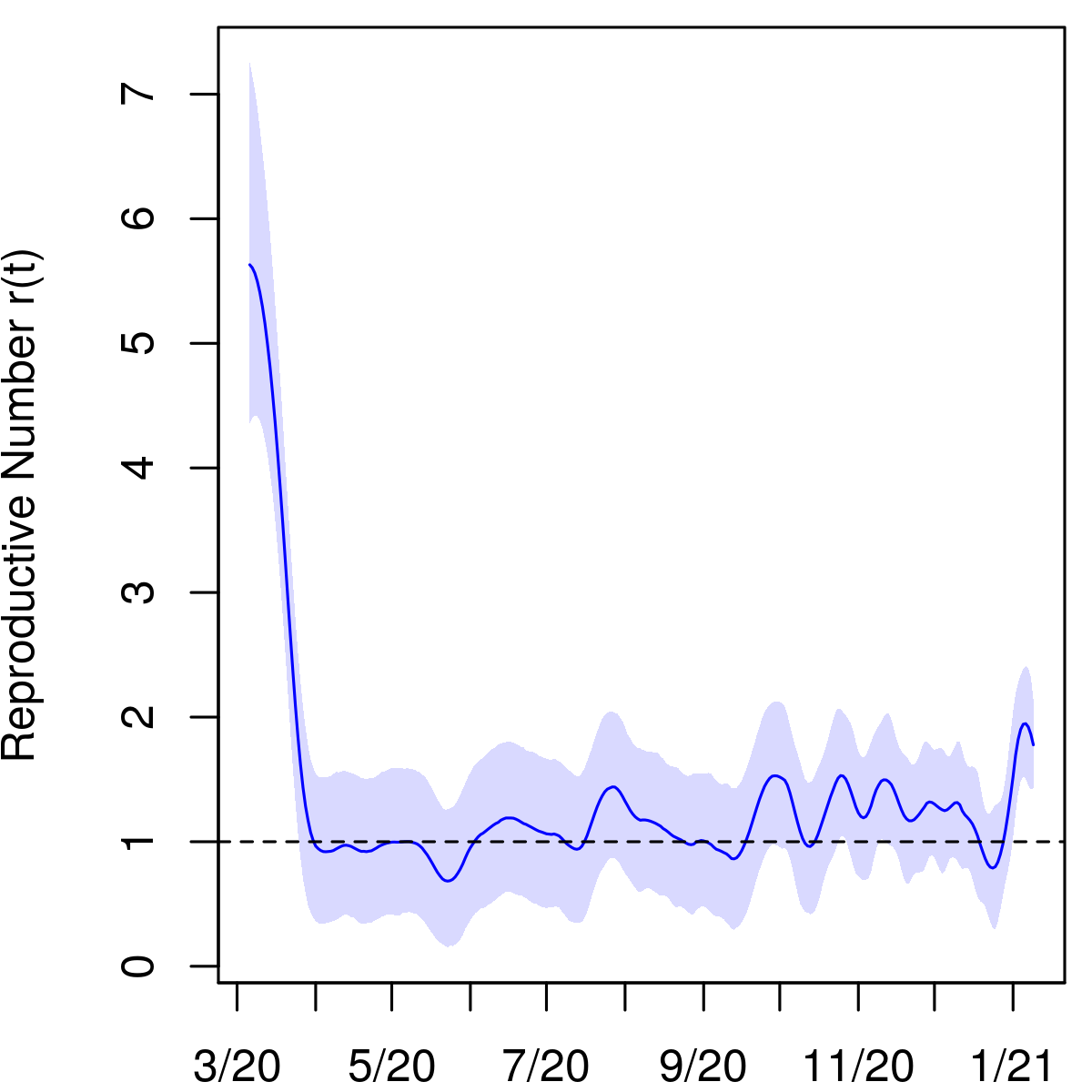}
&
\includegraphics[scale=0.77]{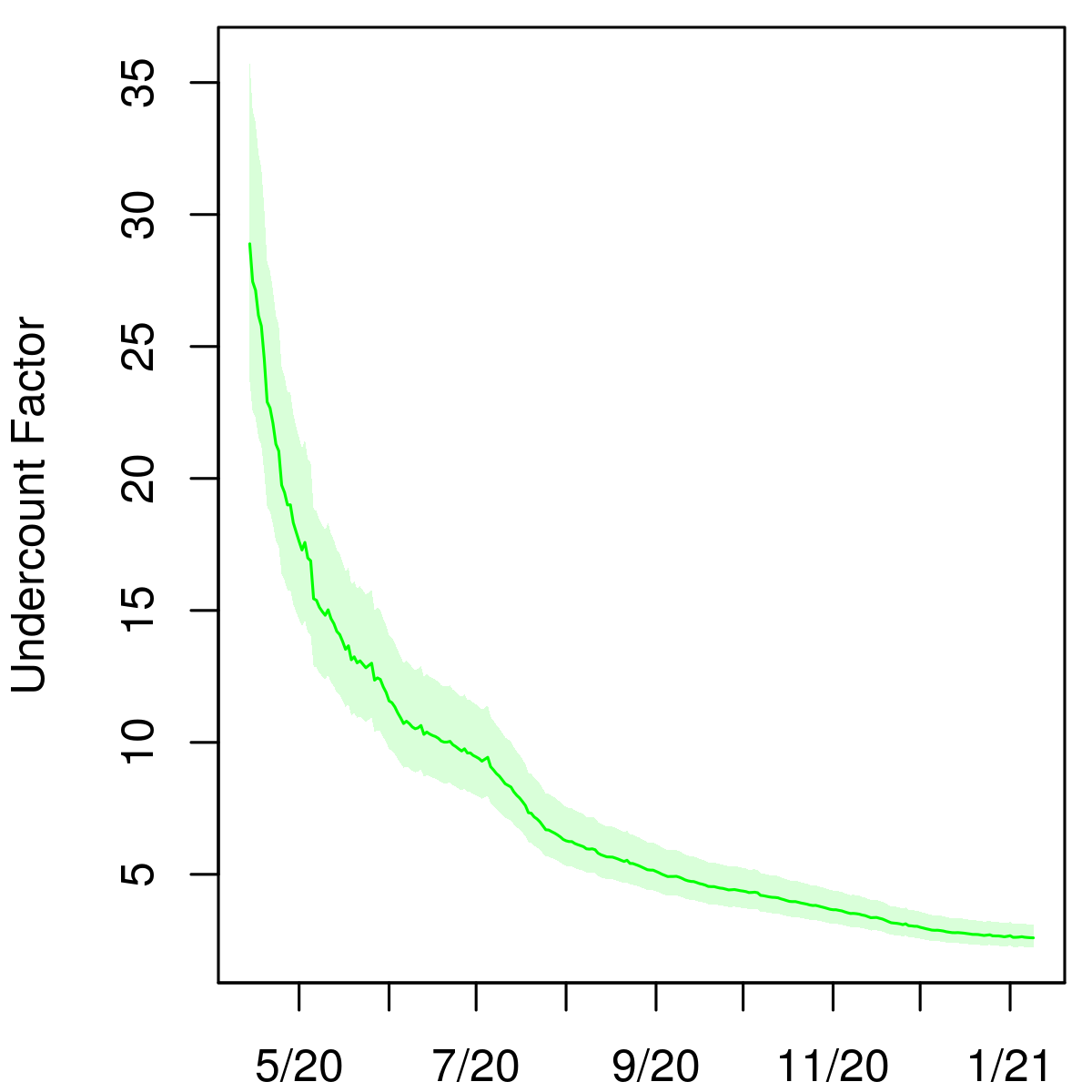} 
\end{tabular}
\caption{Posterior median and middle 95\% intervals for daily new infections, cumulative incidence, $r(t)$, and cumulative undercount from March 2020 to January 2021. In the top left panel, deaths divided by the posterior median IFR are plotted in grey for comparison.}
\end{figure}
\newpage
\begin{figure}[htbp!]
\textbf{Louisiana}
\centering
\begin{tabular}{ll}
\includegraphics[scale=0.77]{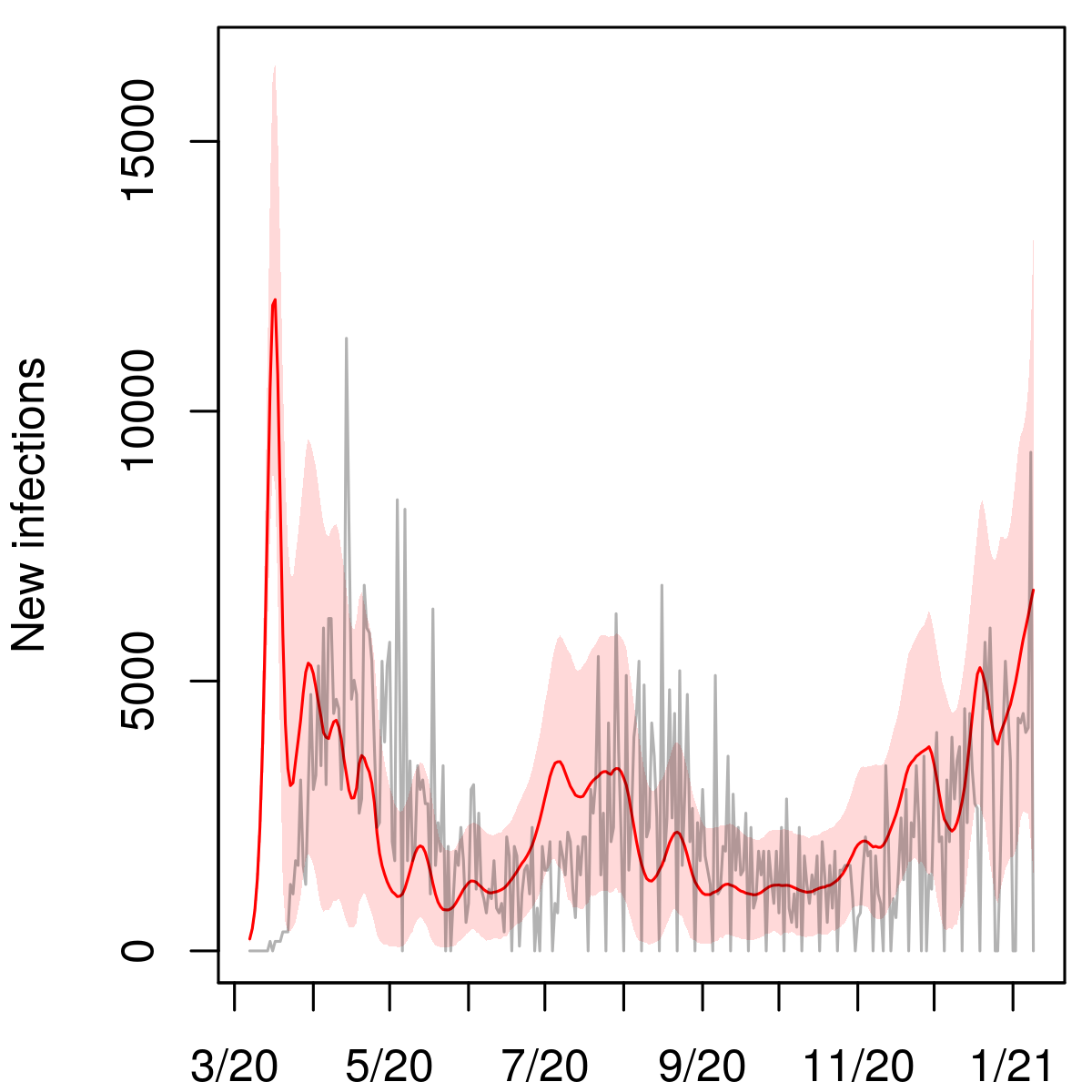}
&
\includegraphics[scale=0.77]{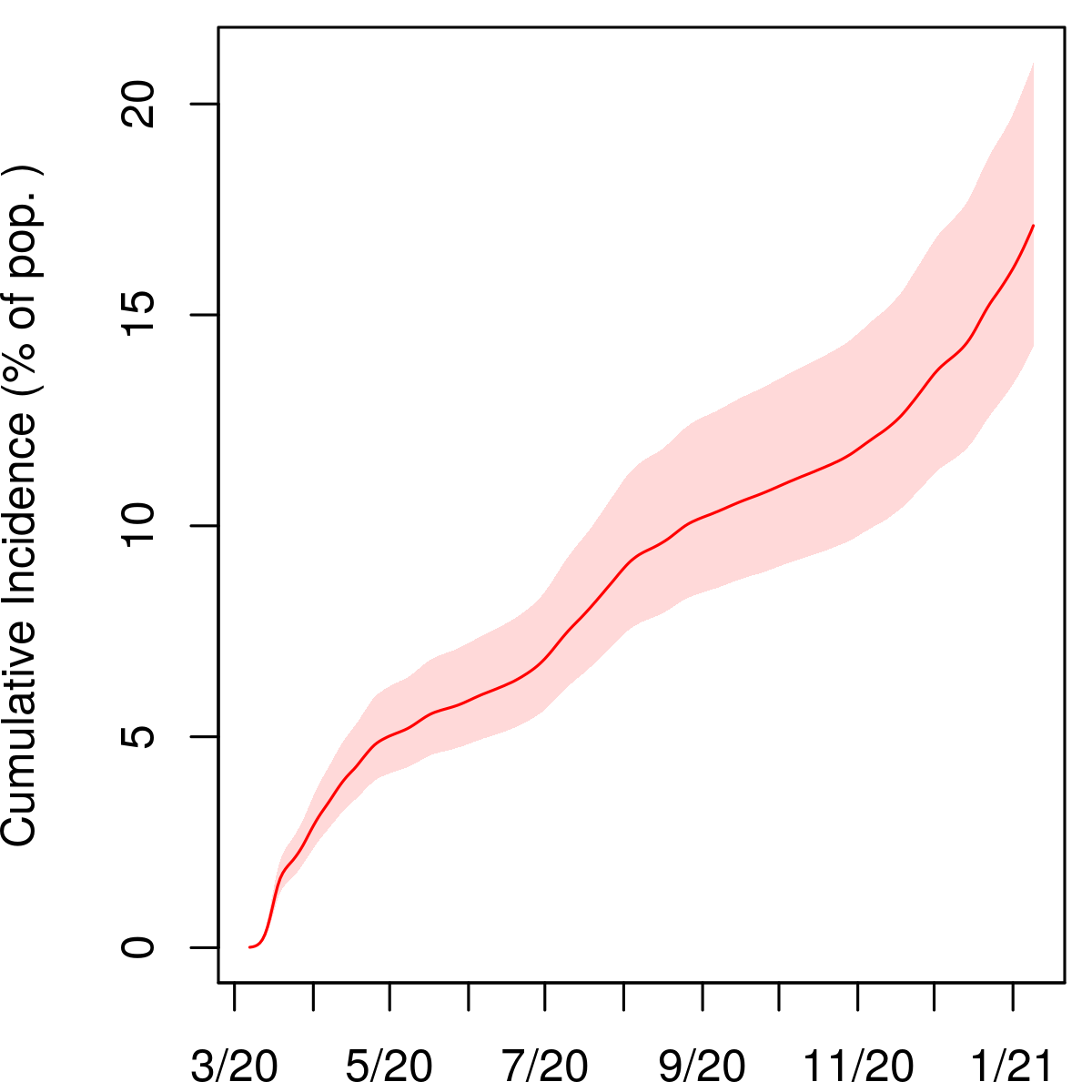} \\
\includegraphics[scale=0.77]{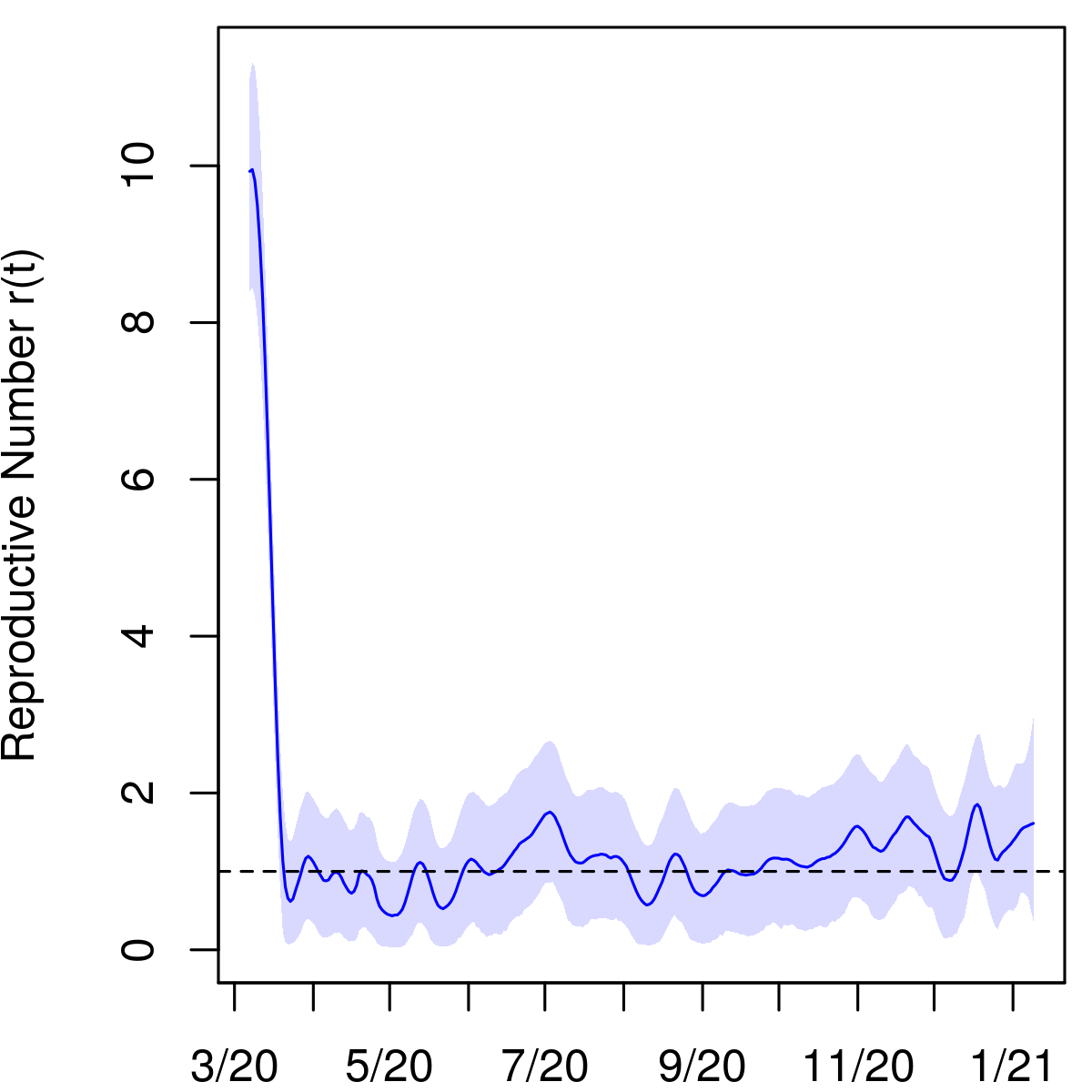}
&
\includegraphics[scale=0.77]{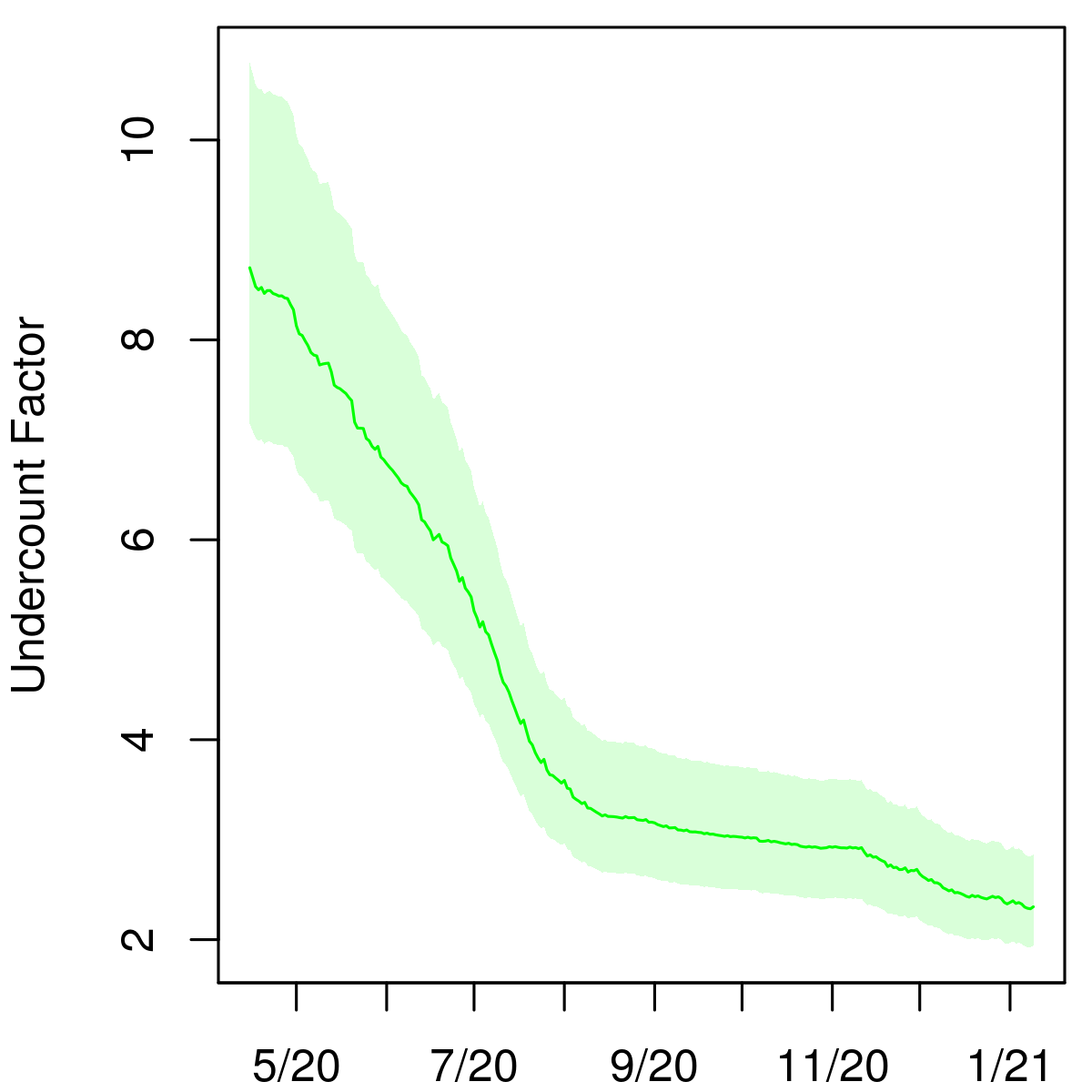} 
\end{tabular}
\caption{Posterior median and middle 95\% intervals for daily new infections, cumulative incidence, $r(t)$, and cumulative undercount from March 2020 to January 2021. In the top left panel, deaths divided by the posterior median IFR are plotted in grey for comparison.}
\end{figure}
\newpage
\begin{figure}[htbp!]
\textbf{Massachusetts}
\centering
\begin{tabular}{ll}
\includegraphics[scale=0.77]{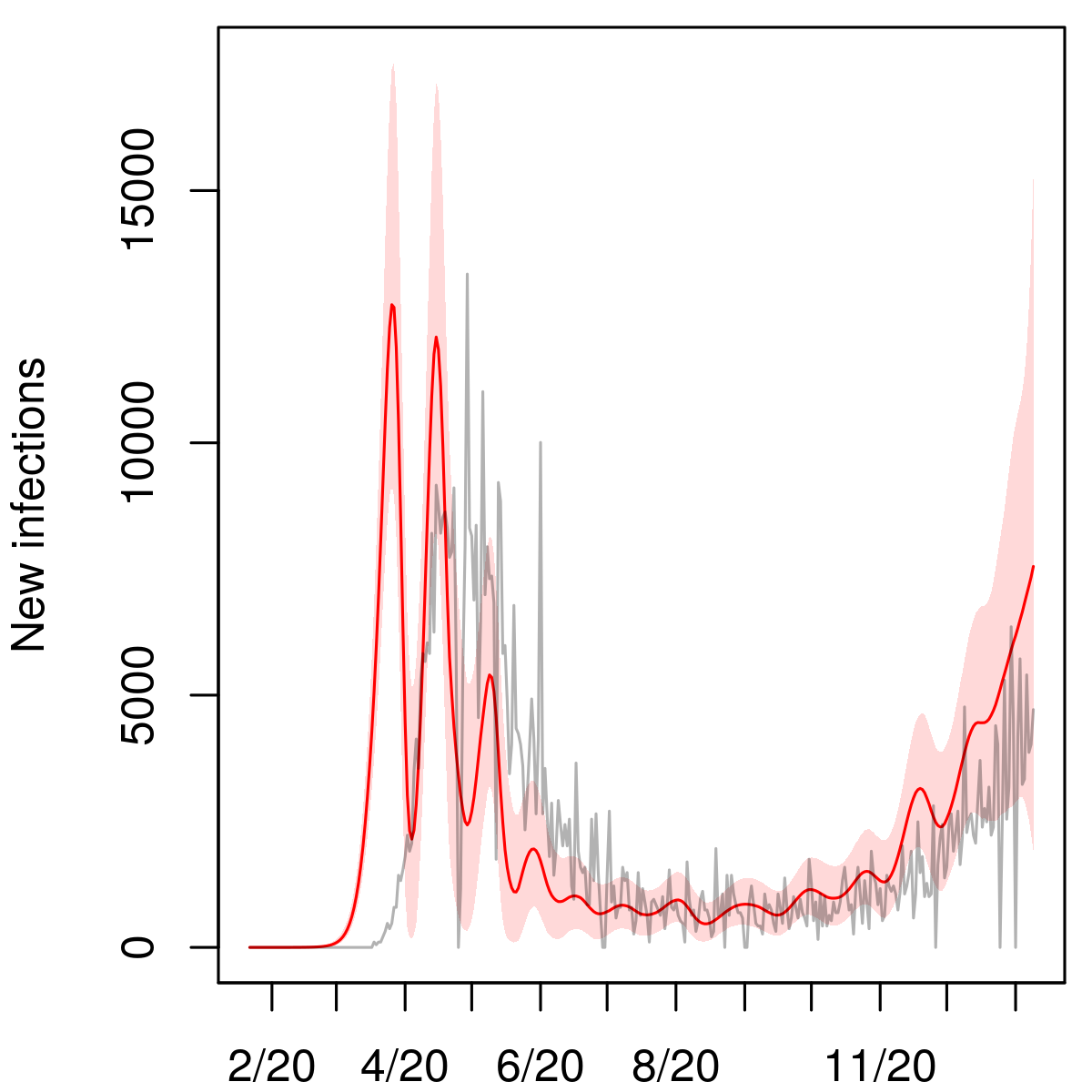}
&
\includegraphics[scale=0.77]{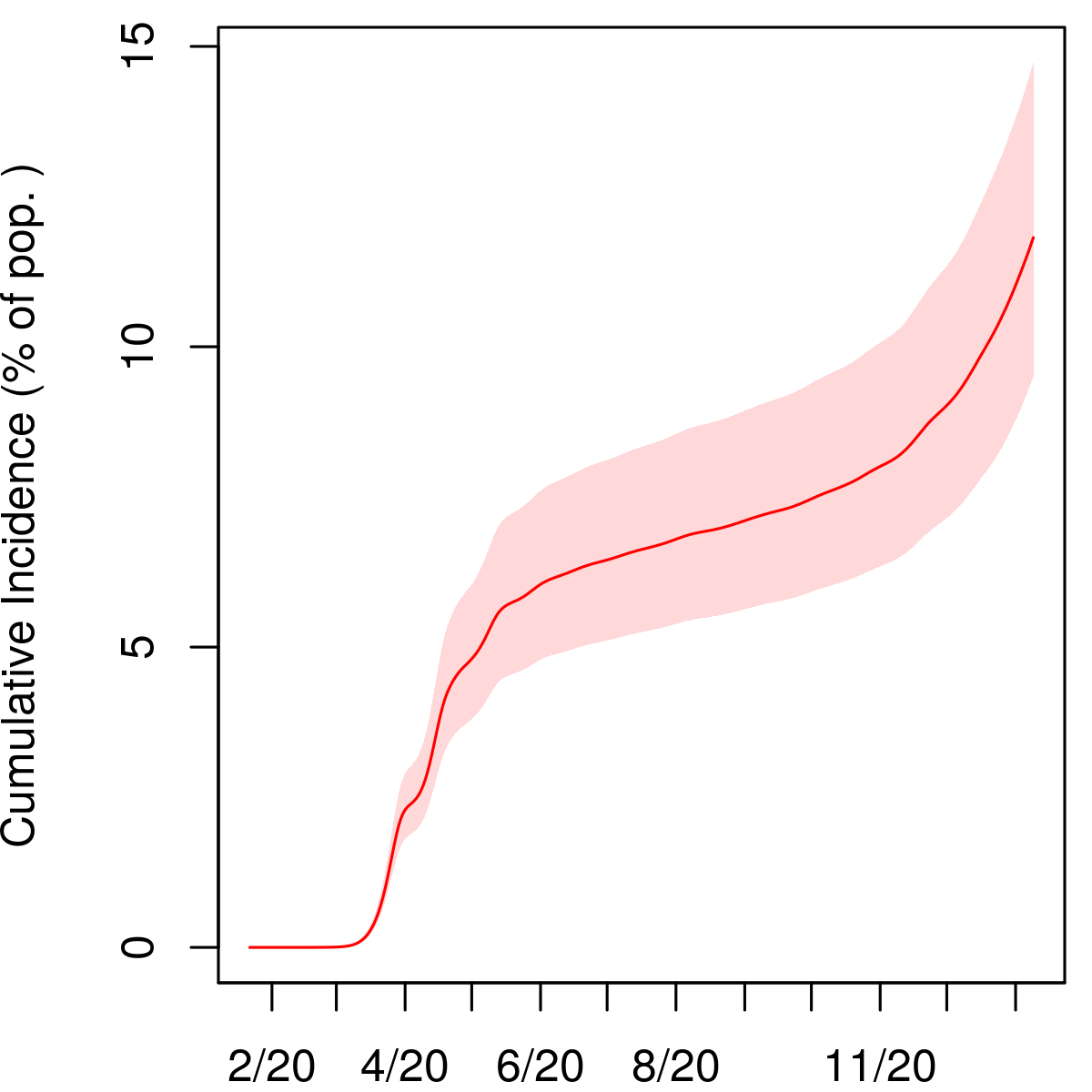} \\
\includegraphics[scale=0.77]{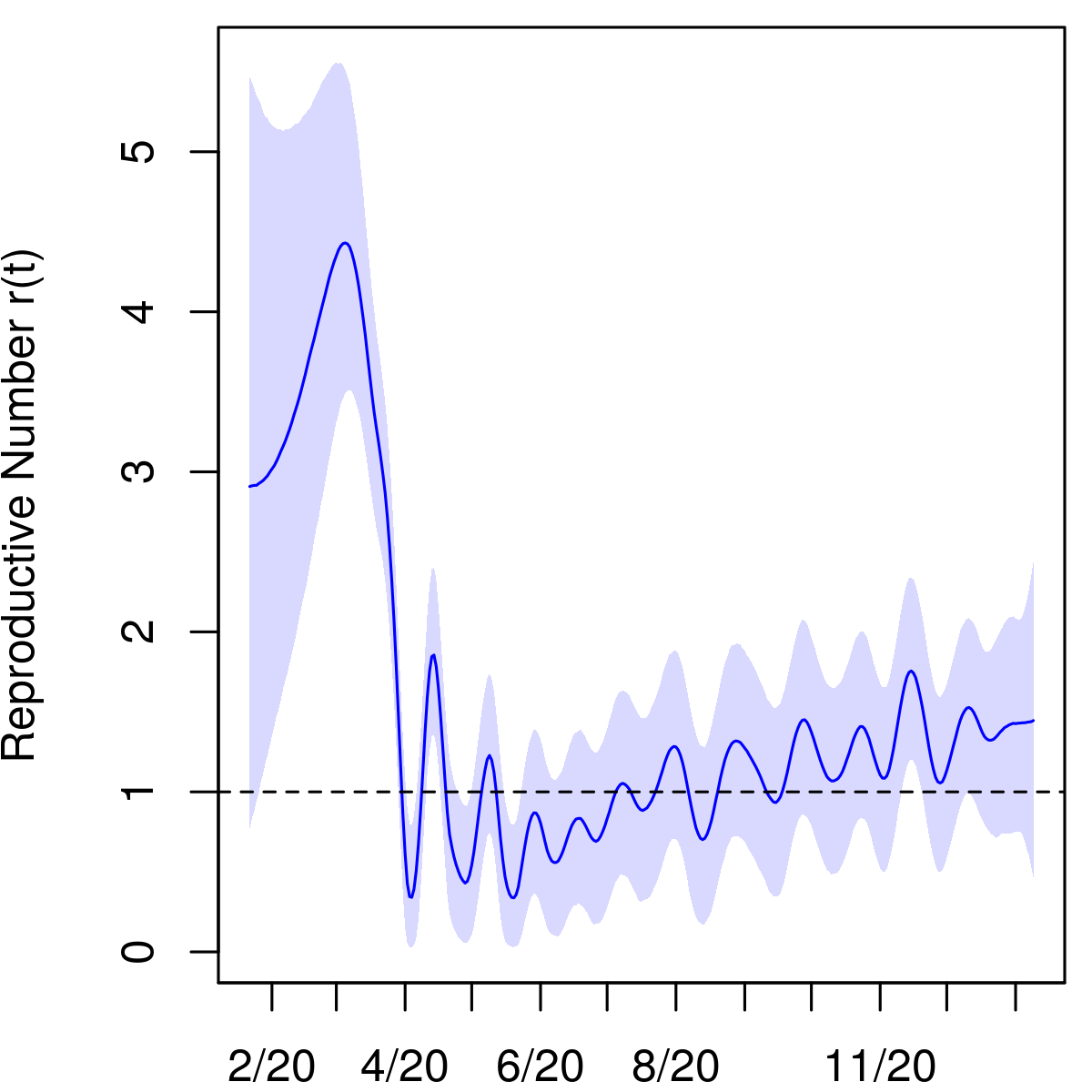}
&
\includegraphics[scale=0.77]{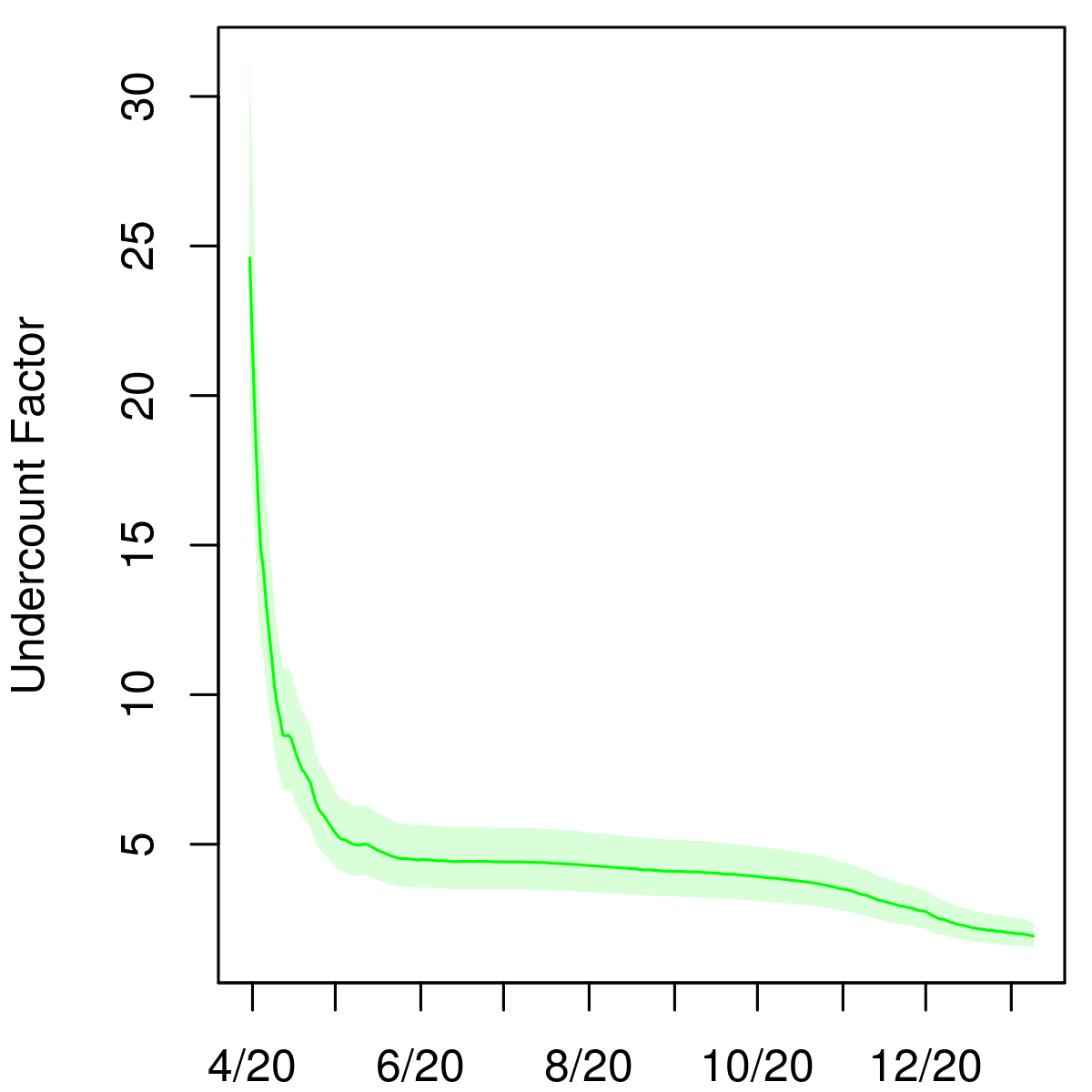} 
\end{tabular}
\caption{Posterior median and middle 95\% intervals for daily new infections, cumulative incidence, $r(t)$, and cumulative undercount from March 2020 to January 2021. In the top left panel, deaths divided by the posterior median IFR are plotted in grey for comparison.}
\end{figure}
\newpage
\begin{figure}[htbp!]
\textbf{Maryland}
\centering
\begin{tabular}{ll}
\includegraphics[scale=0.77]{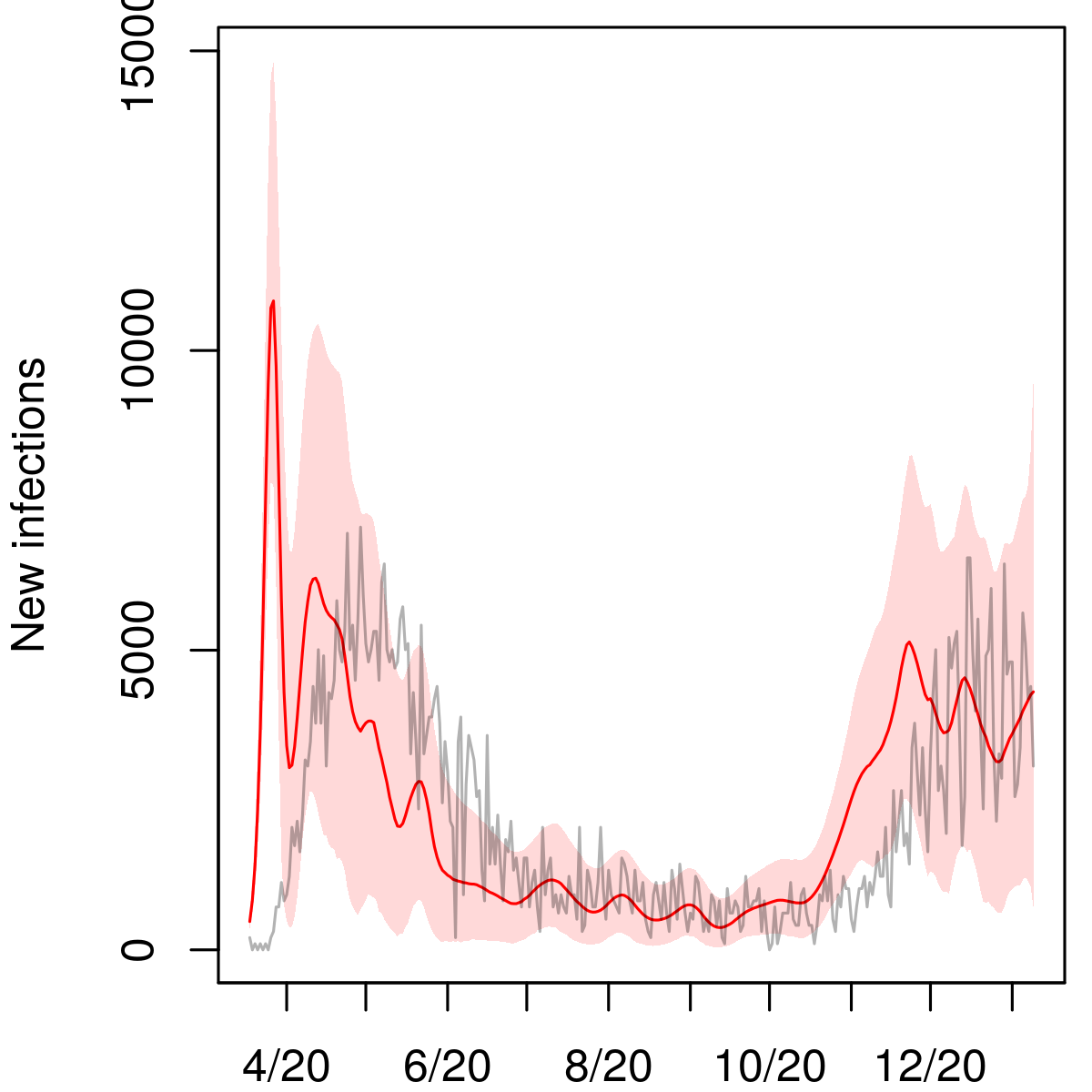}
&
\includegraphics[scale=0.77]{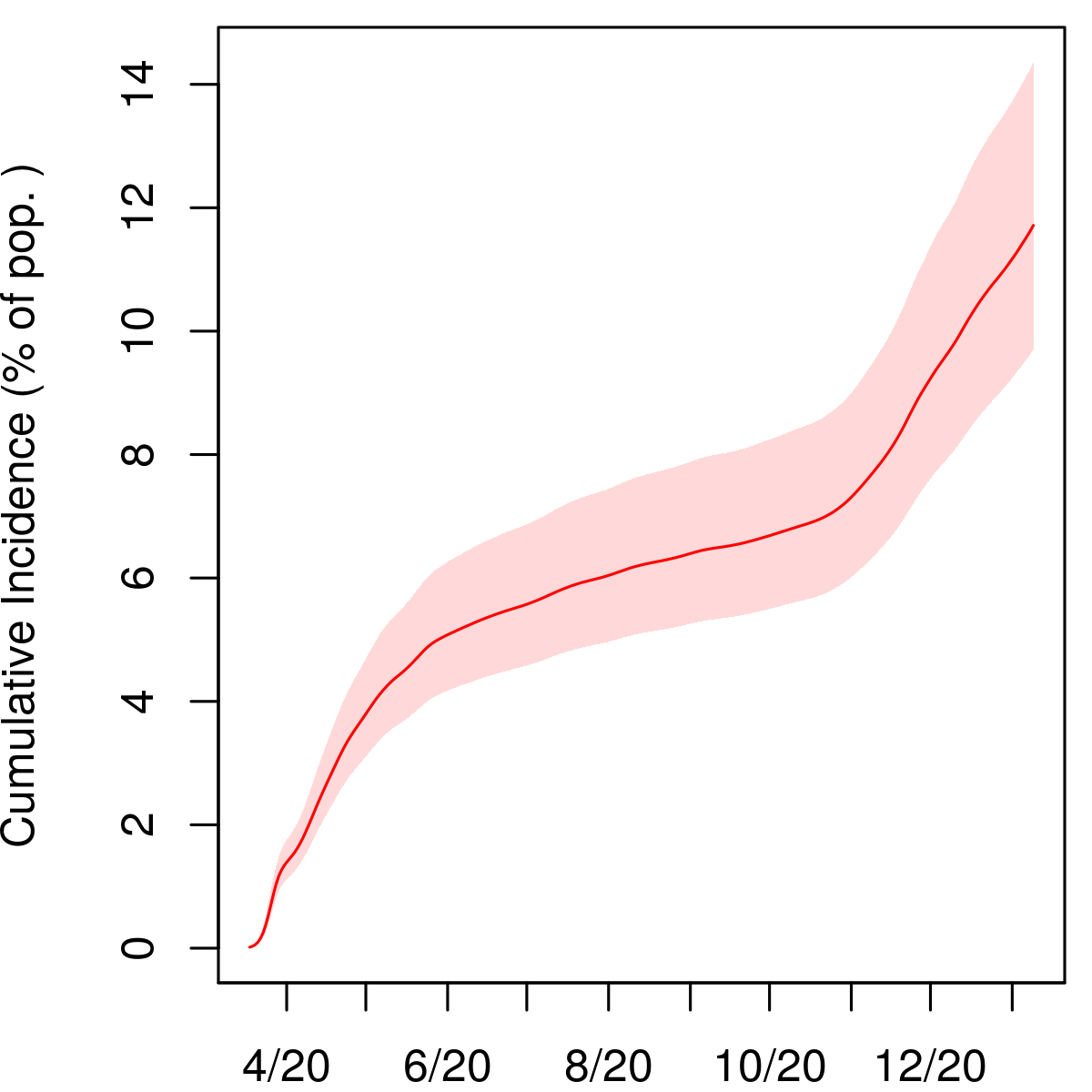} \\
\includegraphics[scale=0.77]{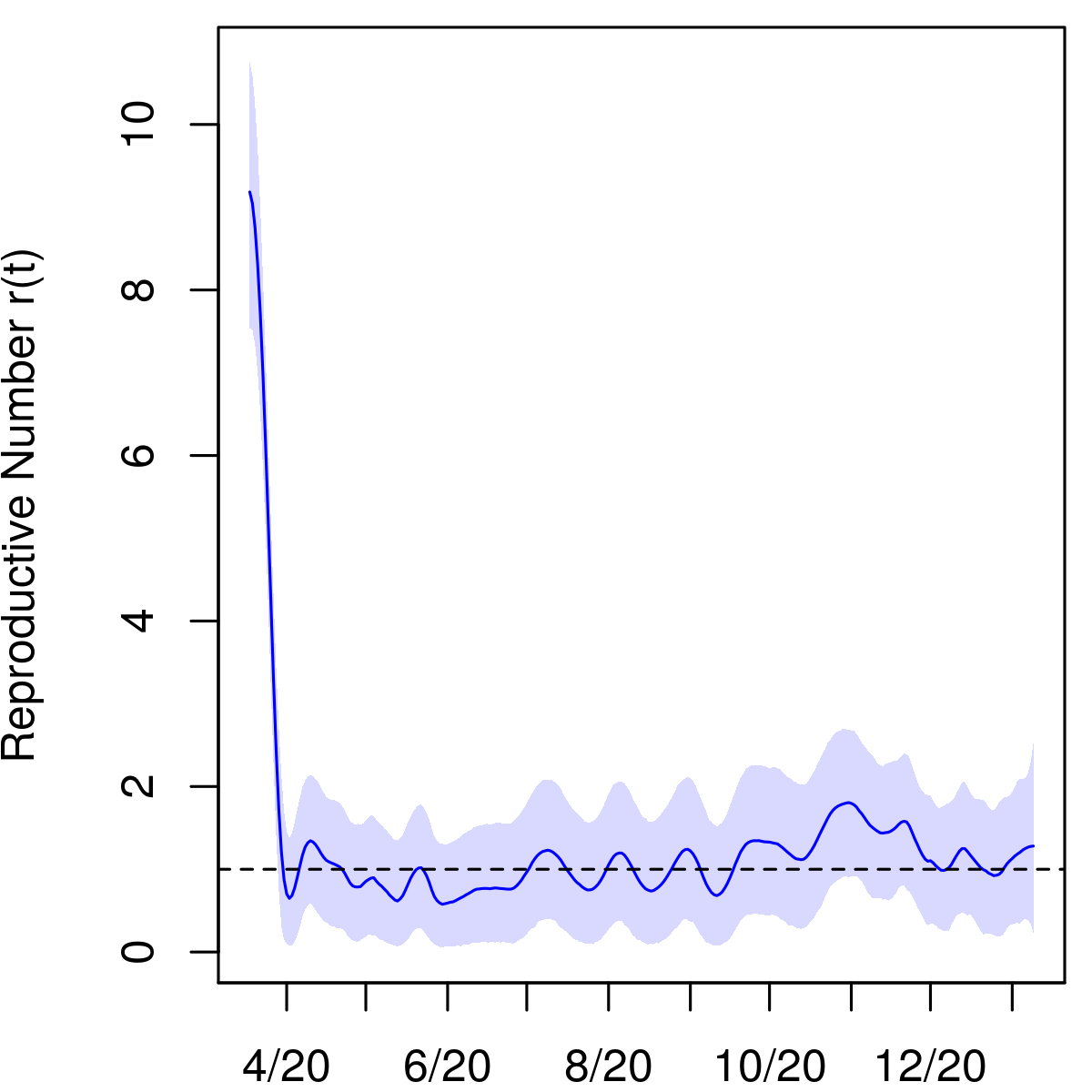}
&
\includegraphics[scale=0.77]{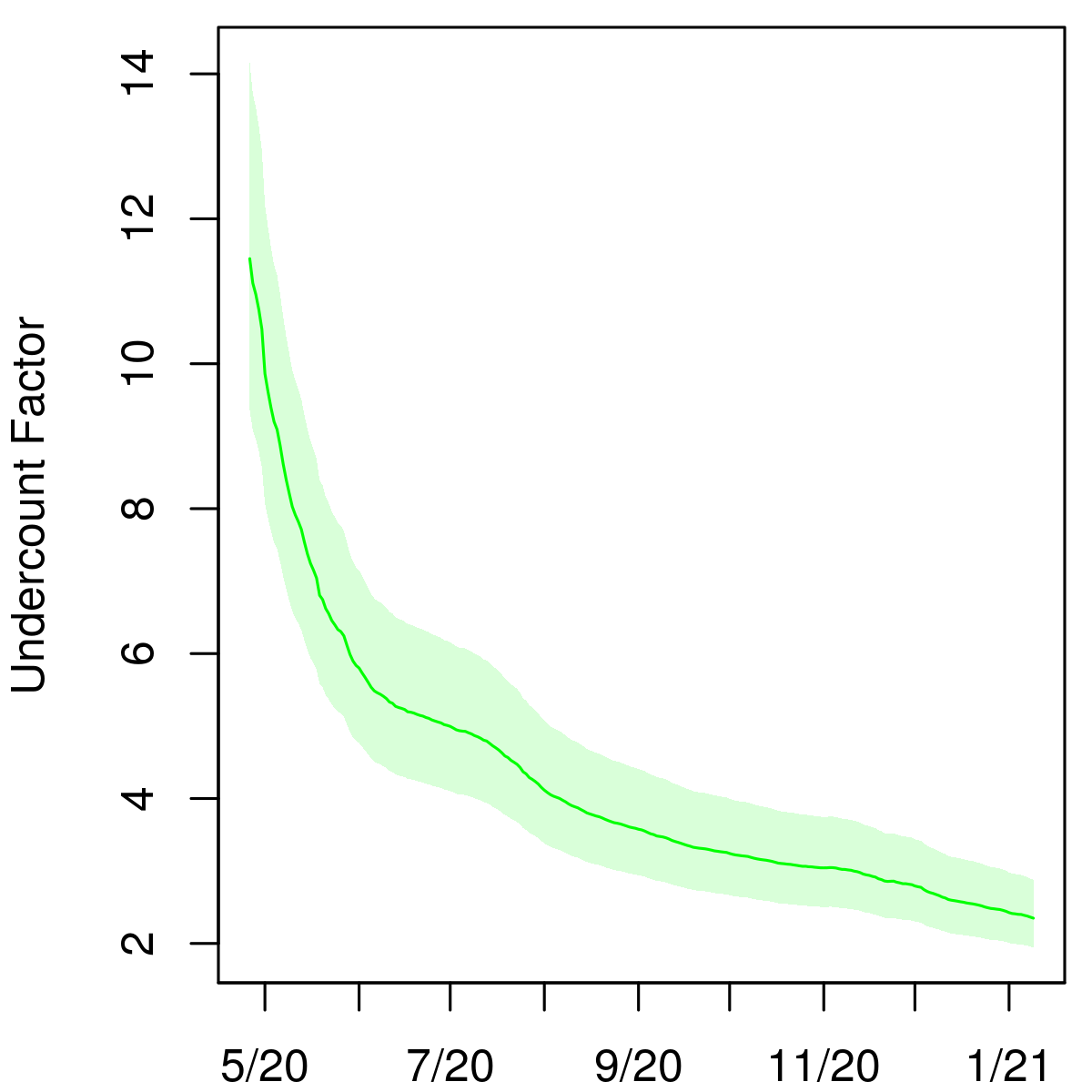} 
\end{tabular}
\caption{Posterior median and middle 95\% intervals for daily new infections, cumulative incidence, $r(t)$, and cumulative undercount from March 2020 to January 2021. In the top left panel, deaths divided by the posterior median IFR are plotted in grey for comparison.}
\end{figure}
\newpage
\begin{figure}[htbp!]
\textbf{Maine}
\centering
\begin{tabular}{ll}
\includegraphics[scale=0.77]{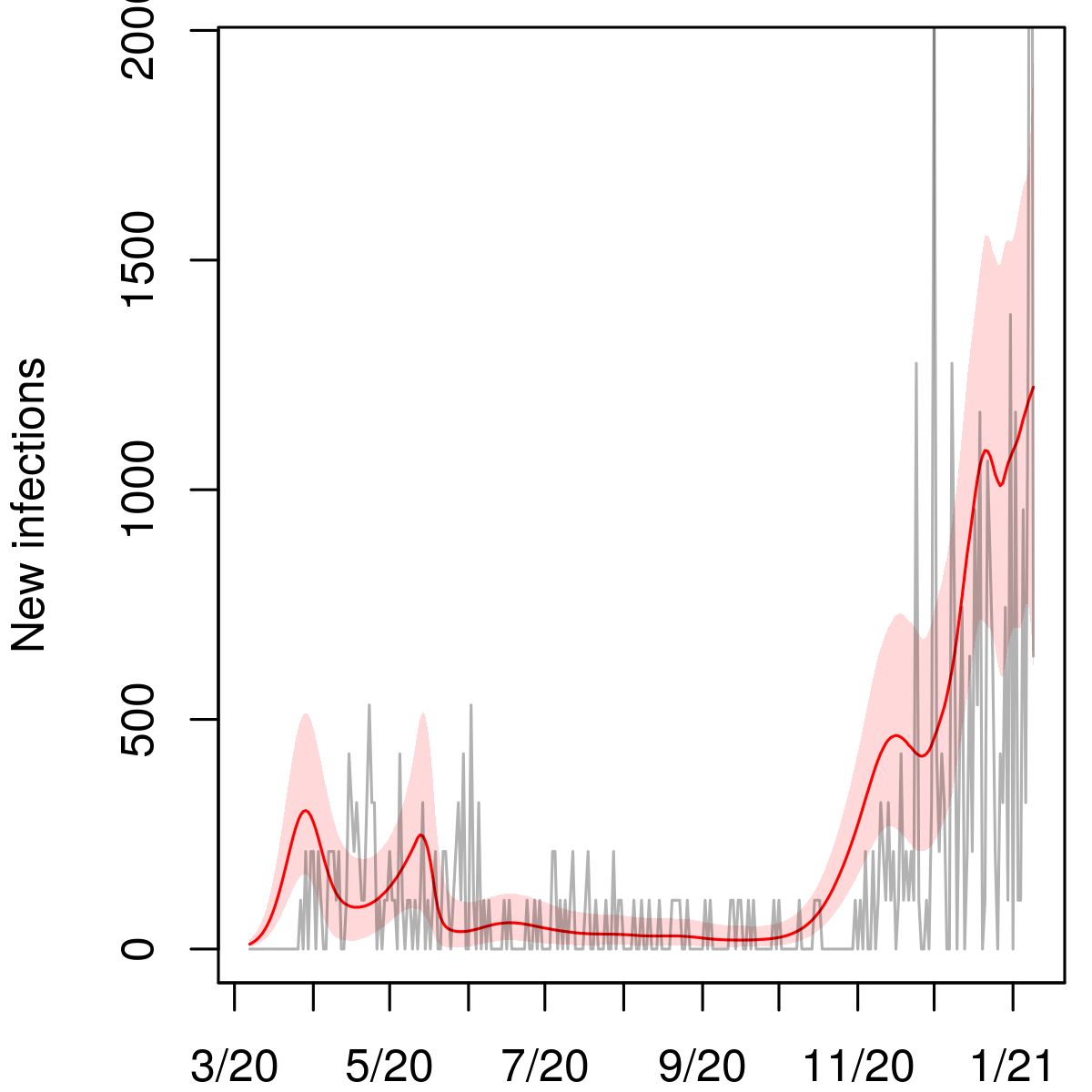}
&
\includegraphics[scale=0.77]{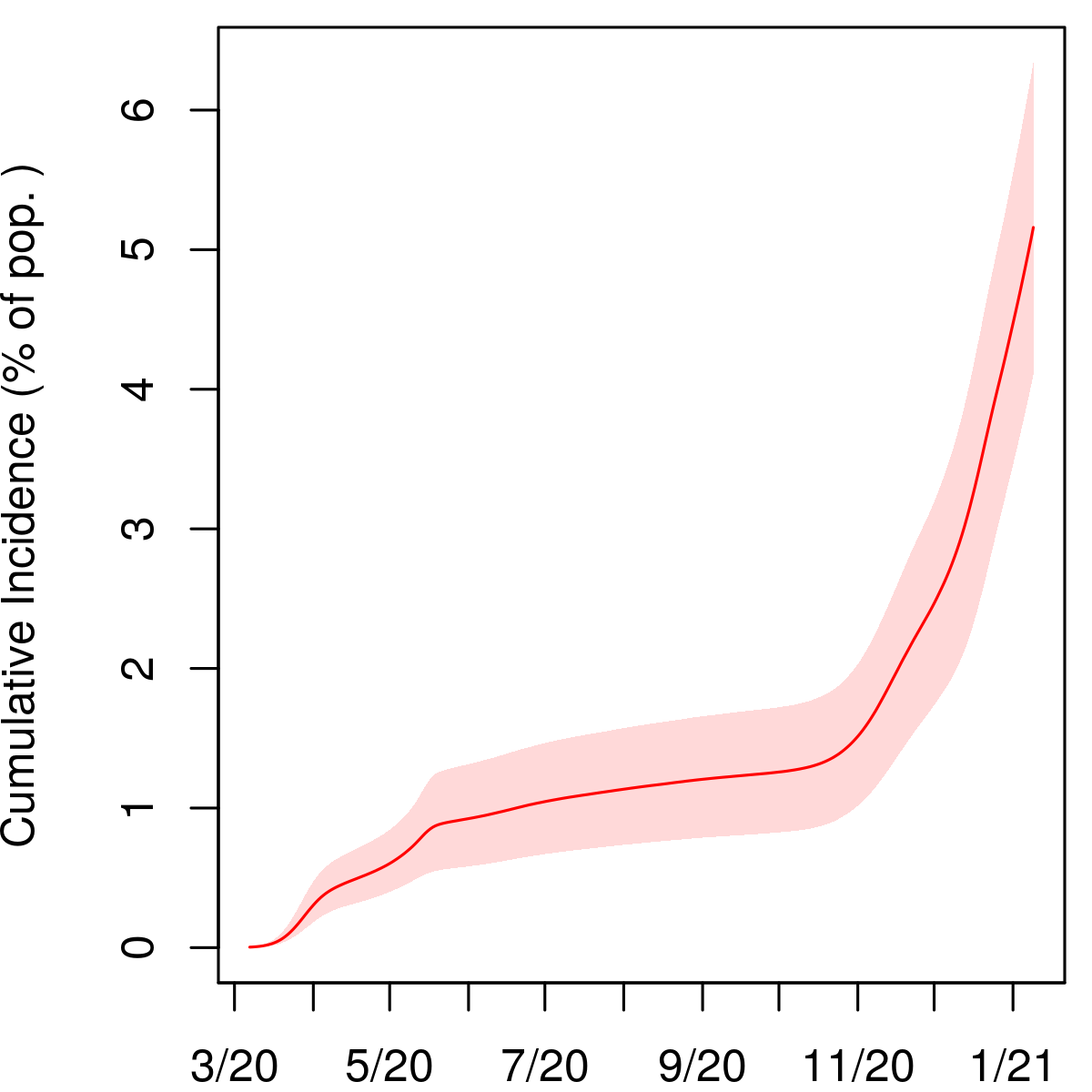} \\
\includegraphics[scale=0.77]{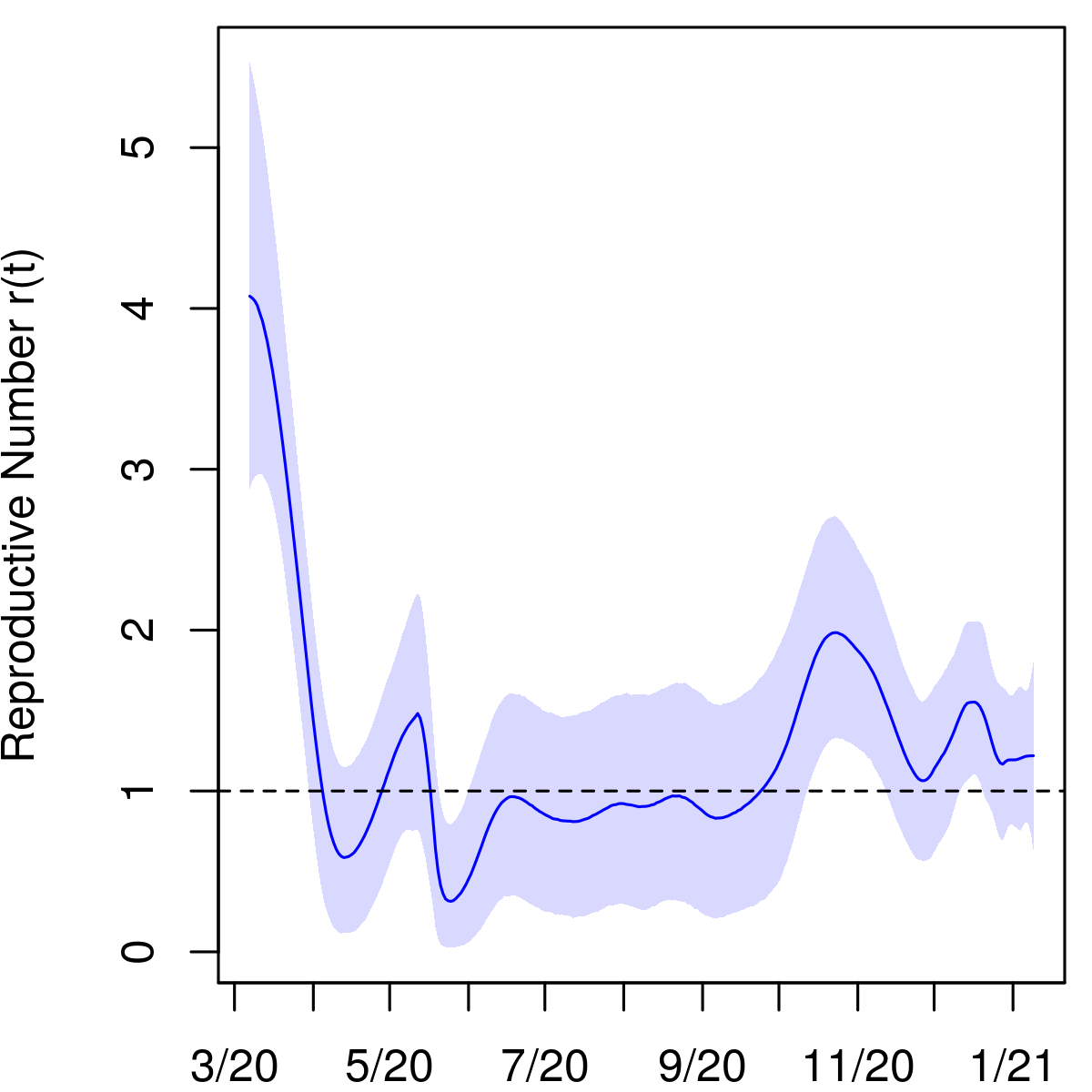}
&
\includegraphics[scale=0.77]{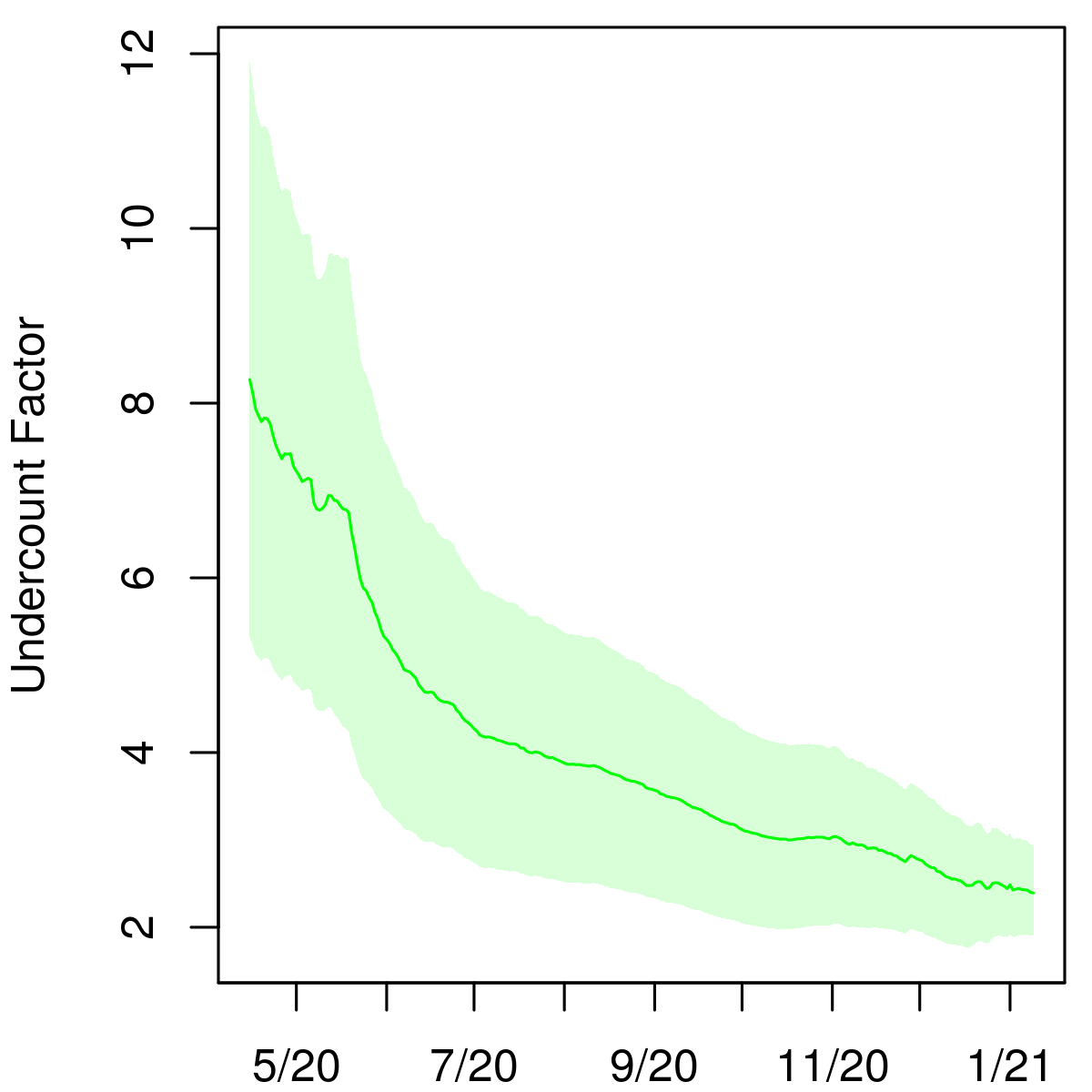} 
\end{tabular}
\caption{Posterior median and middle 95\% intervals for daily new infections, cumulative incidence, $r(t)$, and cumulative undercount from March 2020 to January 2021. In the top left panel, deaths divided by the posterior median IFR are plotted in grey for comparison.}
\end{figure}
\newpage
\begin{figure}[htbp!]
\textbf{Michigan}
\centering
\begin{tabular}{ll}
\includegraphics[scale=0.77]{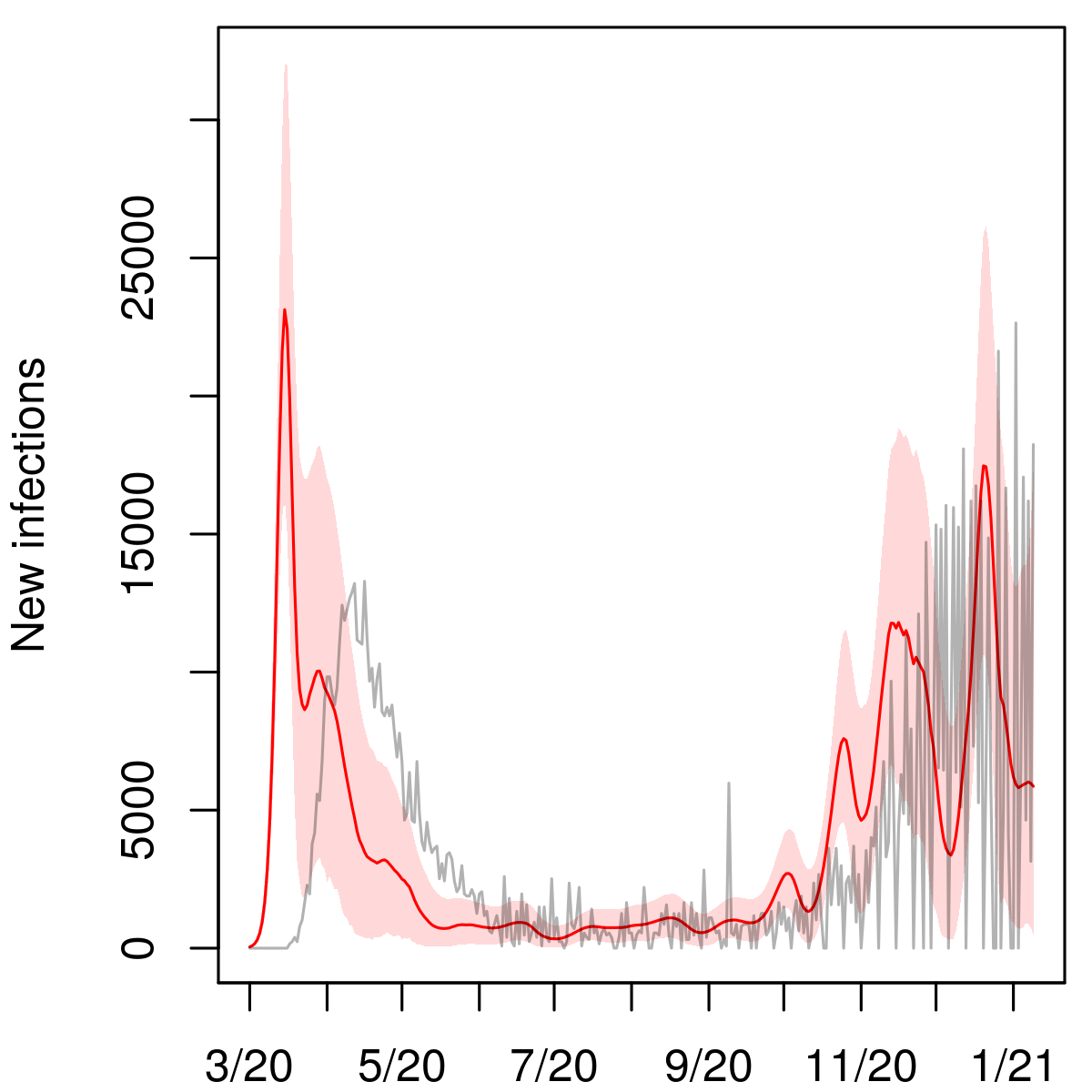}
&
\includegraphics[scale=0.77]{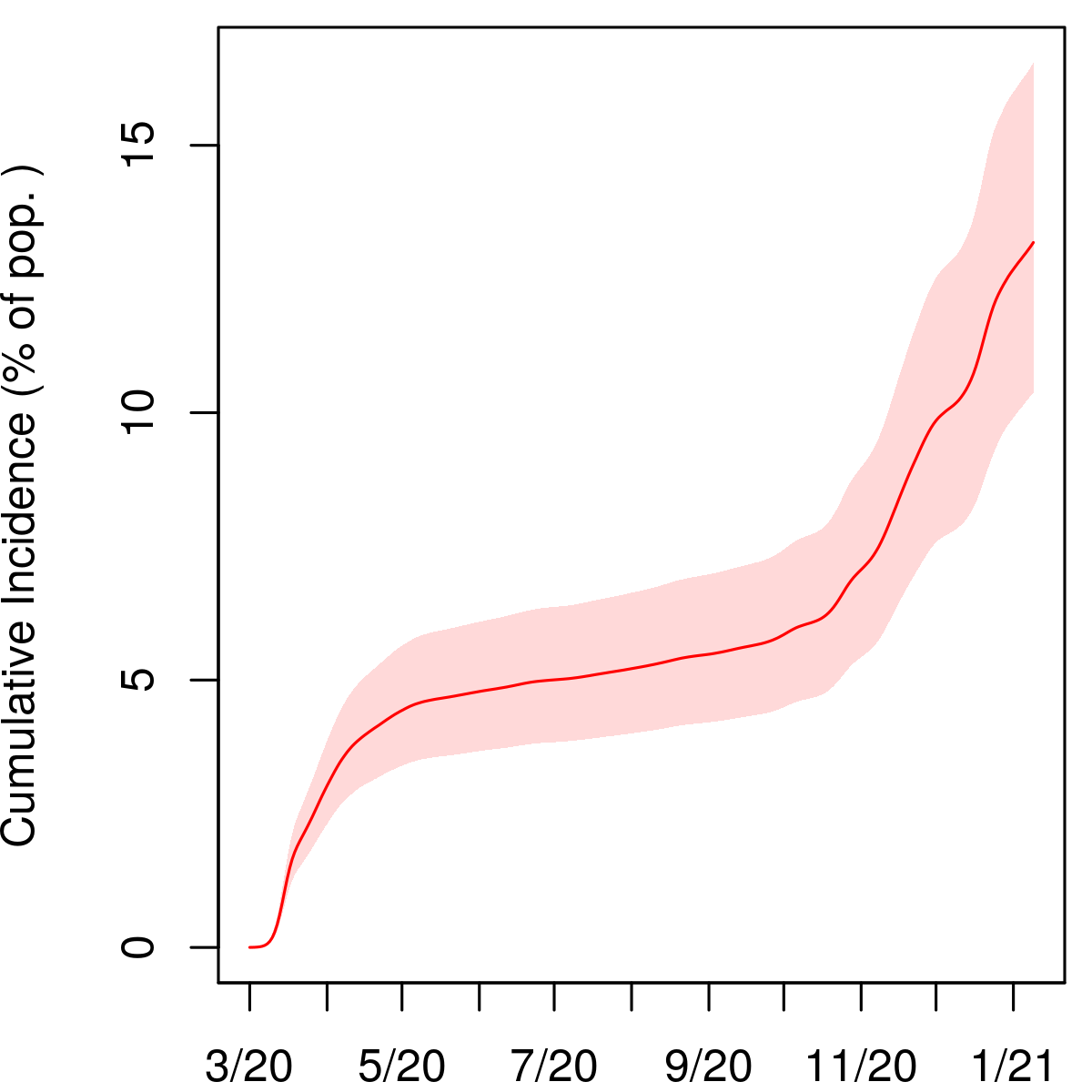} \\
\includegraphics[scale=0.77]{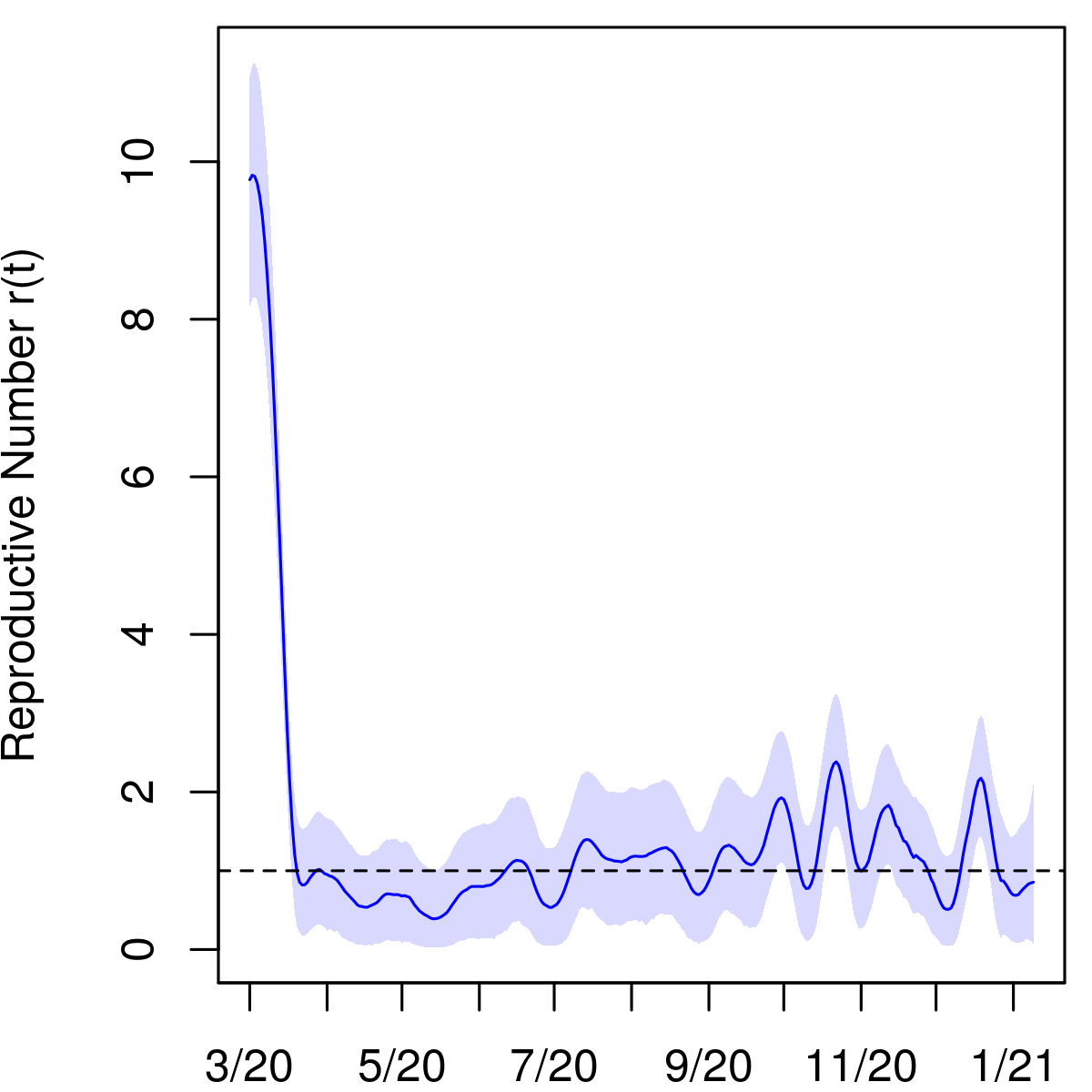}
&
\includegraphics[scale=0.77]{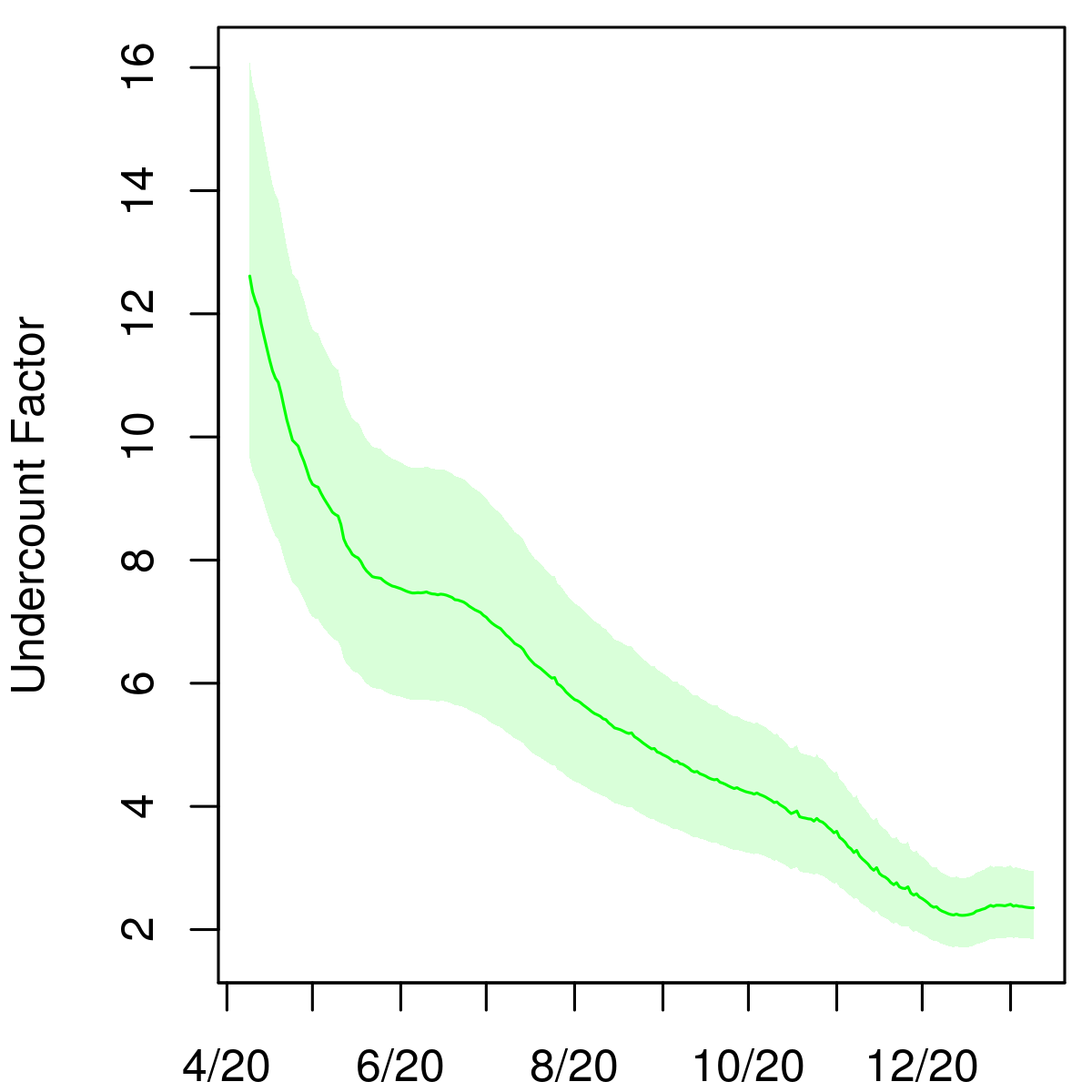} 
\end{tabular}
\caption{Posterior median and middle 95\% intervals for daily new infections, cumulative incidence, $r(t)$, and cumulative undercount from March 2020 to January 2021. In the top left panel, deaths divided by the posterior median IFR are plotted in grey for comparison.}
\end{figure}
\newpage
\begin{figure}[htbp!]
\textbf{Minnesota}
\centering
\begin{tabular}{ll}
\includegraphics[scale=0.77]{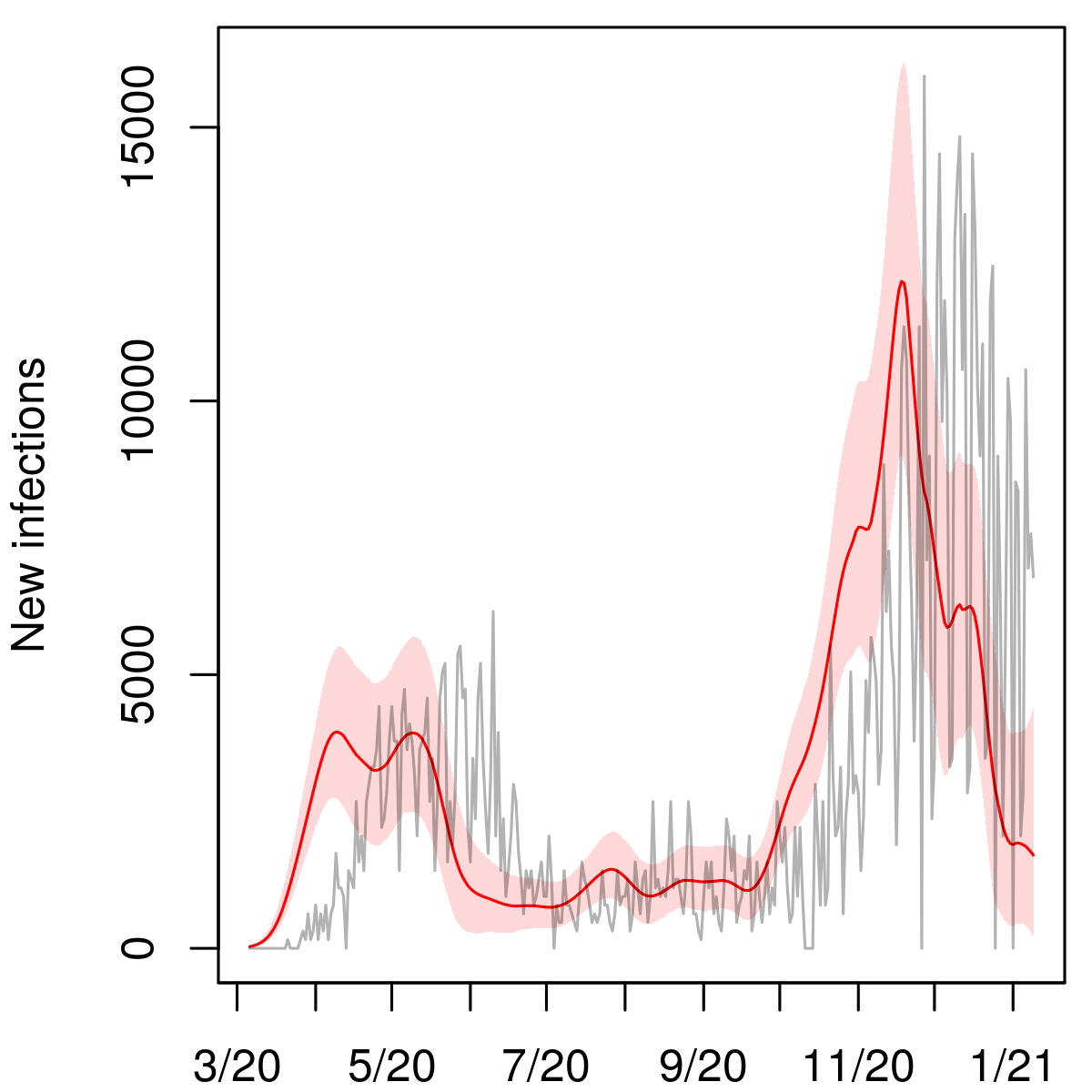}
&
\includegraphics[scale=0.77]{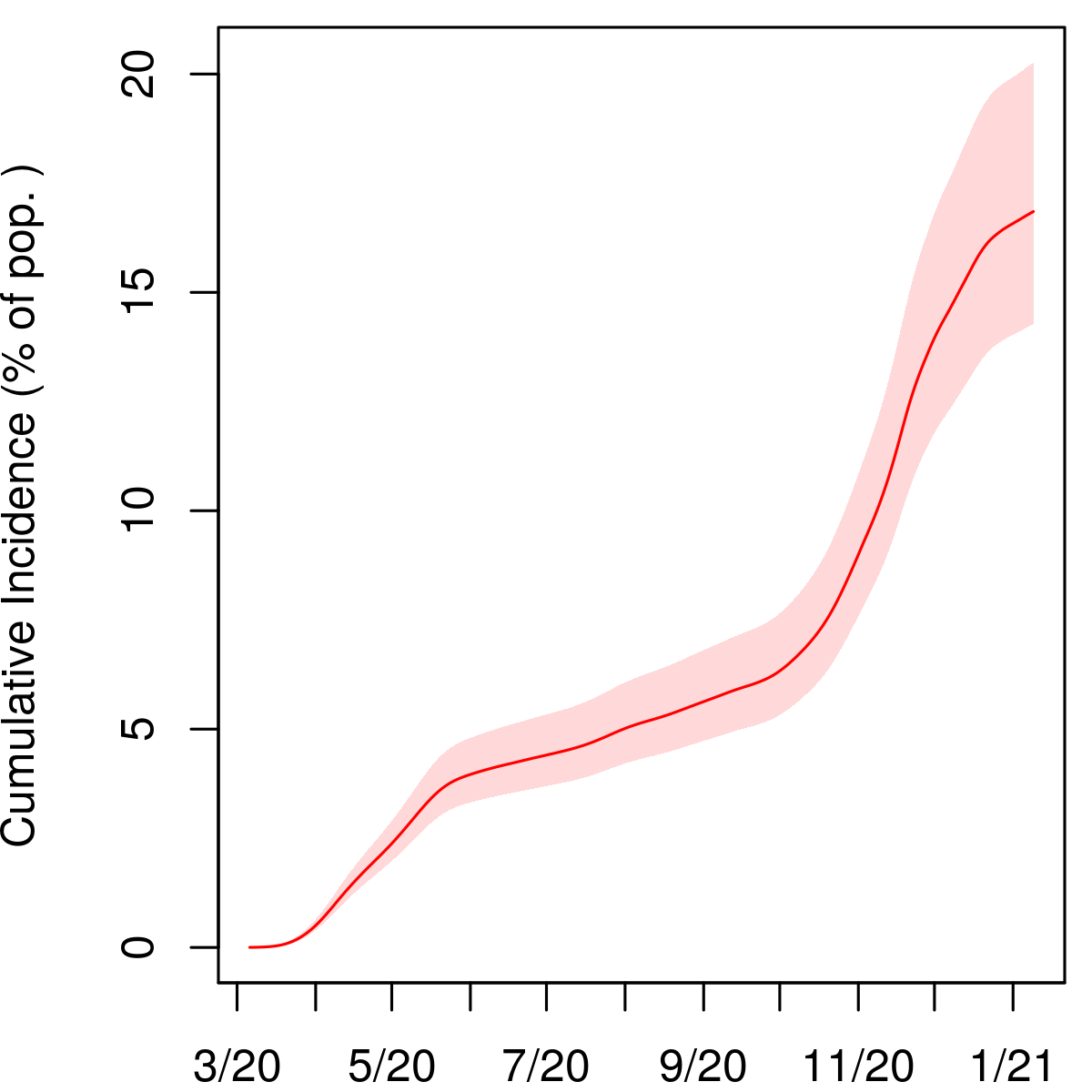} \\
\includegraphics[scale=0.77]{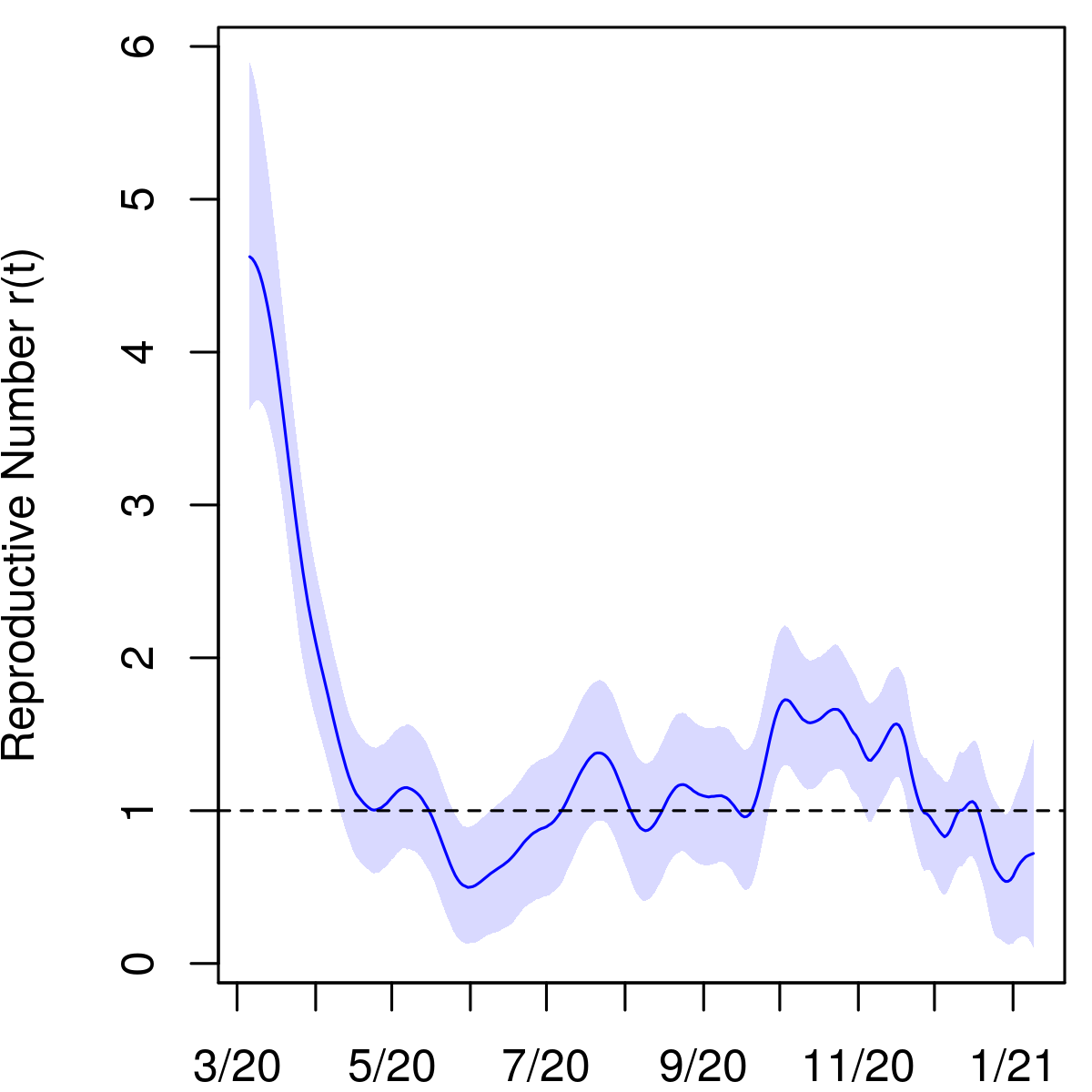}
&
\includegraphics[scale=0.77]{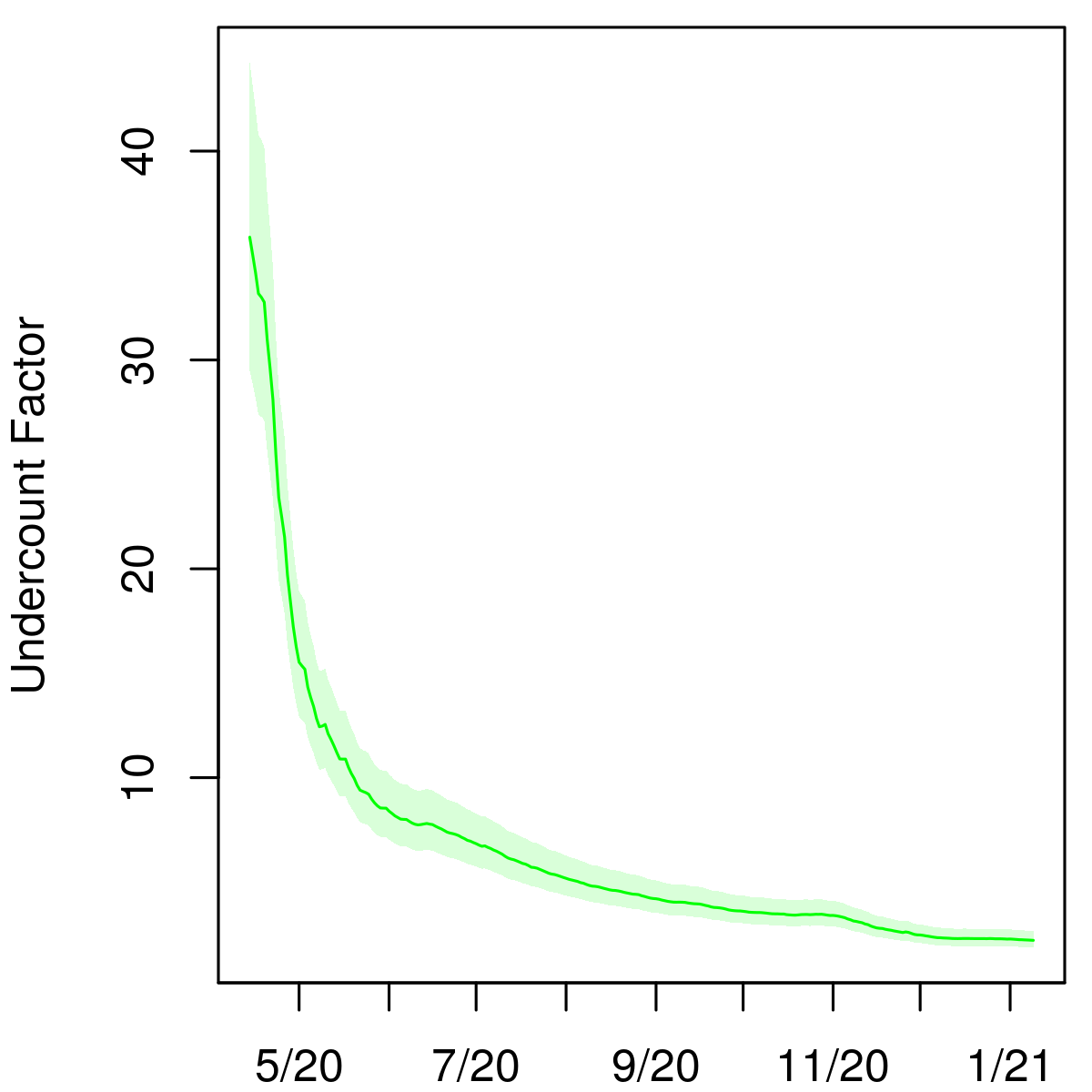} 
\end{tabular}
\caption{Posterior median and middle 95\% intervals for daily new infections, cumulative incidence, $r(t)$, and cumulative undercount from March 2020 to January 2021. In the top left panel, deaths divided by the posterior median IFR are plotted in grey for comparison.}
\end{figure}
\newpage
\begin{figure}[htbp!]
\textbf{Missouri}
\centering
\begin{tabular}{ll}
\includegraphics[scale=0.77]{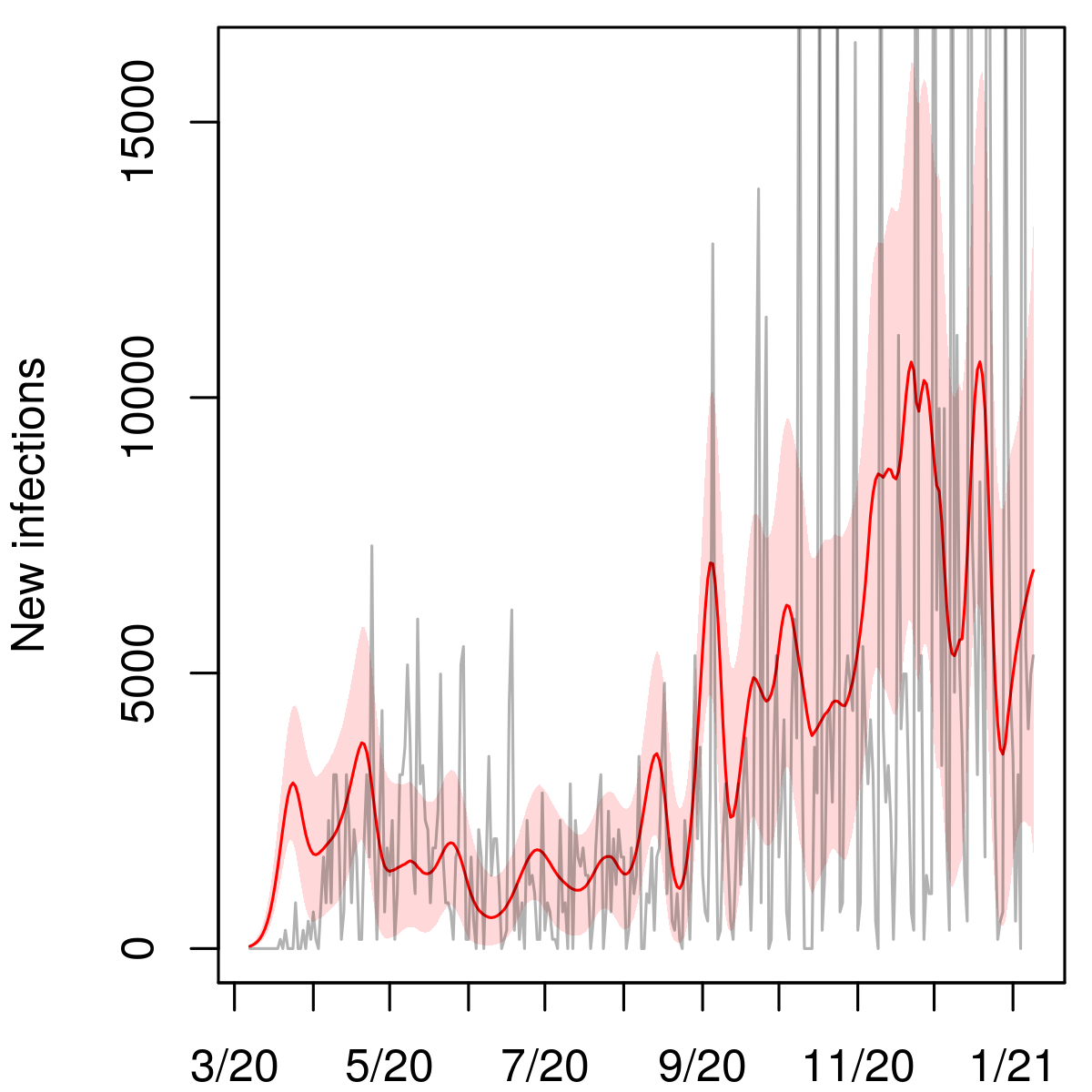}
&
\includegraphics[scale=0.77]{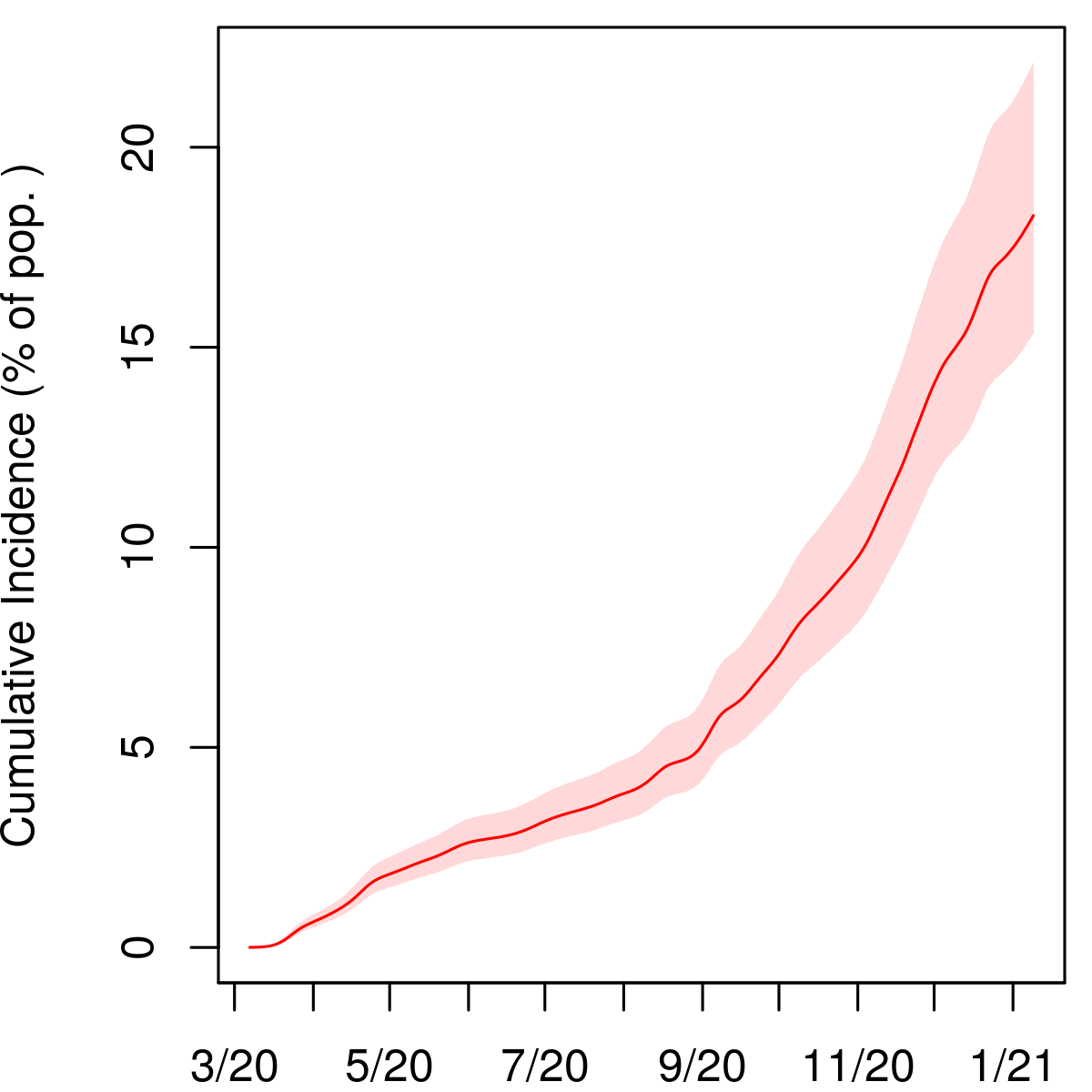} \\
\includegraphics[scale=0.77]{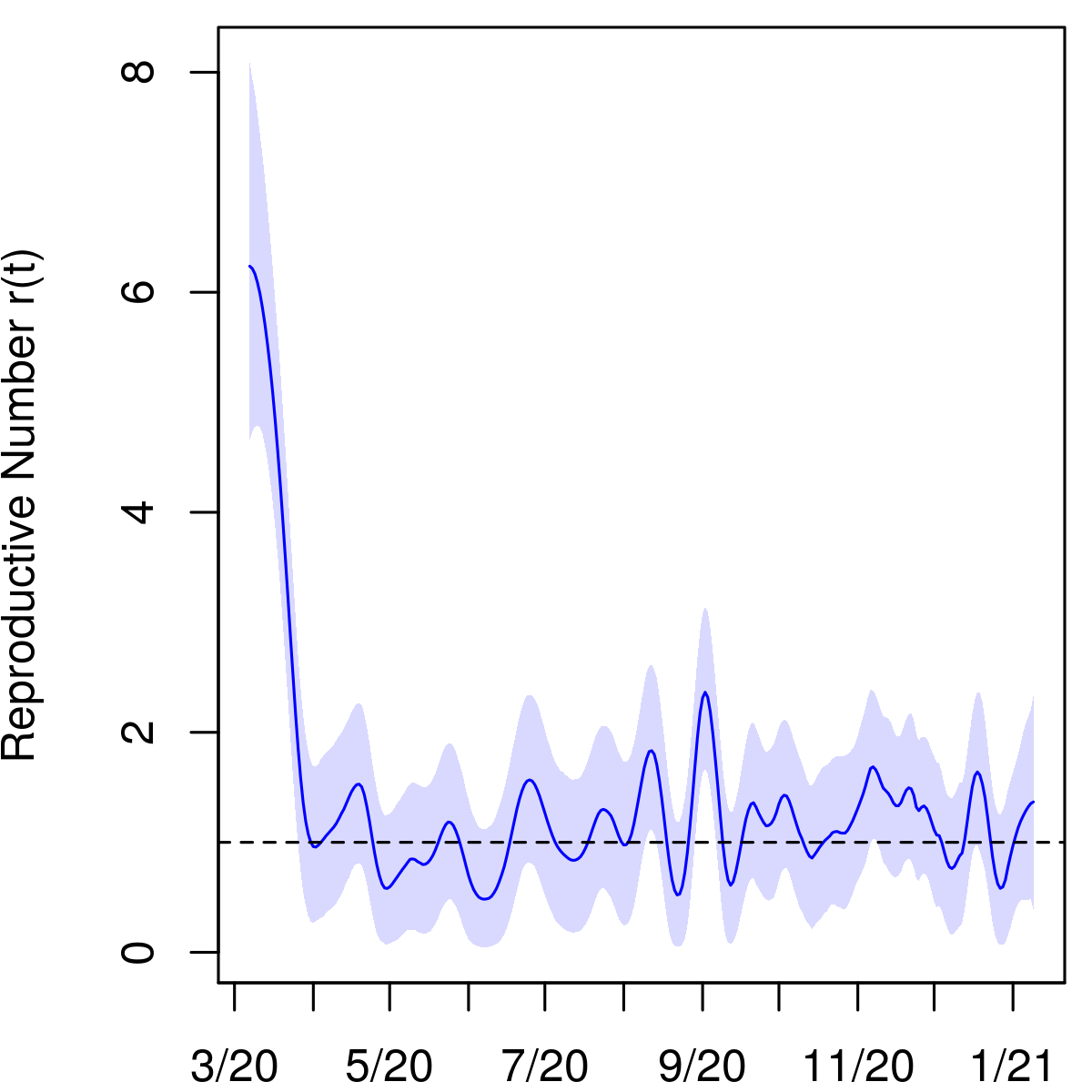}
&
\includegraphics[scale=0.77]{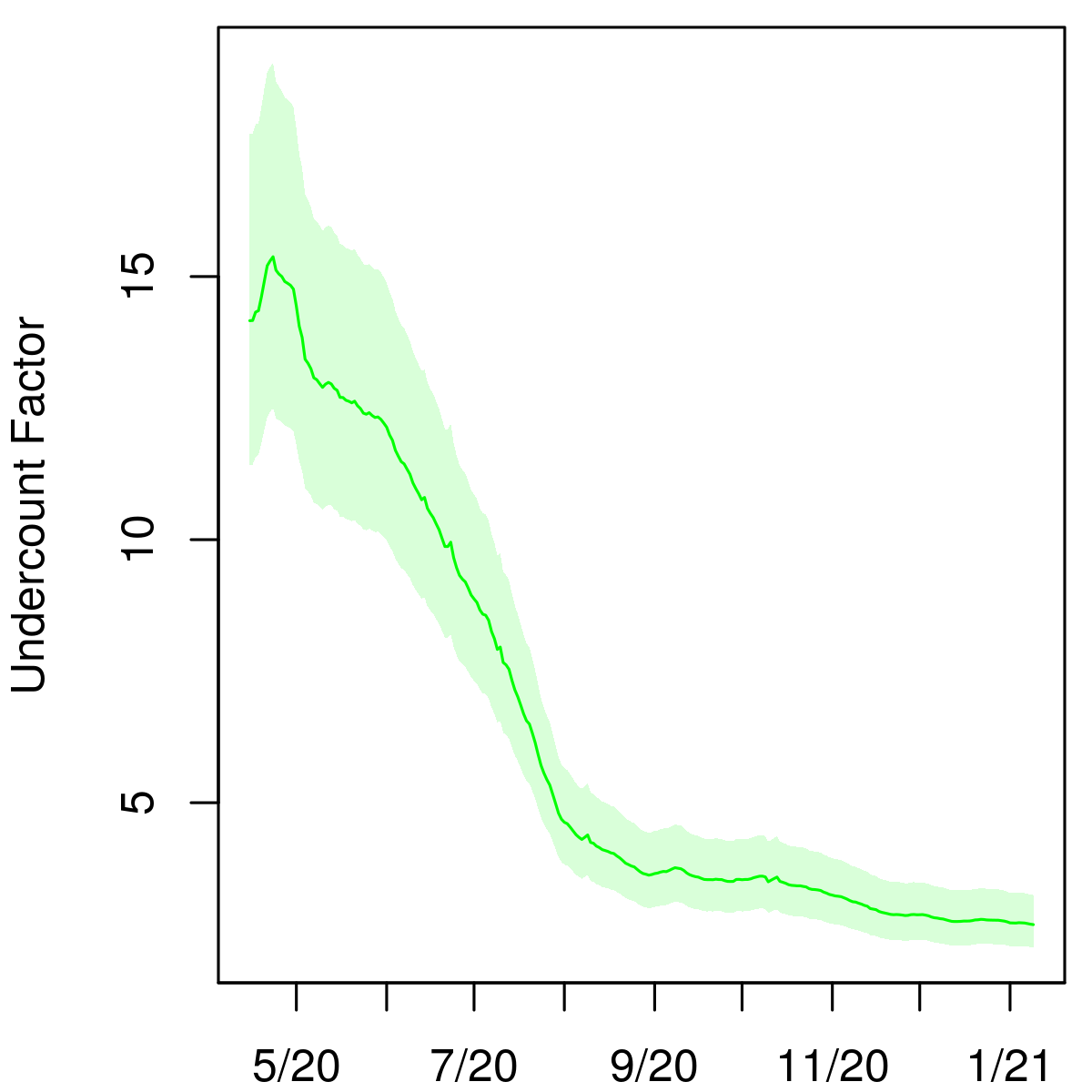} 
\end{tabular}
\caption{Posterior median and middle 95\% intervals for daily new infections, cumulative incidence, $r(t)$, and cumulative undercount from March 2020 to January 2021. In the top left panel, deaths divided by the posterior median IFR are plotted in grey for comparison.}
\end{figure}
\newpage
\begin{figure}[htbp!]
\textbf{Mississippi}
\centering
\begin{tabular}{ll}
\includegraphics[scale=0.77]{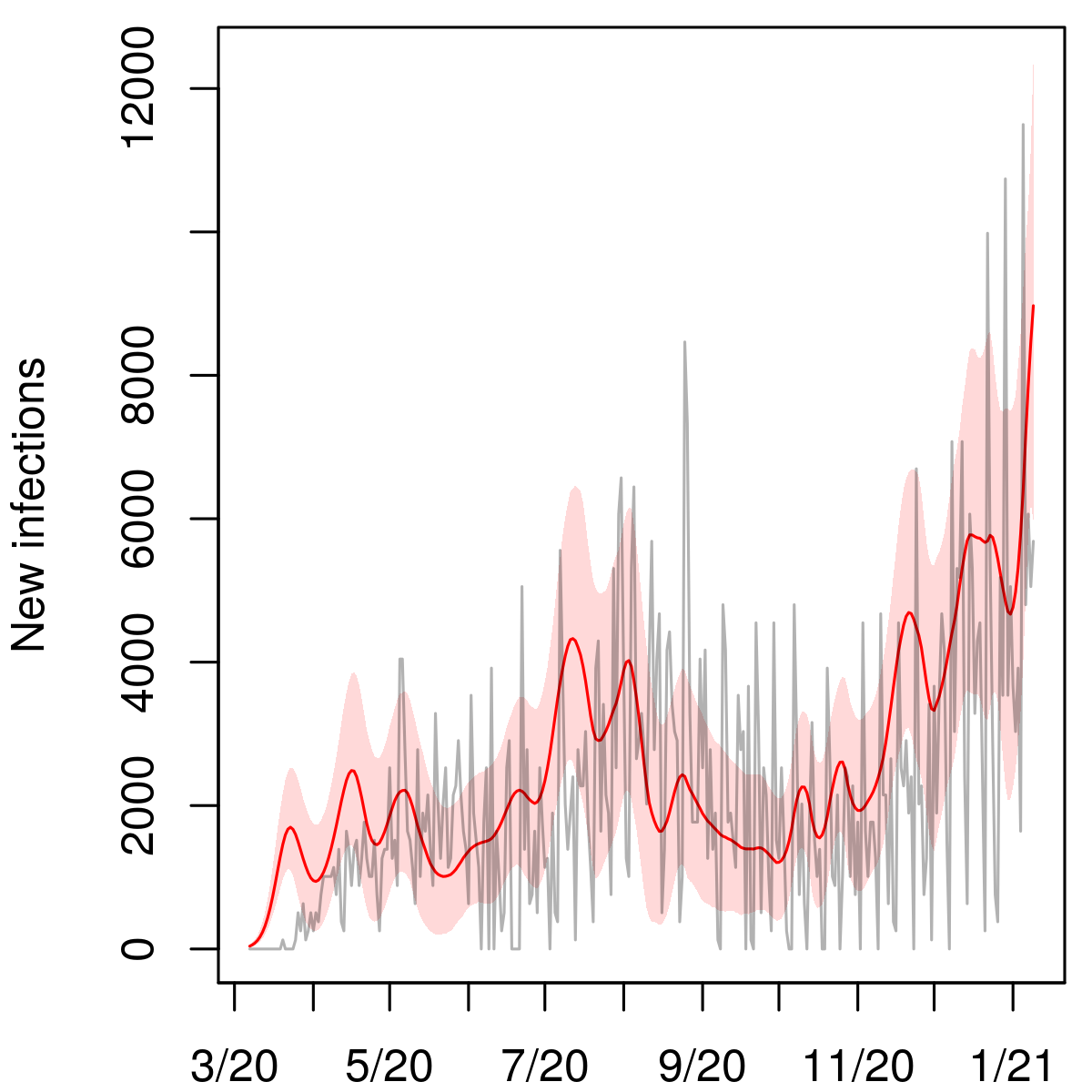}
&
\includegraphics[scale=0.77]{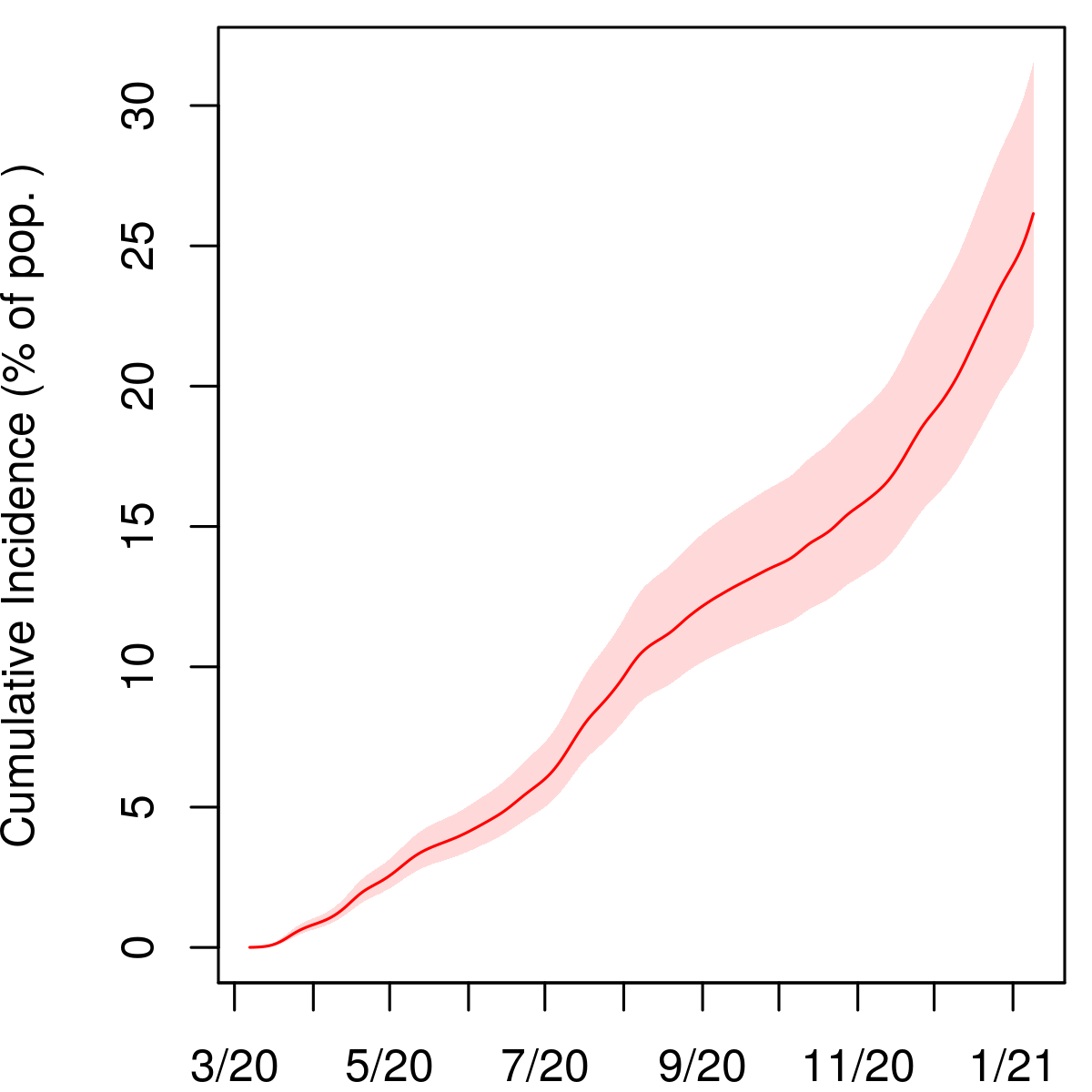} \\
\includegraphics[scale=0.77]{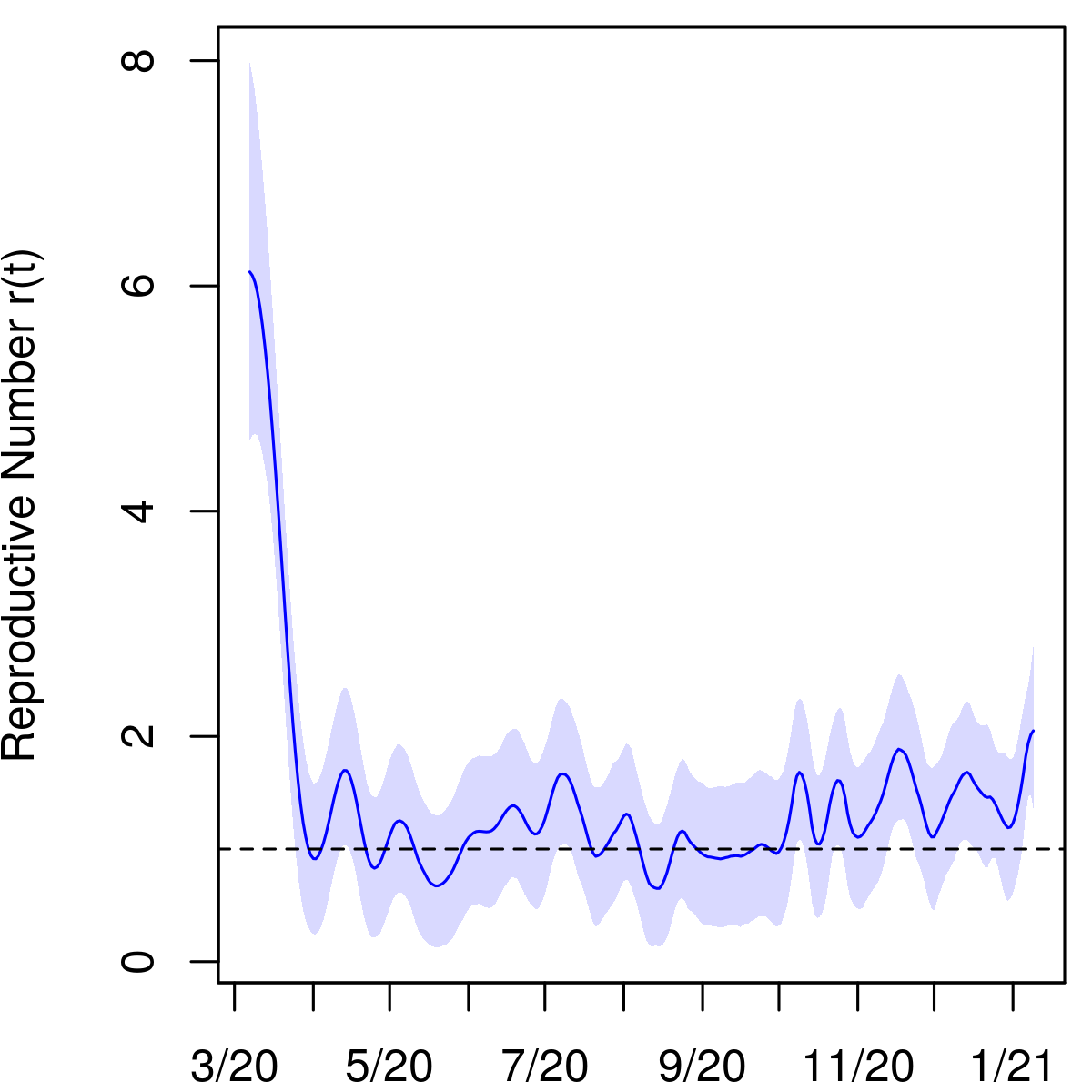}
&
\includegraphics[scale=0.77]{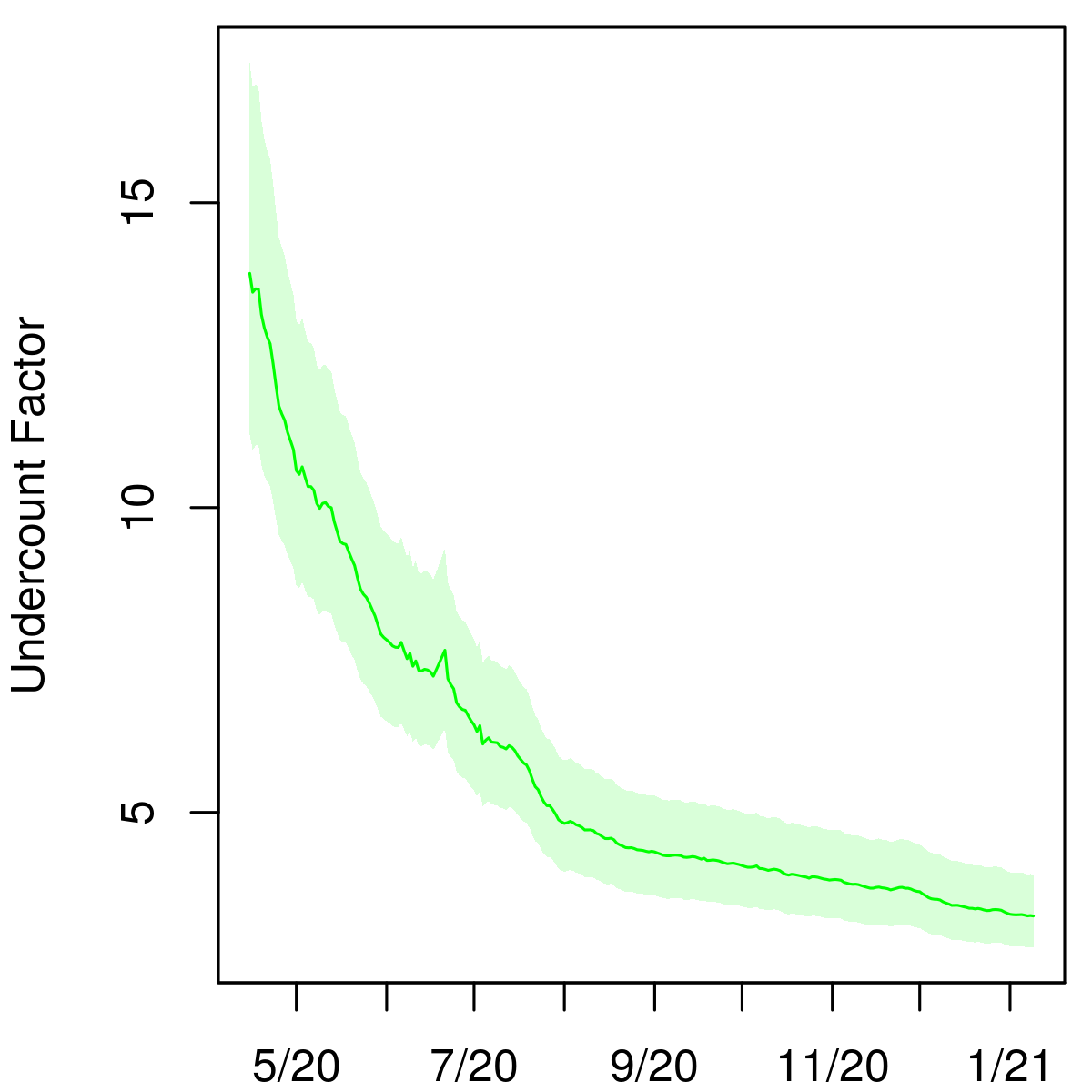} 
\end{tabular}
\caption{Posterior median and middle 95\% intervals for daily new infections, cumulative incidence, $r(t)$, and cumulative undercount from March 2020 to January 2021. In the top left panel, deaths divided by the posterior median IFR are plotted in grey for comparison.}
\end{figure}
\newpage
\begin{figure}[htbp!]
\textbf{Montana}
\centering
\begin{tabular}{ll}
\includegraphics[scale=0.77]{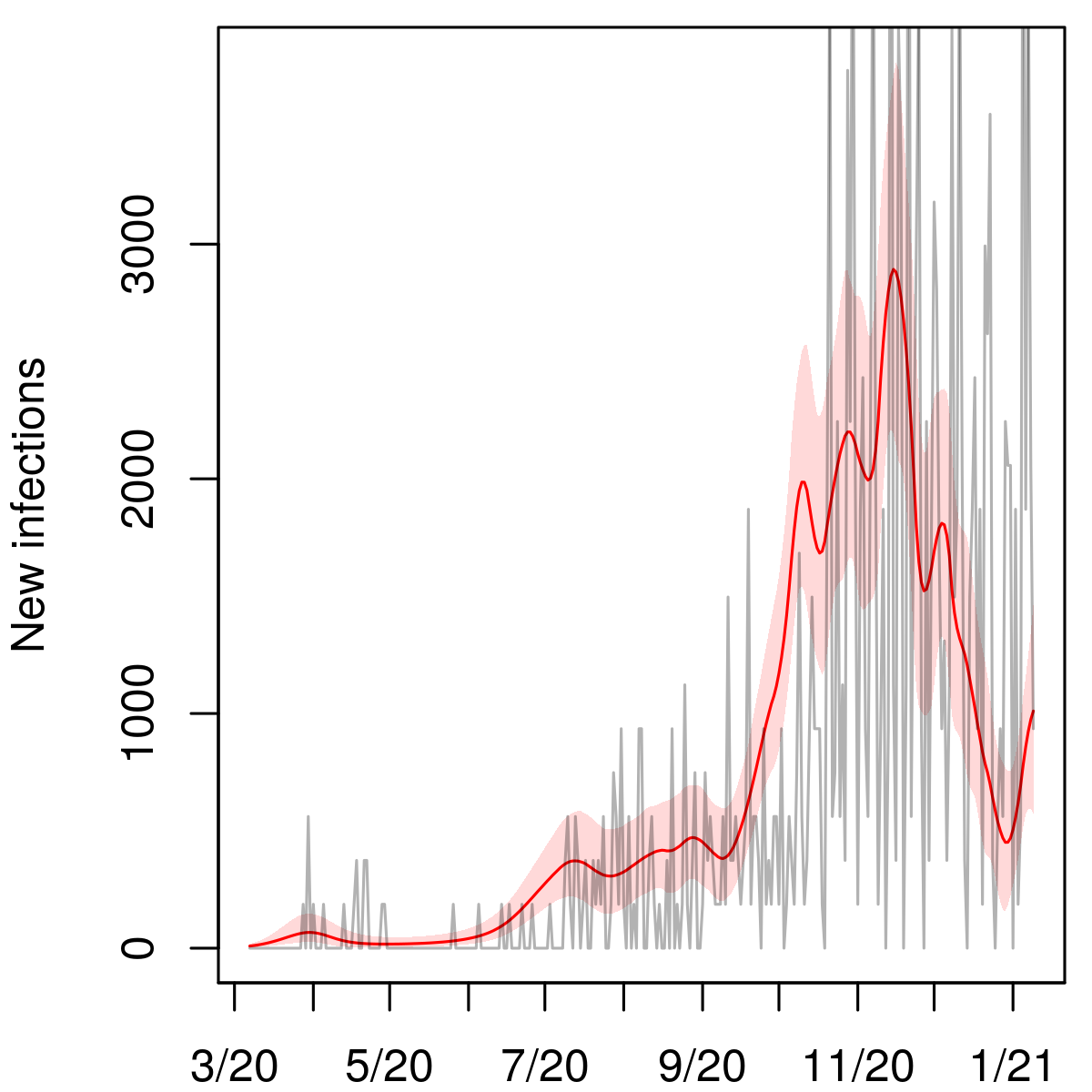}
&
\includegraphics[scale=0.77]{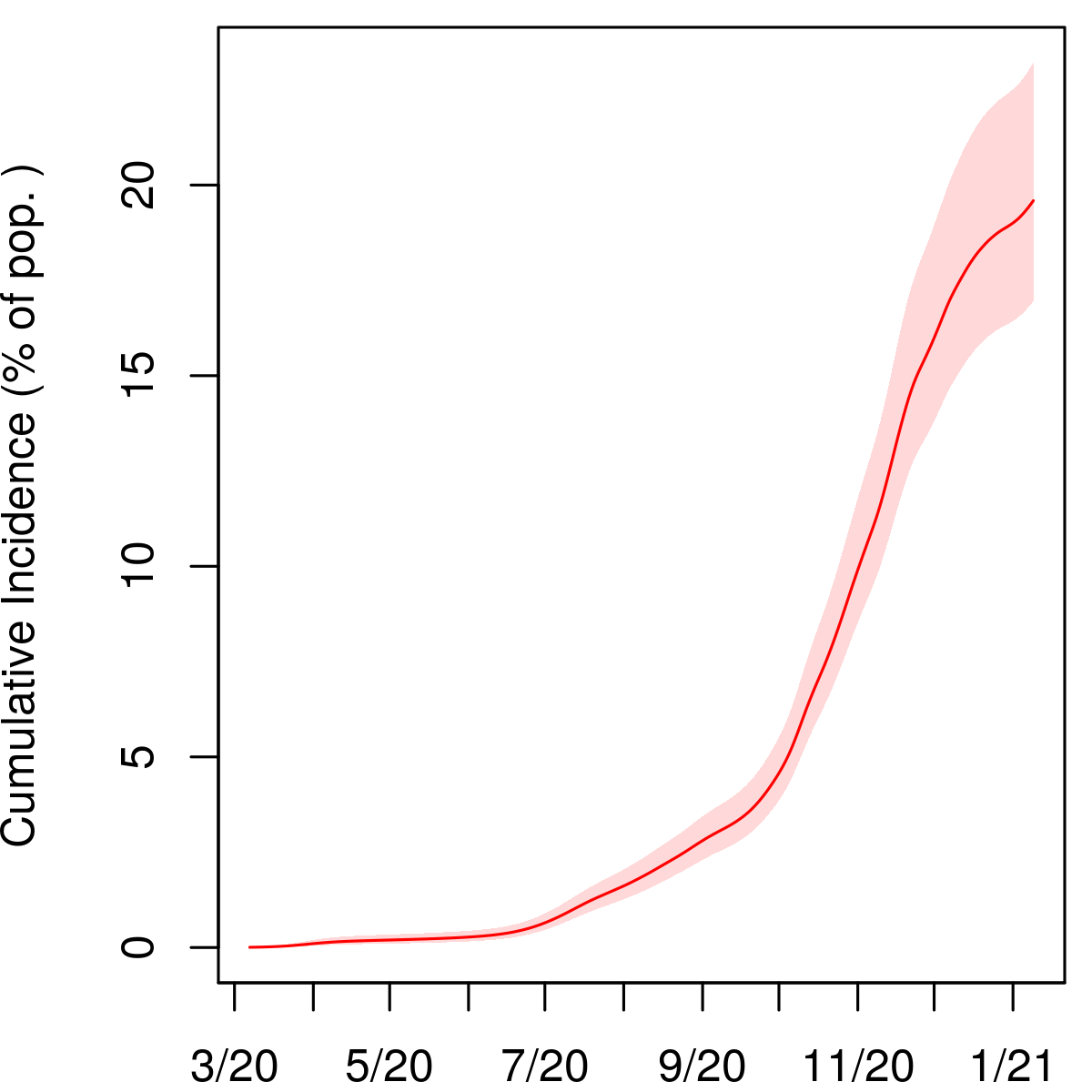} \\
\includegraphics[scale=0.77]{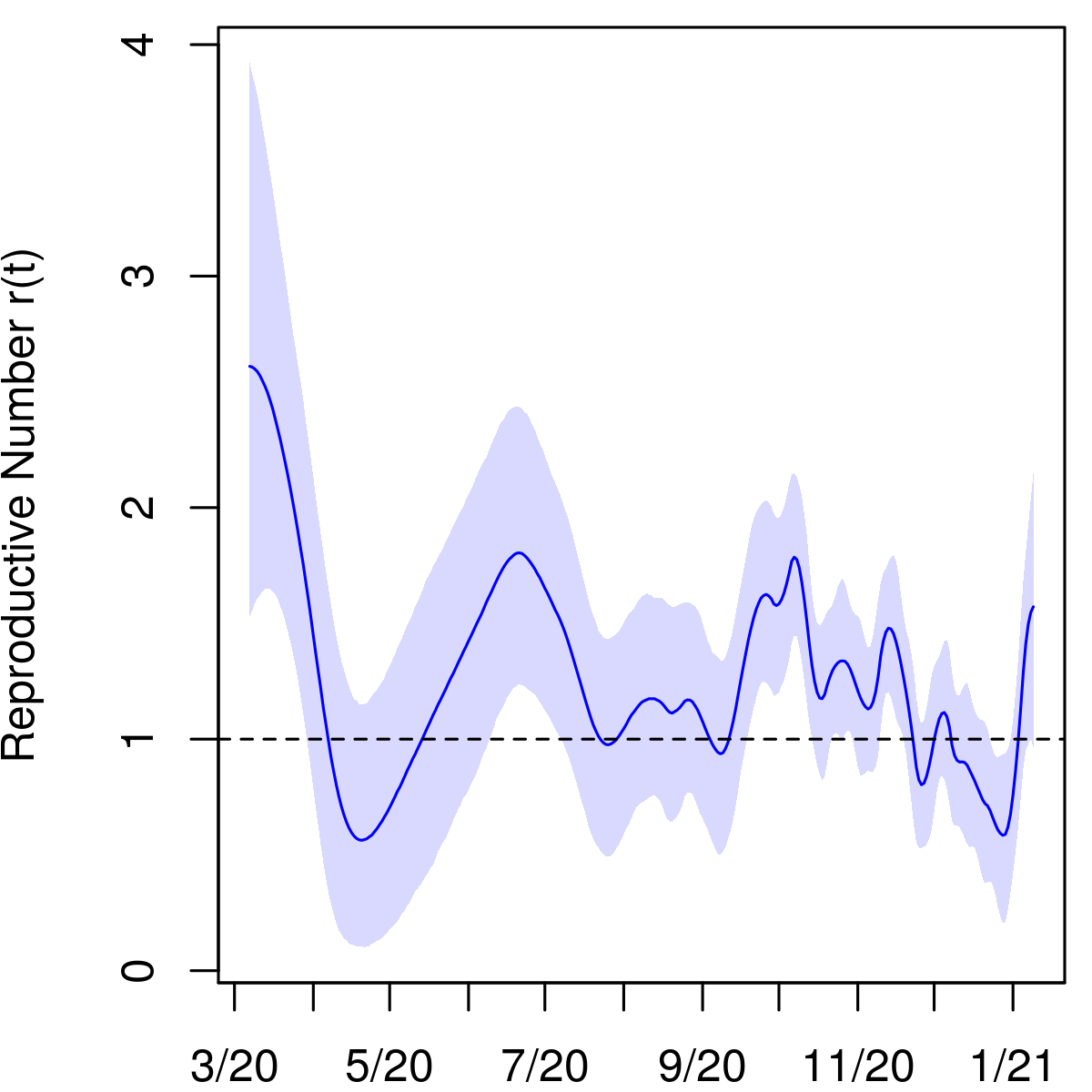}
&
\includegraphics[scale=0.77]{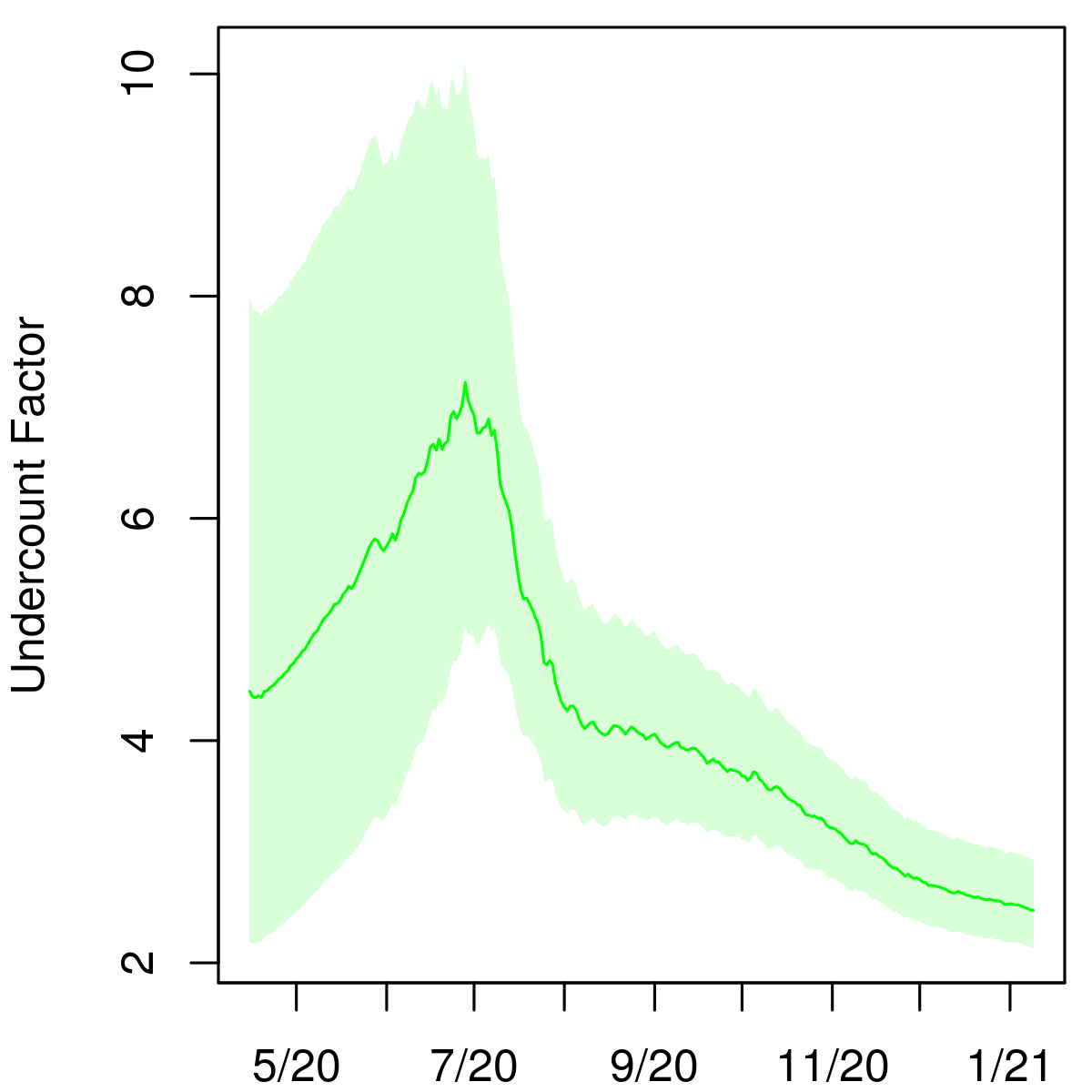} 
\end{tabular}
\caption{Posterior median and middle 95\% intervals for daily new infections, cumulative incidence, $r(t)$, and cumulative undercount from March 2020 to January 2021. In the top left panel, deaths divided by the posterior median IFR are plotted in grey for comparison.}
\end{figure}
\newpage
\begin{figure}[htbp!]
\textbf{North Carolina}
\centering
\begin{tabular}{ll}
\includegraphics[scale=0.77]{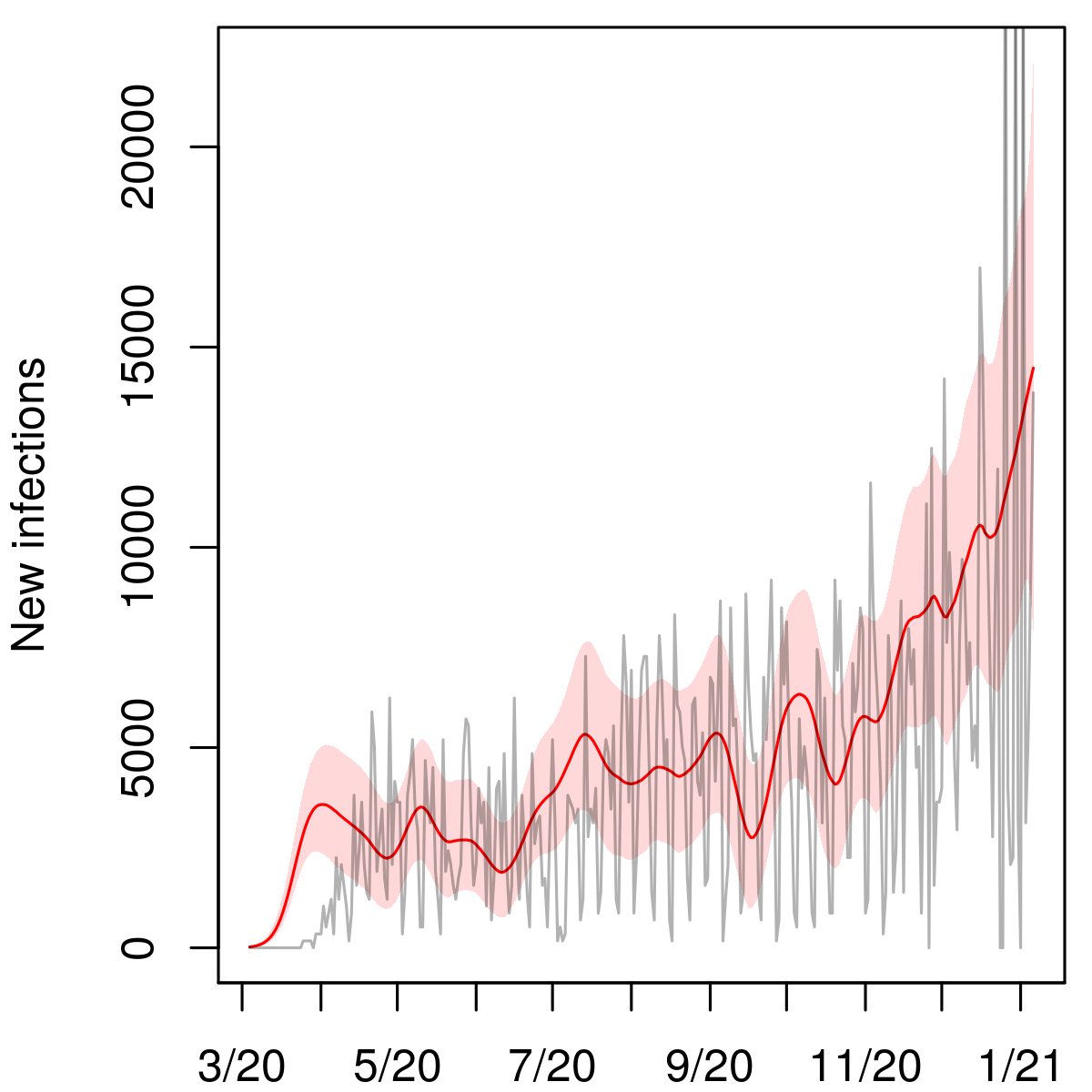}
&
\includegraphics[scale=0.77]{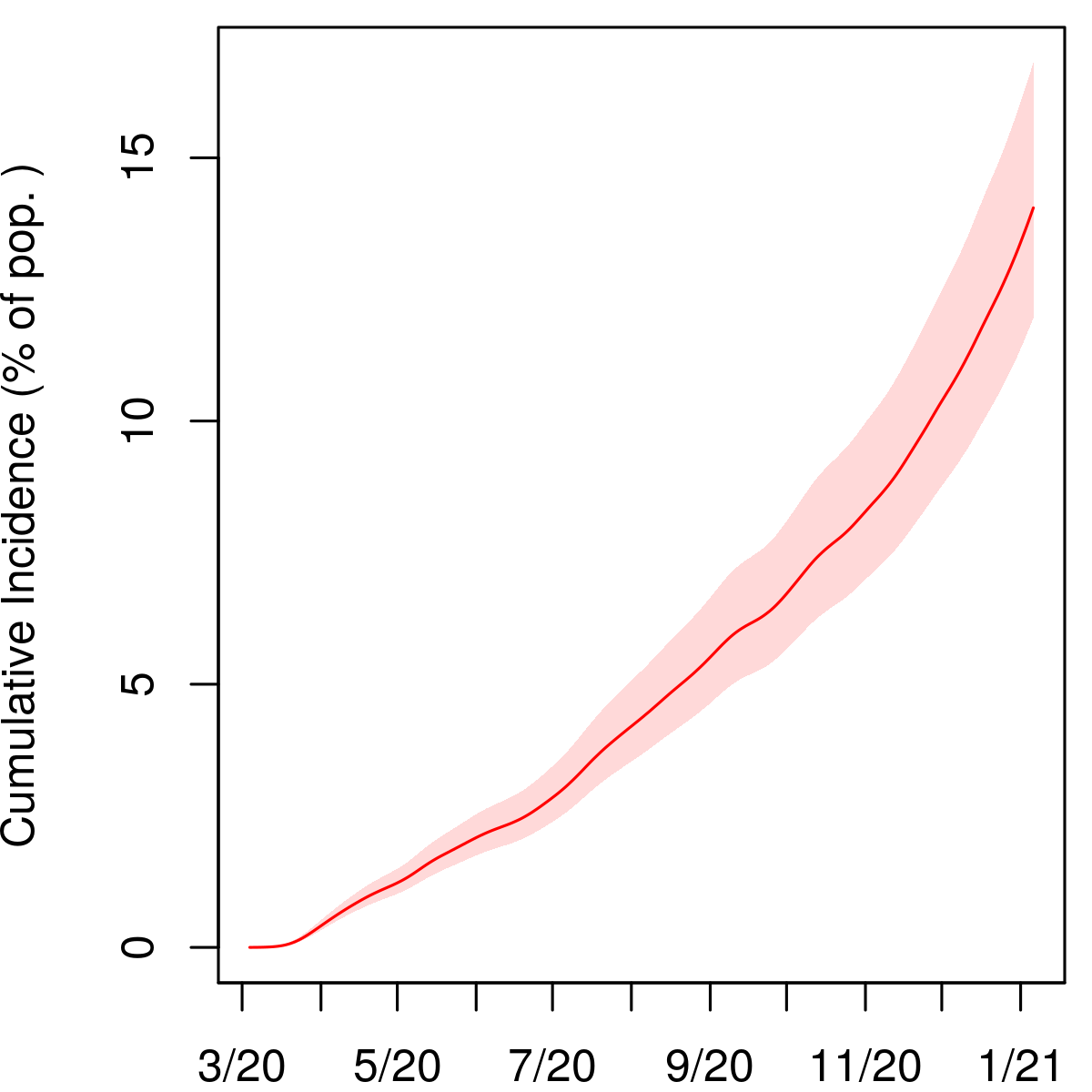} \\
\includegraphics[scale=0.77]{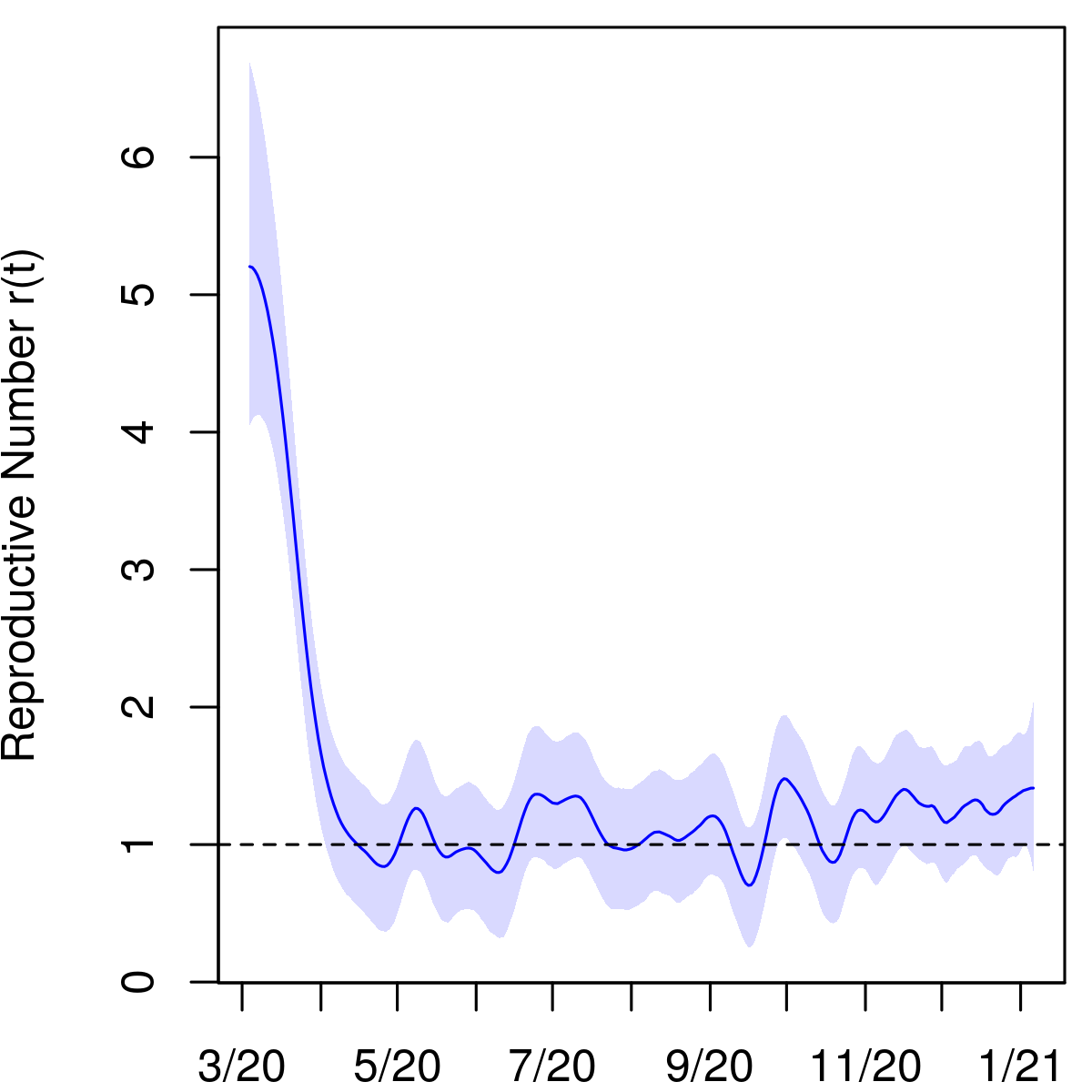}
&
\includegraphics[scale=0.77]{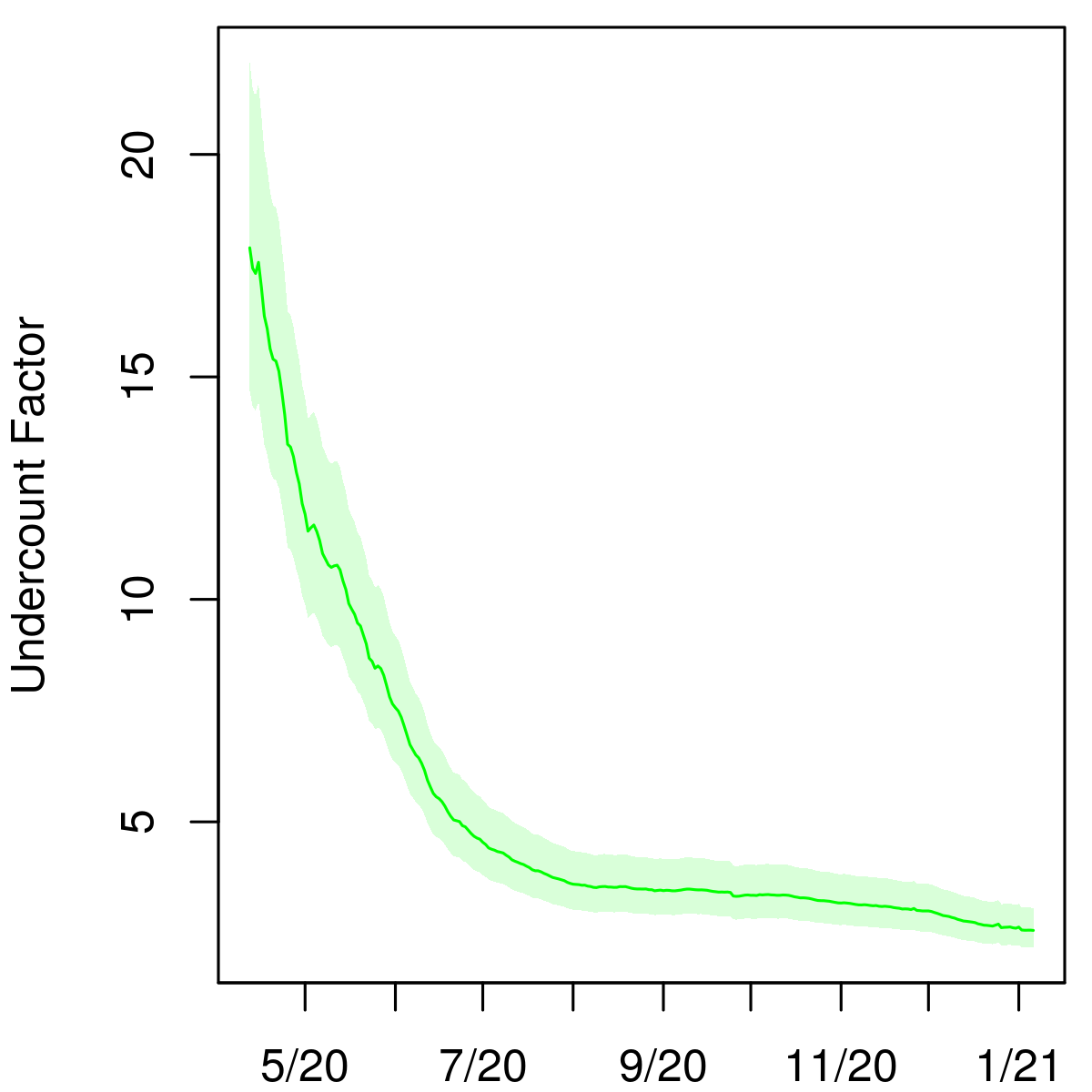} 
\end{tabular}
\caption{Posterior median and middle 95\% intervals for daily new infections, cumulative incidence, $r(t)$, and cumulative undercount from March 2020 to January 2021. In the top left panel, deaths divided by the posterior median IFR are plotted in grey for comparison.}
\end{figure}
\newpage
\begin{figure}[htbp!]
\textbf{North Dakota}
\centering
\begin{tabular}{ll}
\includegraphics[scale=0.77]{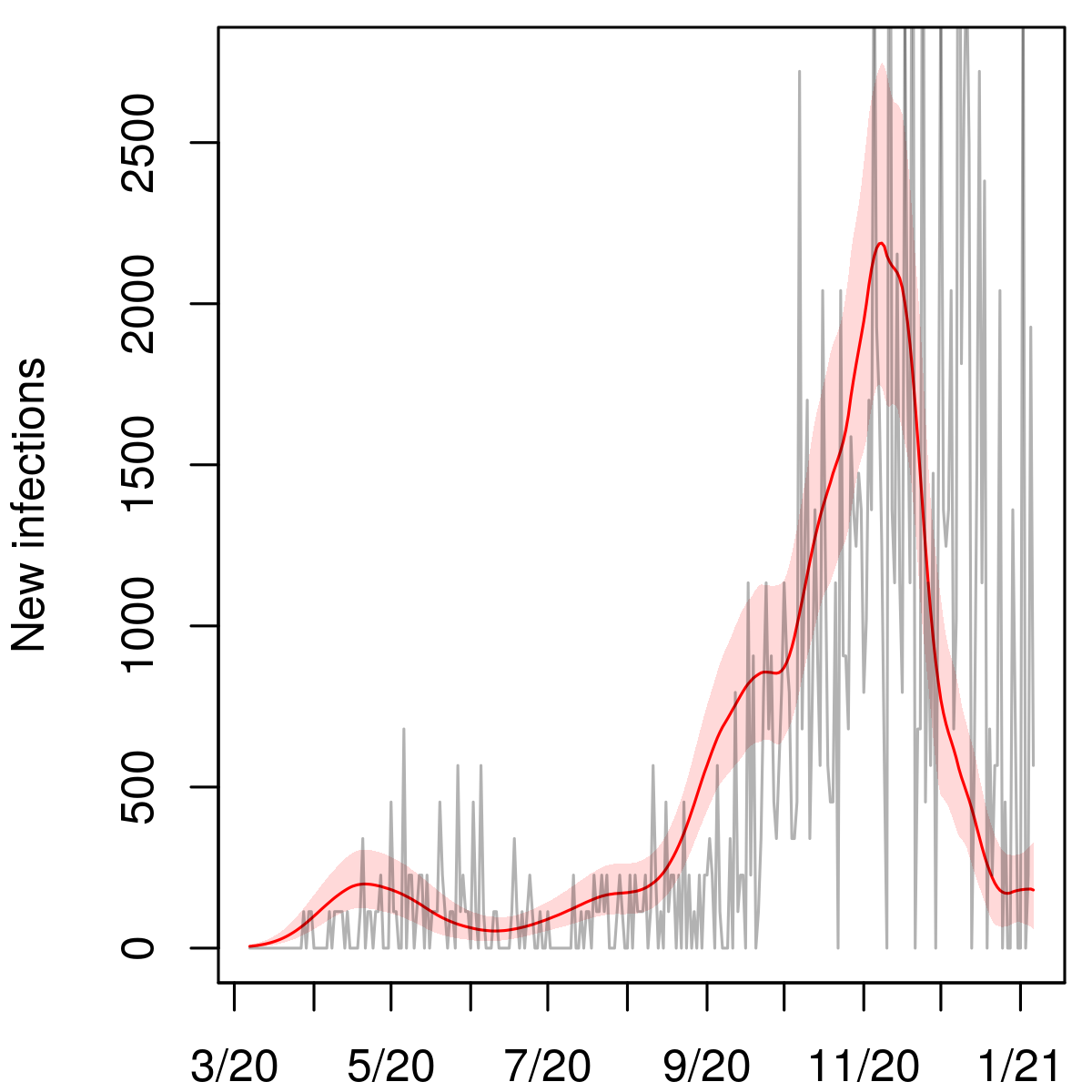}
&
\includegraphics[scale=0.77]{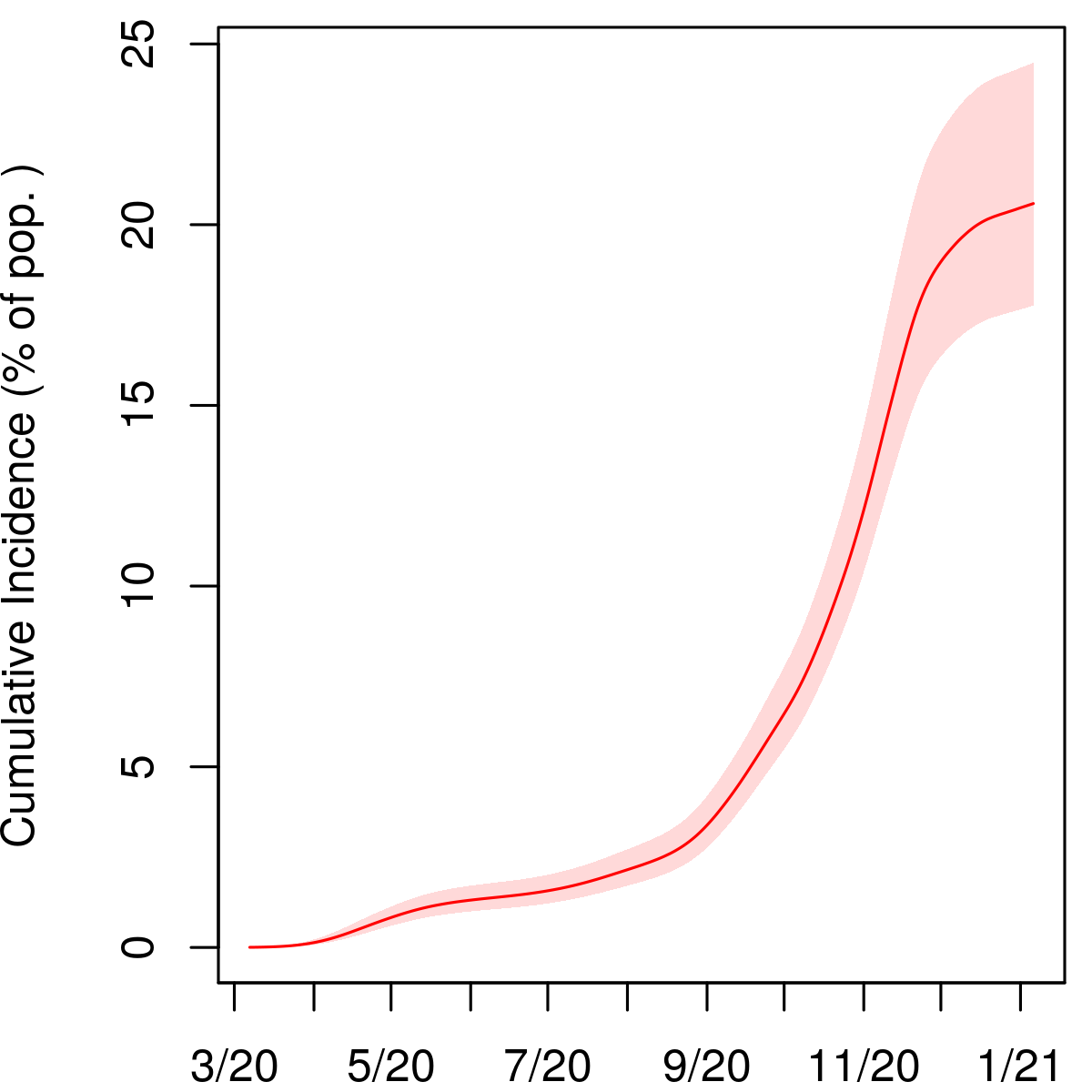} \\
\includegraphics[scale=0.77]{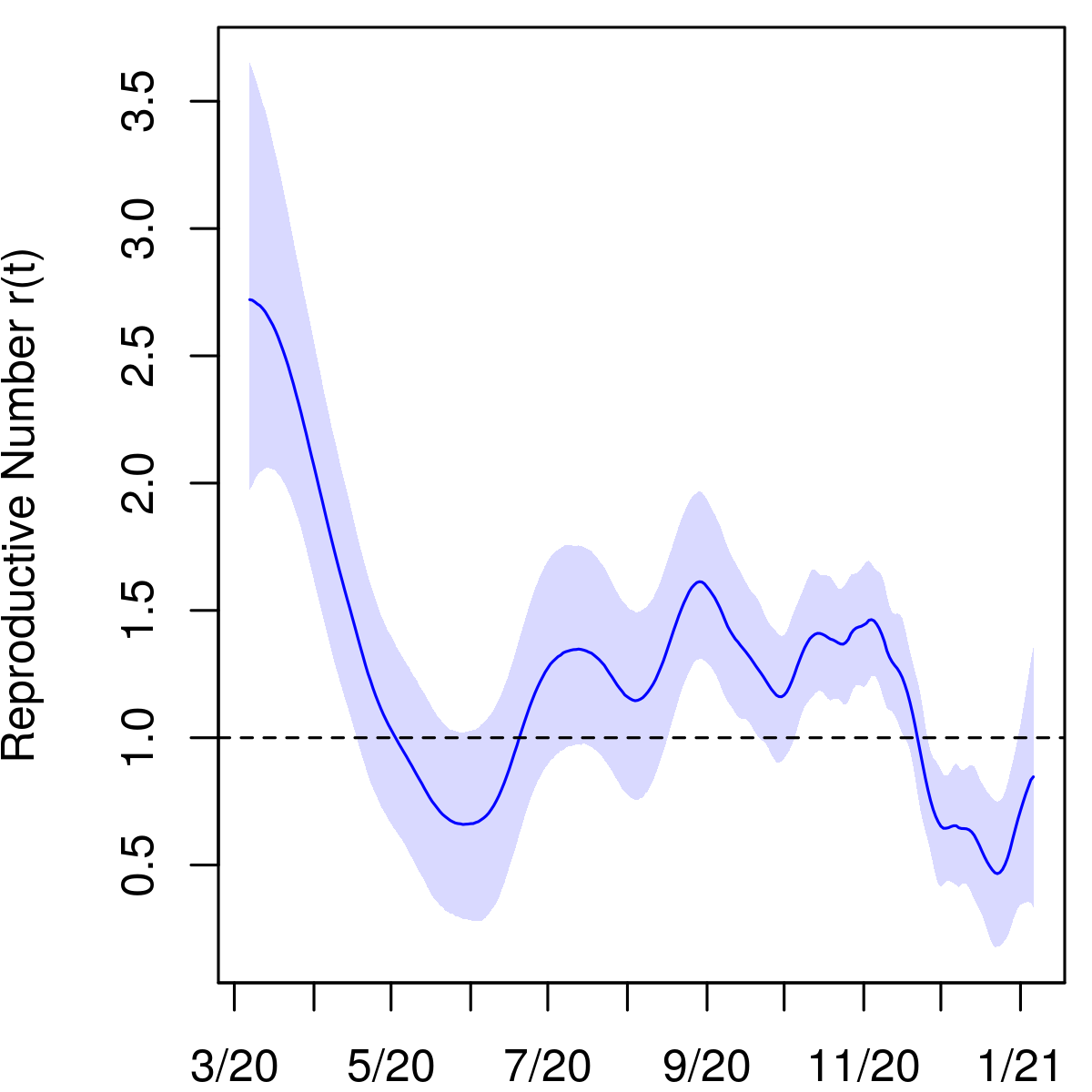}
&
\includegraphics[scale=0.77]{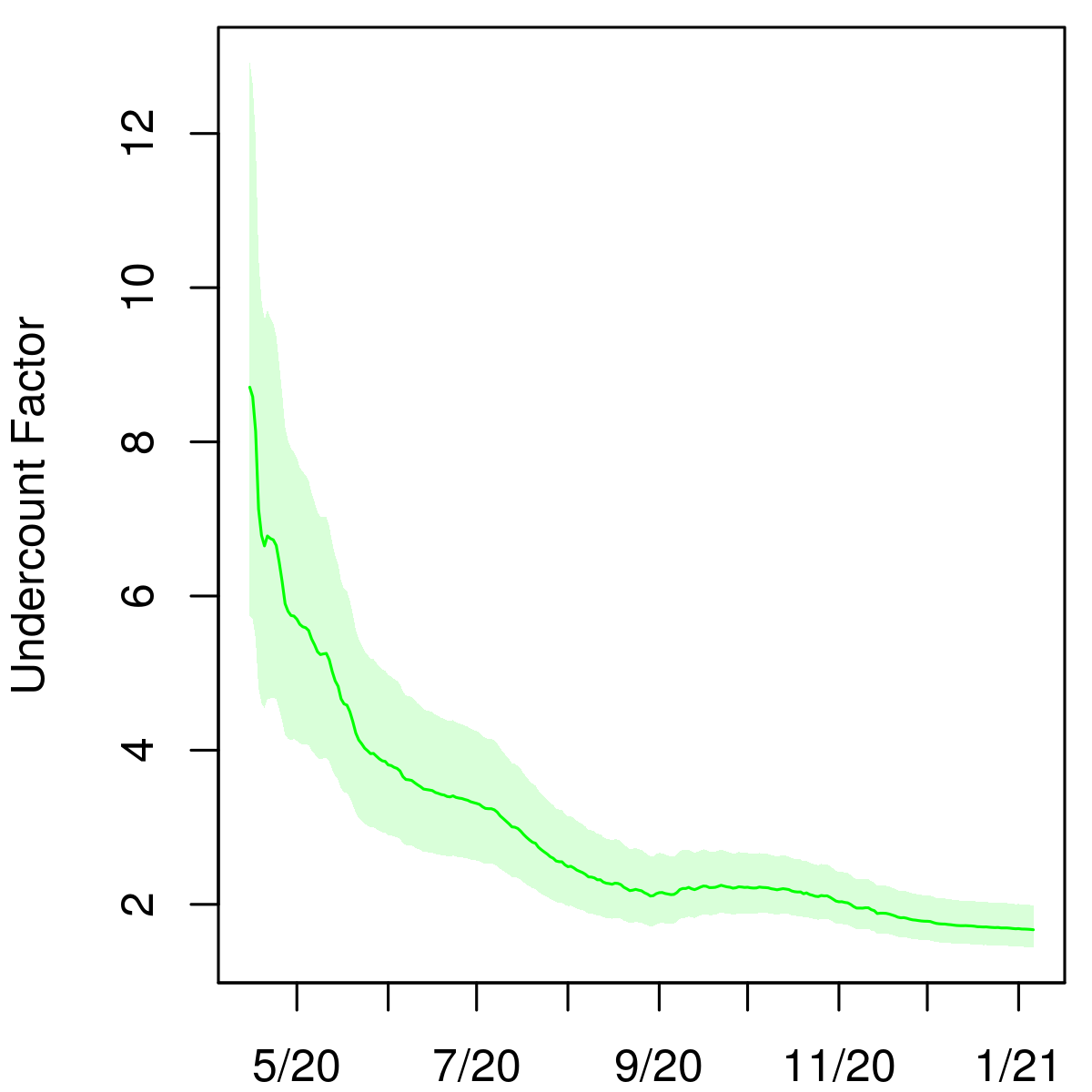} 
\end{tabular}
\caption{Posterior median and middle 95\% intervals for daily new infections, cumulative incidence, $r(t)$, and cumulative undercount from March 2020 to January 2021. In the top left panel, deaths divided by the posterior median IFR are plotted in grey for comparison.}
\end{figure}
\newpage
\begin{figure}[htbp!]
\textbf{Nebraska}
\centering
\begin{tabular}{ll}
\includegraphics[scale=0.77]{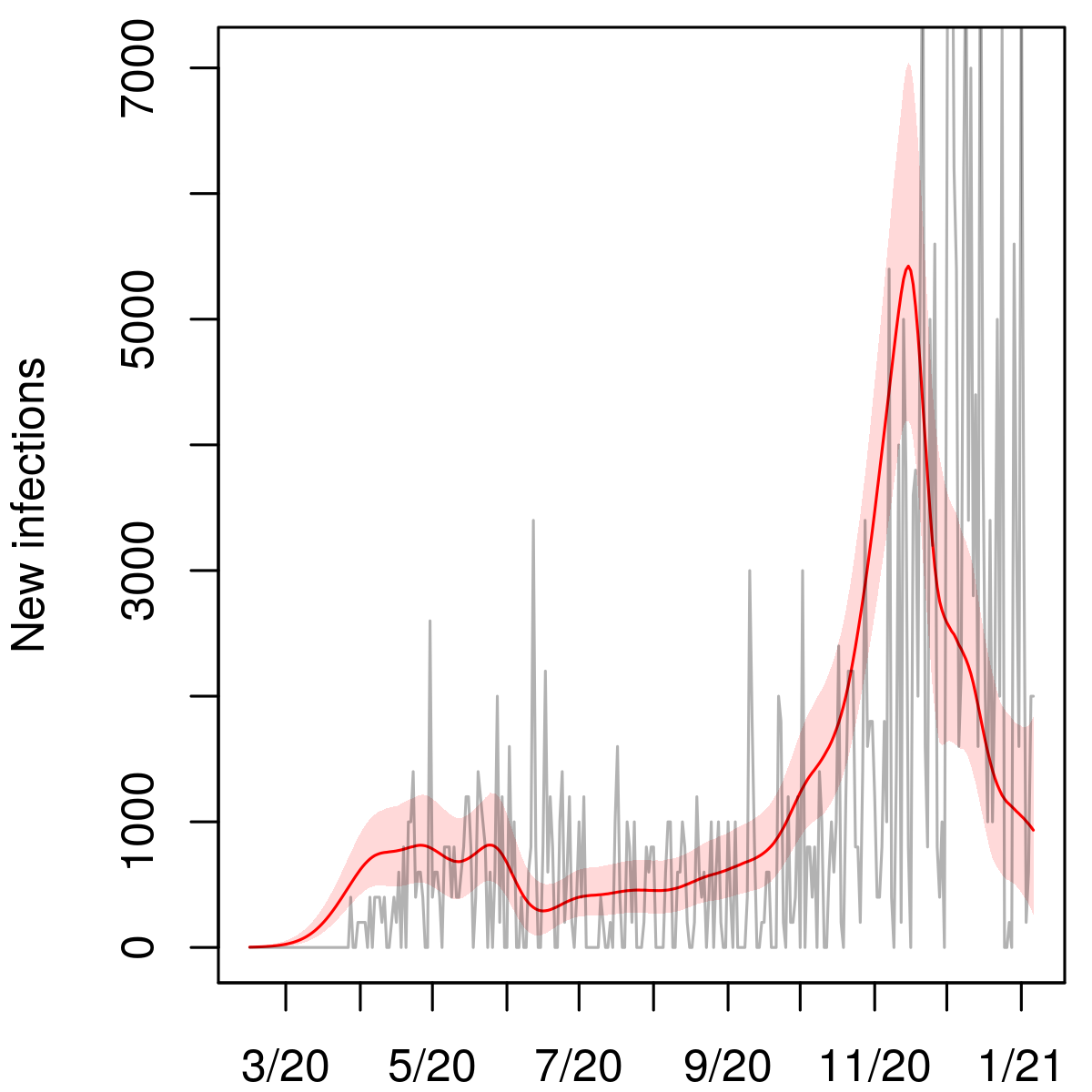}
&
\includegraphics[scale=0.77]{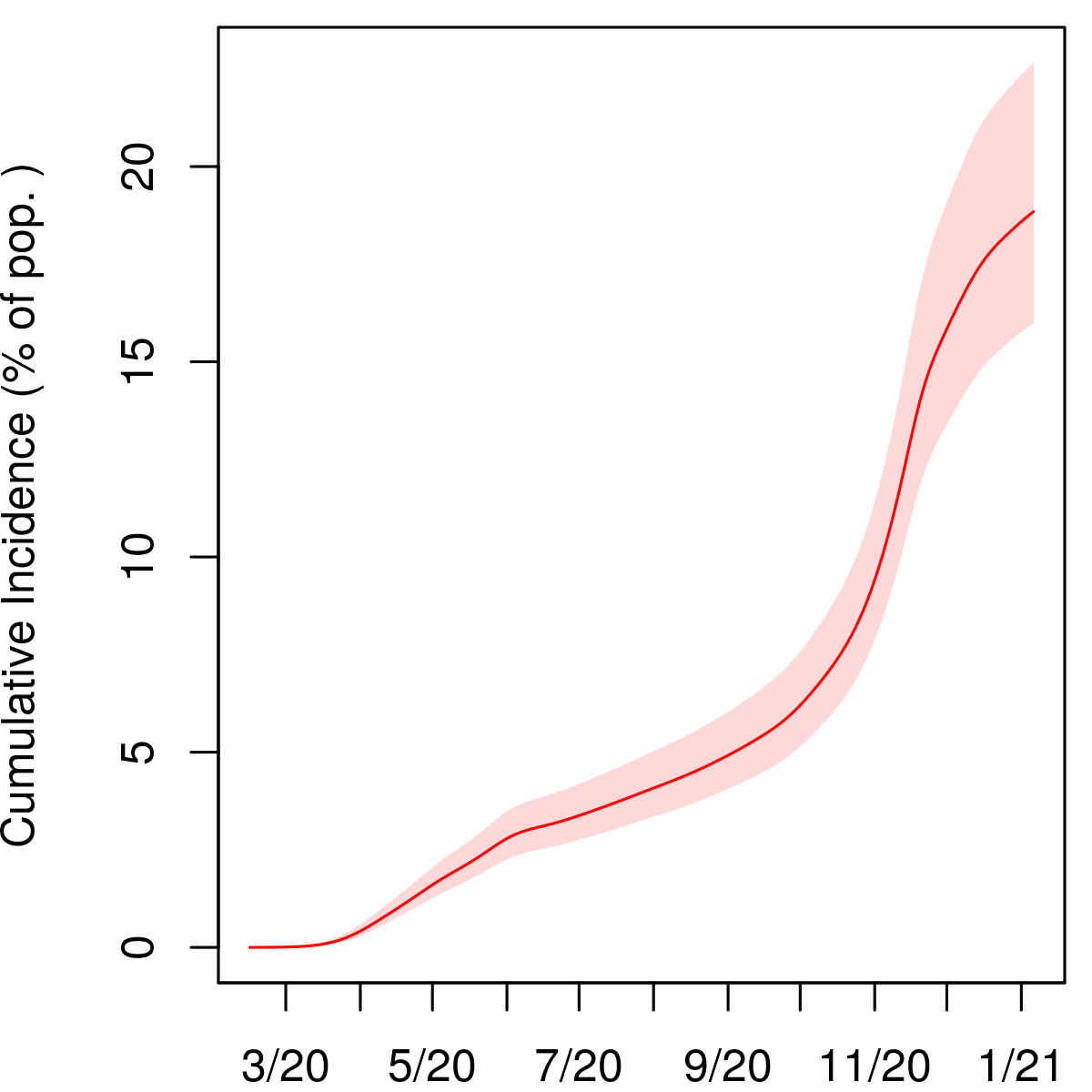} \\
\includegraphics[scale=0.77]{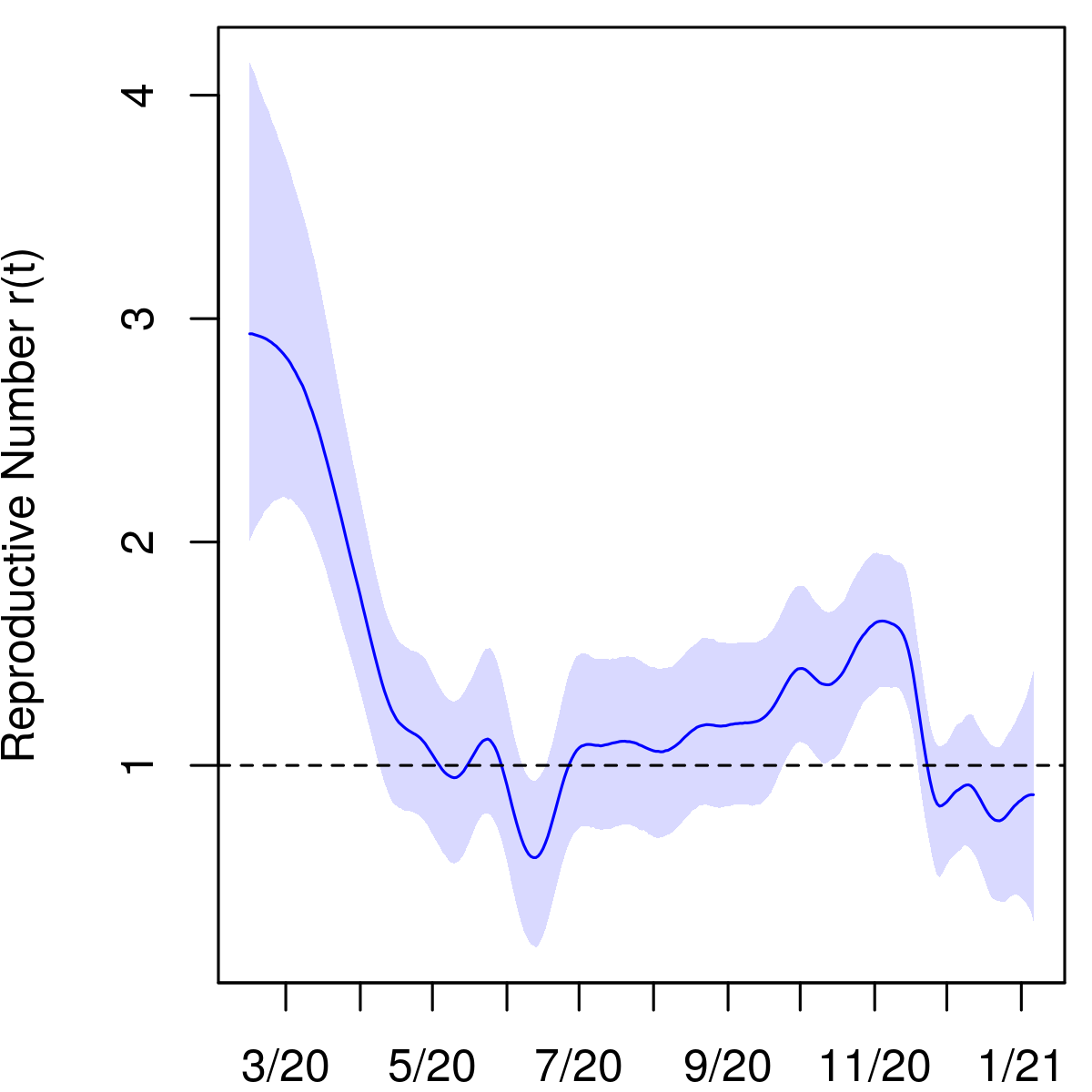}
&
\includegraphics[scale=0.77]{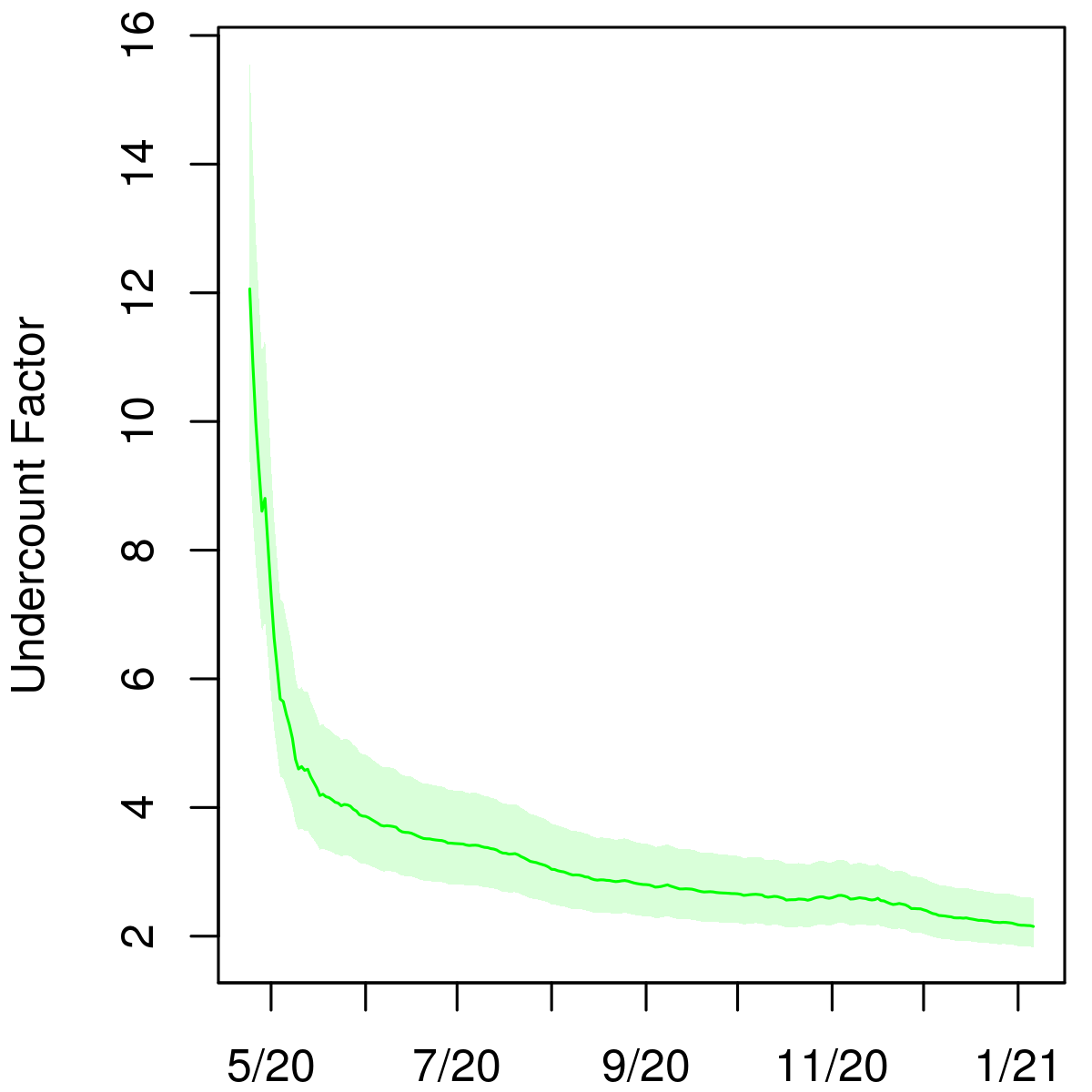} 
\end{tabular}
\caption{Posterior median and middle 95\% intervals for daily new infections, cumulative incidence, $r(t)$, and cumulative undercount from March 2020 to January 2021. In the top left panel, deaths divided by the posterior median IFR are plotted in grey for comparison.}
\end{figure}
\newpage
\begin{figure}[htbp!]
\textbf{New Hampshire}
\centering
\begin{tabular}{ll}
\includegraphics[scale=0.77]{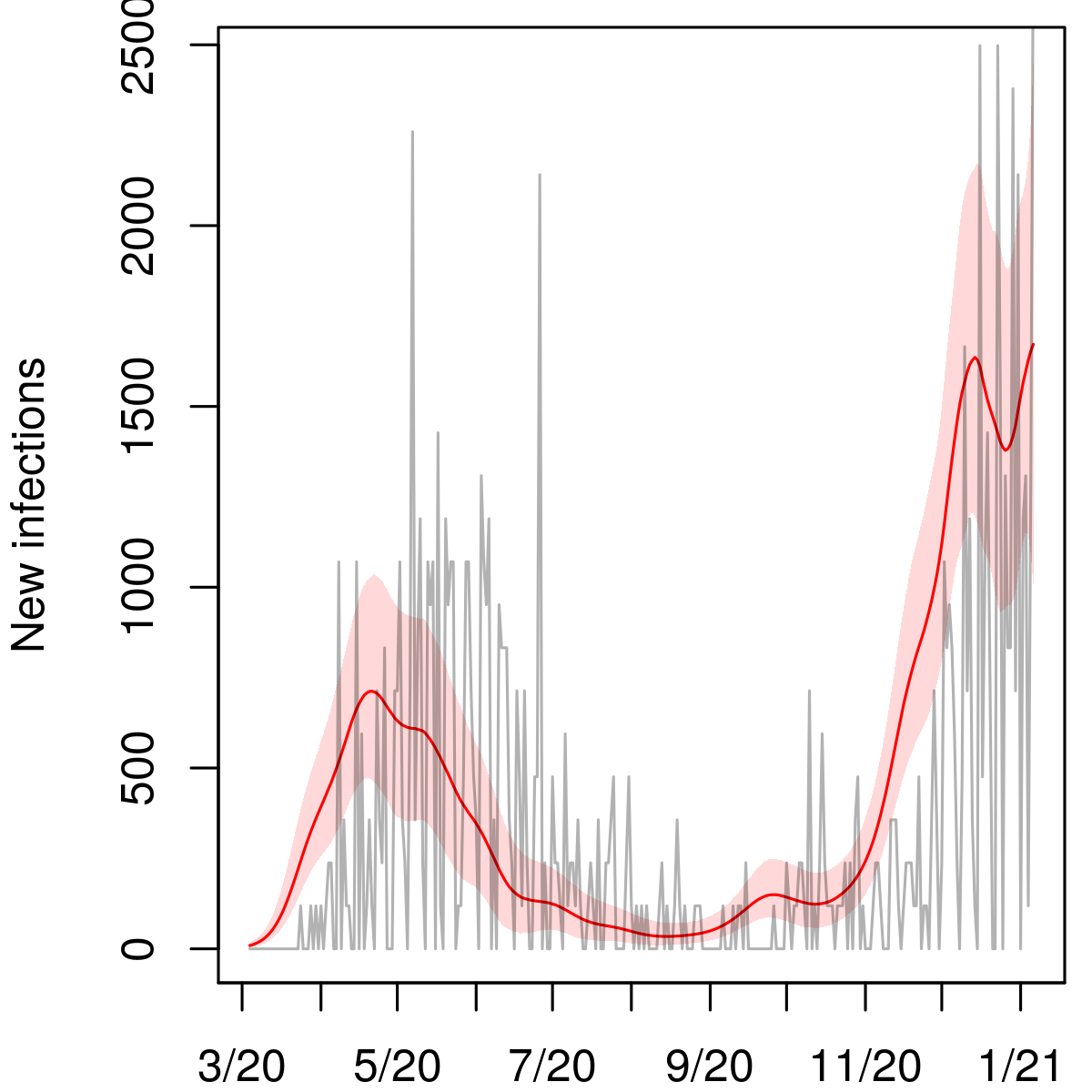}
&
\includegraphics[scale=0.77]{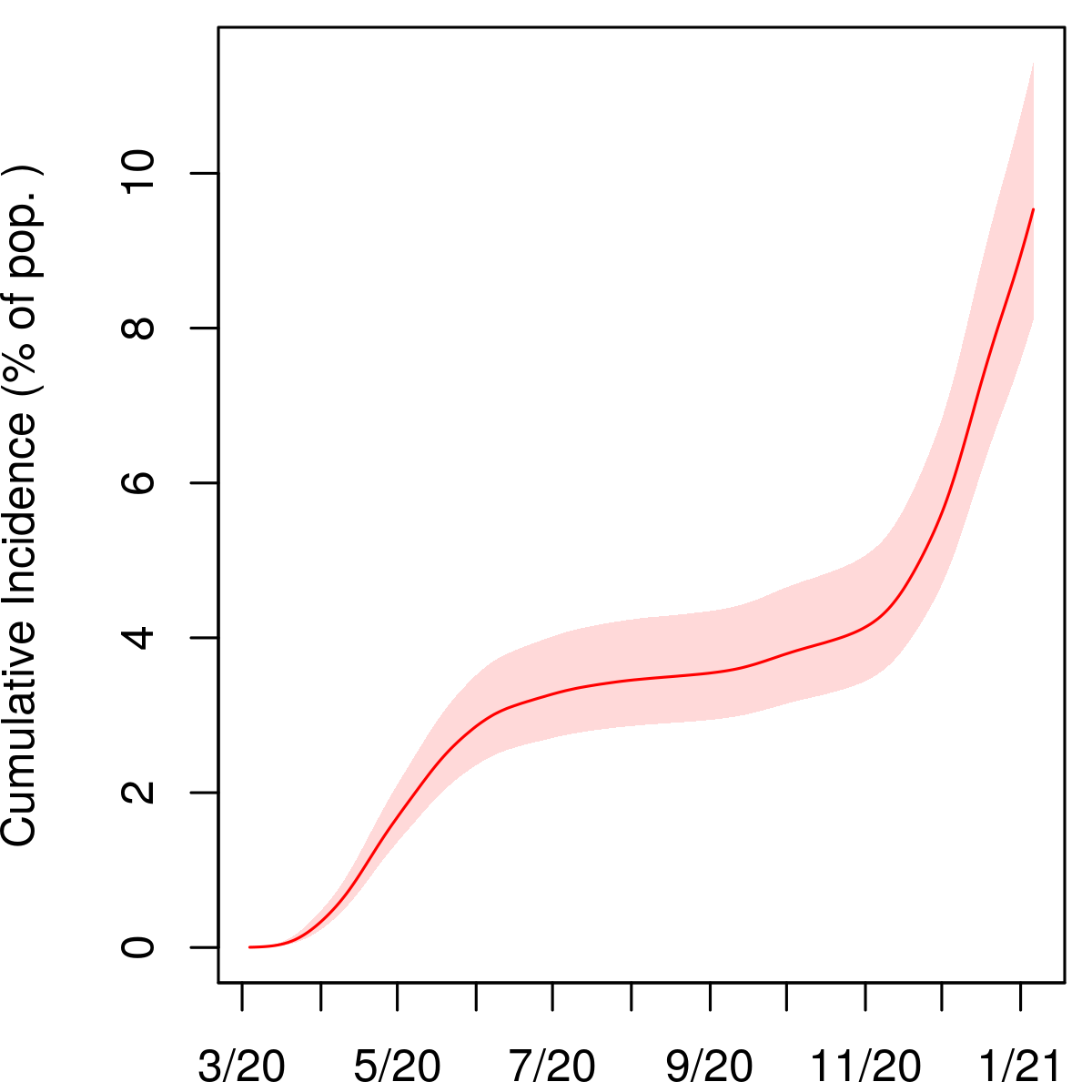} \\
\includegraphics[scale=0.77]{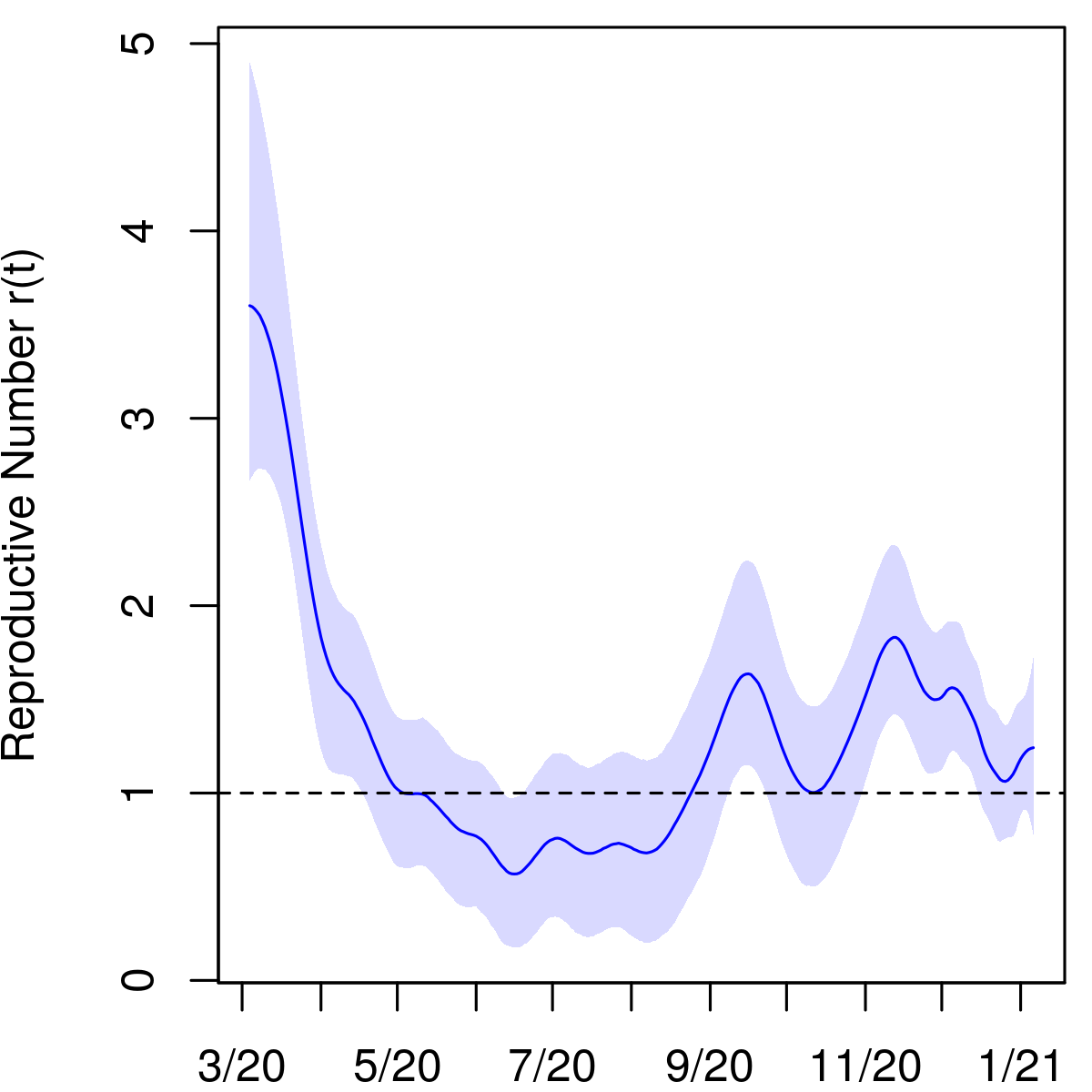}
&
\includegraphics[scale=0.77]{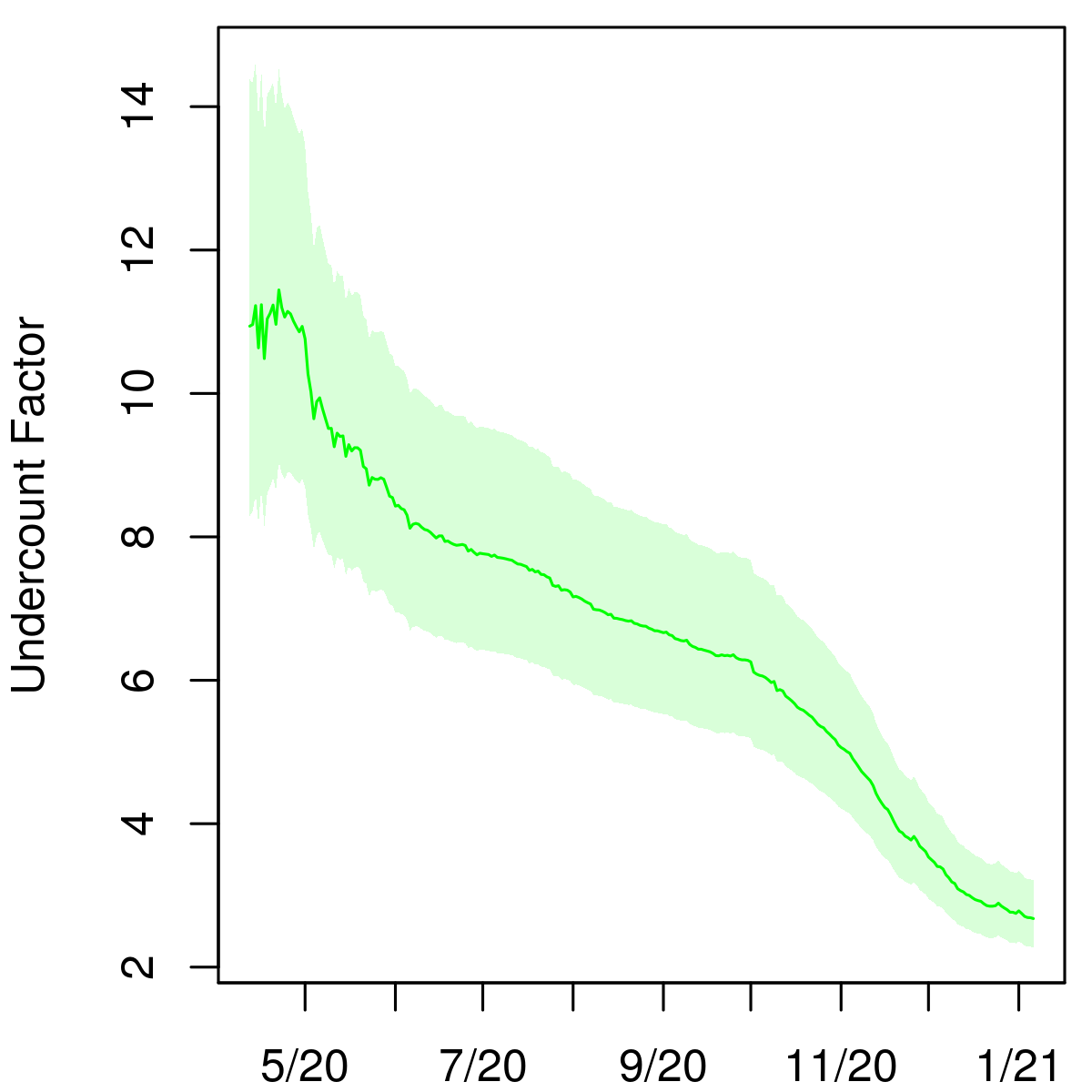} 
\end{tabular}
\caption{Posterior median and middle 95\% intervals for daily new infections, cumulative incidence, $r(t)$, and cumulative undercount from March 2020 to January 2021. In the top left panel, deaths divided by the posterior median IFR are plotted in grey for comparison.}
\end{figure}
\newpage
\begin{figure}[htbp!]
\textbf{New Jersey}
\centering
\begin{tabular}{ll}
\includegraphics[scale=0.77]{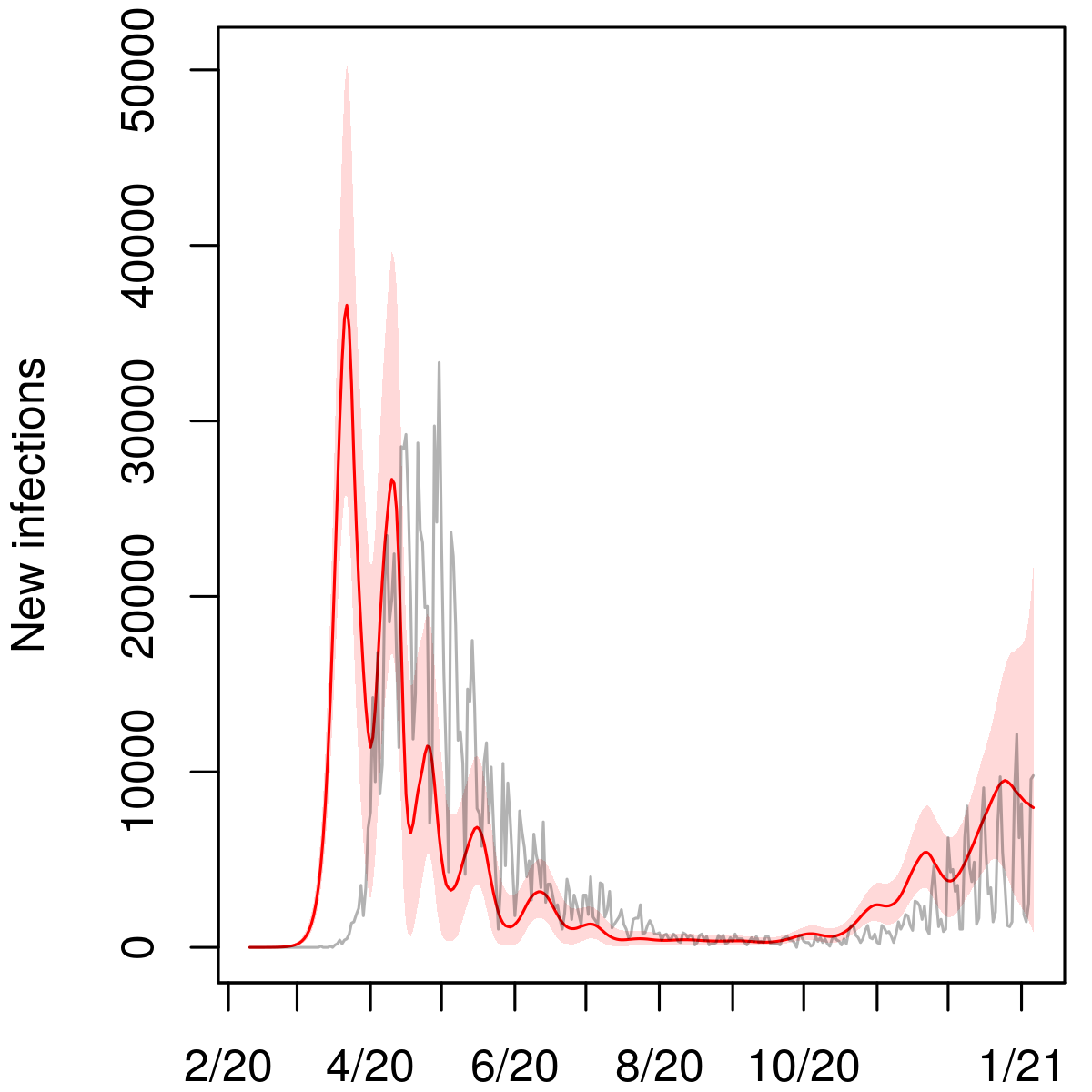}
&
\includegraphics[scale=0.77]{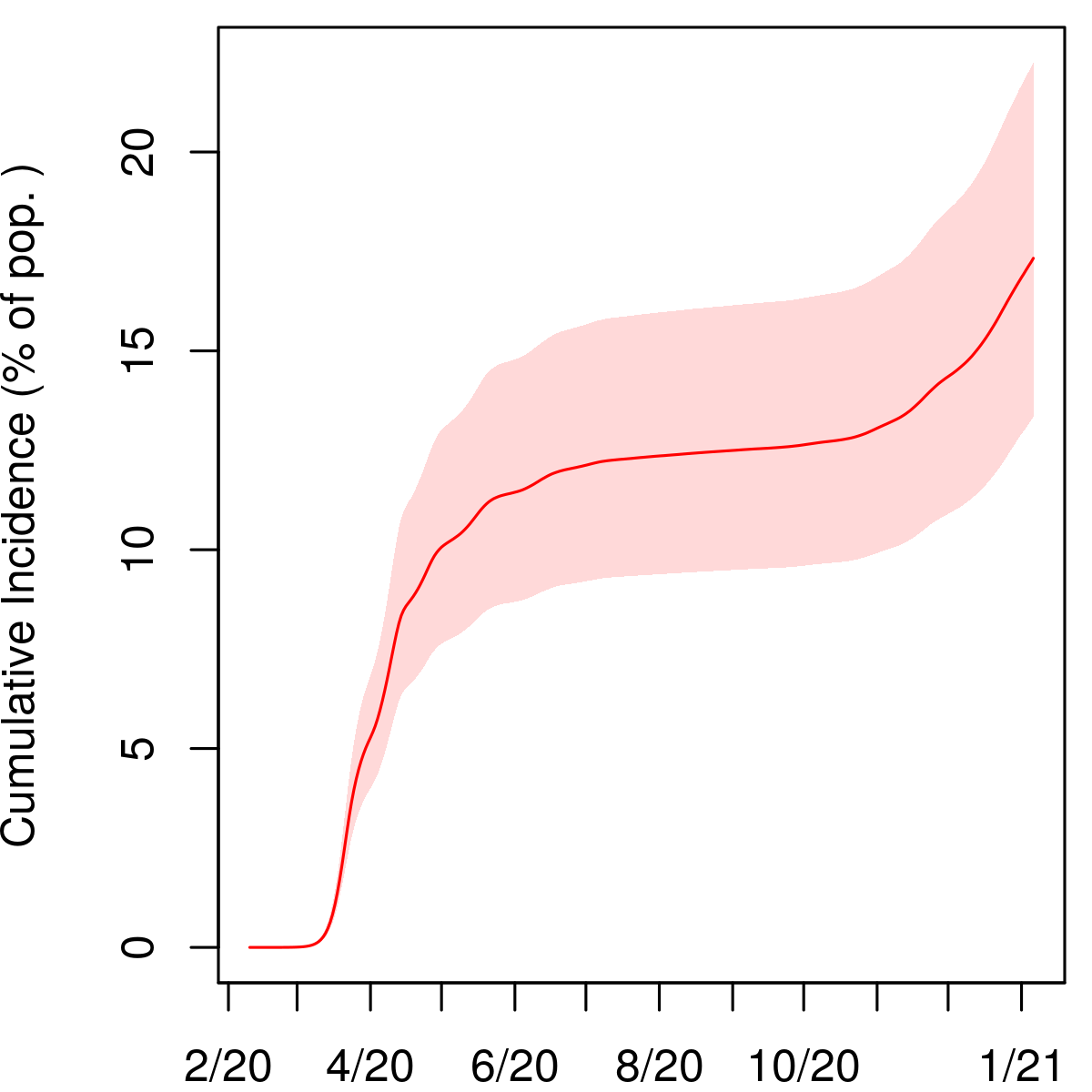} \\
\includegraphics[scale=0.77]{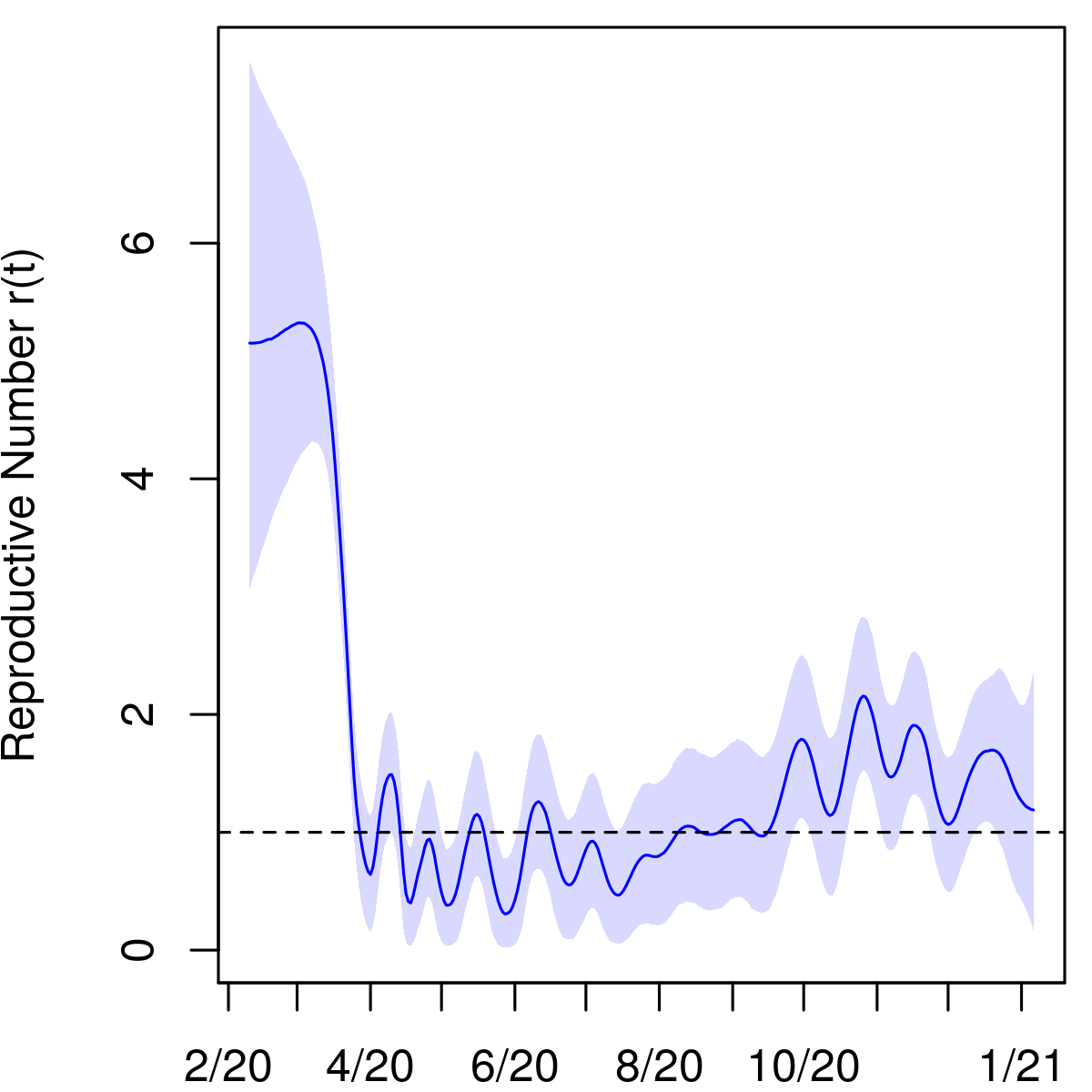}
&
\includegraphics[scale=0.77]{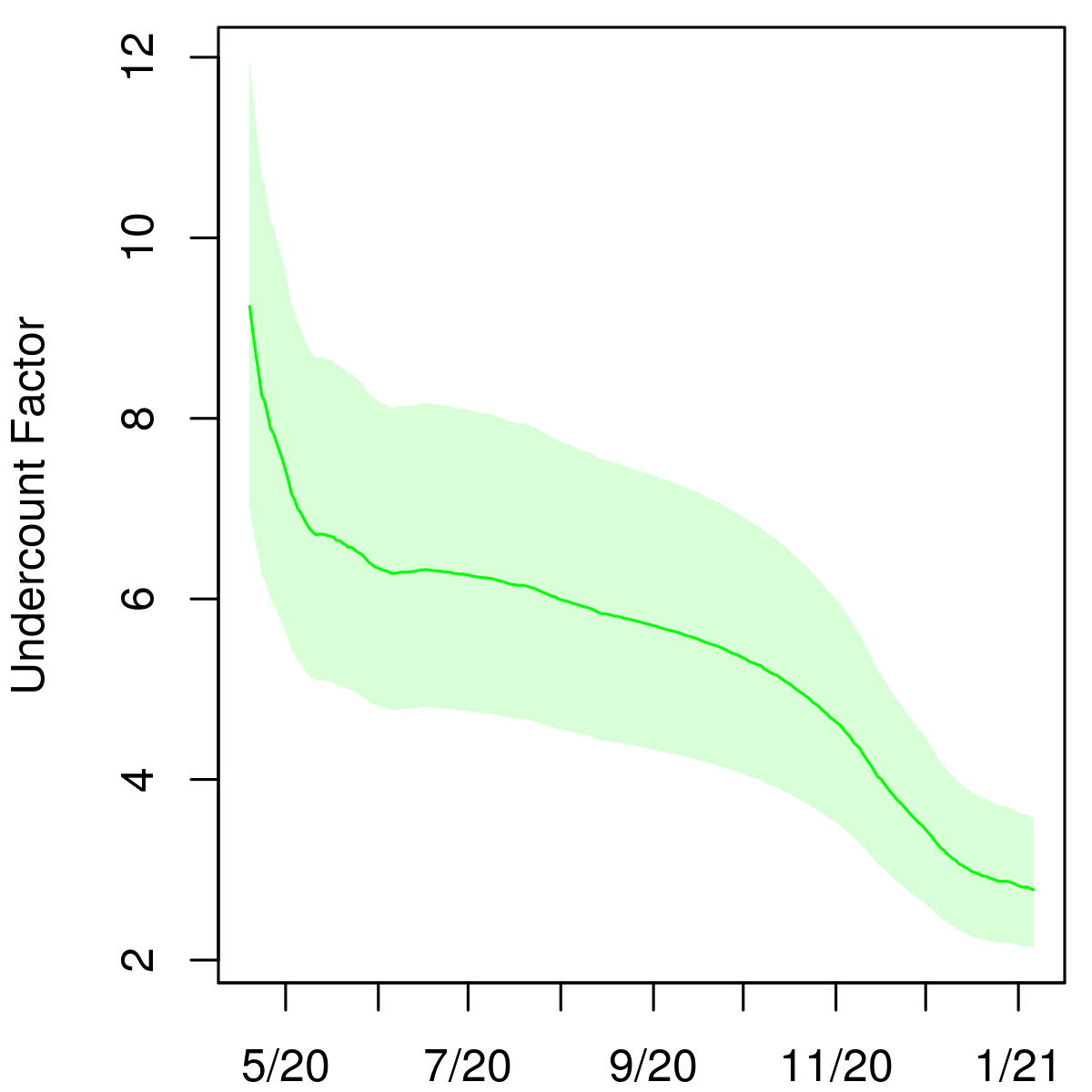} 
\end{tabular}
\caption{Posterior median and middle 95\% intervals for daily new infections, cumulative incidence, $r(t)$, and cumulative undercount from March 2020 to January 2021. In the top left panel, deaths divided by the posterior median IFR are plotted in grey for comparison.}
\end{figure}
\newpage
\begin{figure}[htbp!]
\textbf{New Mexico}
\centering
\begin{tabular}{ll}
\includegraphics[scale=0.77]{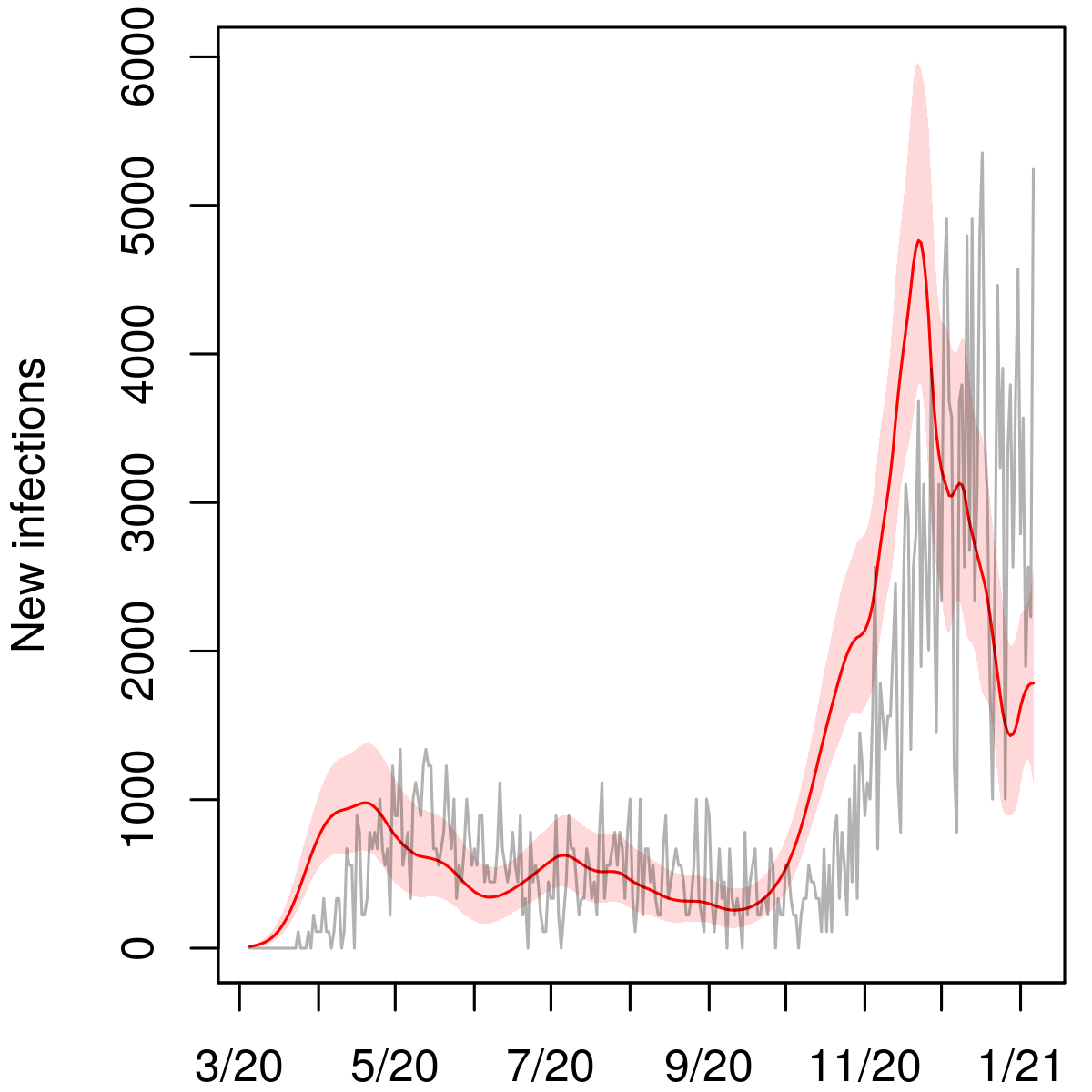}
&
\includegraphics[scale=0.77]{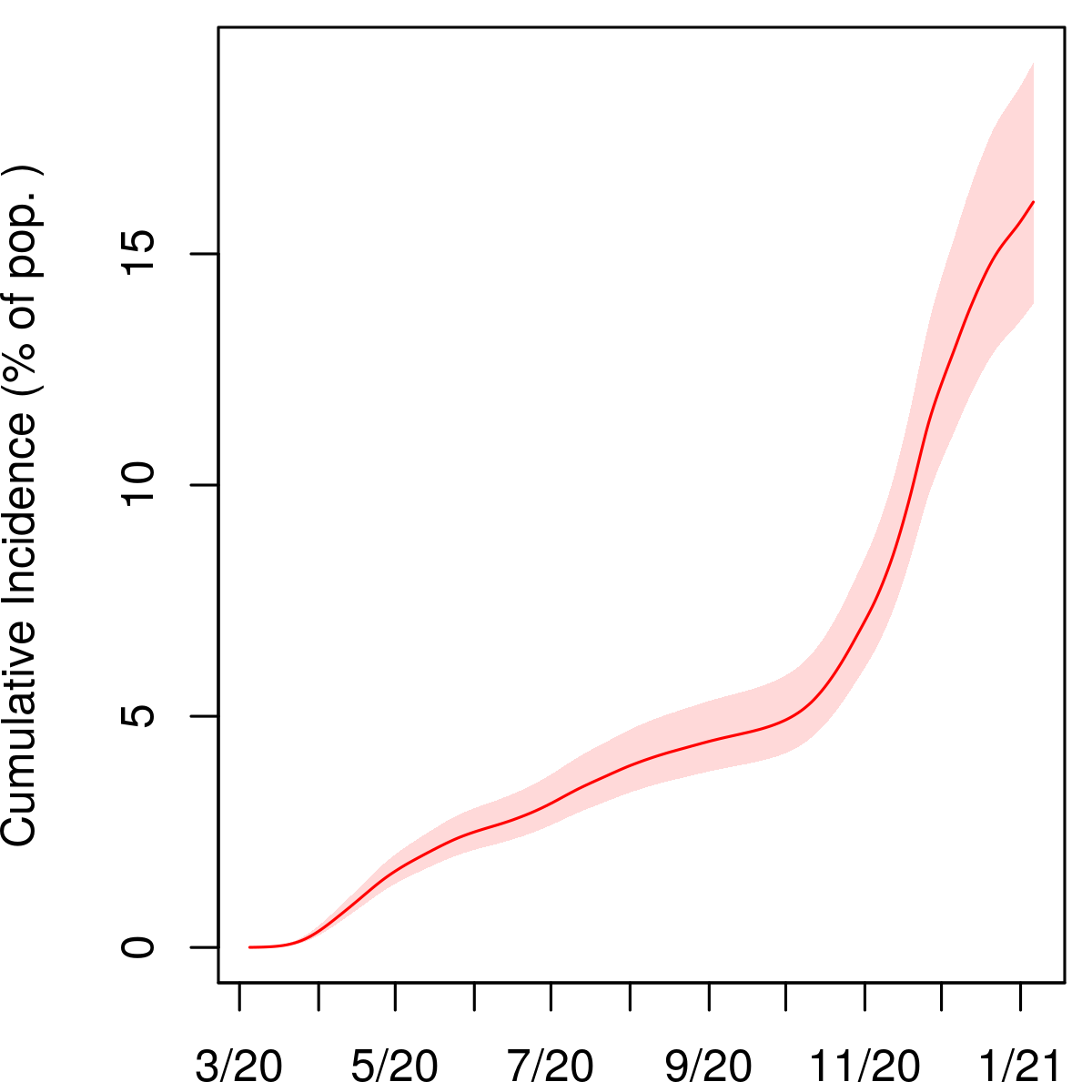} \\
\includegraphics[scale=0.77]{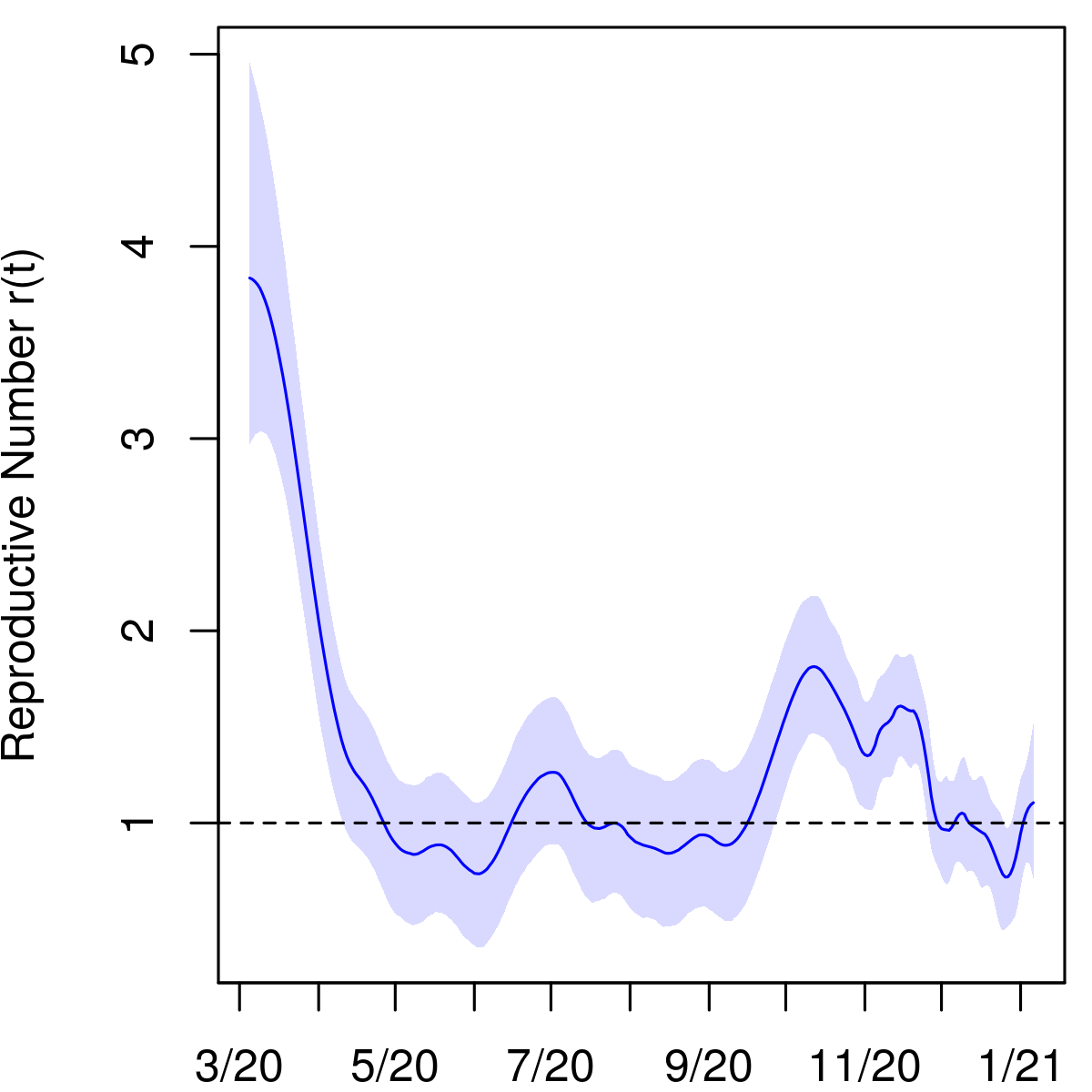}
&
\includegraphics[scale=0.77]{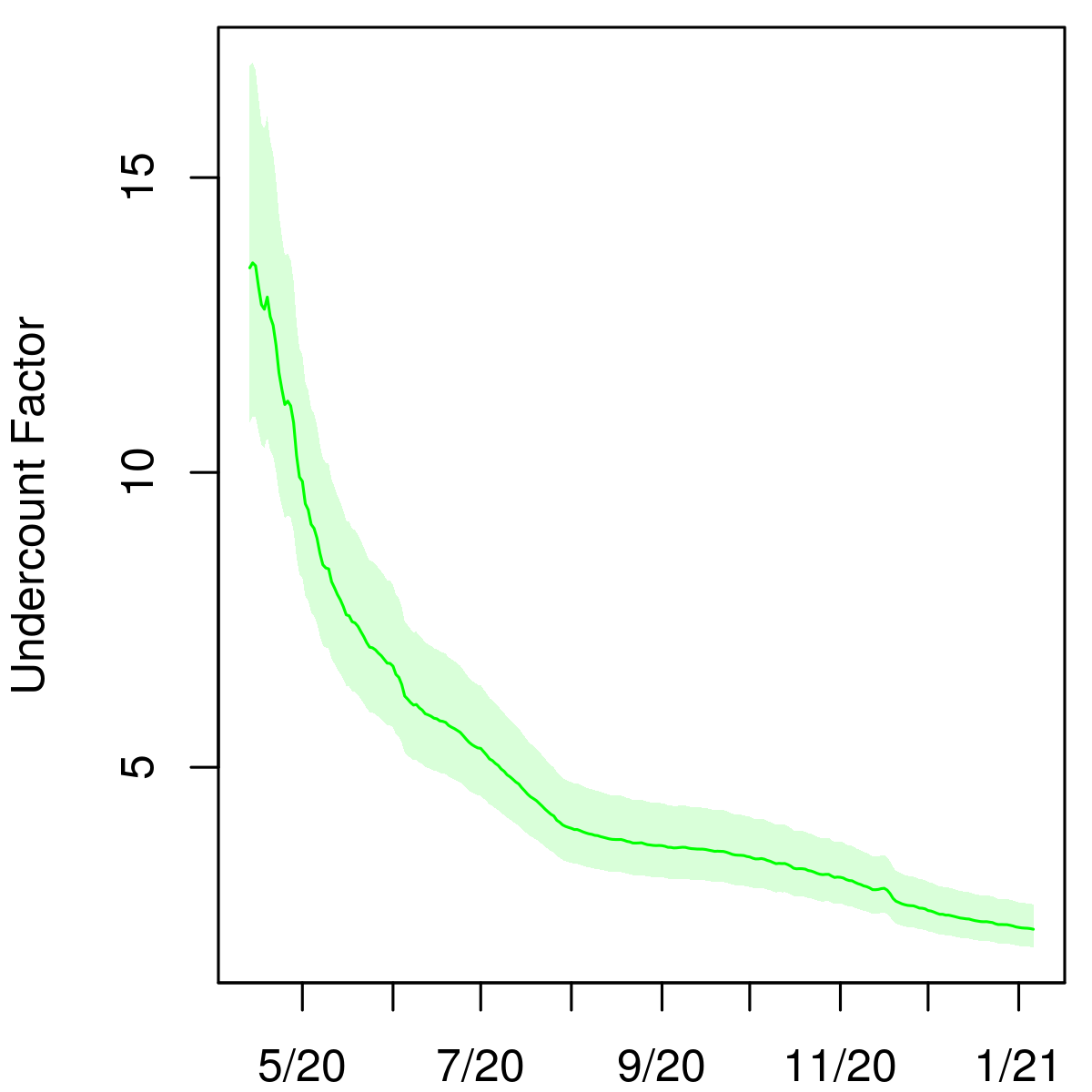} 
\end{tabular}
\caption{Posterior median and middle 95\% intervals for daily new infections, cumulative incidence, $r(t)$, and cumulative undercount from March 2020 to January 2021. In the top left panel, deaths divided by the posterior median IFR are plotted in grey for comparison.}
\end{figure}
\newpage
\begin{figure}[htbp!]
\textbf{Nevada}
\centering
\begin{tabular}{ll}
\includegraphics[scale=0.77]{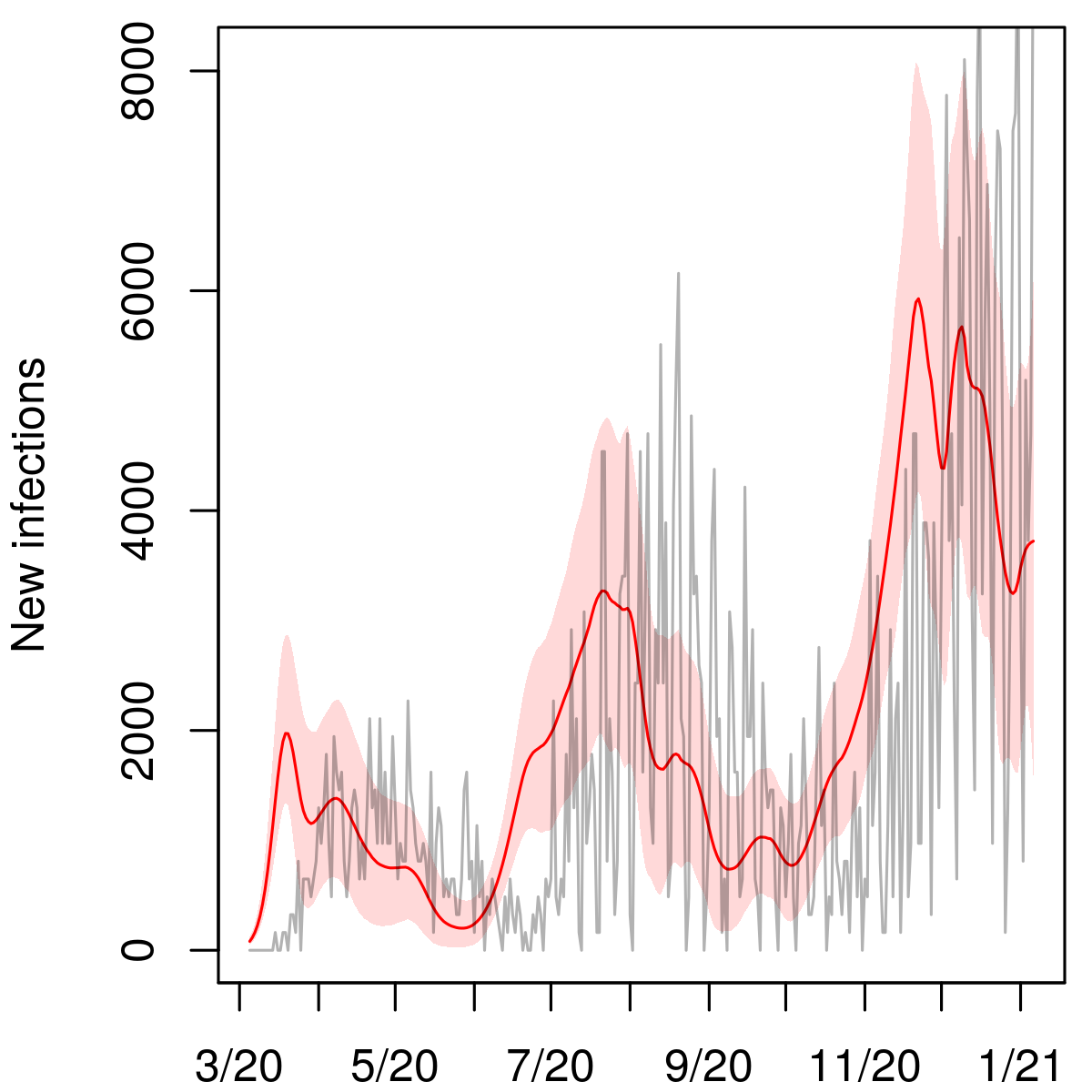}
&
\includegraphics[scale=0.77]{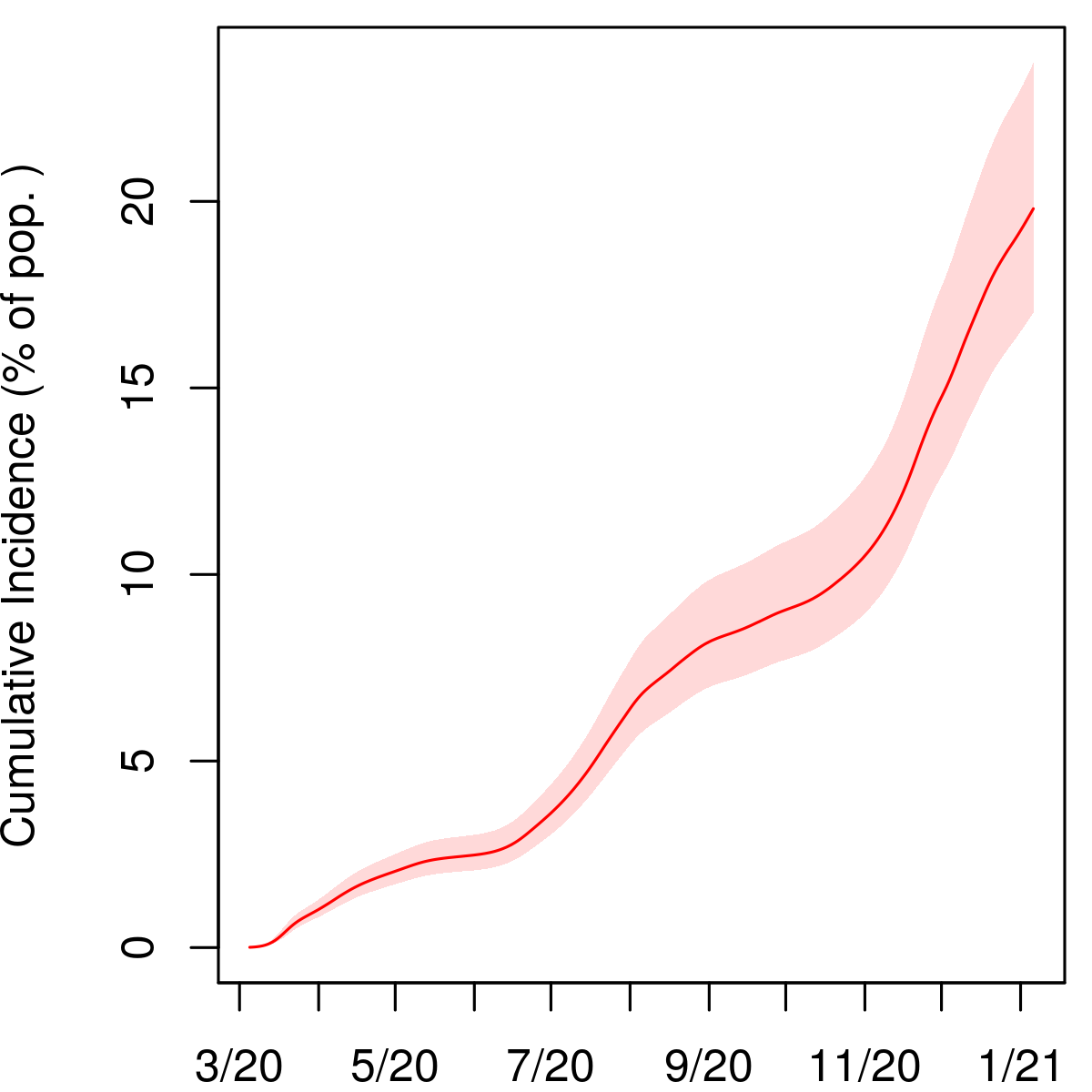} \\
\includegraphics[scale=0.77]{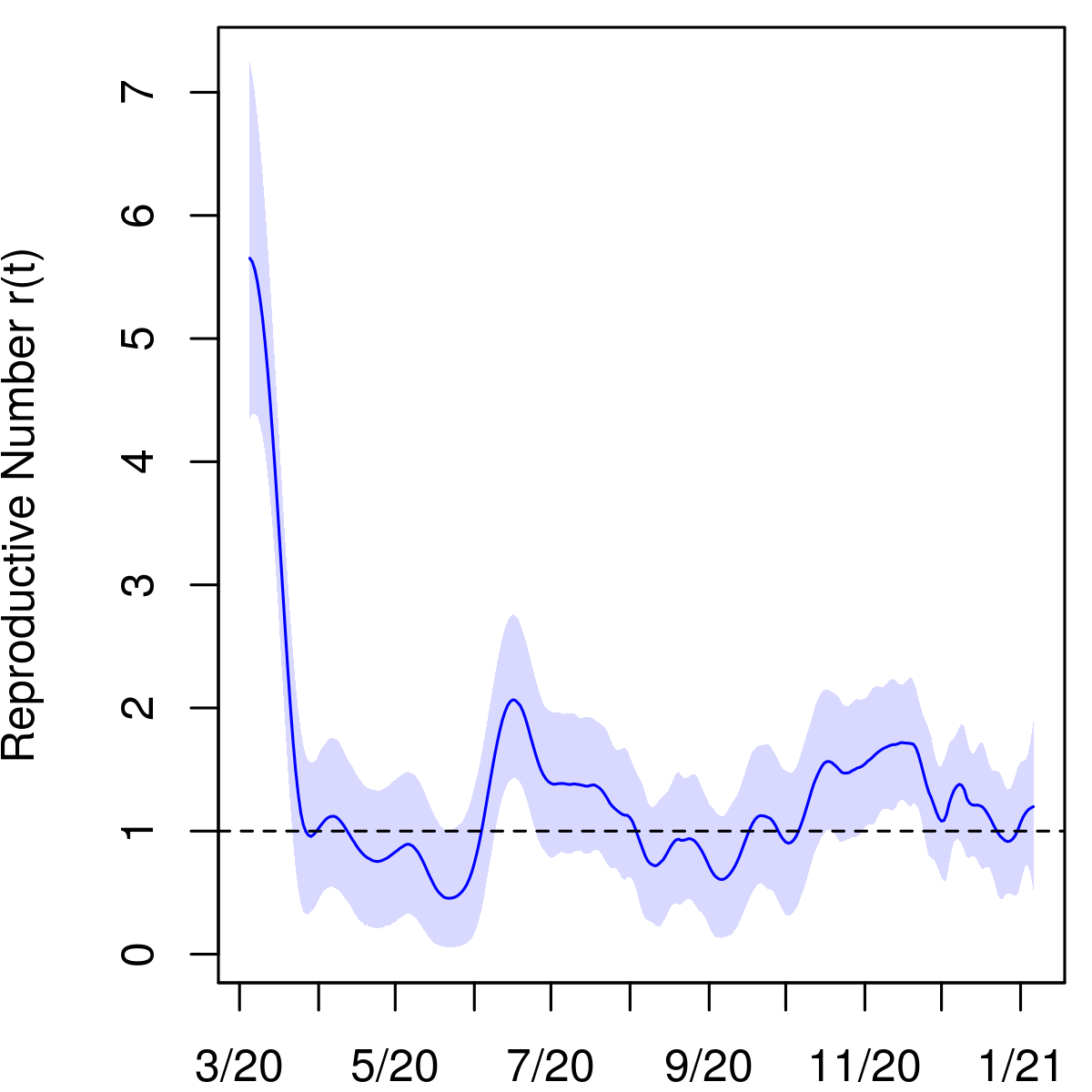}
&
\includegraphics[scale=0.77]{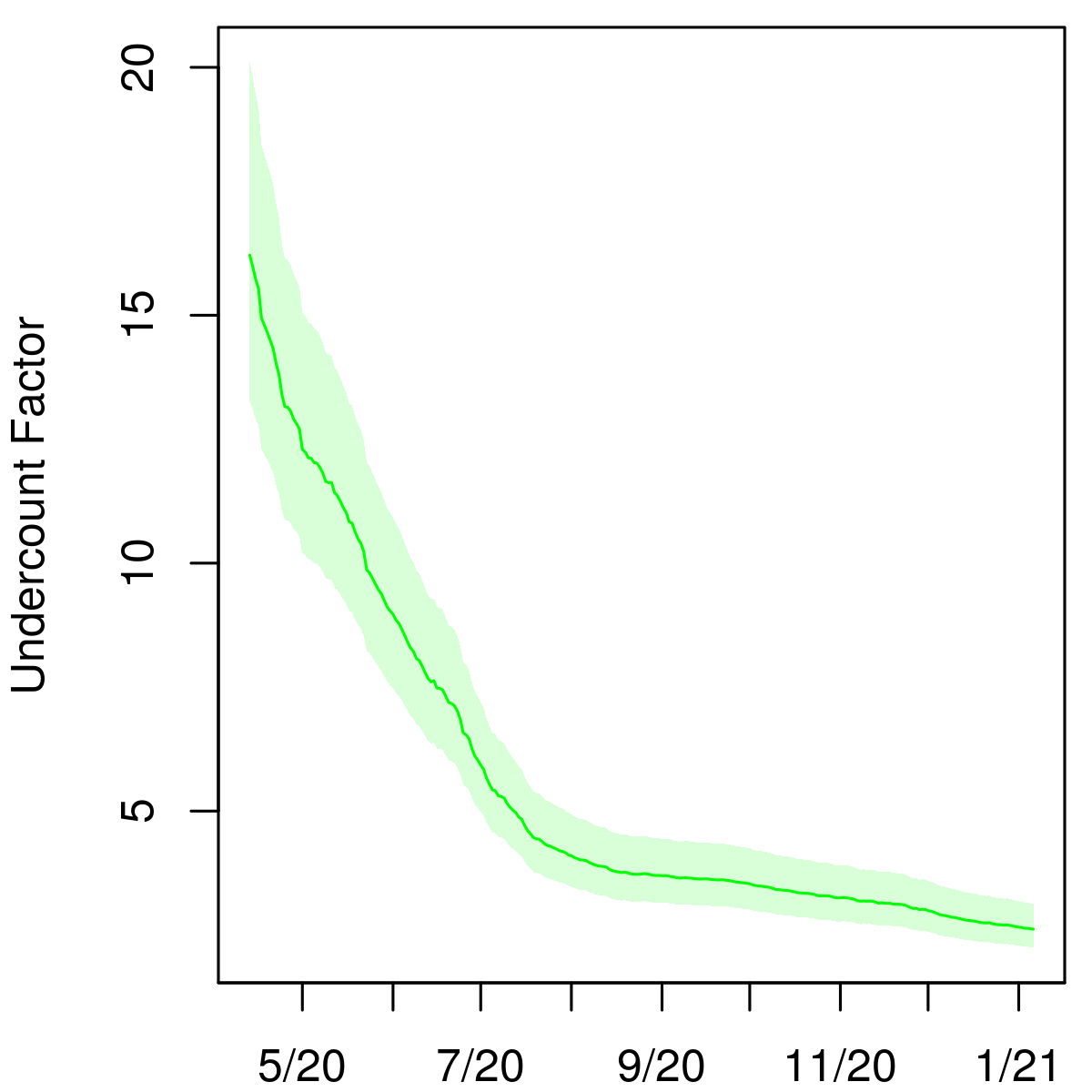} 
\end{tabular}
\caption{Posterior median and middle 95\% intervals for daily new infections, cumulative incidence, $r(t)$, and cumulative undercount from March 2020 to January 2021. In the top left panel, deaths divided by the posterior median IFR are plotted in grey for comparison.}
\end{figure}
\newpage
\begin{figure}[htbp!]
\textbf{New York}
\centering
\begin{tabular}{ll}
\includegraphics[scale=0.77]{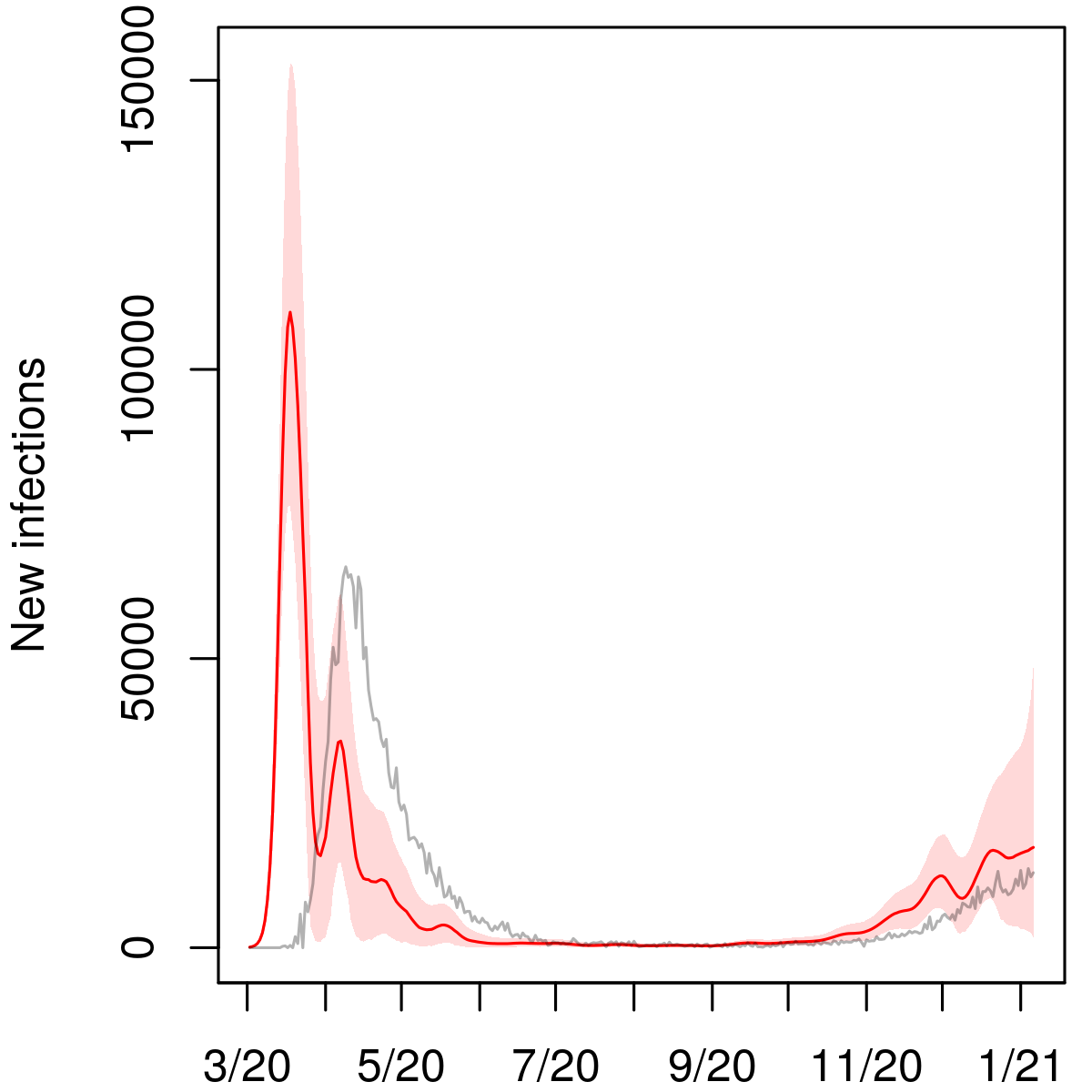}
&
\includegraphics[scale=0.77]{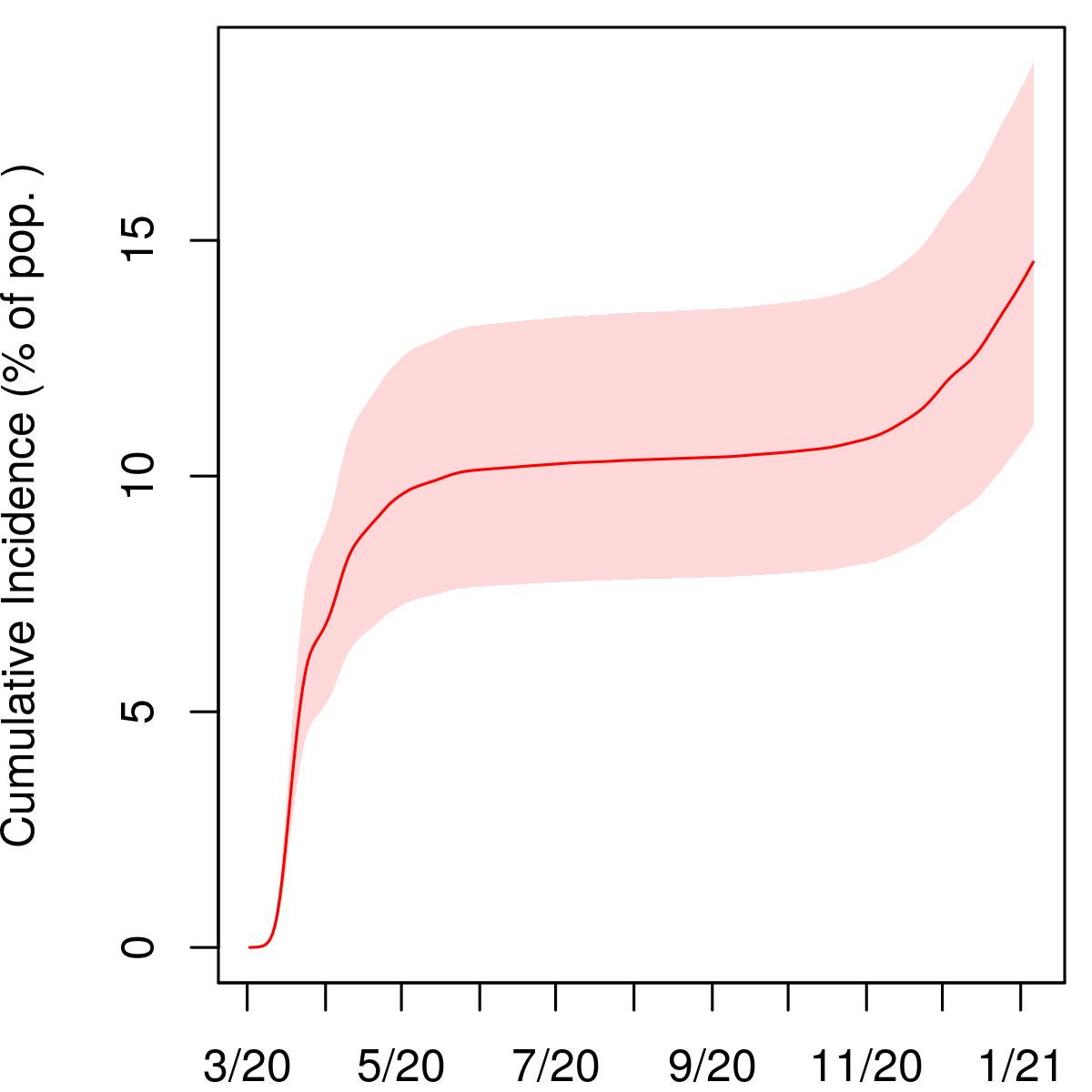} \\
\includegraphics[scale=0.77]{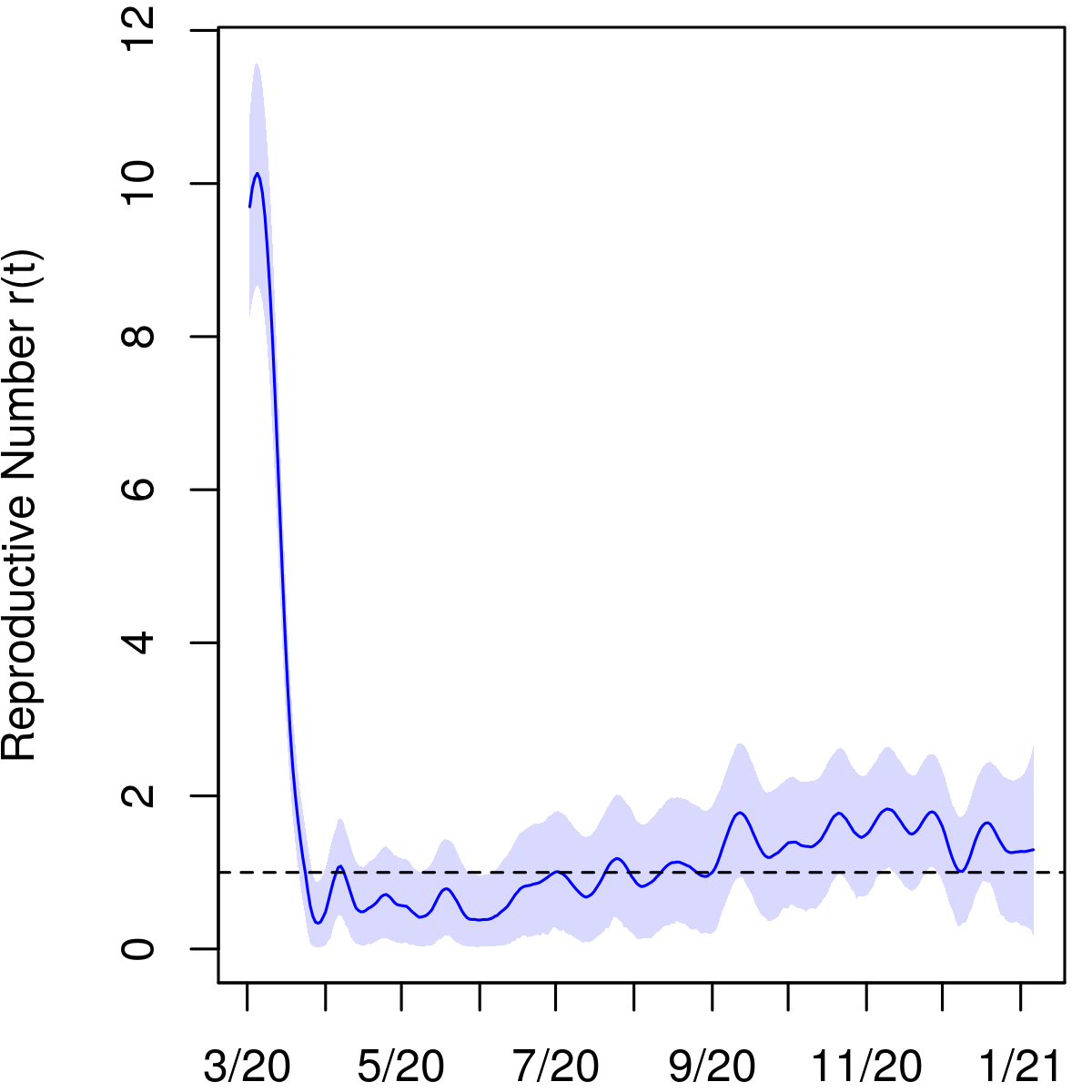}
&
\includegraphics[scale=0.77]{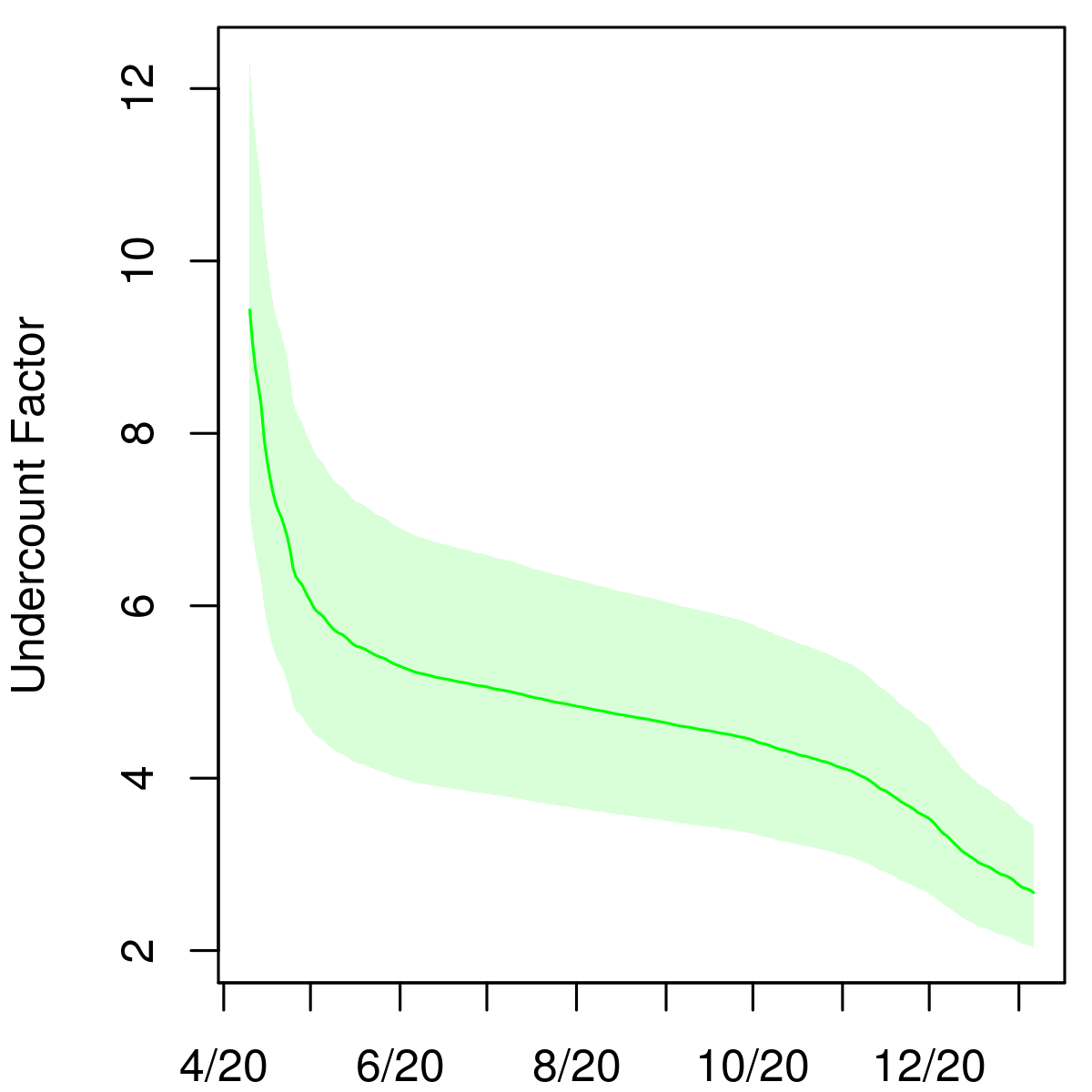} 
\end{tabular}
\caption{Posterior median and middle 95\% intervals for daily new infections, cumulative incidence, $r(t)$, and cumulative undercount from March 2020 to January 2021. In the top left panel, deaths divided by the posterior median IFR are plotted in grey for comparison.}
\end{figure}
\newpage
\begin{figure}[htbp!]
\textbf{Ohio}
\centering
\begin{tabular}{ll}
\includegraphics[scale=0.185]{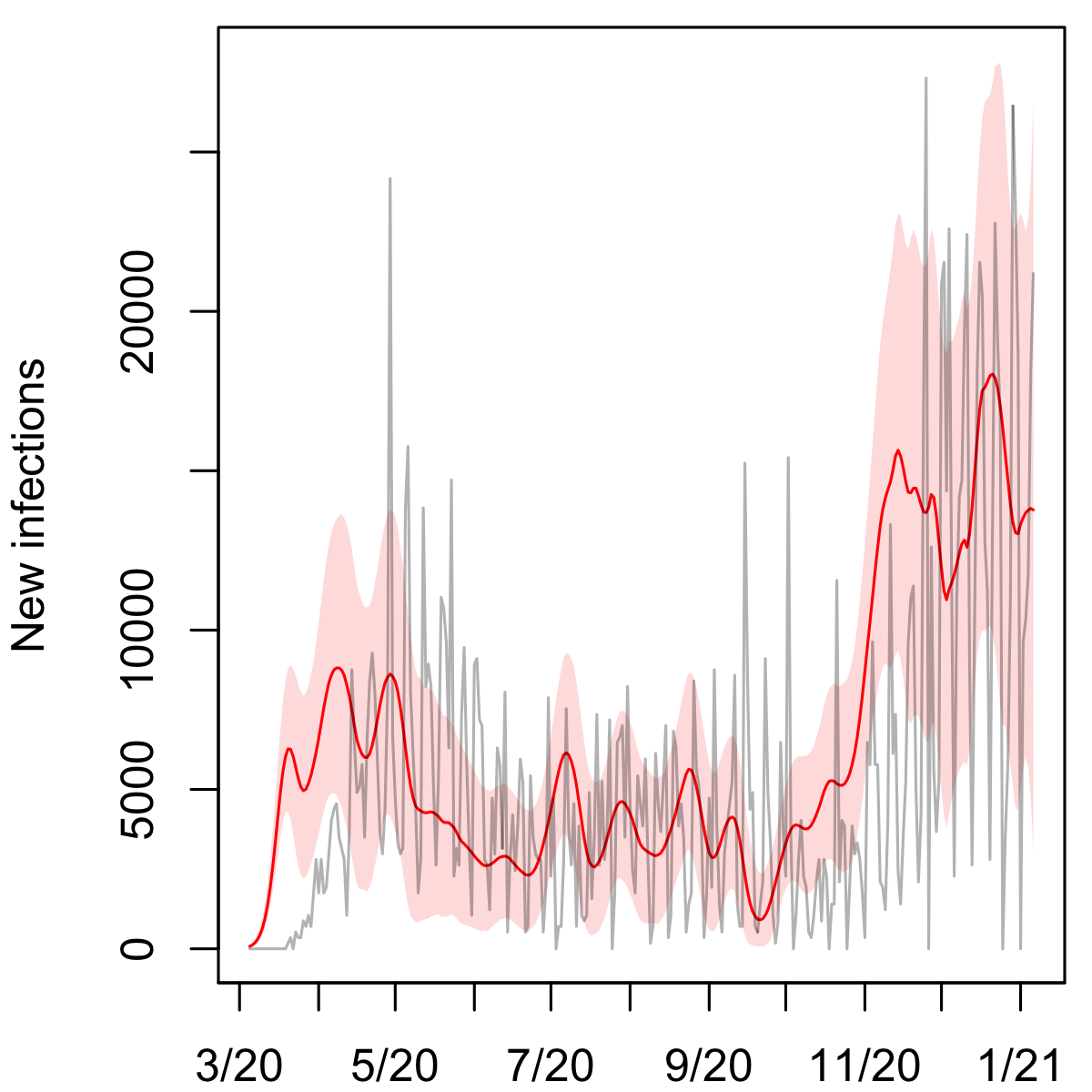}
&
\includegraphics[scale=0.185]{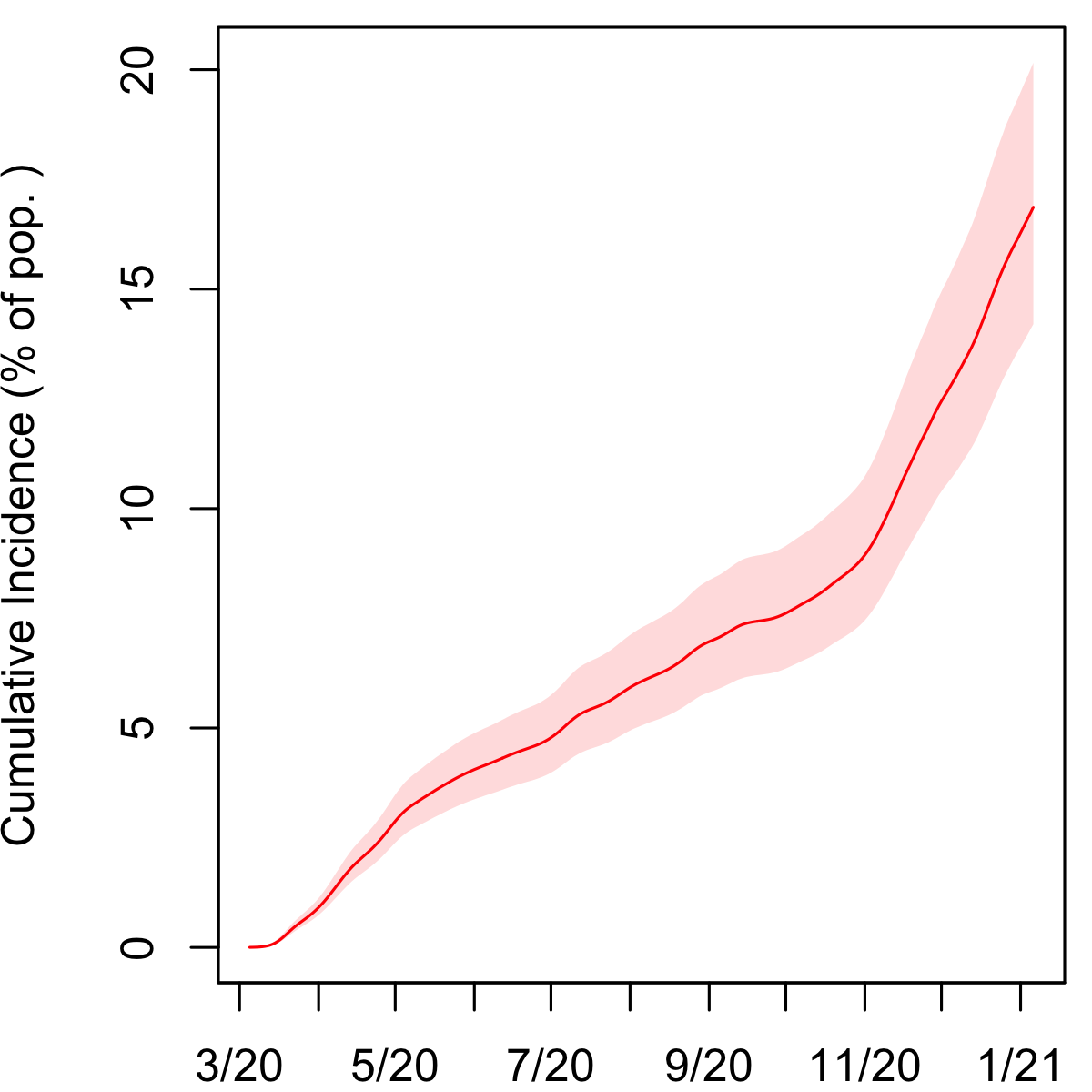} \\
\includegraphics[scale=0.185]{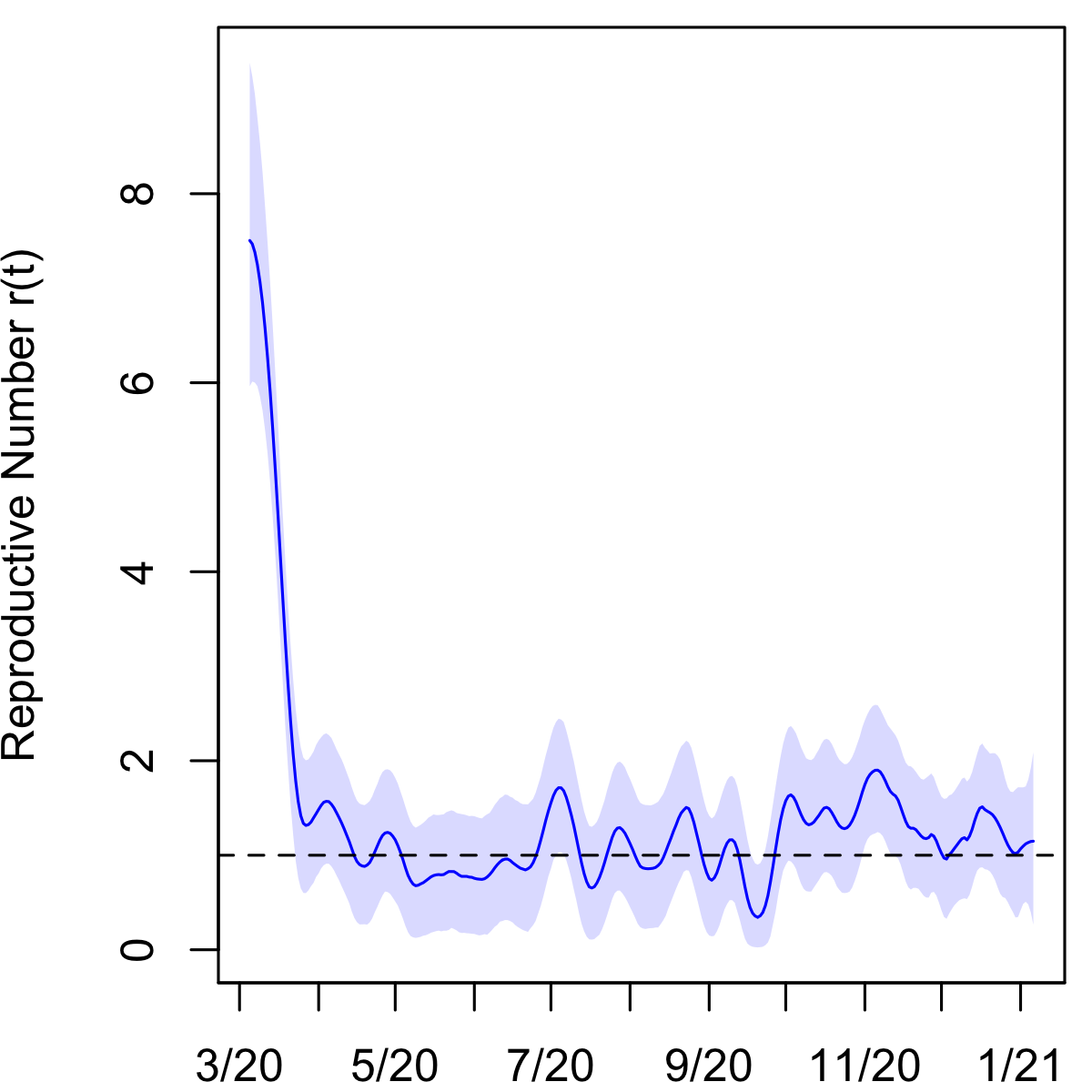}
&
\includegraphics[scale=0.185]{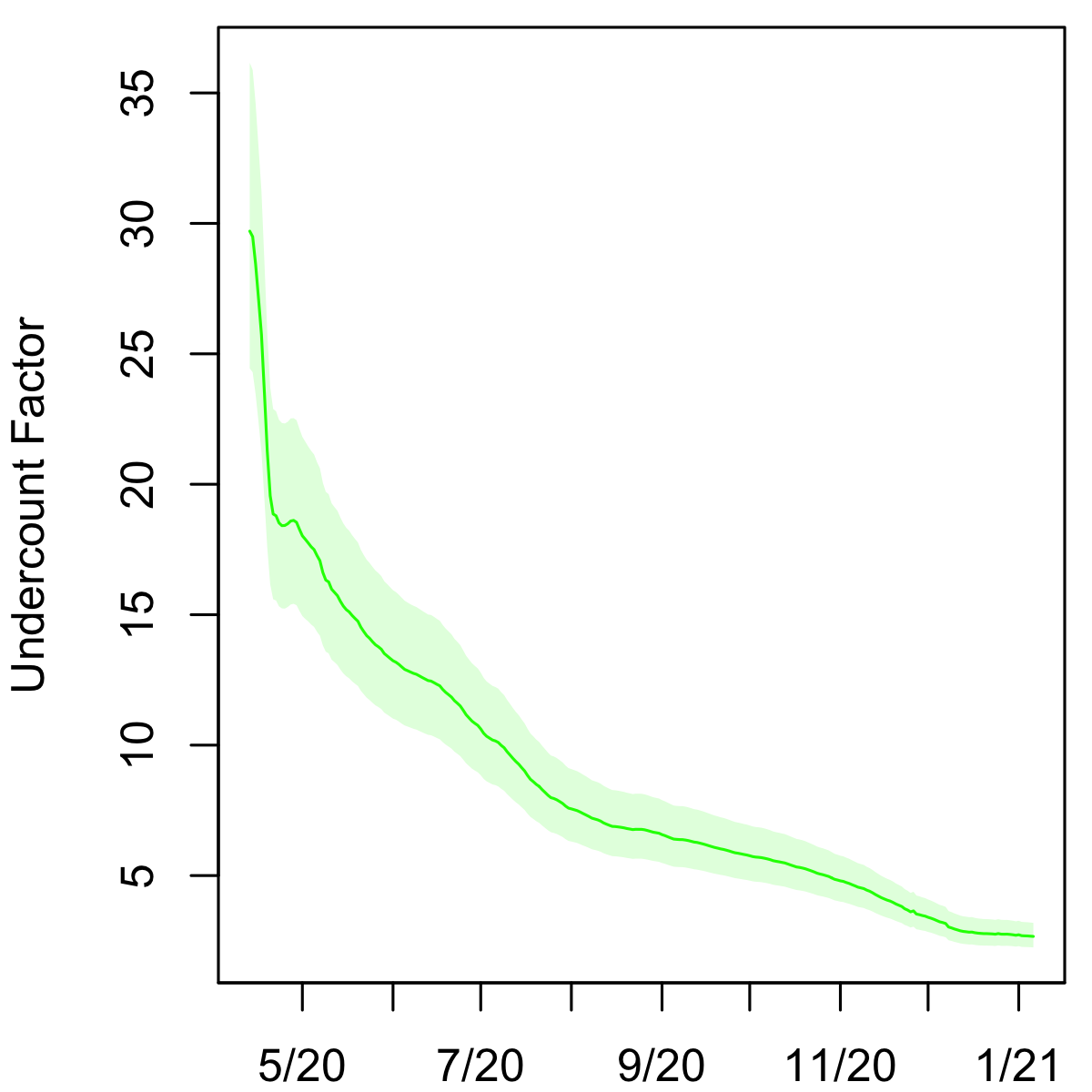} 
\end{tabular}
\caption{Posterior median and middle 95\% intervals for daily new infections, cumulative incidence, $r(t)$, and cumulative undercount from March 2020 to January 2021. In the top left panel, deaths divided by the posterior median IFR are plotted in grey for comparison.}
\end{figure}
\newpage
\begin{figure}[htbp!]
\textbf{Oklahoma}
\centering
\begin{tabular}{ll}
\includegraphics[scale=0.77]{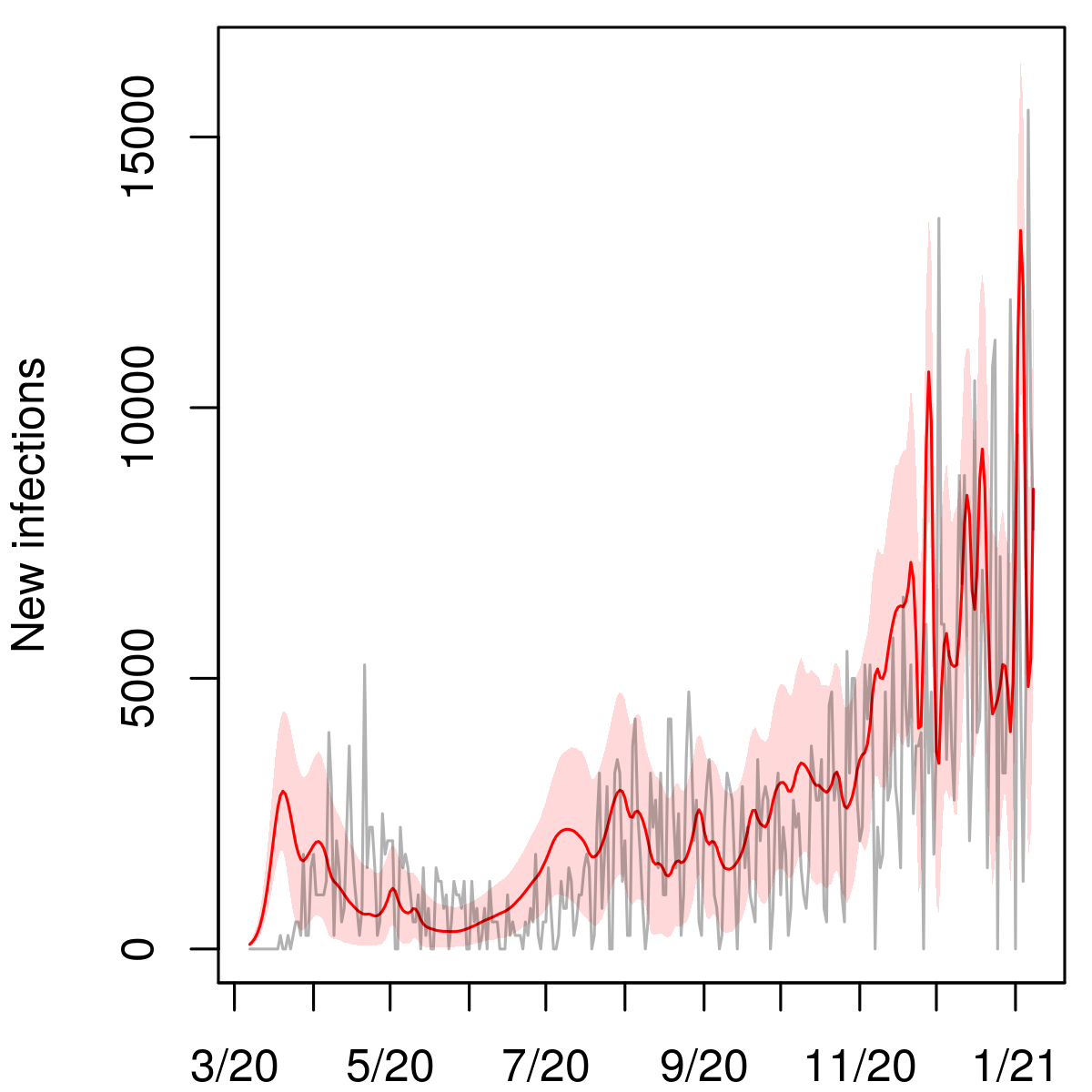}
&
\includegraphics[scale=0.77]{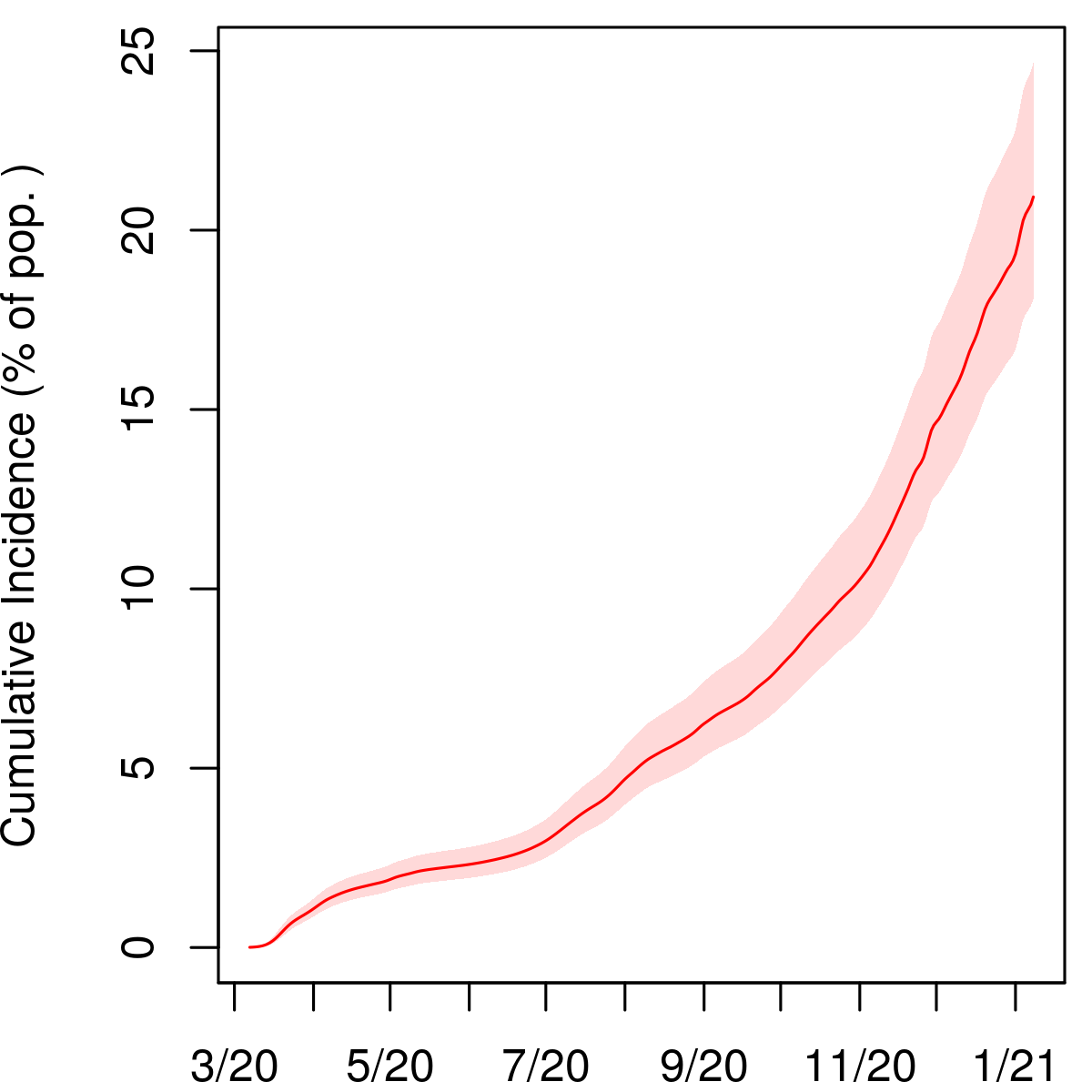} \\
\includegraphics[scale=0.77]{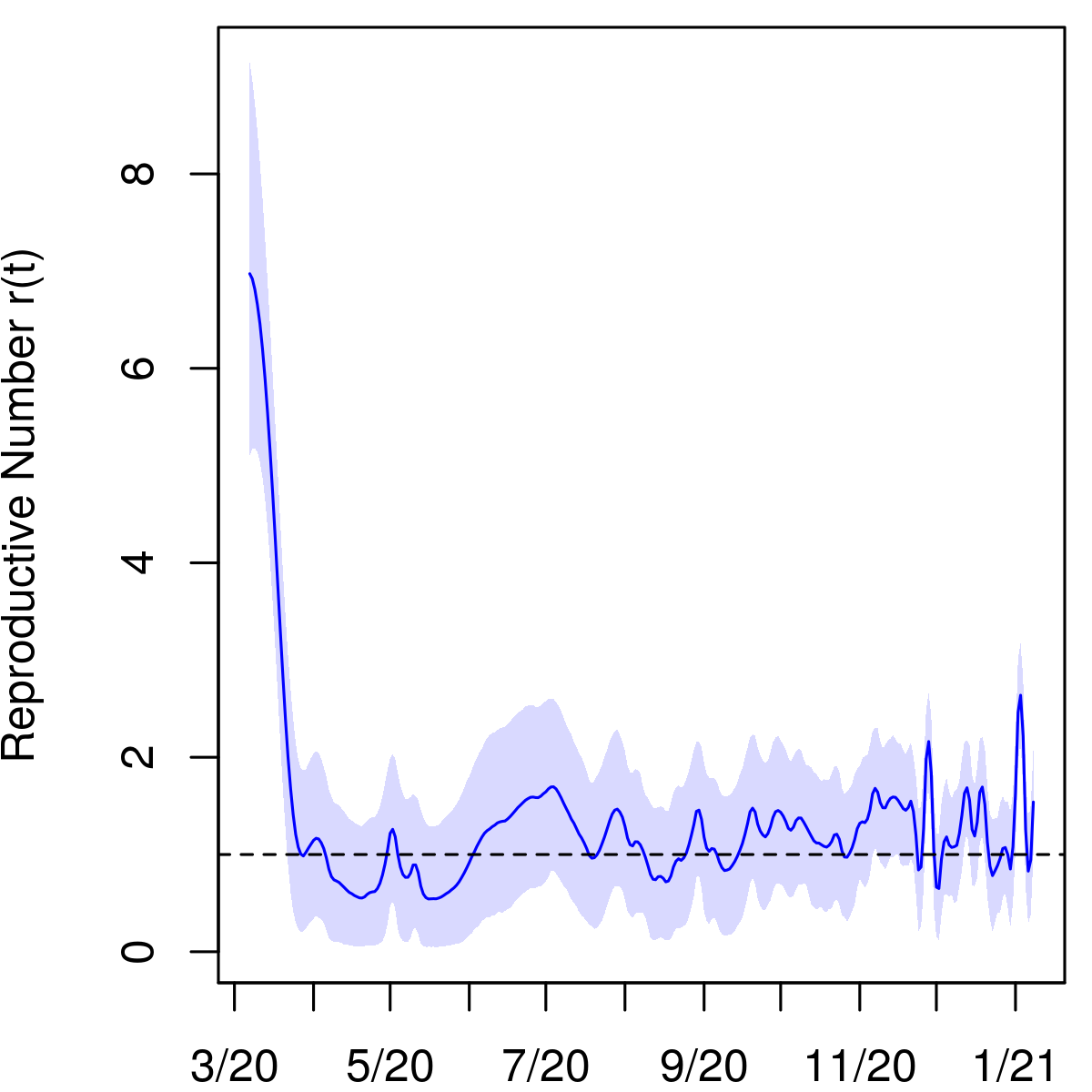}
&
\includegraphics[scale=0.77]{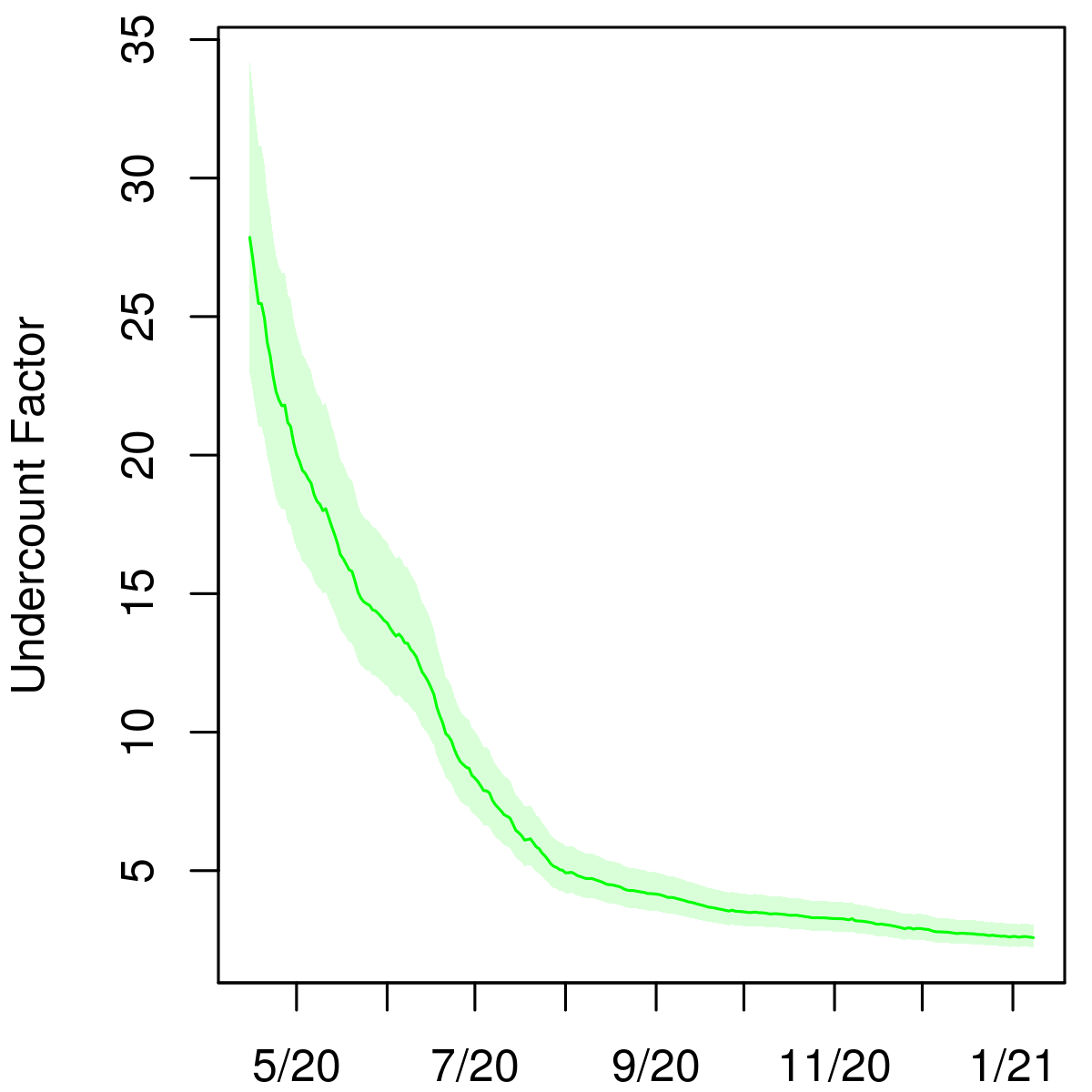} 
\end{tabular}
\caption{Posterior median and middle 95\% intervals for daily new infections, cumulative incidence, $r(t)$, and cumulative undercount from March 2020 to January 2021. In the top left panel, deaths divided by the posterior median IFR are plotted in grey for comparison.}
\end{figure}
\newpage
\begin{figure}[htbp!]
\textbf{Oregon}
\centering
\begin{tabular}{ll}
\includegraphics[scale=0.77]{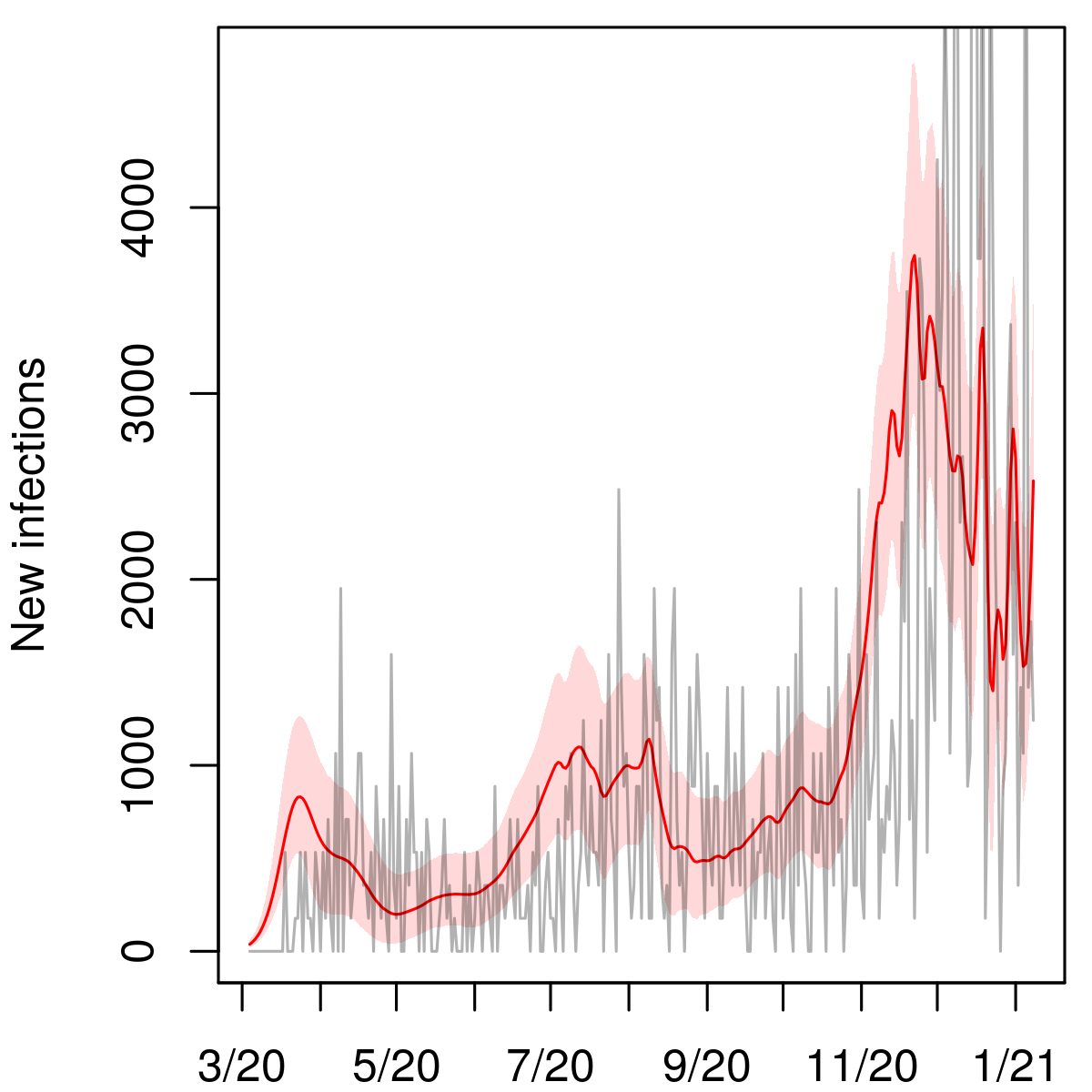}
&
\includegraphics[scale=0.77]{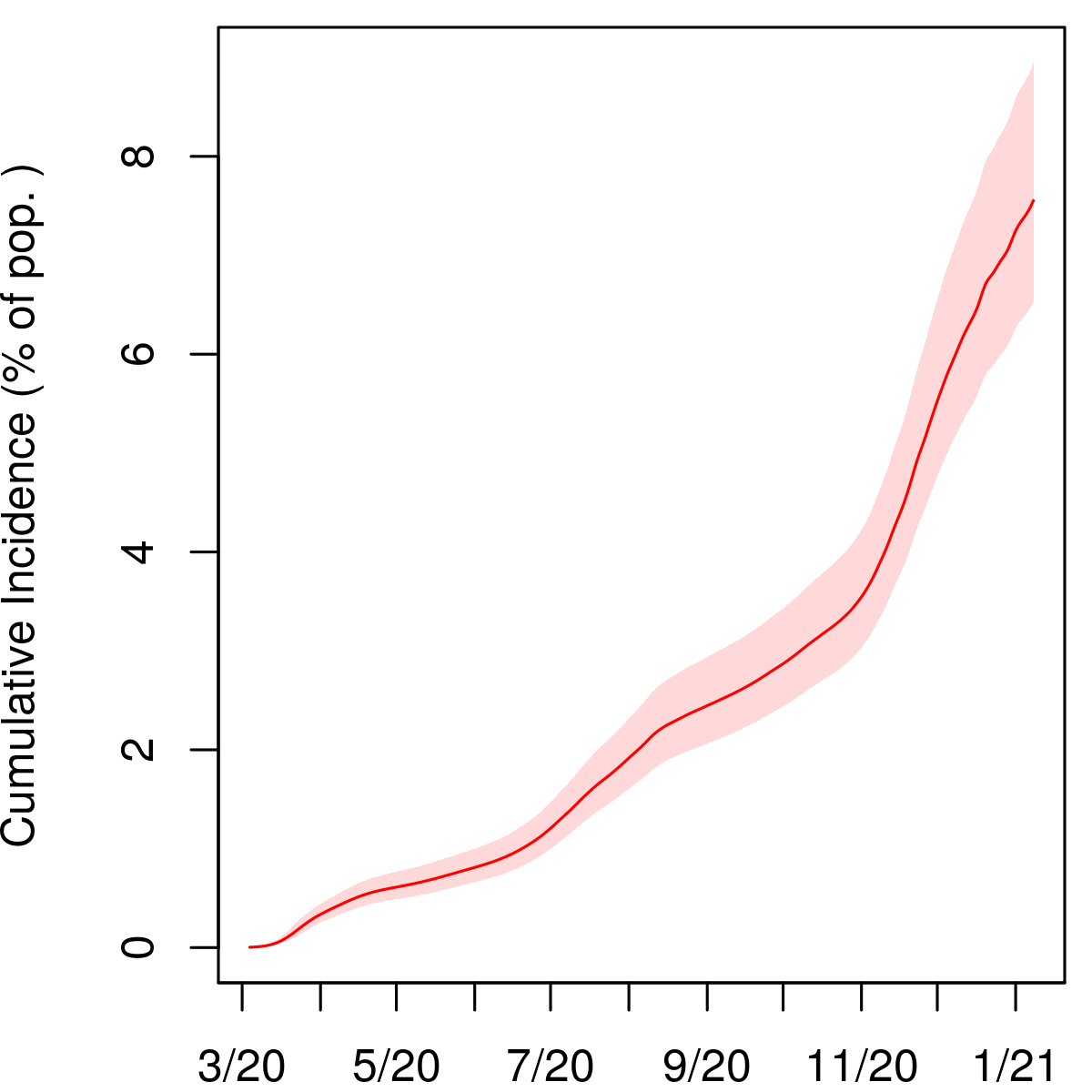} \\
\includegraphics[scale=0.77]{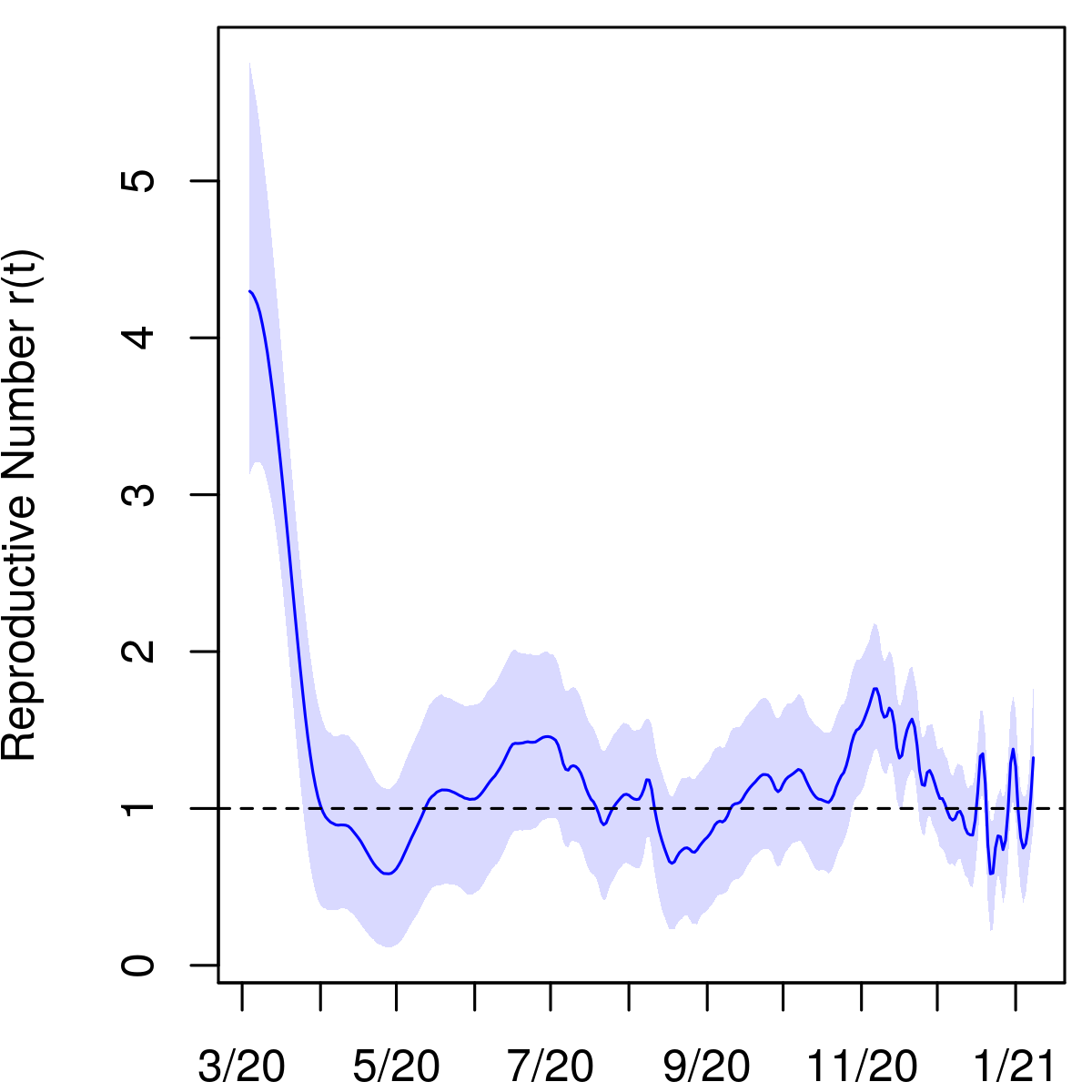}
&
\includegraphics[scale=0.77]{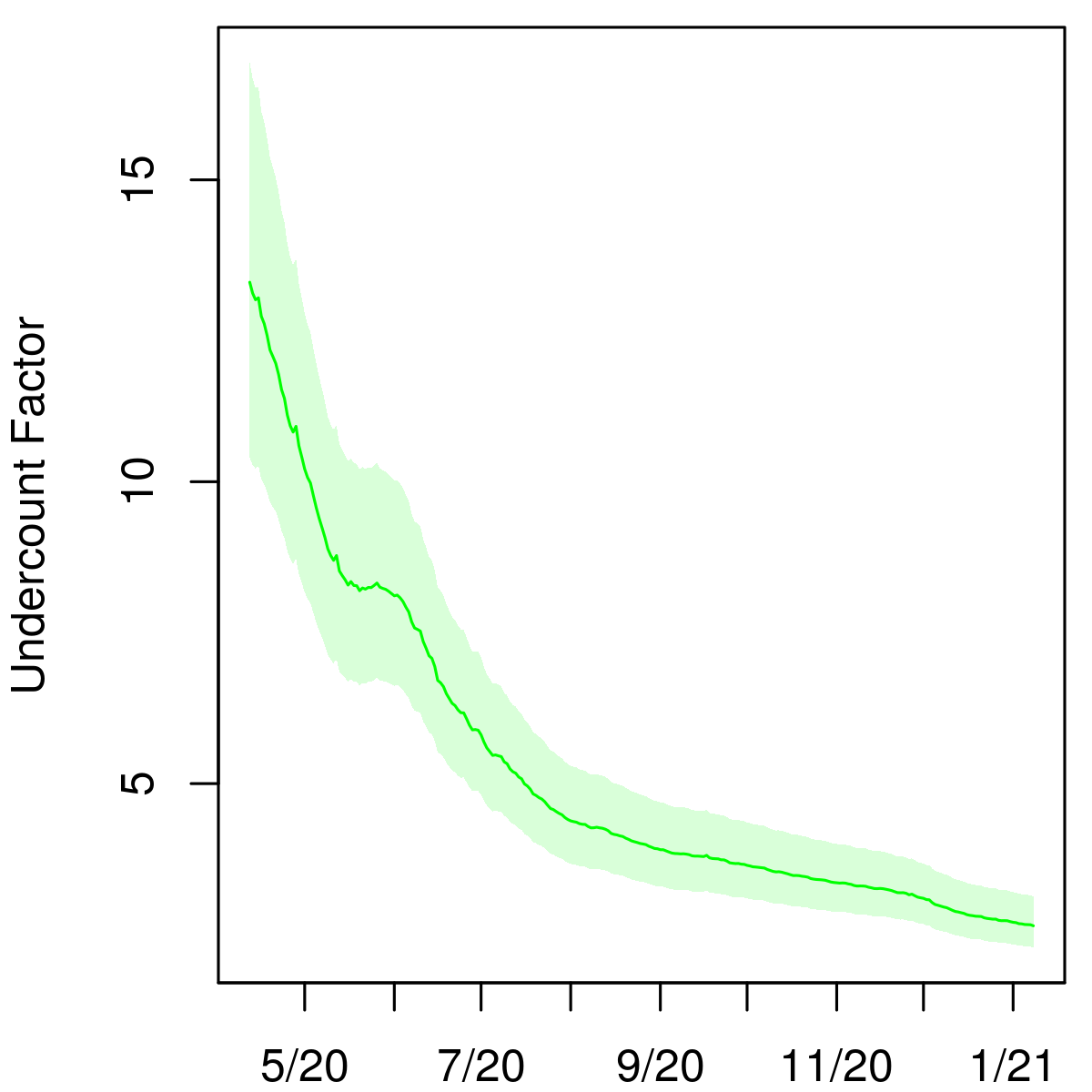} 
\end{tabular}
\caption{Posterior median and middle 95\% intervals for daily new infections, cumulative incidence, $r(t)$, and cumulative undercount from March 2020 to January 2021. In the top left panel, deaths divided by the posterior median IFR are plotted in grey for comparison.}
\end{figure}
\newpage
\begin{figure}[htbp!]
\textbf{Pennsylvania}
\centering
\begin{tabular}{ll}
\includegraphics[scale=0.77]{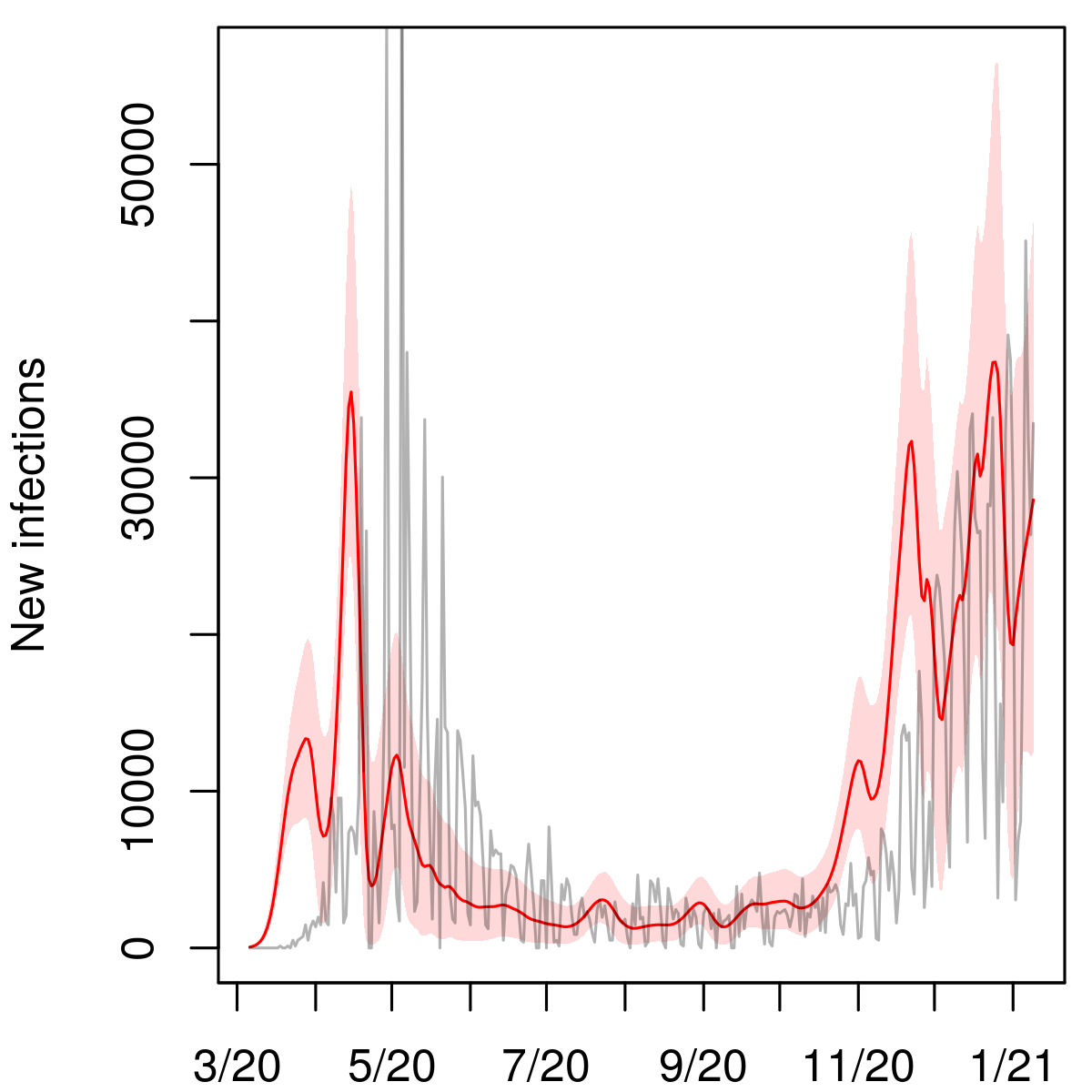}
&
\includegraphics[scale=0.77]{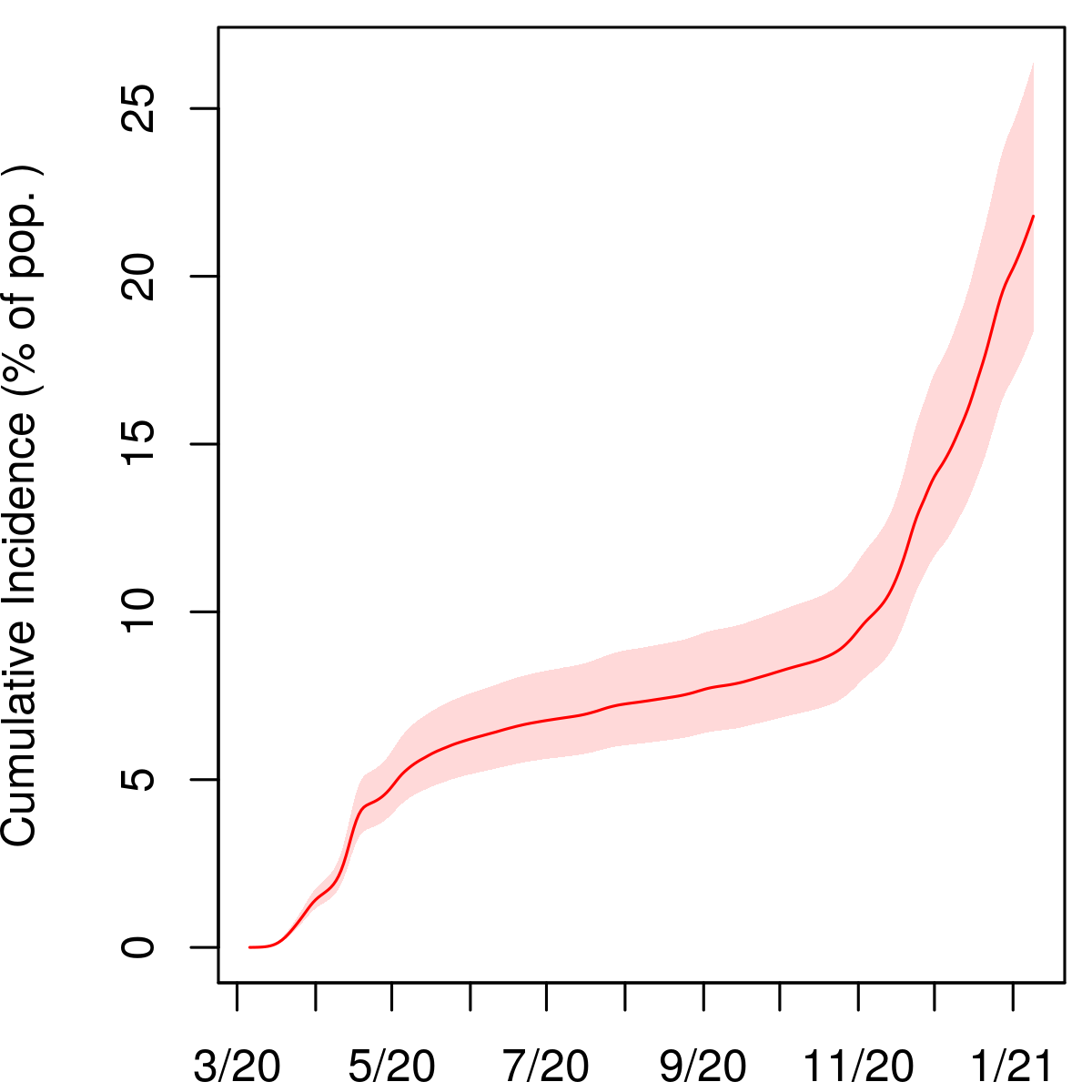} \\
\includegraphics[scale=0.77]{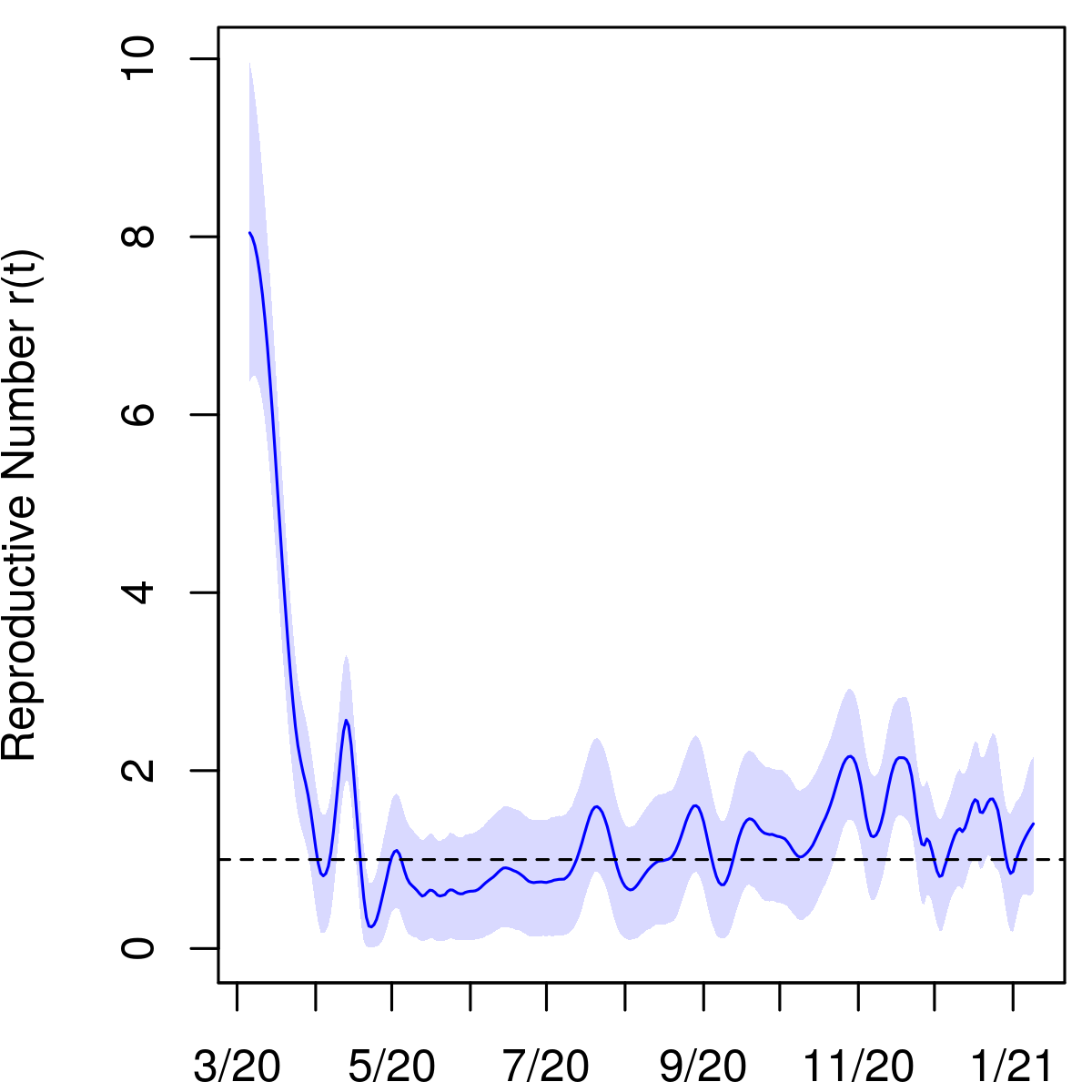}
&
\includegraphics[scale=0.77]{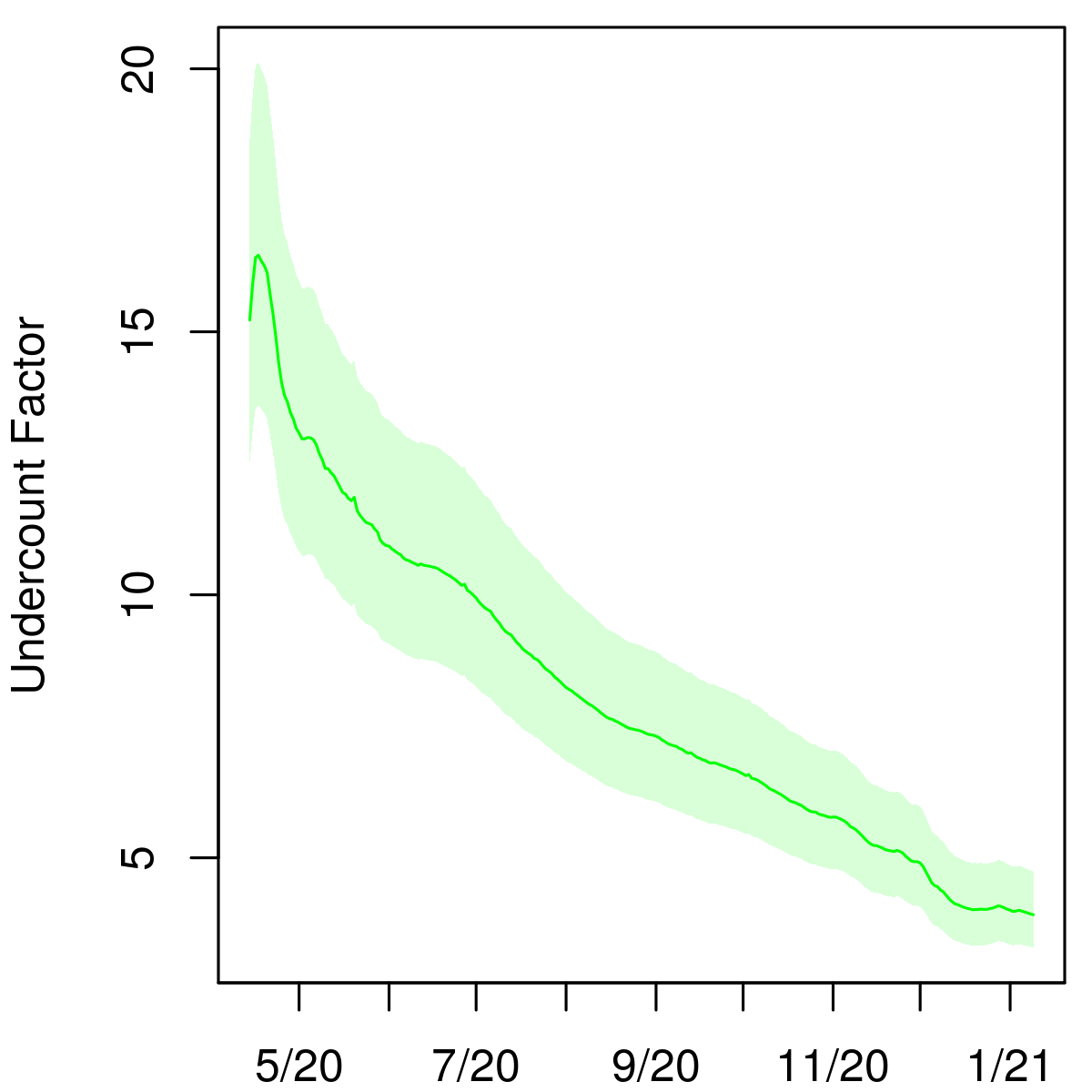} 
\end{tabular}
\caption{Posterior median and middle 95\% intervals for daily new infections, cumulative incidence, $r(t)$, and cumulative undercount from March 2020 to January 2021. In the top left panel, deaths divided by the posterior median IFR are plotted in grey for comparison.}
\end{figure}
\newpage
\begin{figure}[htbp!]
\textbf{Rhode Island}
\centering
\begin{tabular}{ll}
\includegraphics[scale=0.77]{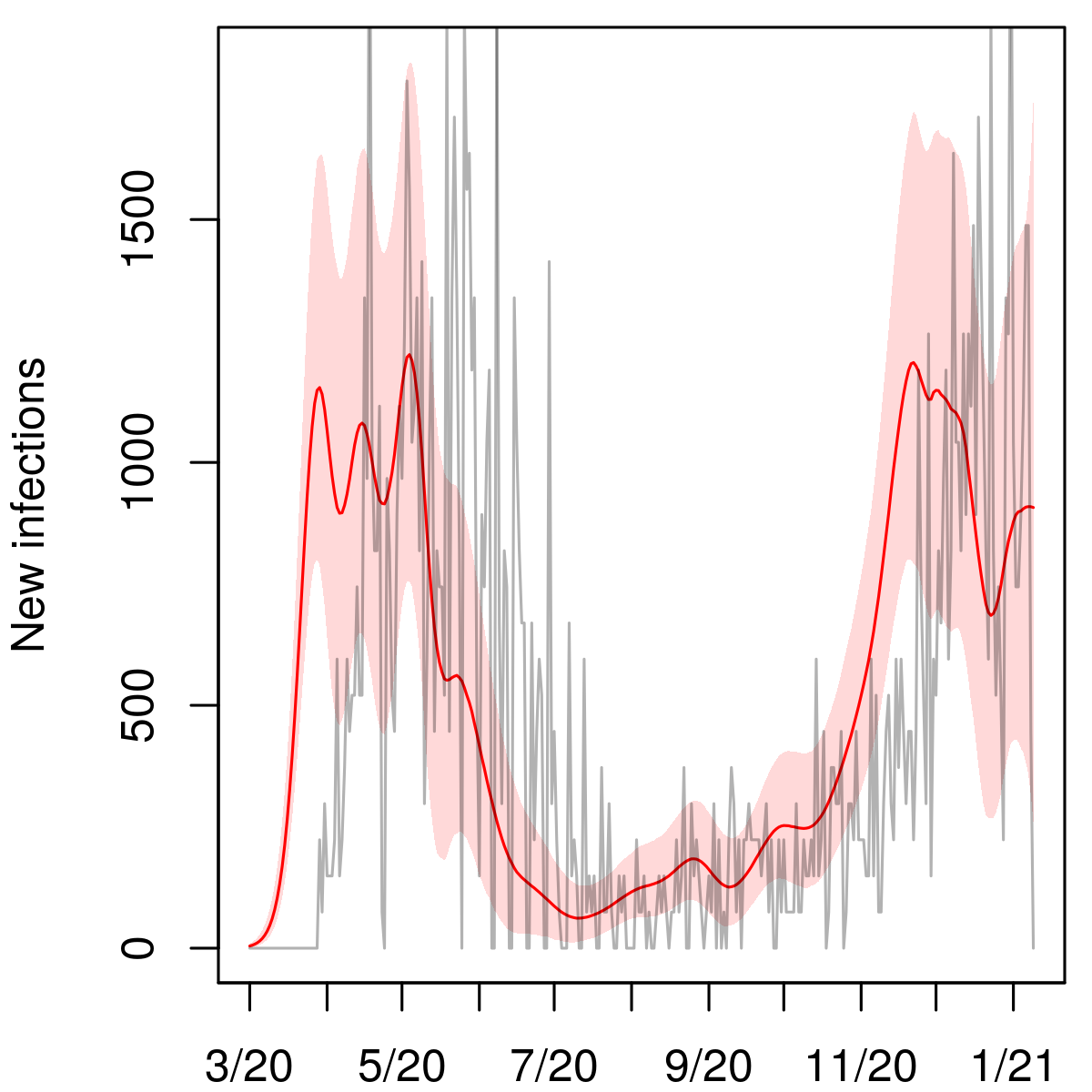}
&
\includegraphics[scale=0.77]{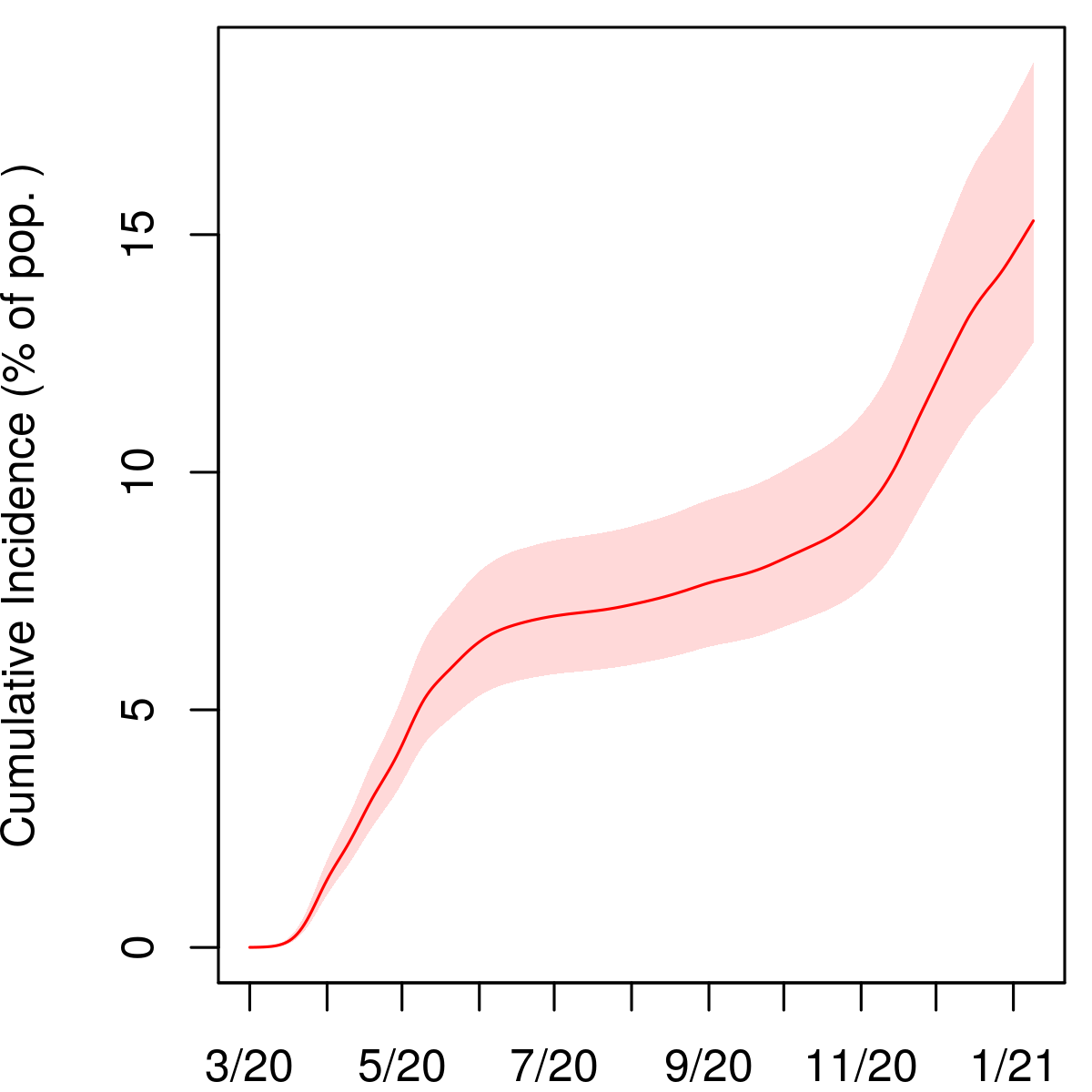} \\
\includegraphics[scale=0.77]{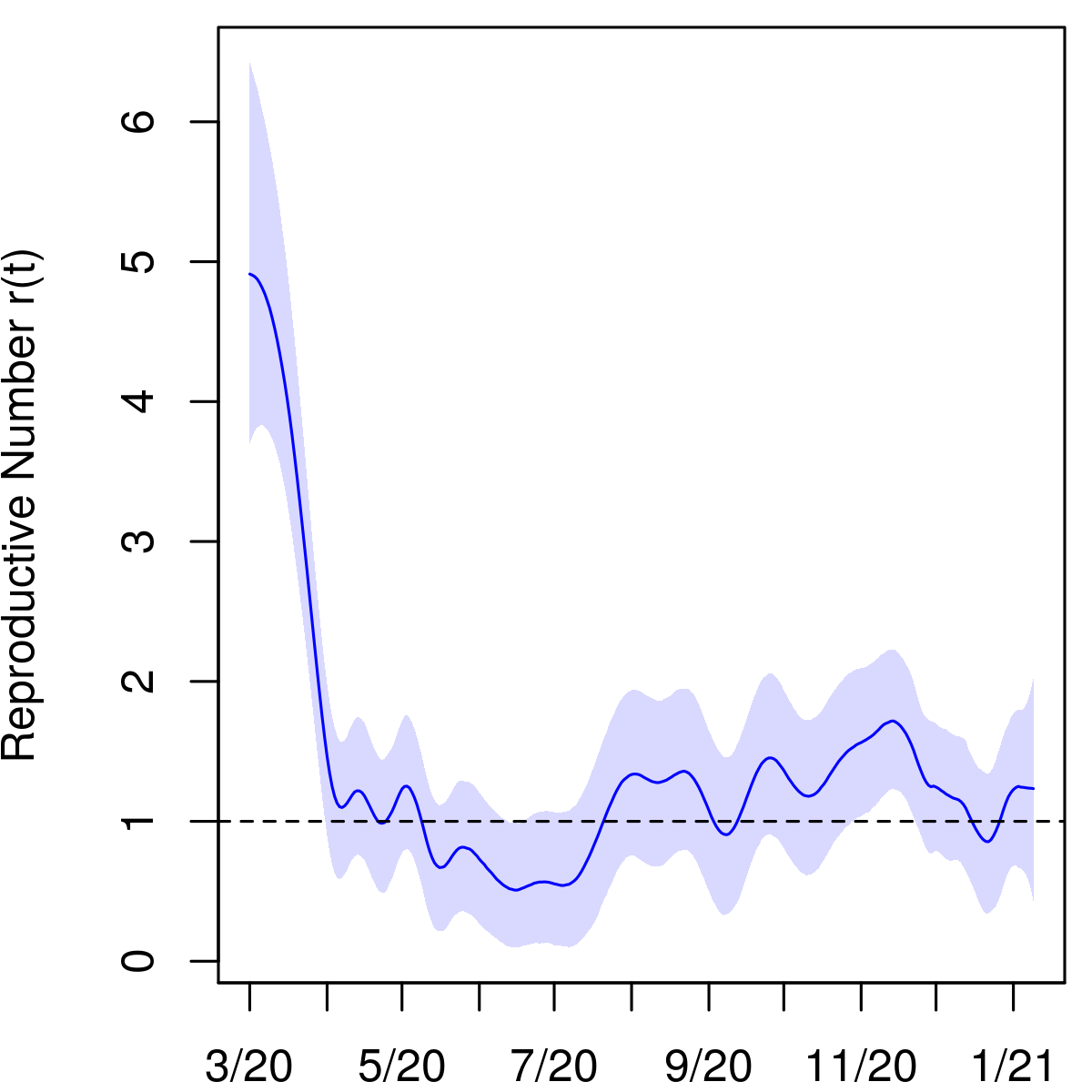}
&
\includegraphics[scale=0.77]{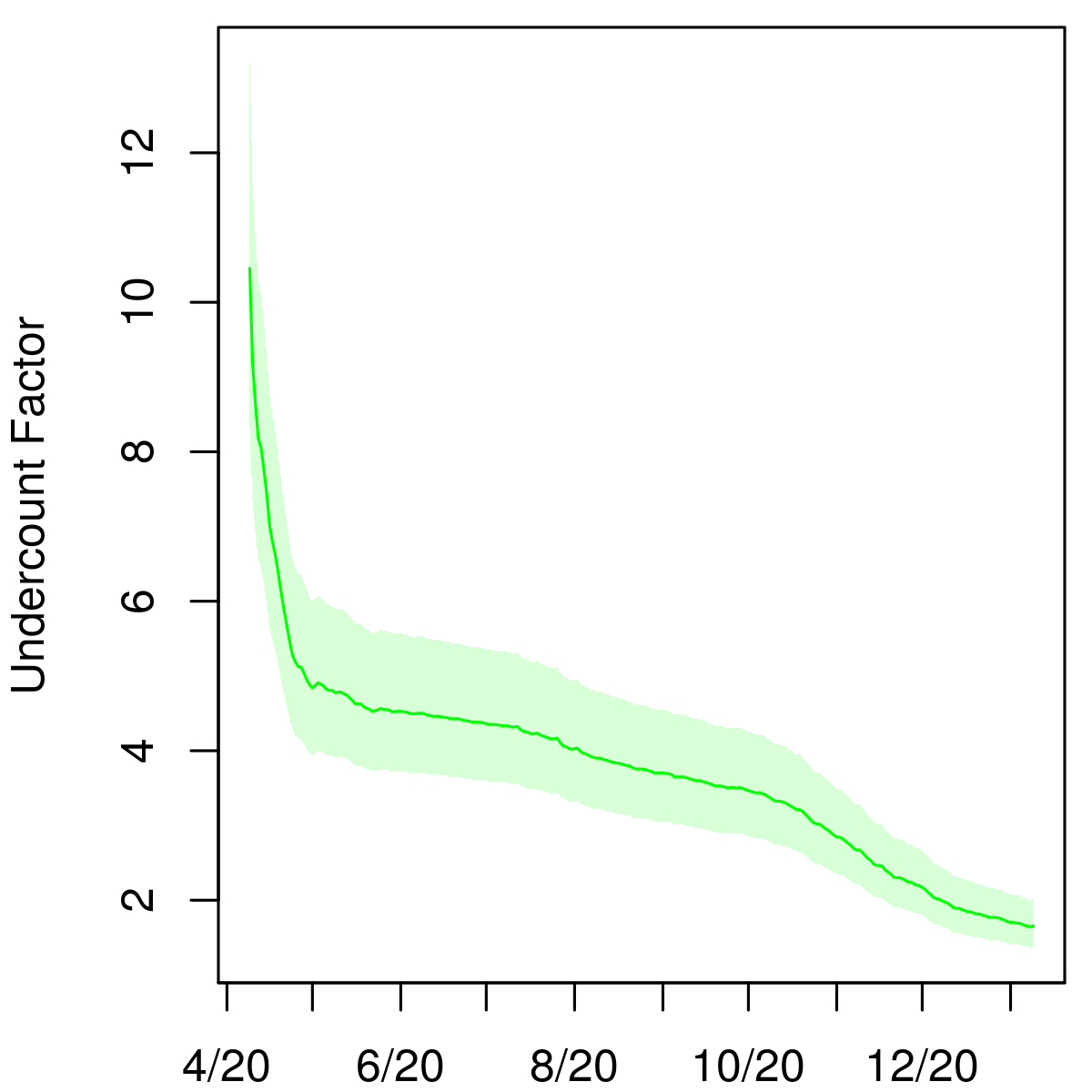} 
\end{tabular}
\caption{Posterior median and middle 95\% intervals for daily new infections, cumulative incidence, $r(t)$, and cumulative undercount from March 2020 to January 2021. In the top left panel, deaths divided by the posterior median IFR are plotted in grey for comparison.}
\end{figure}
\newpage
\begin{figure}[htbp!]
\textbf{South Carolina}
\centering
\begin{tabular}{ll}
\includegraphics[scale=0.77]{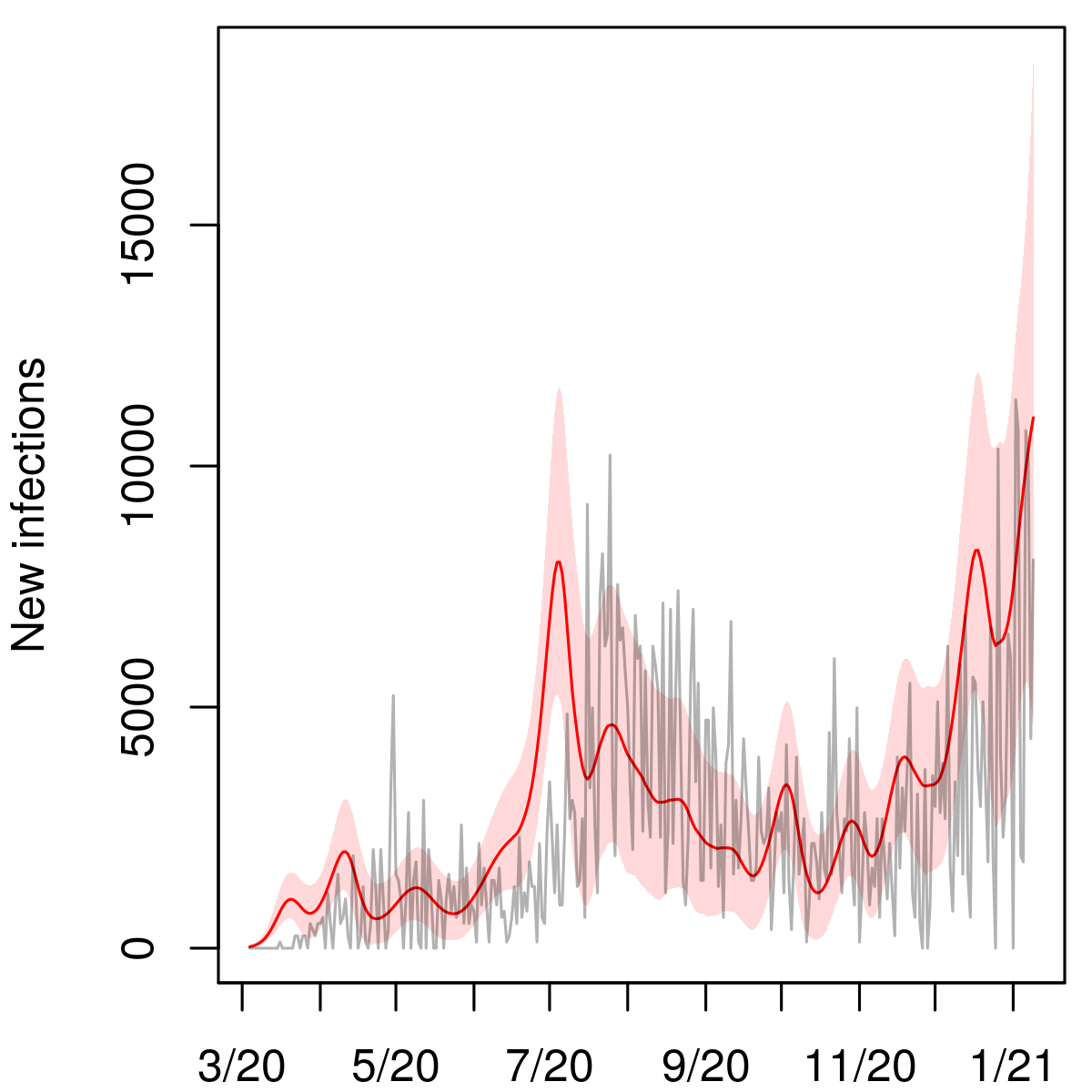}
&
\includegraphics[scale=0.77]{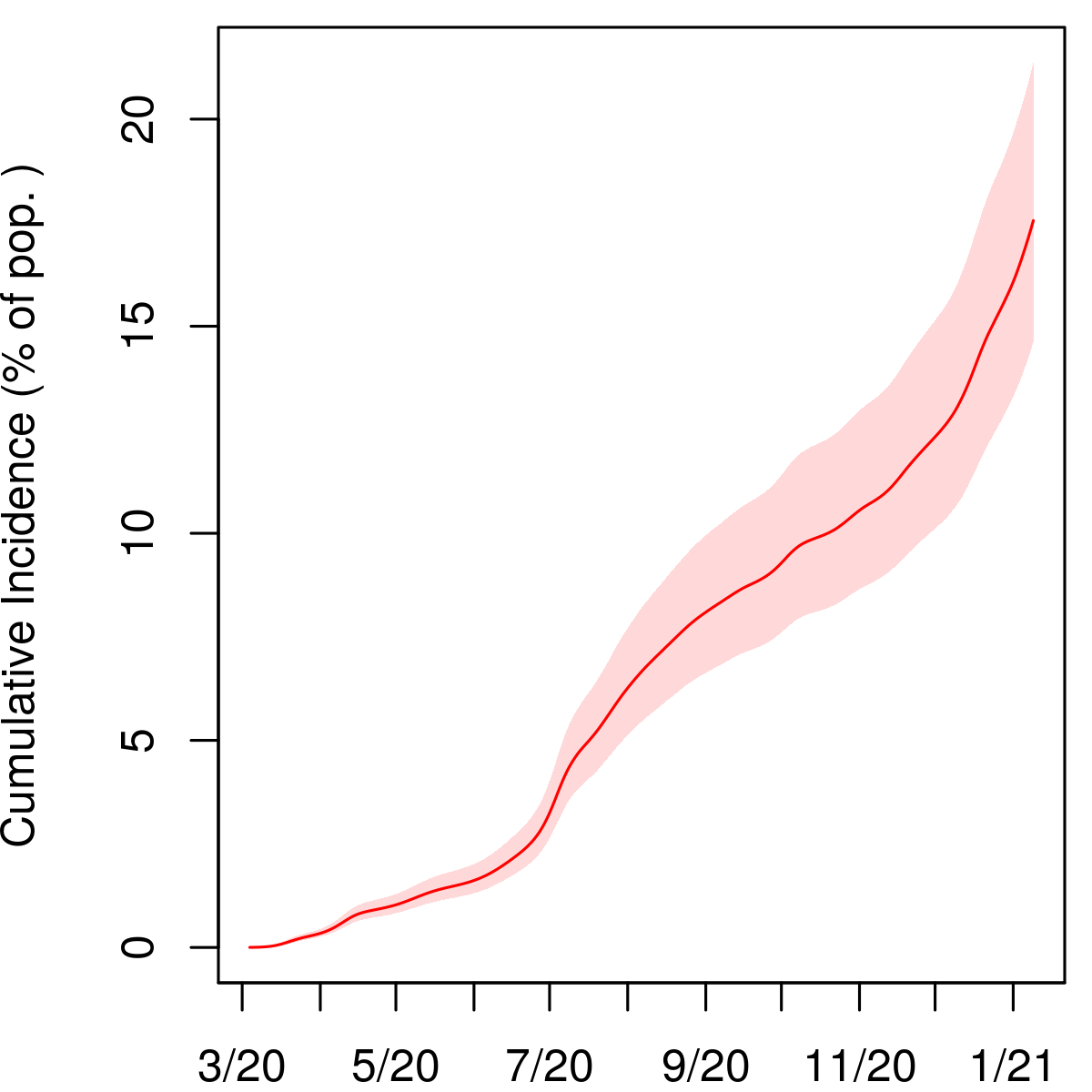} \\
\includegraphics[scale=0.77]{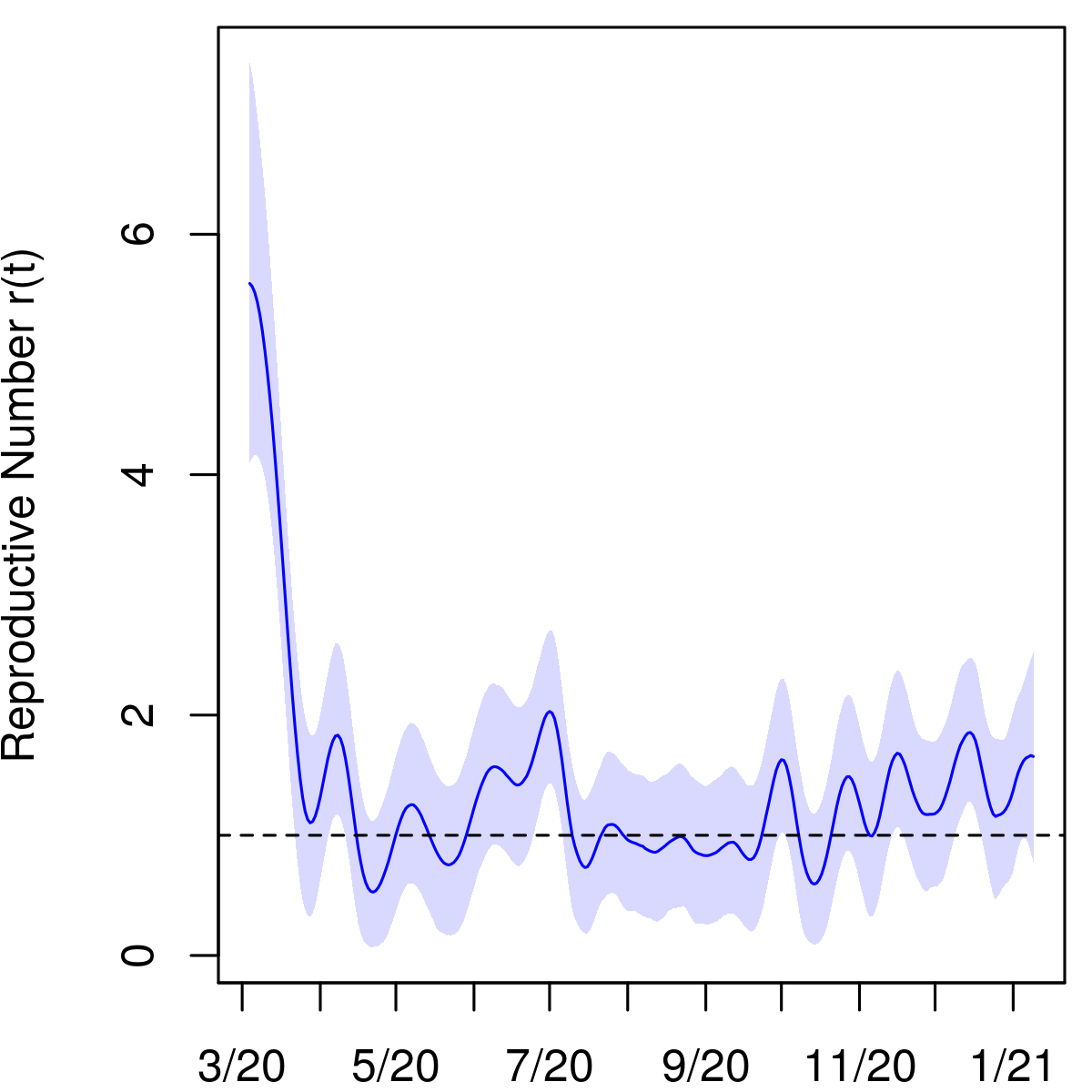}
&
\includegraphics[scale=0.77]{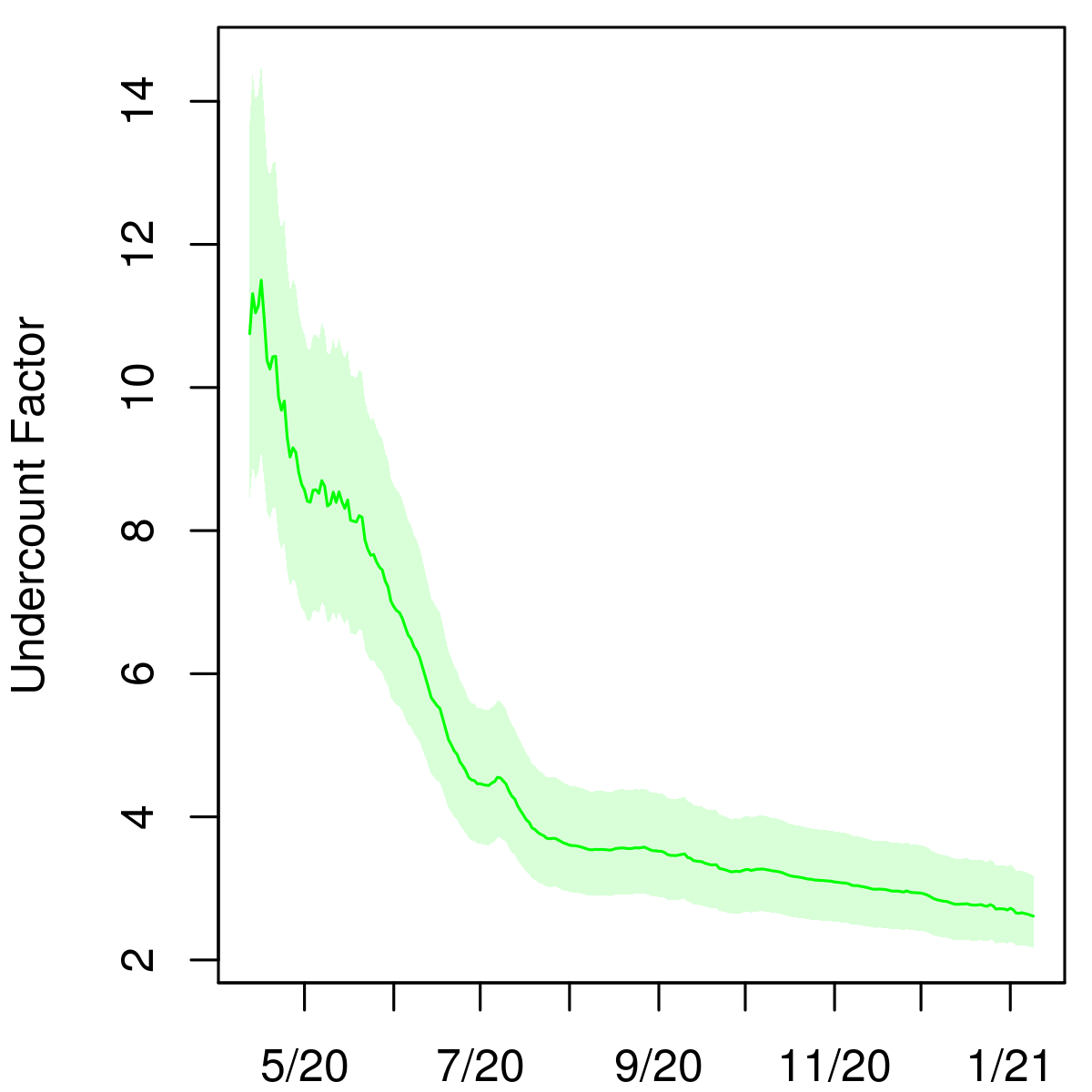} 
\end{tabular}
\caption{Posterior median and middle 95\% intervals for daily new infections, cumulative incidence, $r(t)$, and cumulative undercount from March 2020 to January 2021. In the top left panel, deaths divided by the posterior median IFR are plotted in grey for comparison.}
\end{figure}
\newpage
\begin{figure}[htbp!]
\textbf{South Dakota}
\centering
\begin{tabular}{ll}
\includegraphics[scale=0.77]{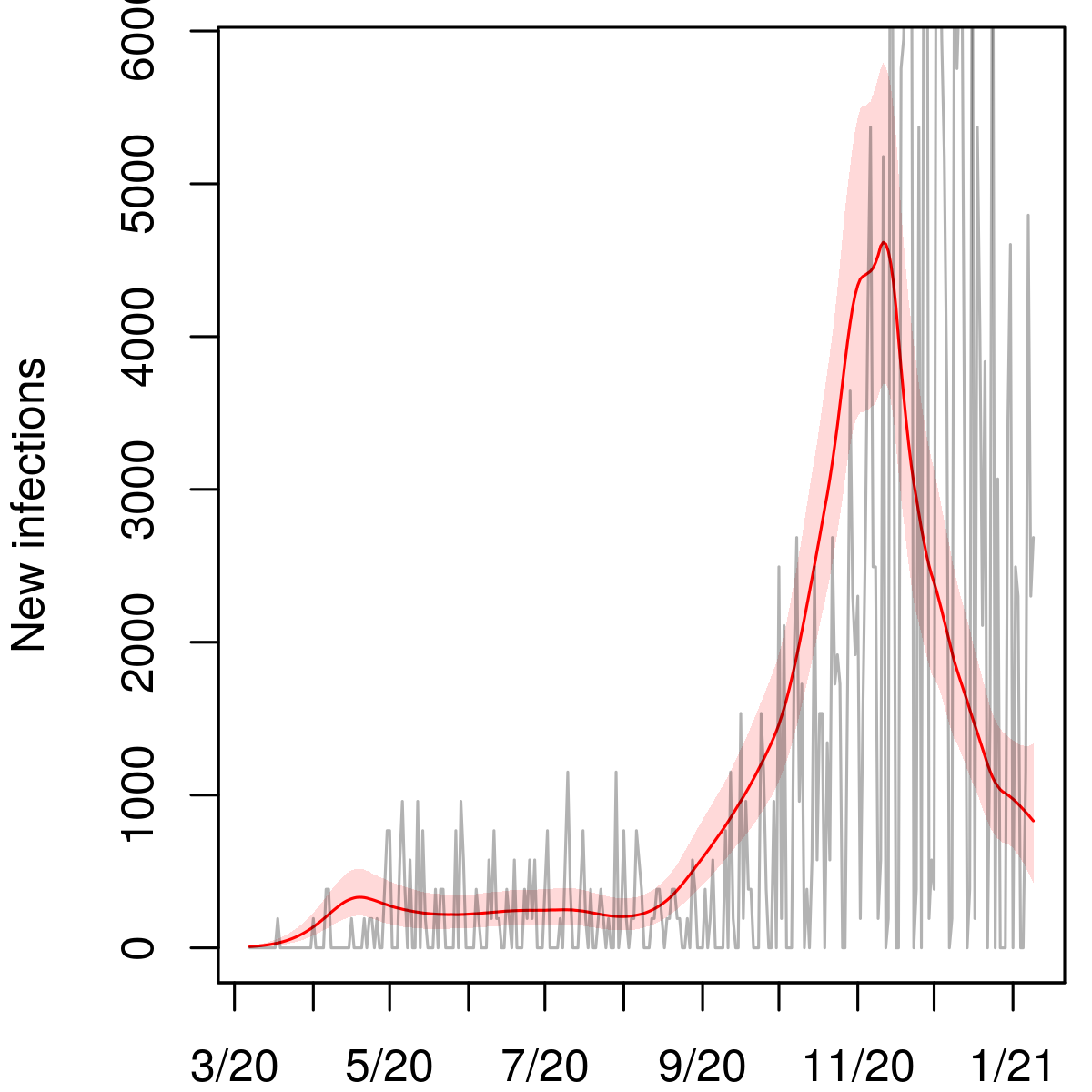}
&
\includegraphics[scale=0.77]{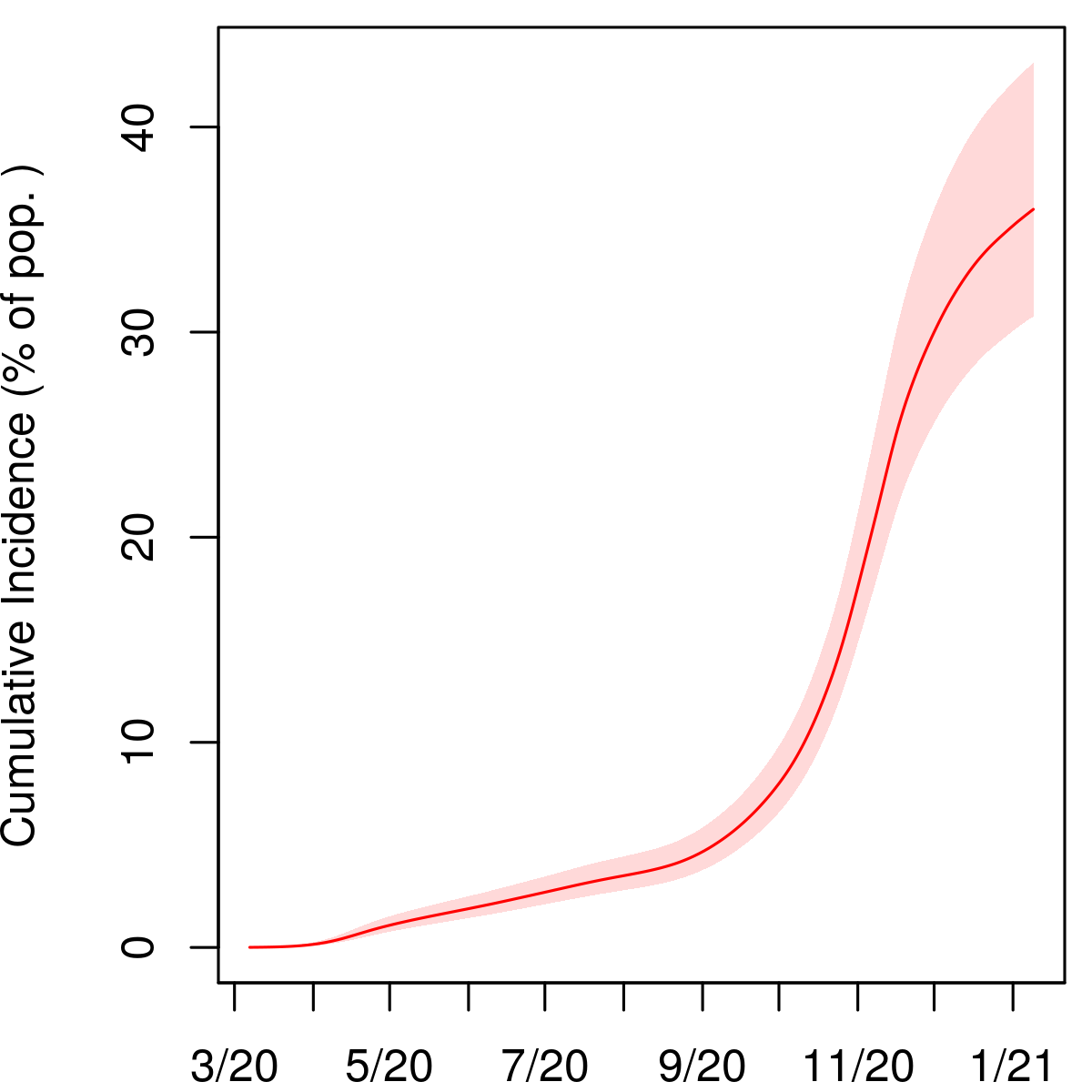} \\
\includegraphics[scale=0.77]{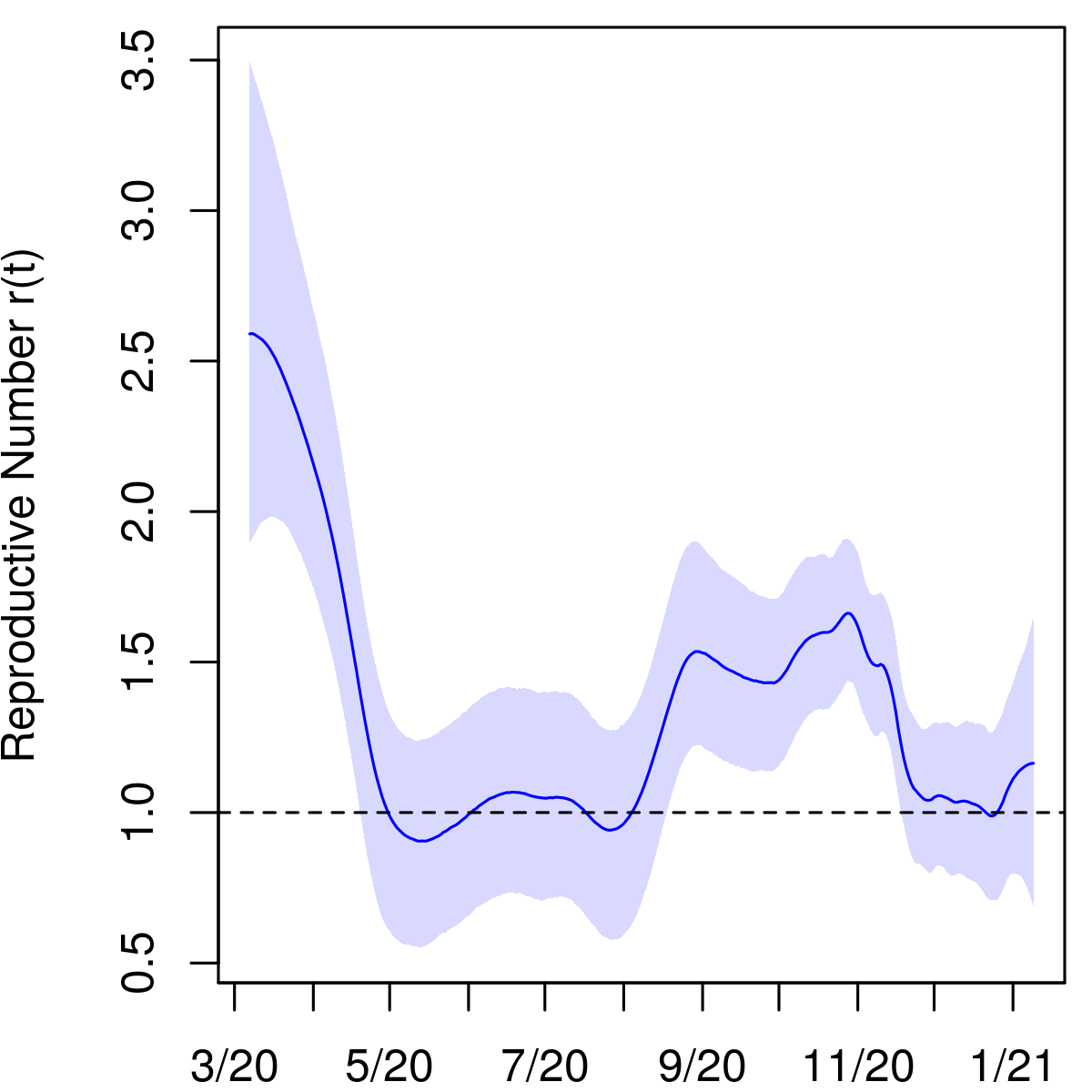}
&
\includegraphics[scale=0.77]{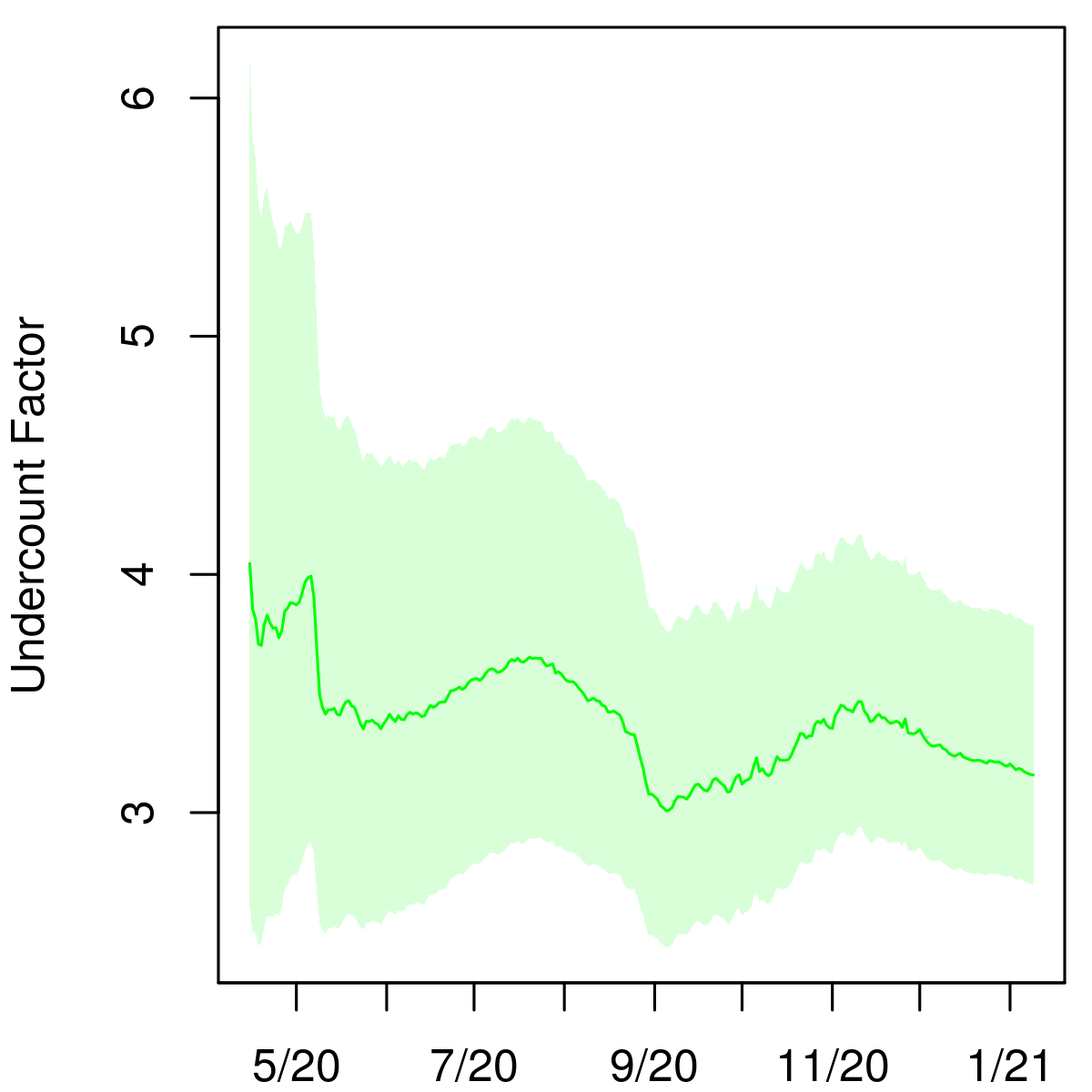} 
\end{tabular}
\caption{Posterior median and middle 95\% intervals for daily new infections, cumulative incidence, $r(t)$, and cumulative undercount from March 2020 to January 2021. In the top left panel, deaths divided by the posterior median IFR are plotted in grey for comparison.}
\end{figure}
\newpage
\begin{figure}[htbp!]
\textbf{Tennessee}
\centering
\begin{tabular}{ll}
\includegraphics[scale=0.77]{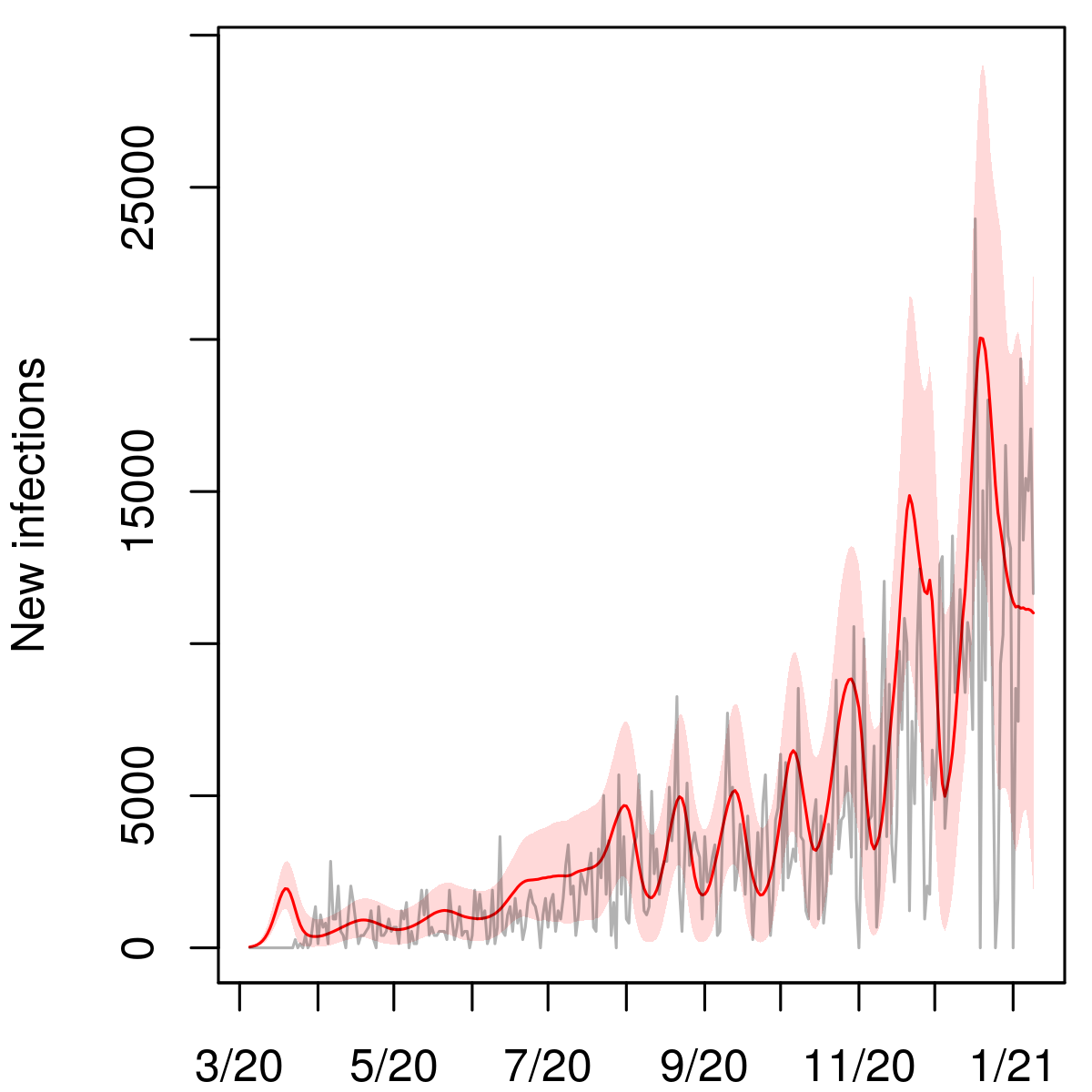}
&
\includegraphics[scale=0.77]{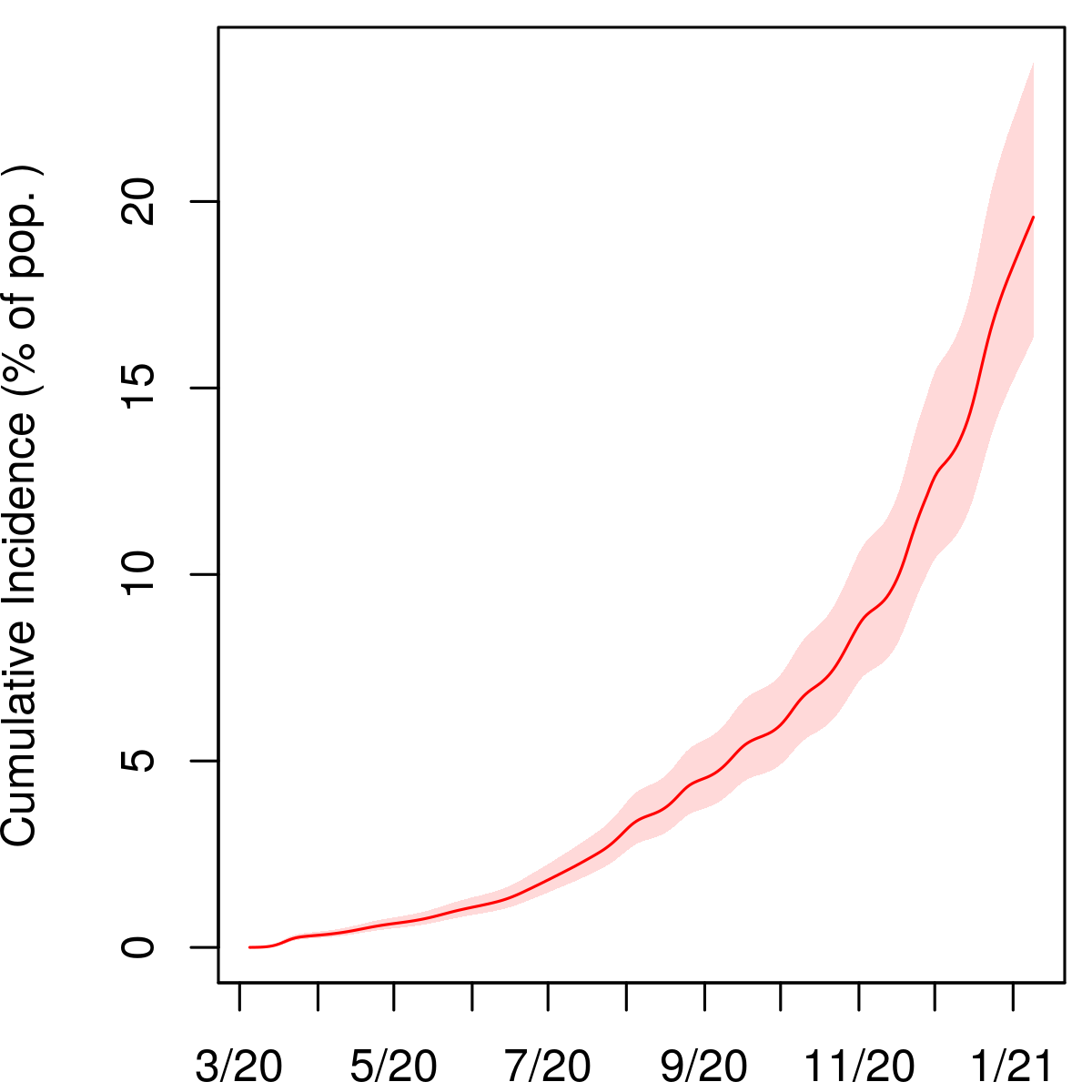} \\
\includegraphics[scale=0.77]{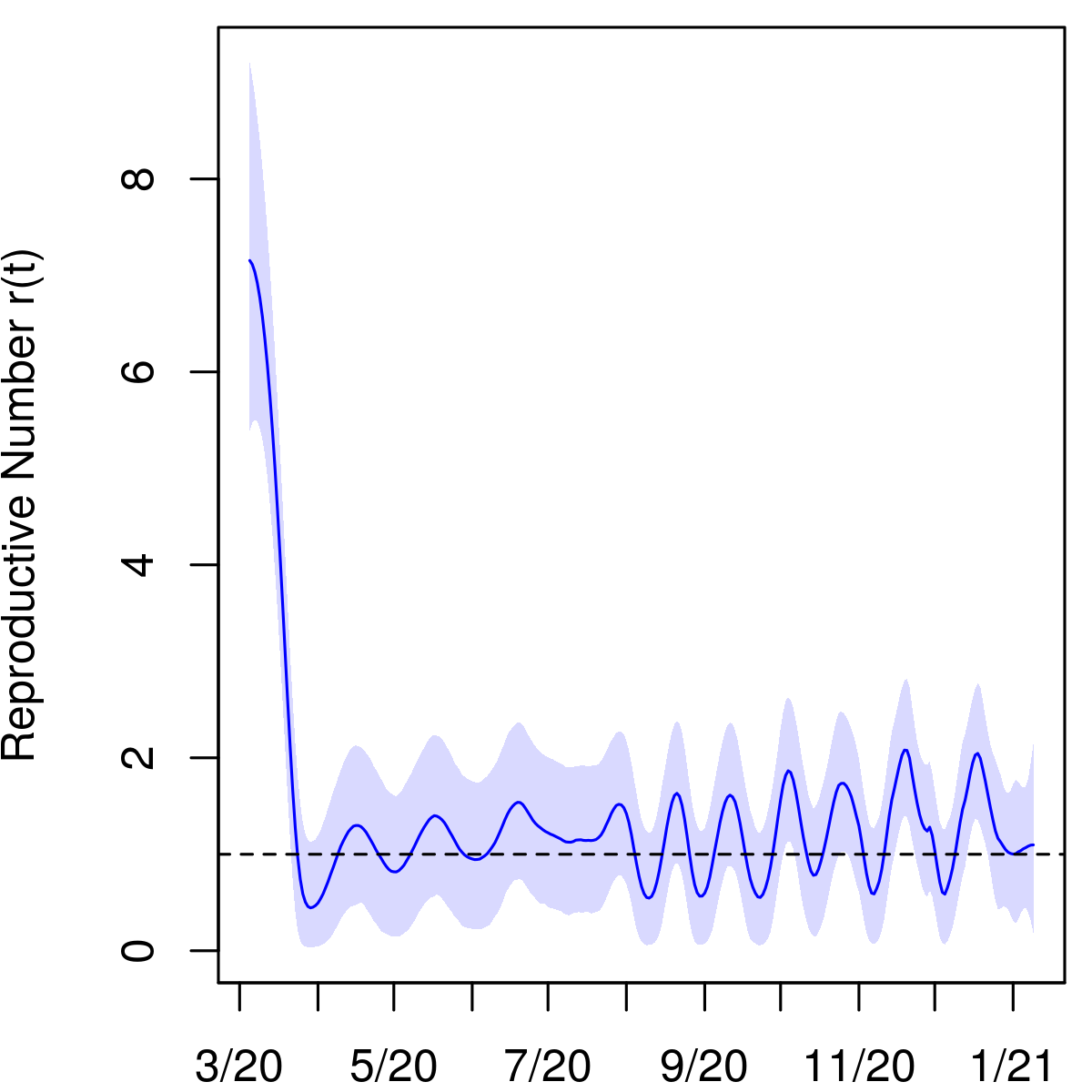}
&
\includegraphics[scale=0.77]{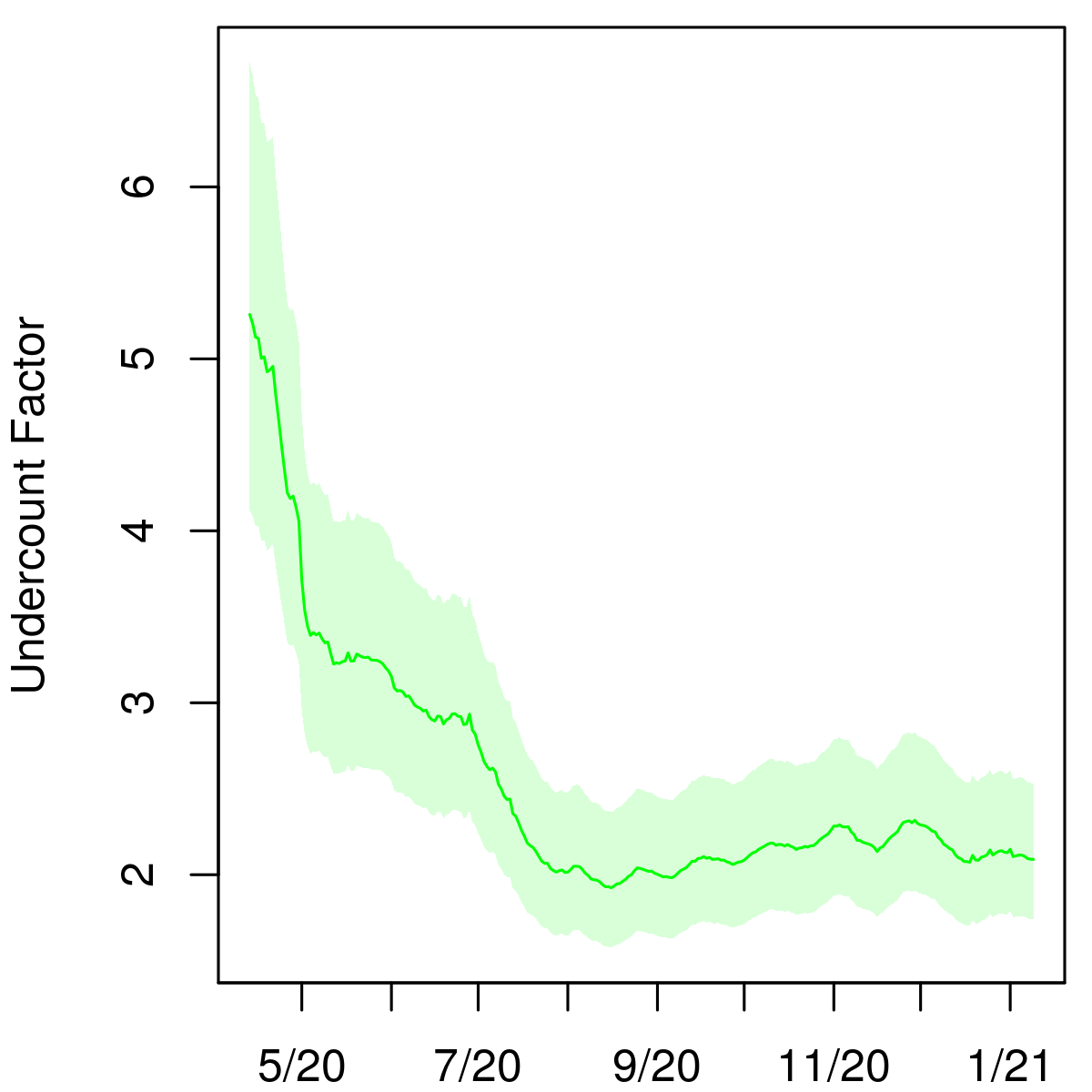} 
\end{tabular}
\caption{Posterior median and middle 95\% intervals for daily new infections, cumulative incidence, $r(t)$, and cumulative undercount from March 2020 to January 2021. In the top left panel, deaths divided by the posterior median IFR are plotted in grey for comparison.}
\end{figure}
\newpage
\begin{figure}[htbp!]
\textbf{Texas}
\centering
\begin{tabular}{ll}
\includegraphics[scale=0.77]{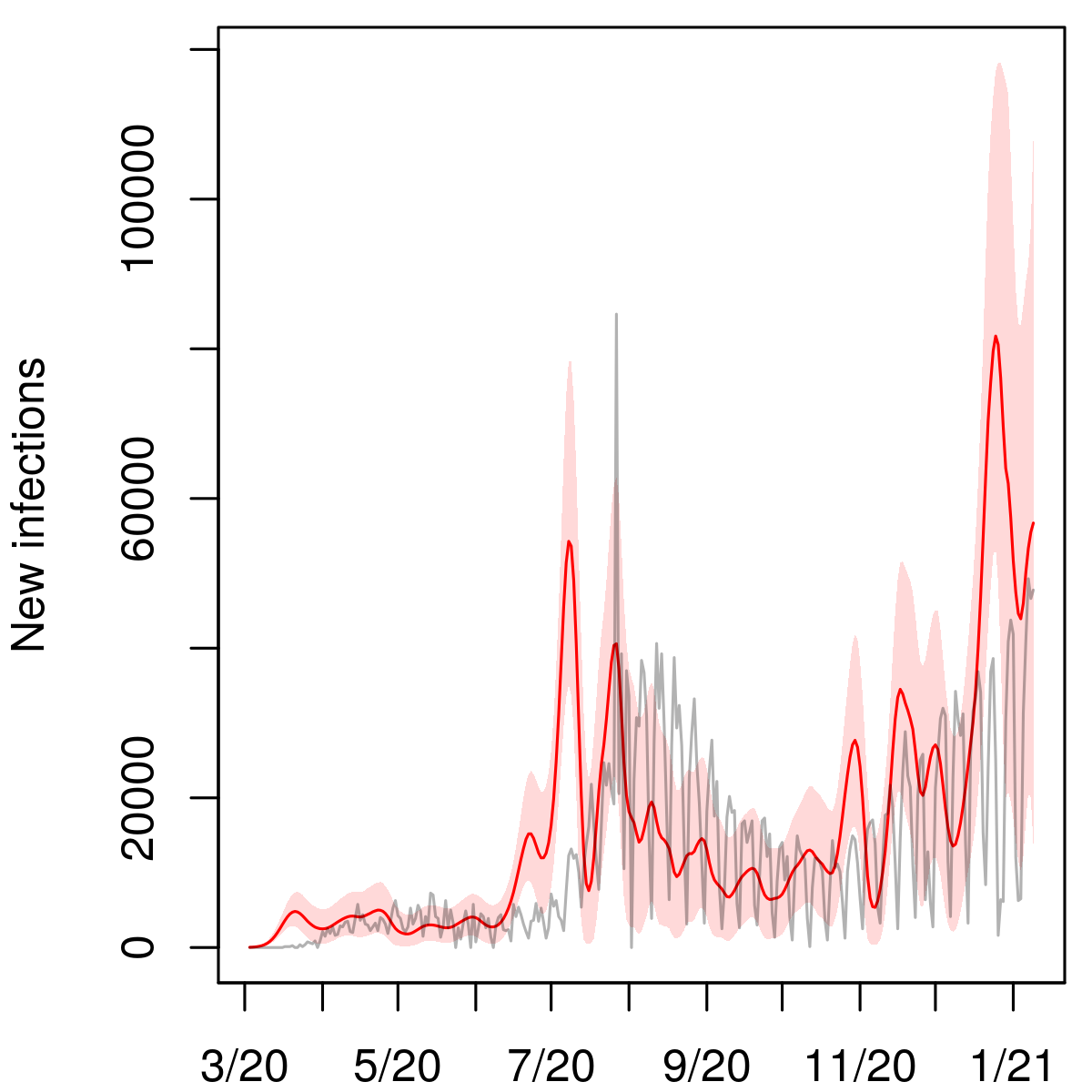}
&
\includegraphics[scale=0.77]{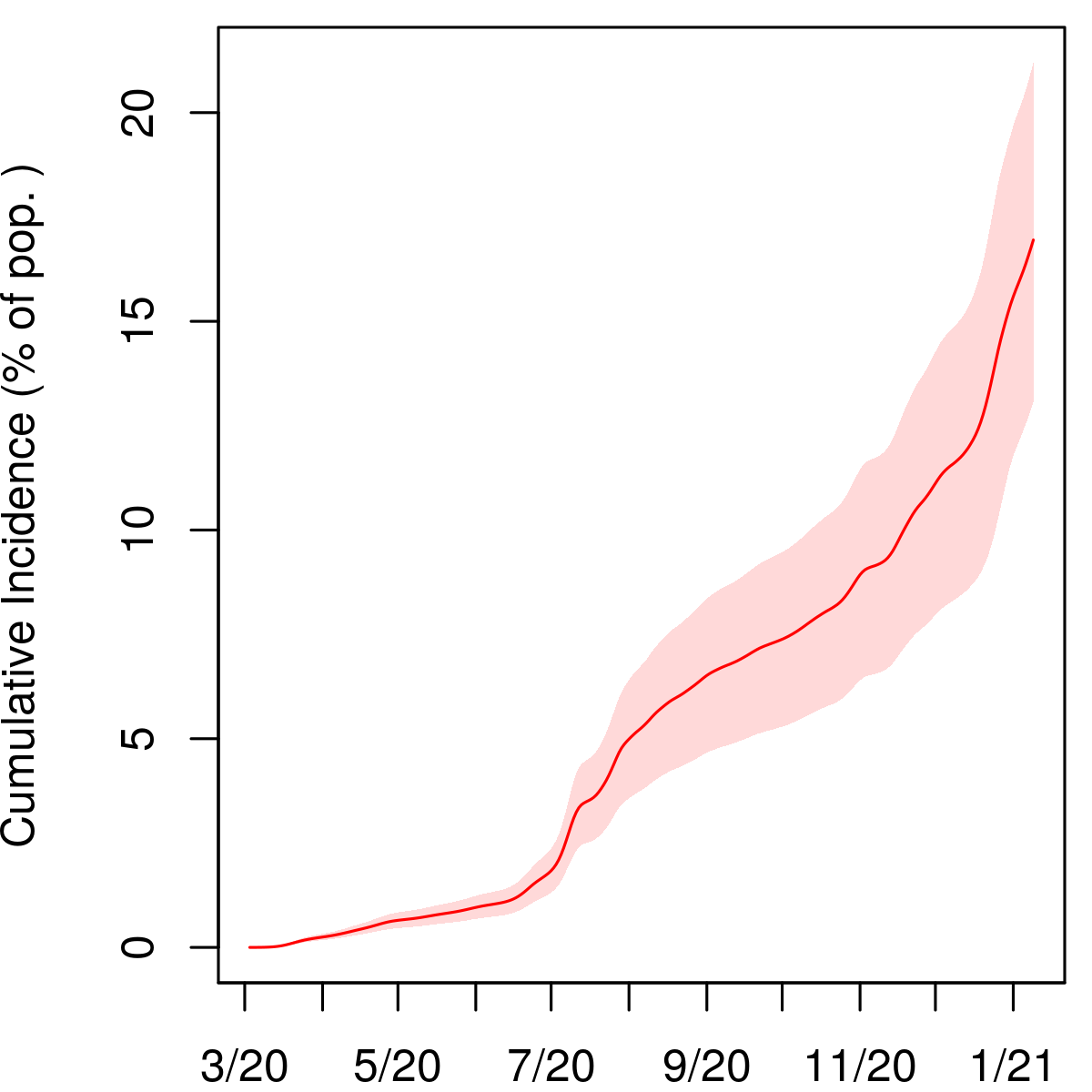} \\
\includegraphics[scale=0.77]{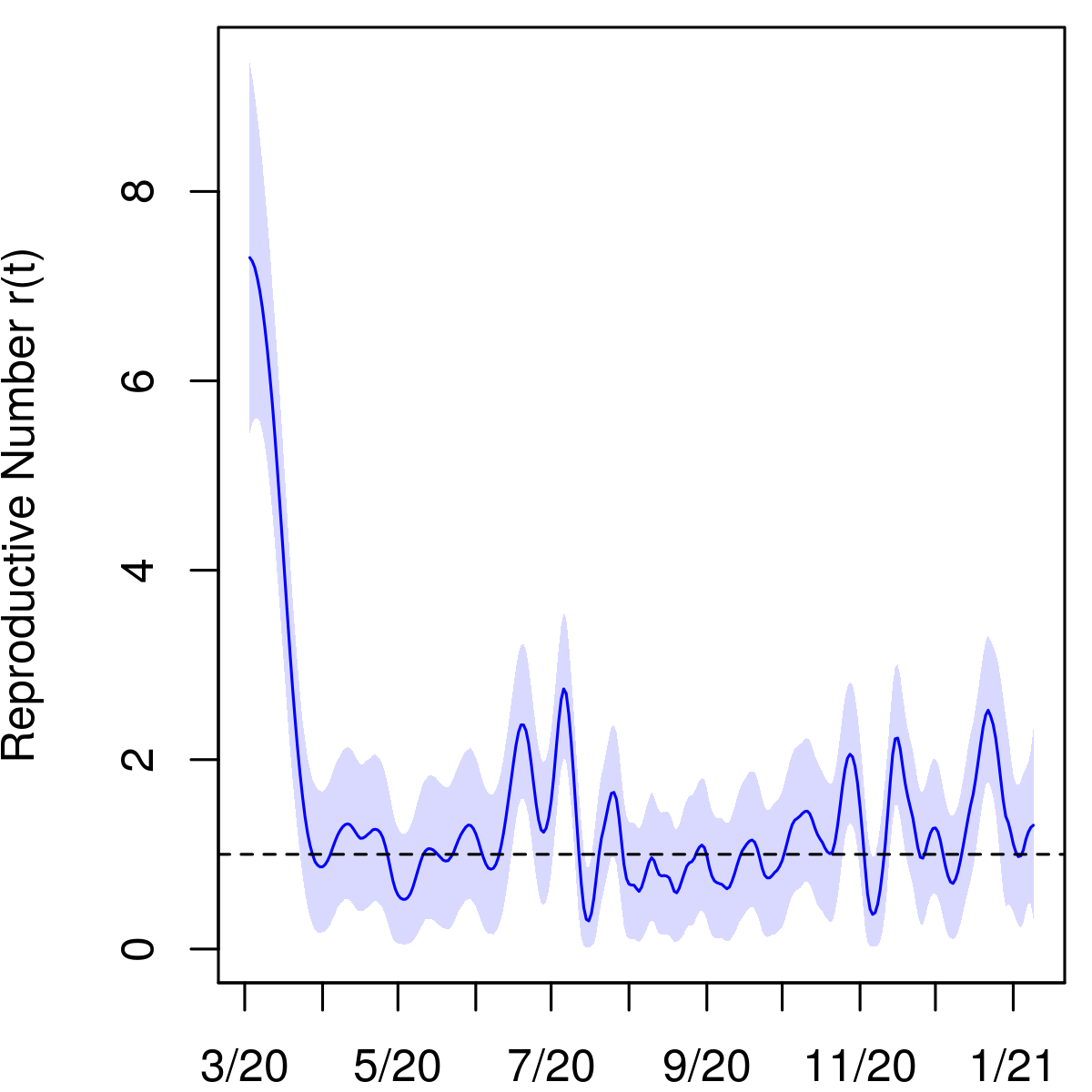}
&
\includegraphics[scale=0.77]{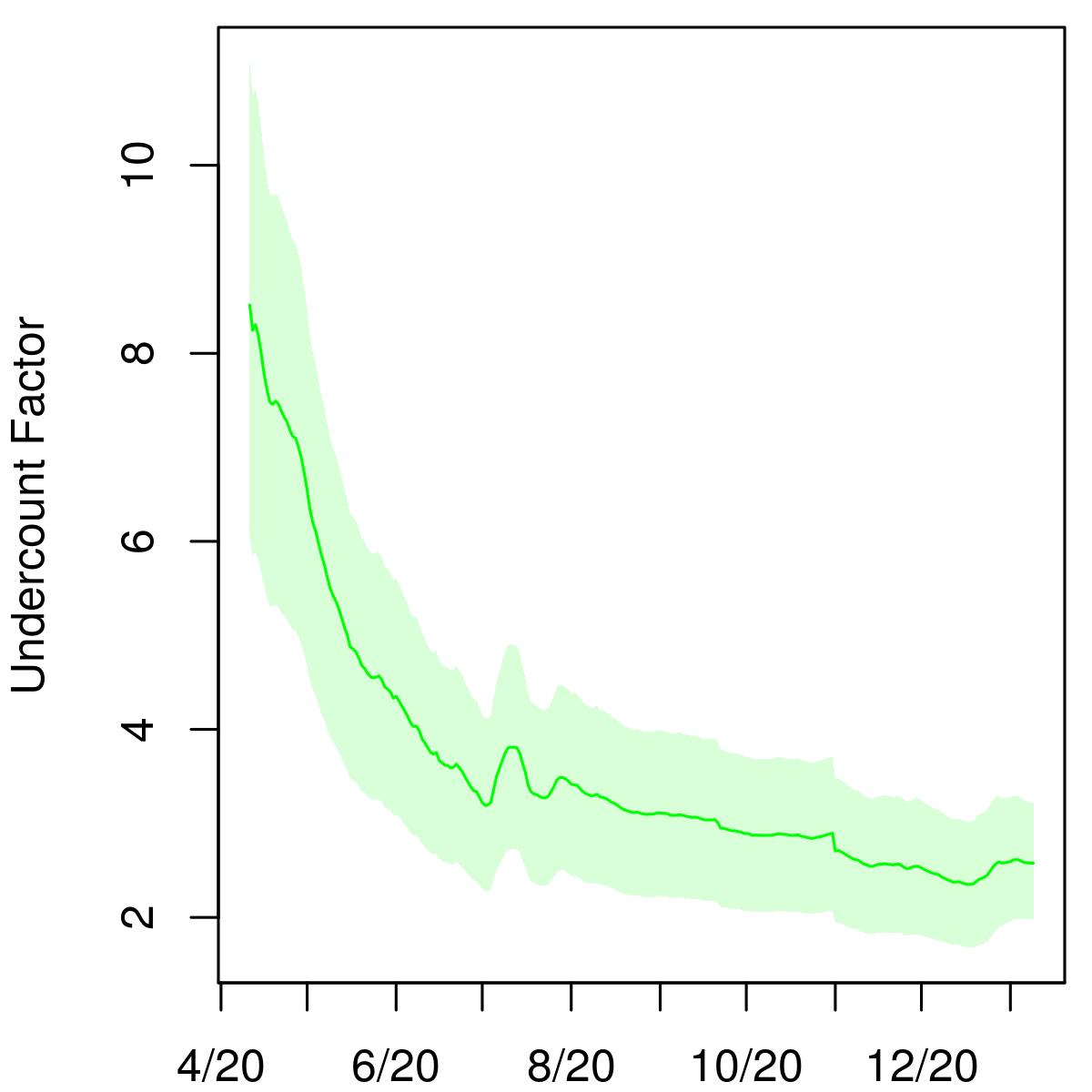} 
\end{tabular}
\caption{Posterior median and middle 95\% intervals for daily new infections, cumulative incidence, $r(t)$, and cumulative undercount from March 2020 to January 2021. In the top left panel, deaths divided by the posterior median IFR are plotted in grey for comparison.}
\end{figure}
\newpage
\begin{figure}[htbp!]
\textbf{Utah}
\centering
\begin{tabular}{ll}
\includegraphics[scale=0.77]{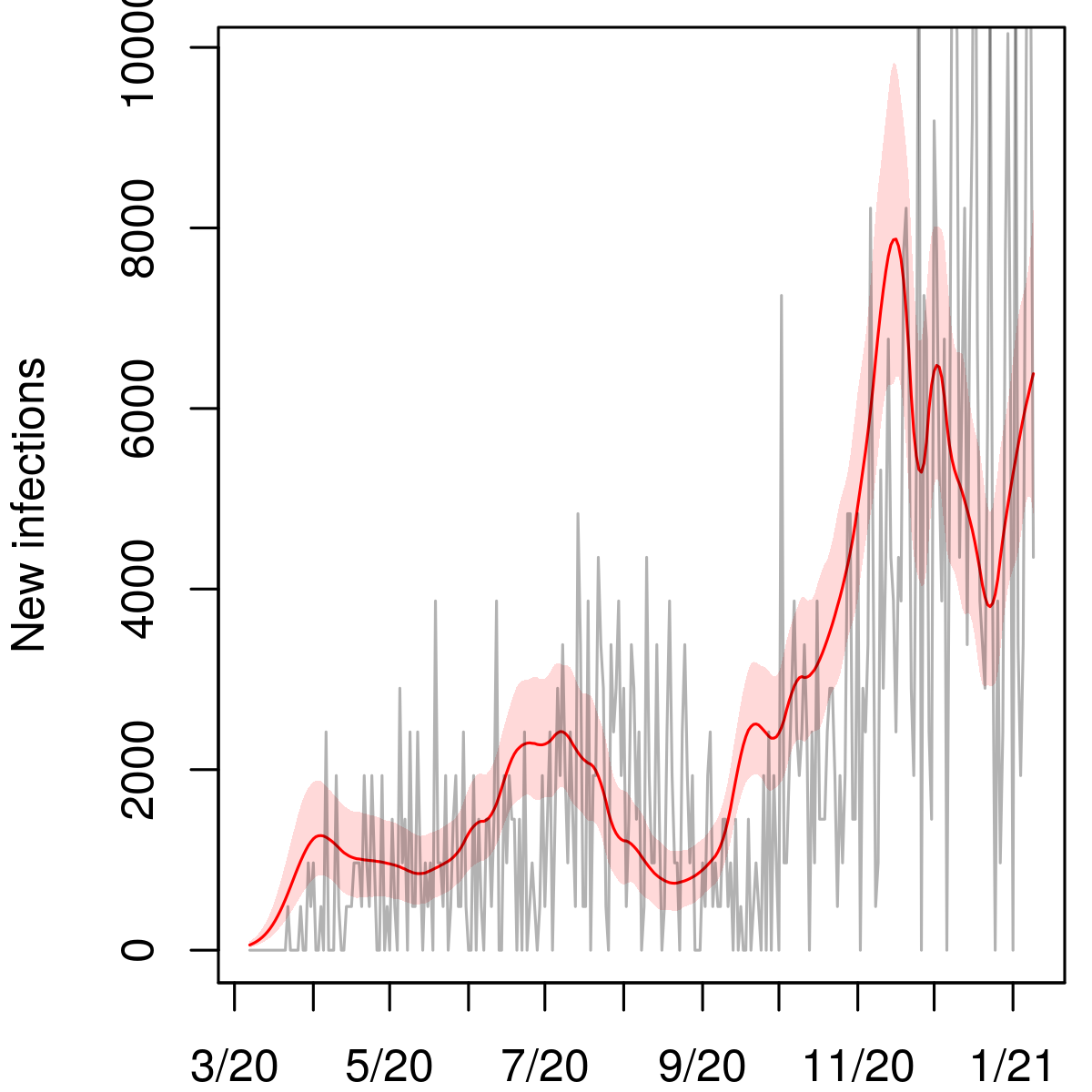}
&
\includegraphics[scale=0.77]{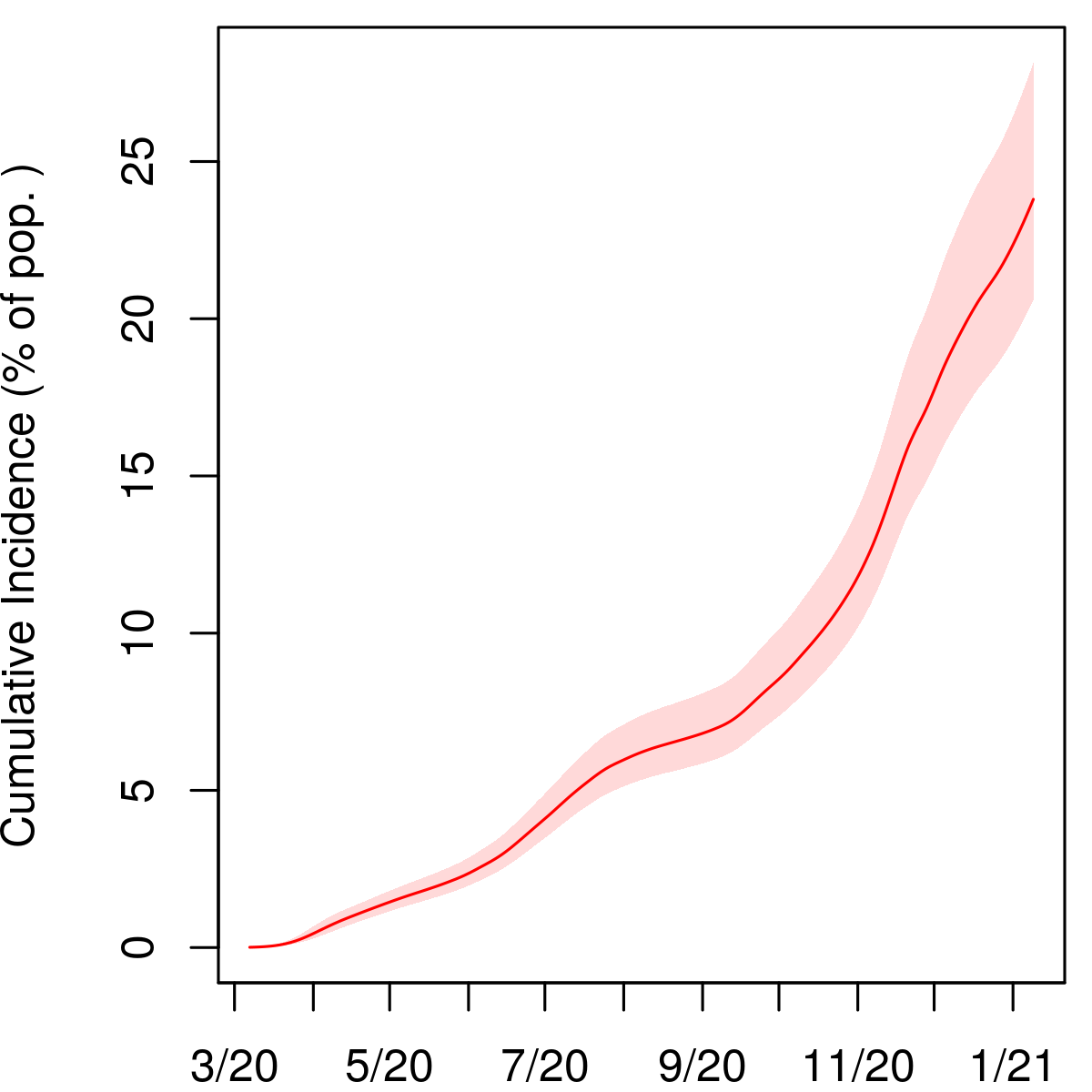} \\
\includegraphics[scale=0.77]{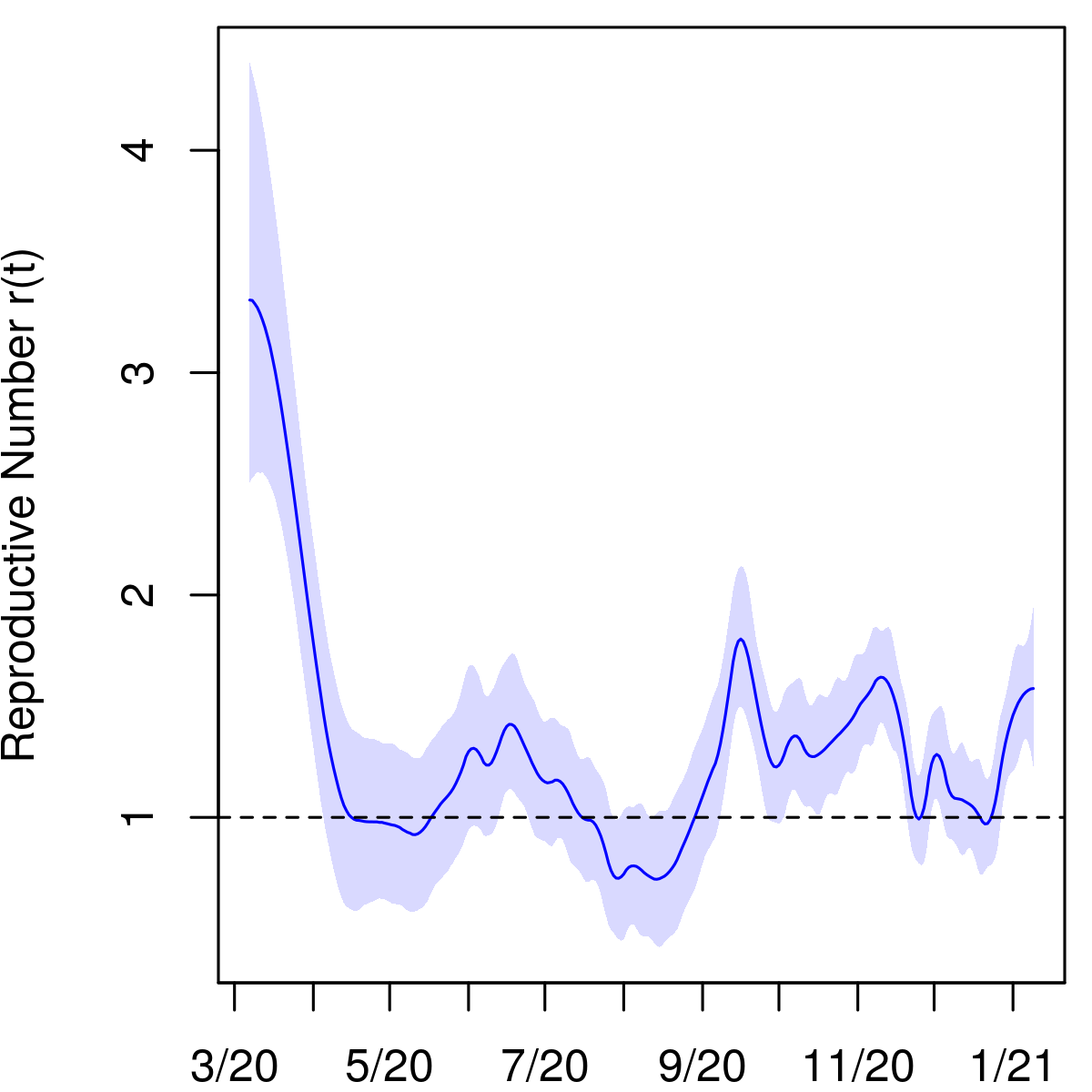}
&
\includegraphics[scale=0.77]{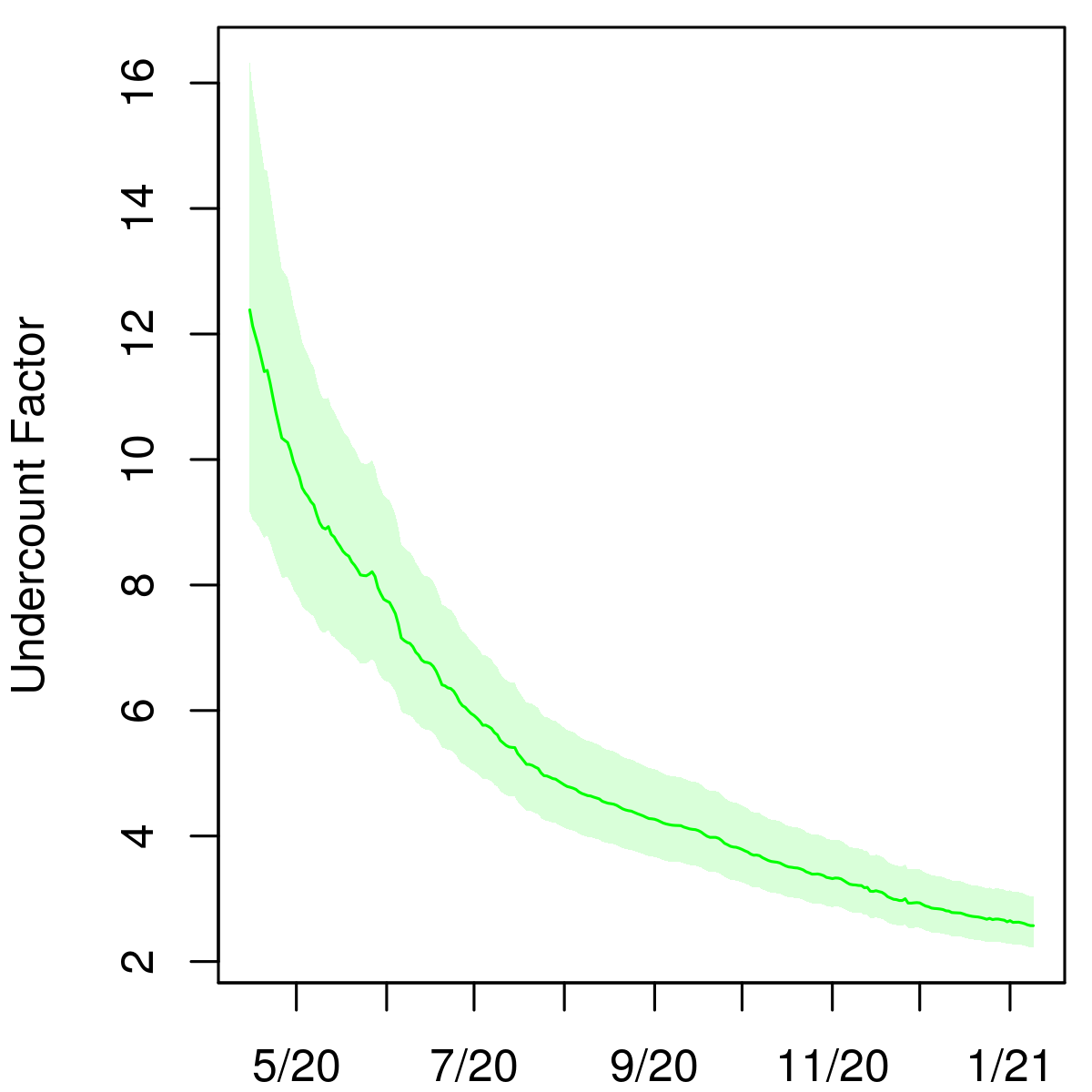} 
\end{tabular}
\caption{Posterior median and middle 95\% intervals for daily new infections, cumulative incidence, $r(t)$, and cumulative undercount from March 2020 to January 2021. In the top left panel, deaths divided by the posterior median IFR are plotted in grey for comparison.}
\end{figure}
\newpage
\begin{figure}[htbp!]
\textbf{Virginia}
\centering
\begin{tabular}{ll}
\includegraphics[scale=0.77]{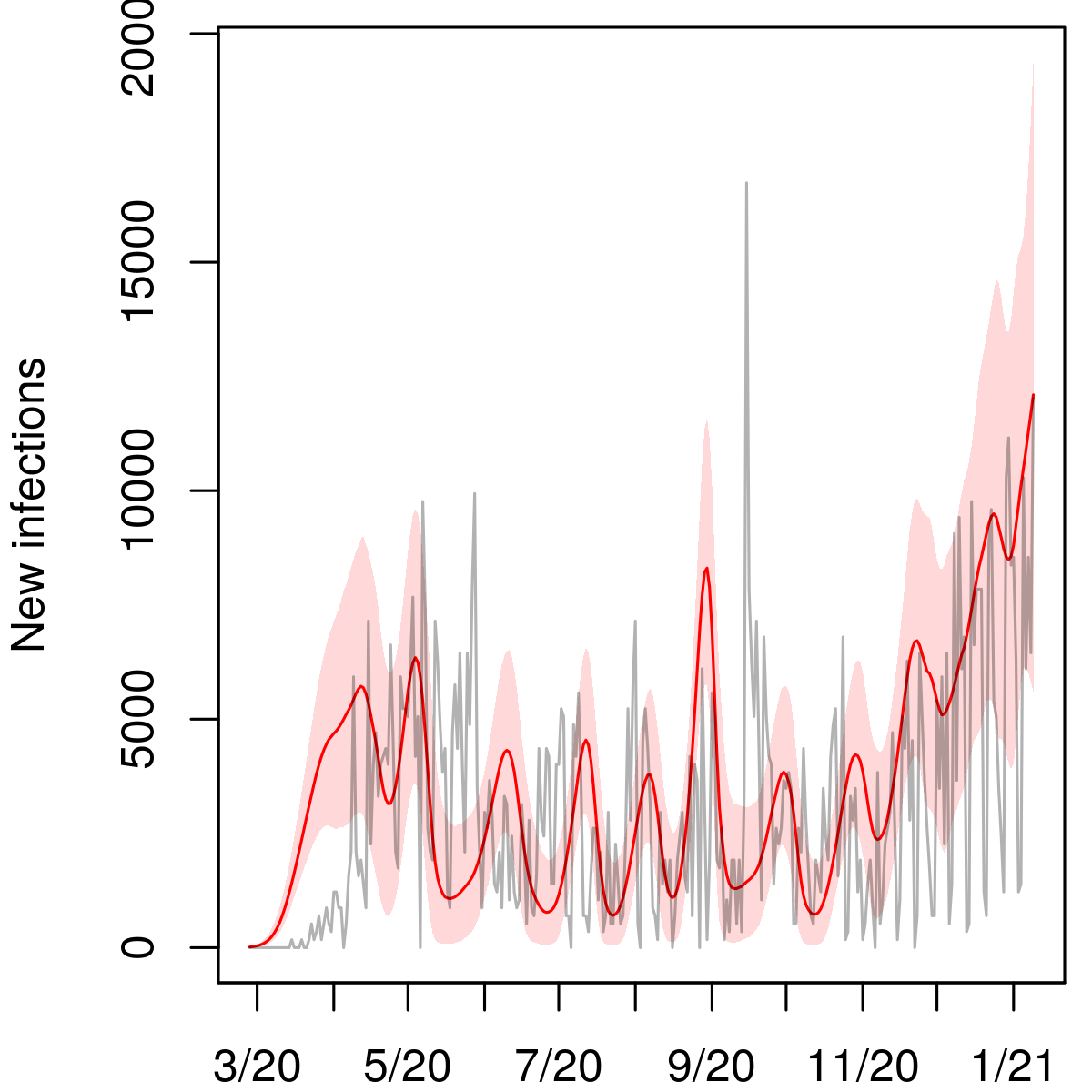}
&
\includegraphics[scale=0.77]{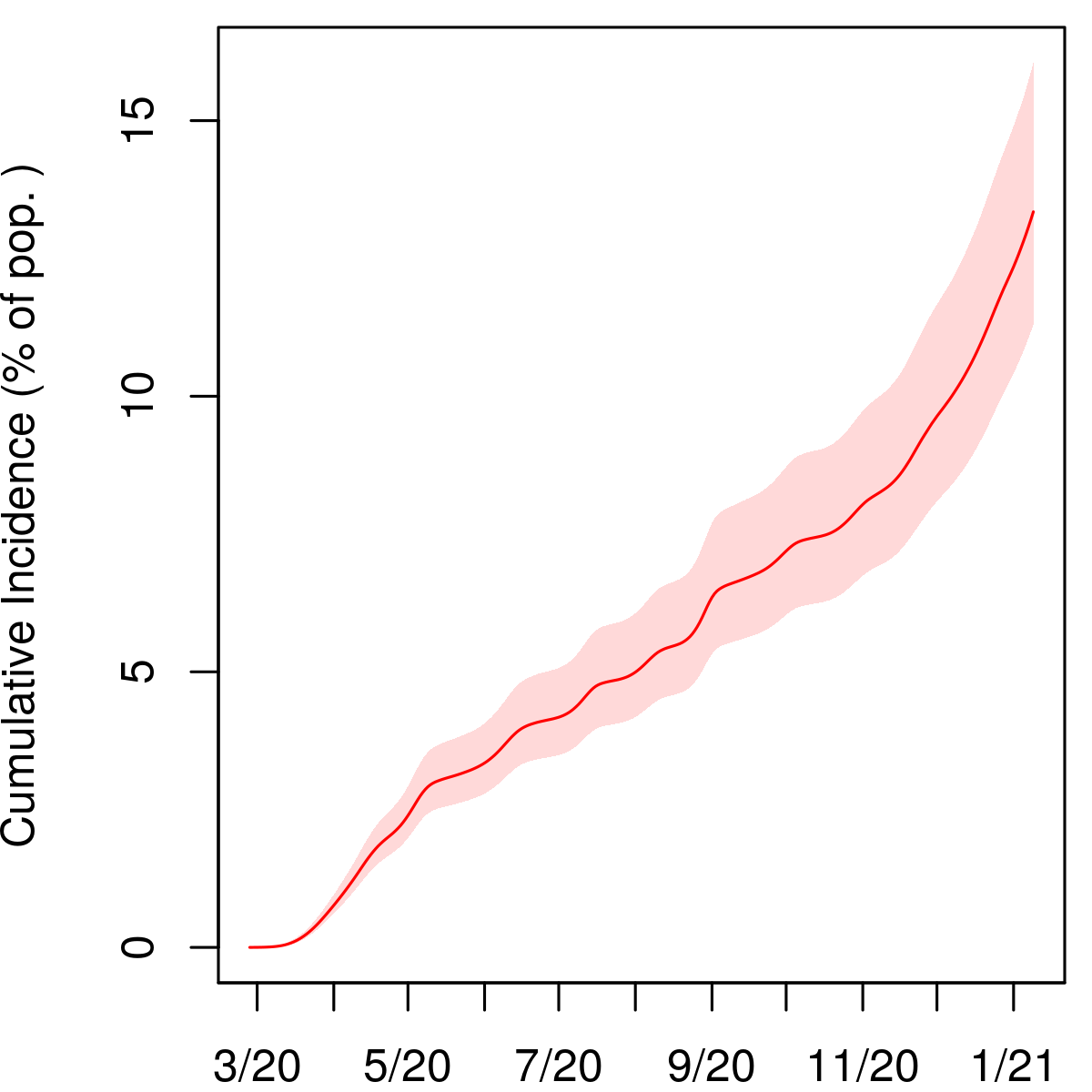} \\
\includegraphics[scale=0.77]{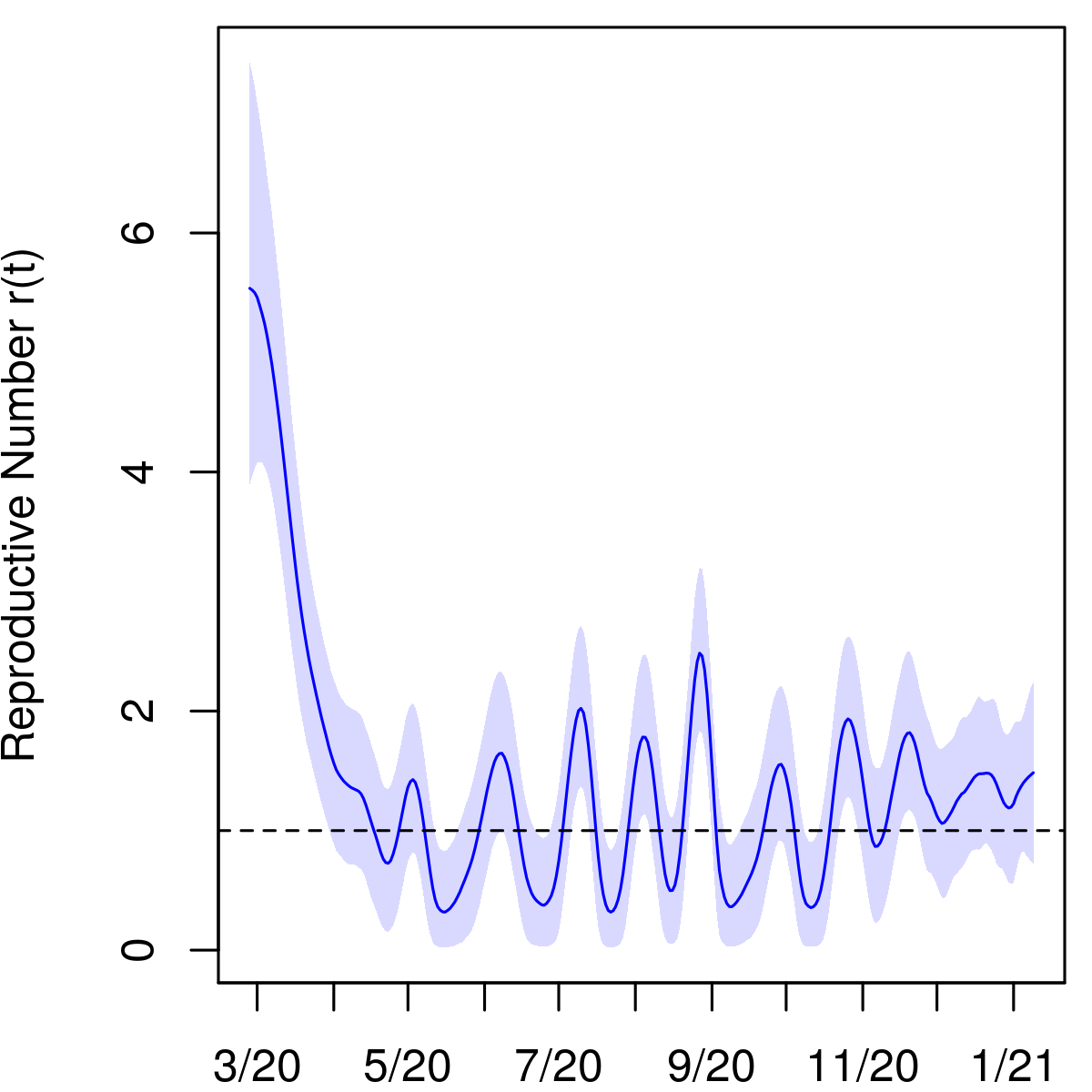}
&
\includegraphics[scale=0.77]{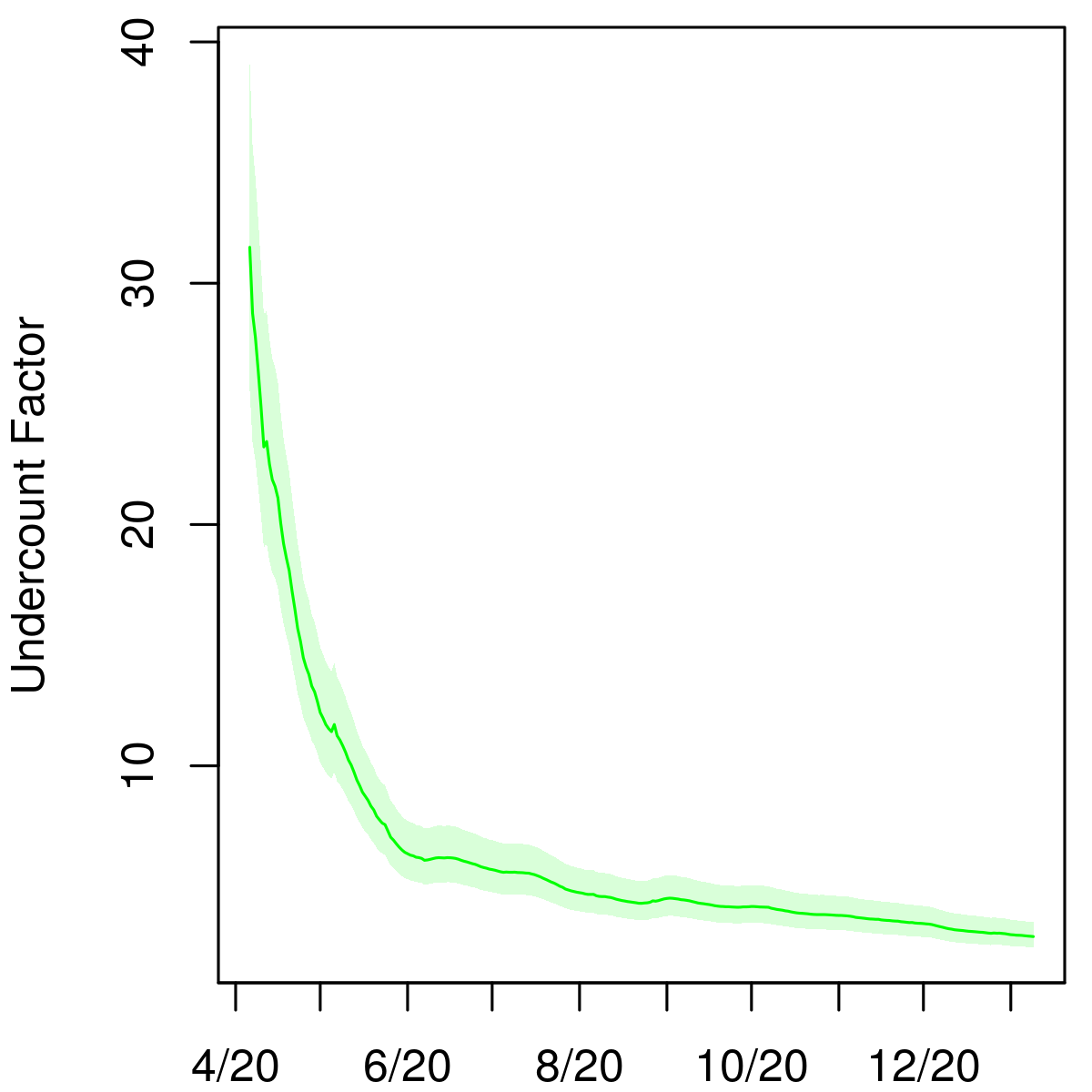} 
\end{tabular}
\caption{Posterior median and middle 95\% intervals for daily new infections, cumulative incidence, $r(t)$, and cumulative undercount from March 2020 to January 2021. In the top left panel, deaths divided by the posterior median IFR are plotted in grey for comparison.}
\end{figure}
\newpage
\begin{figure}[htbp!]
\textbf{Vermont}
\centering
\begin{tabular}{ll}
\includegraphics[scale=0.77]{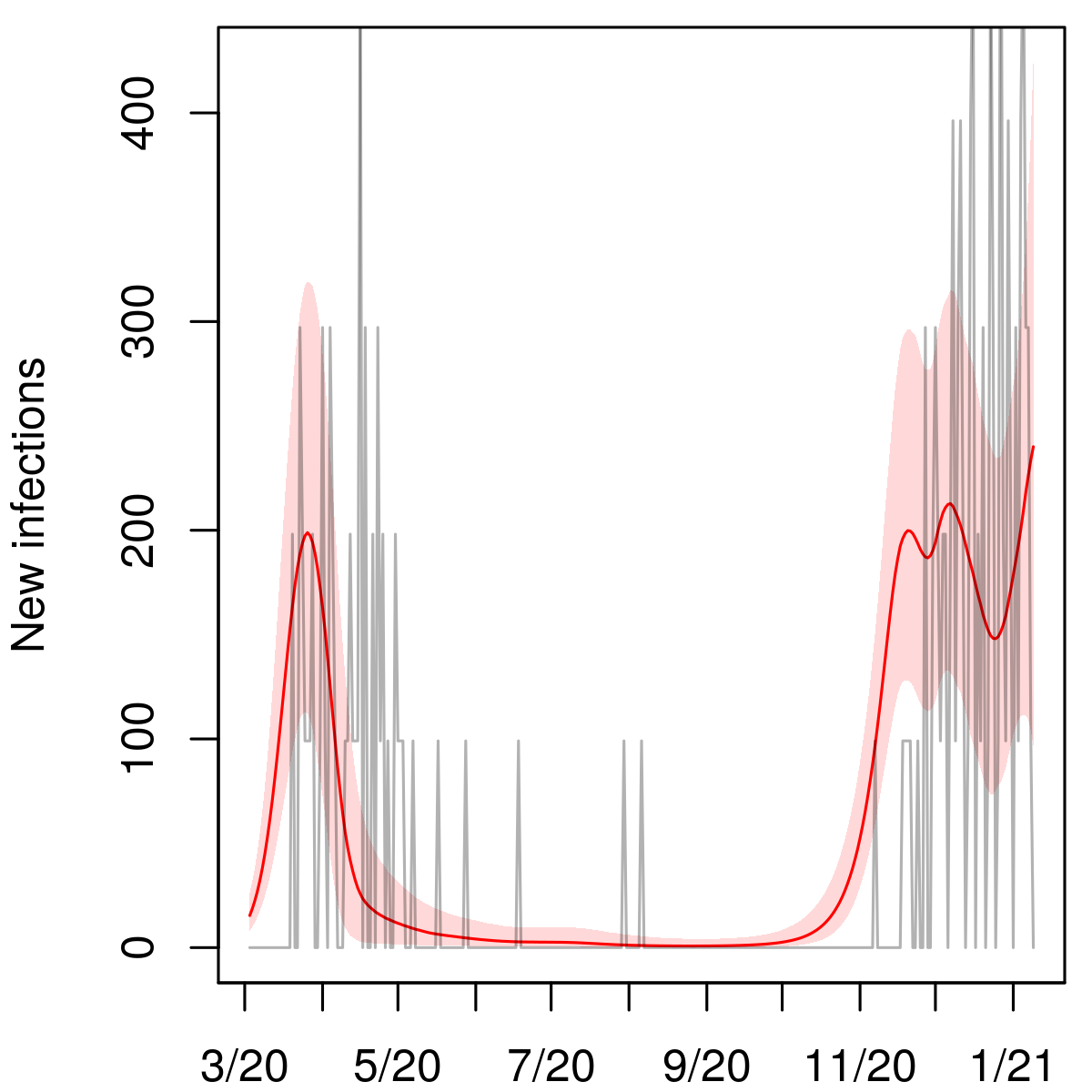}
&
\includegraphics[scale=0.77]{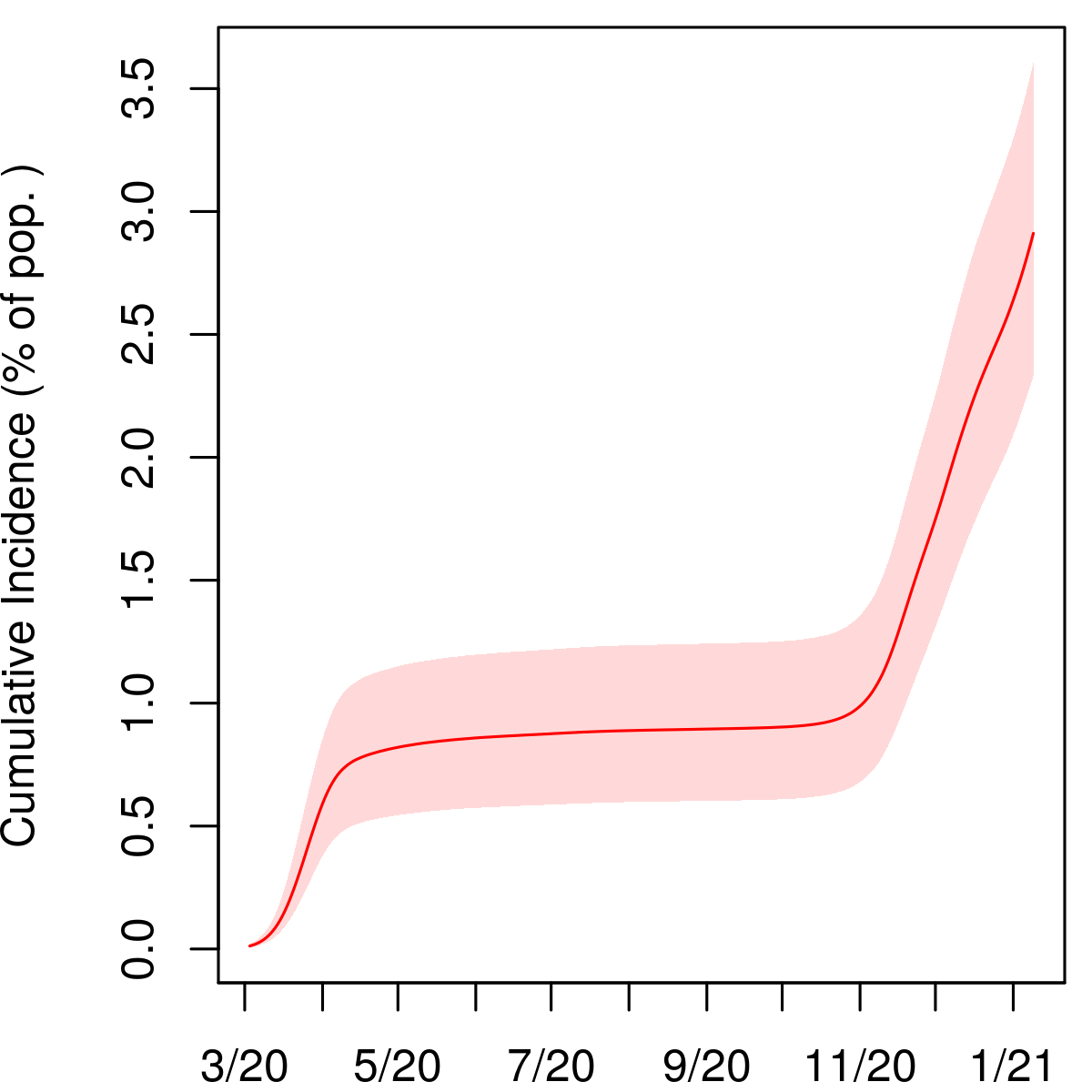} \\
\includegraphics[scale=0.77]{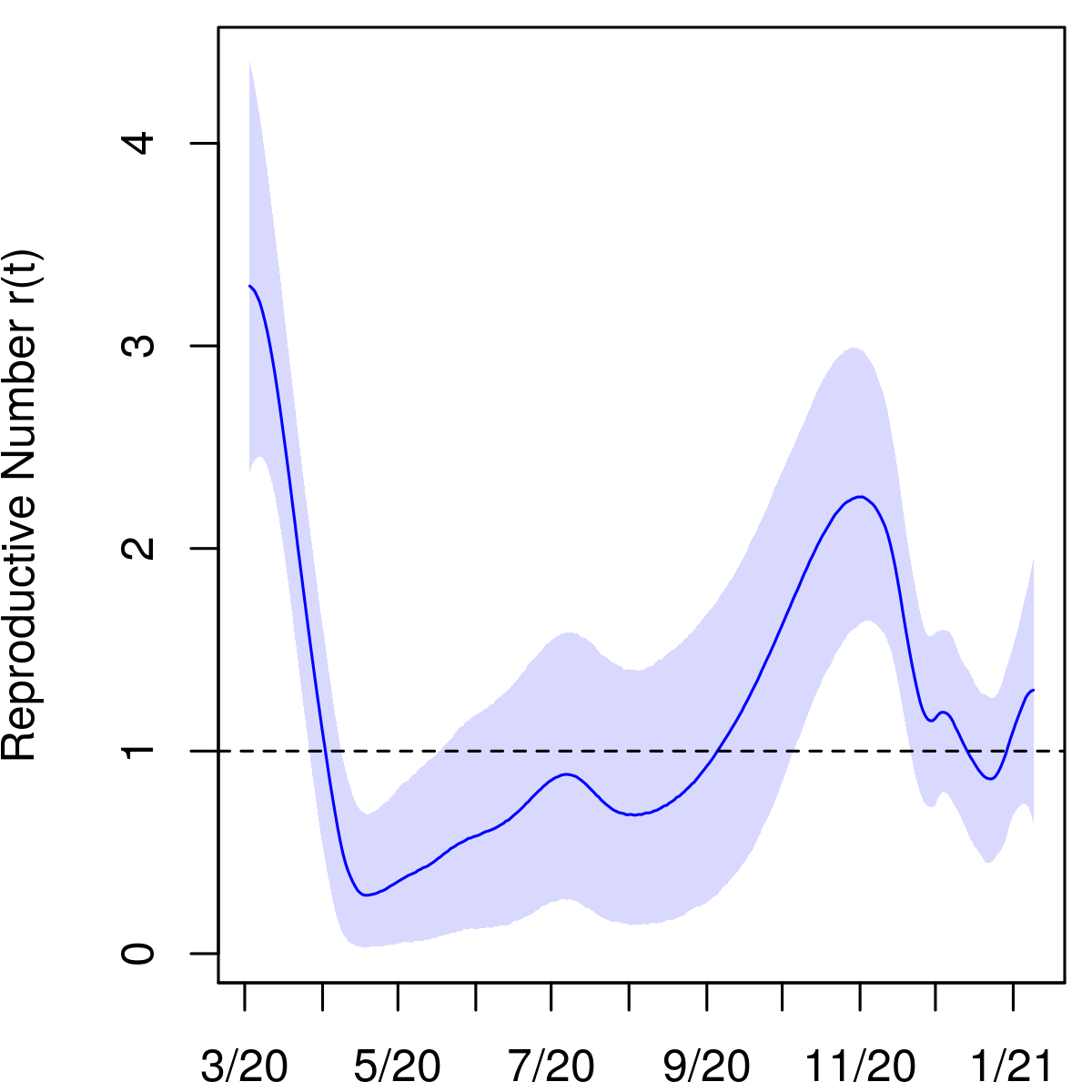}
&
\includegraphics[scale=0.77]{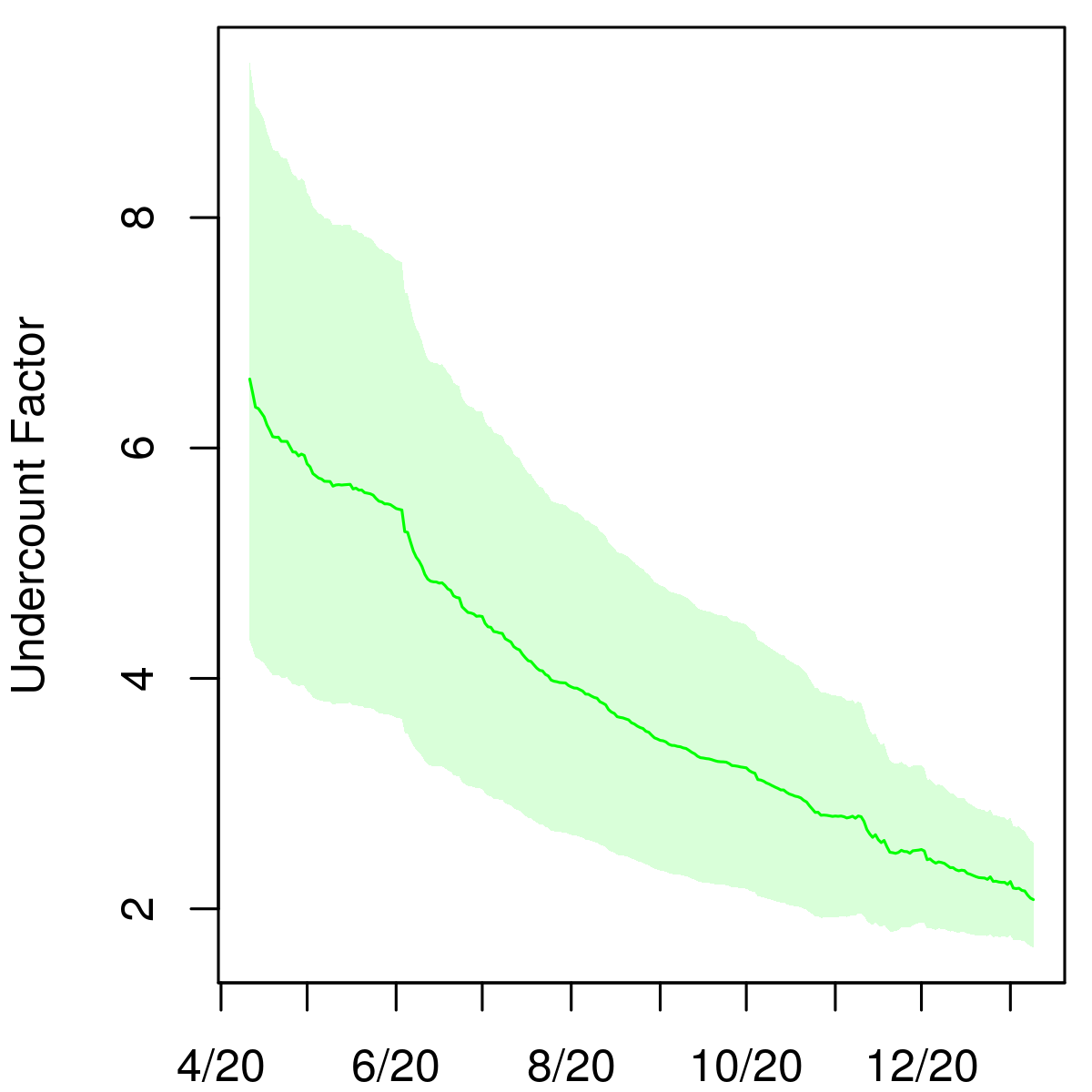} 
\end{tabular}
\caption{Posterior median and middle 95\% intervals for daily new infections, cumulative incidence, $r(t)$, and cumulative undercount from March 2020 to January 2021. In the top left panel, deaths divided by the posterior median IFR are plotted in grey for comparison.}
\end{figure}
\newpage
\begin{figure}[htbp!]
\textbf{Washington}
\centering
\begin{tabular}{ll}
\includegraphics[scale=0.77]{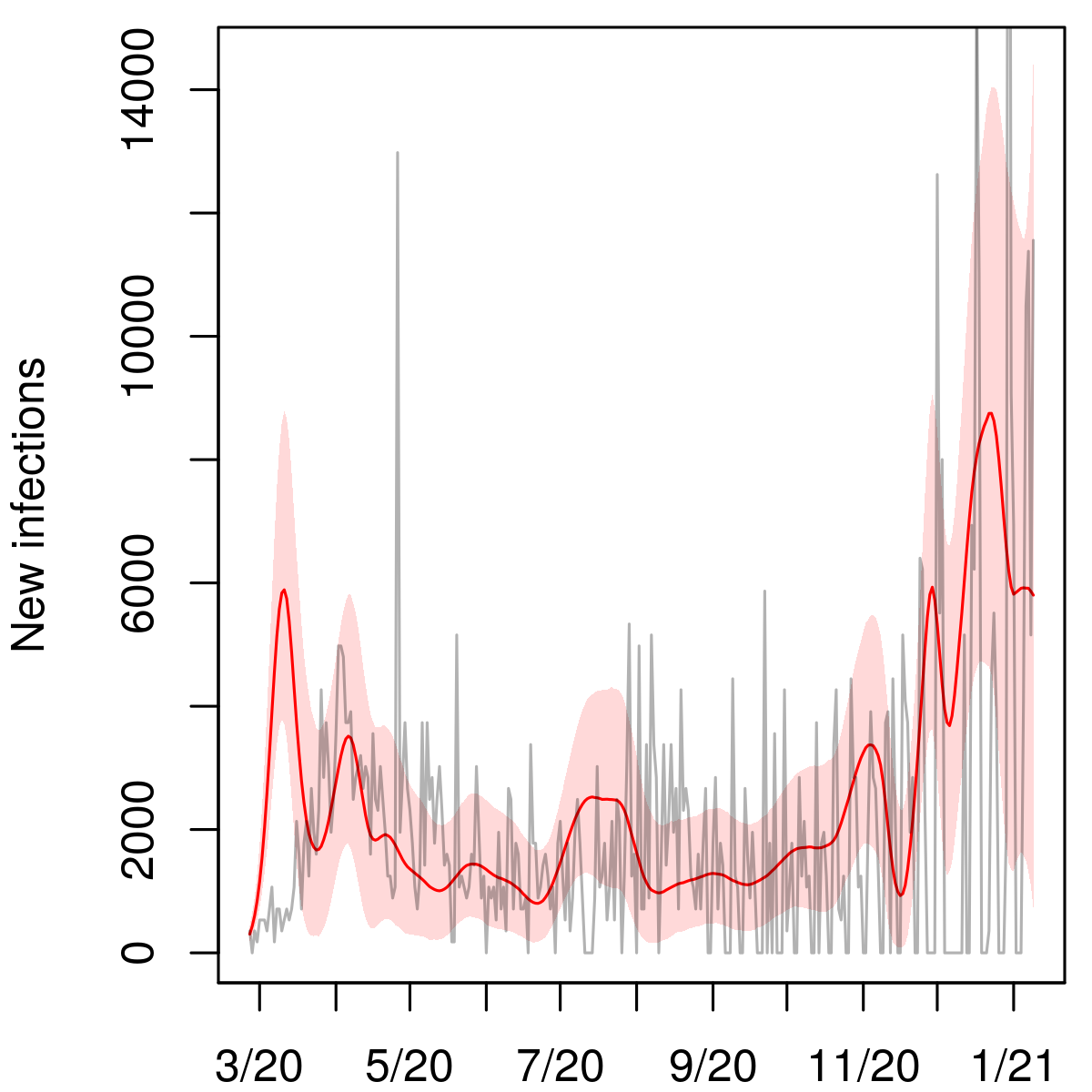}
&
\includegraphics[scale=0.77]{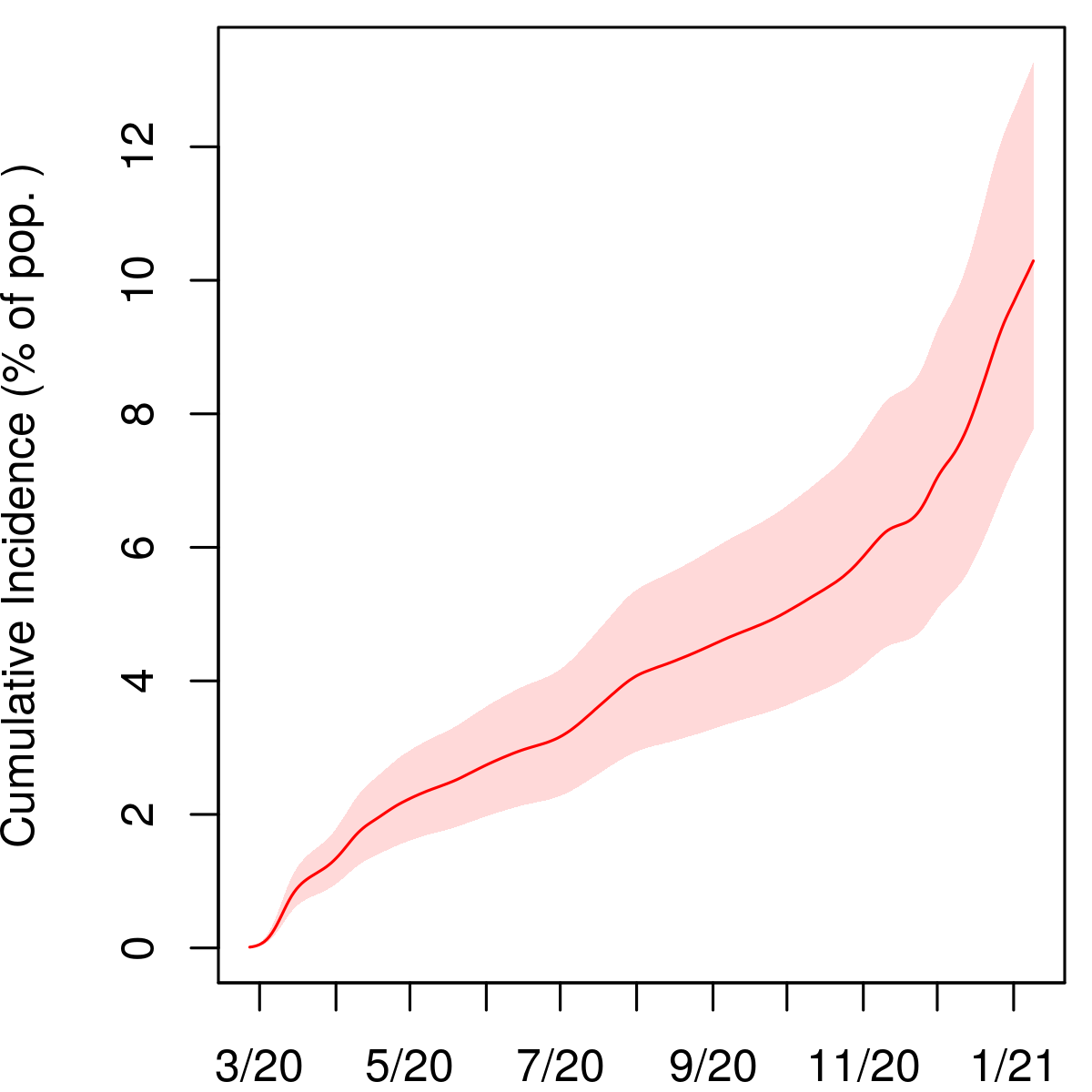} \\
\includegraphics[scale=0.77]{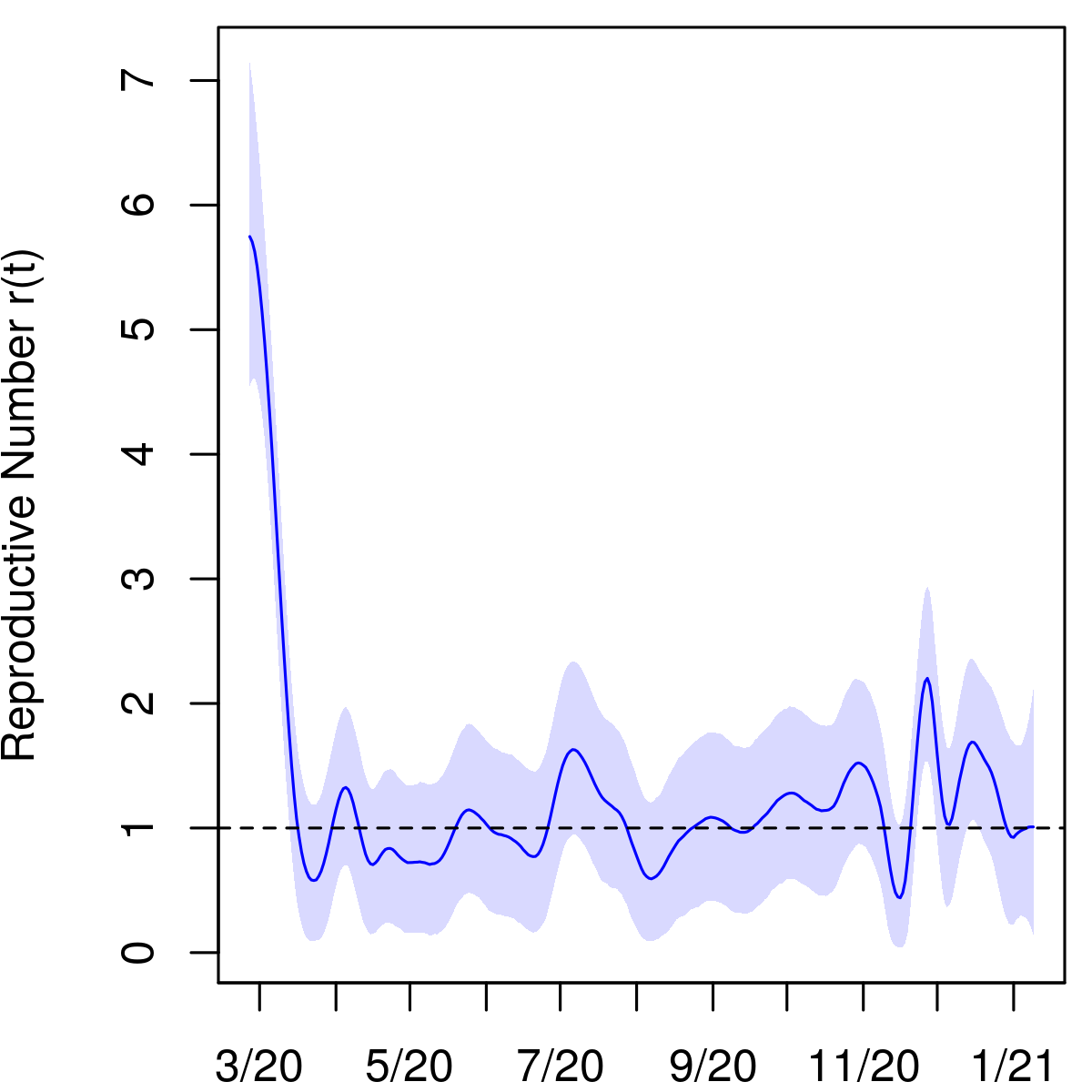}
&
\includegraphics[scale=0.77]{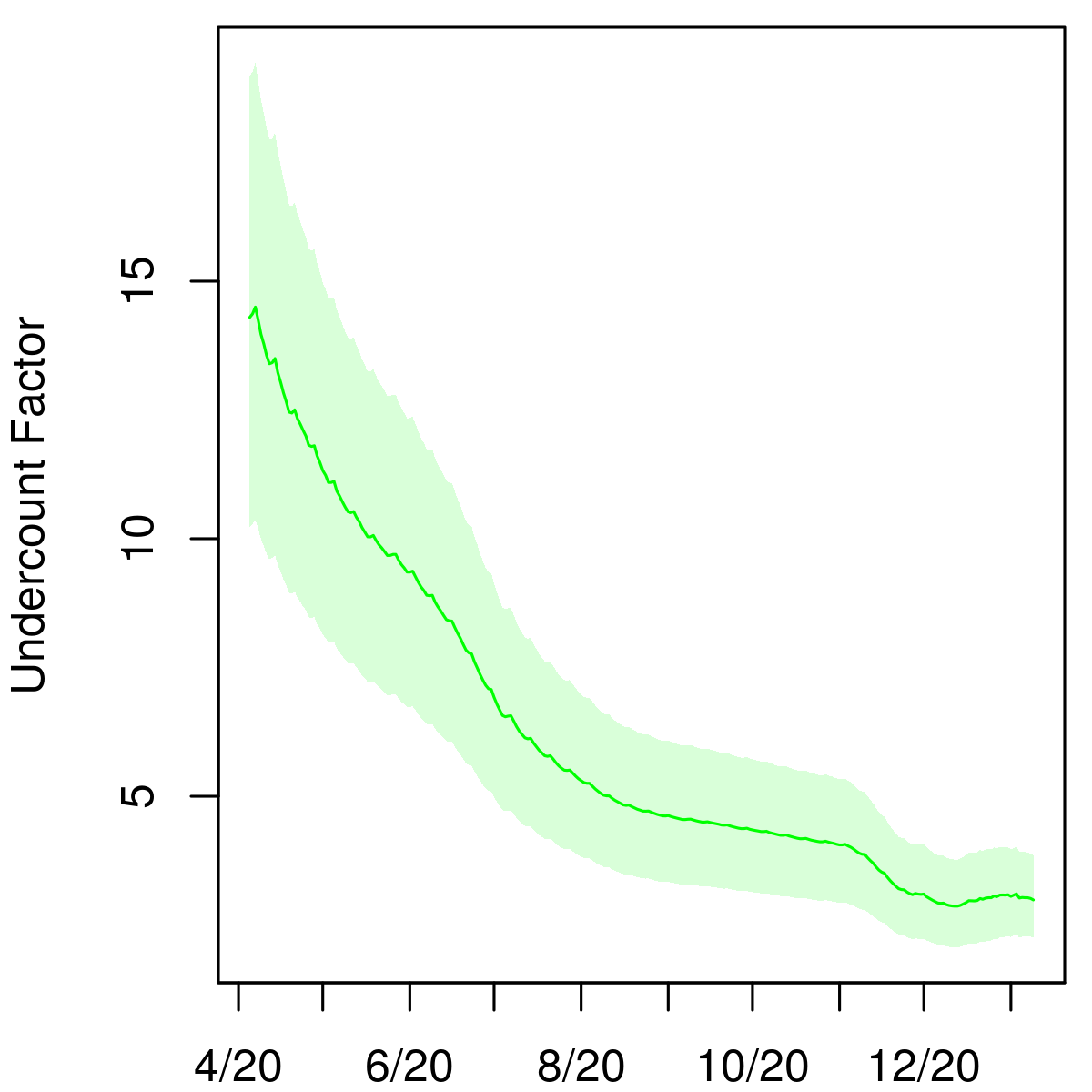} 
\end{tabular}
\caption{Posterior median and middle 95\% intervals for daily new infections, cumulative incidence, $r(t)$, and cumulative undercount from March 2020 to January 2021. In the top left panel, deaths divided by the posterior median IFR are plotted in grey for comparison.}
\end{figure}
\newpage
\begin{figure}[htbp!]
\textbf{Wisconsin}
\centering
\begin{tabular}{ll}
\includegraphics[scale=0.77]{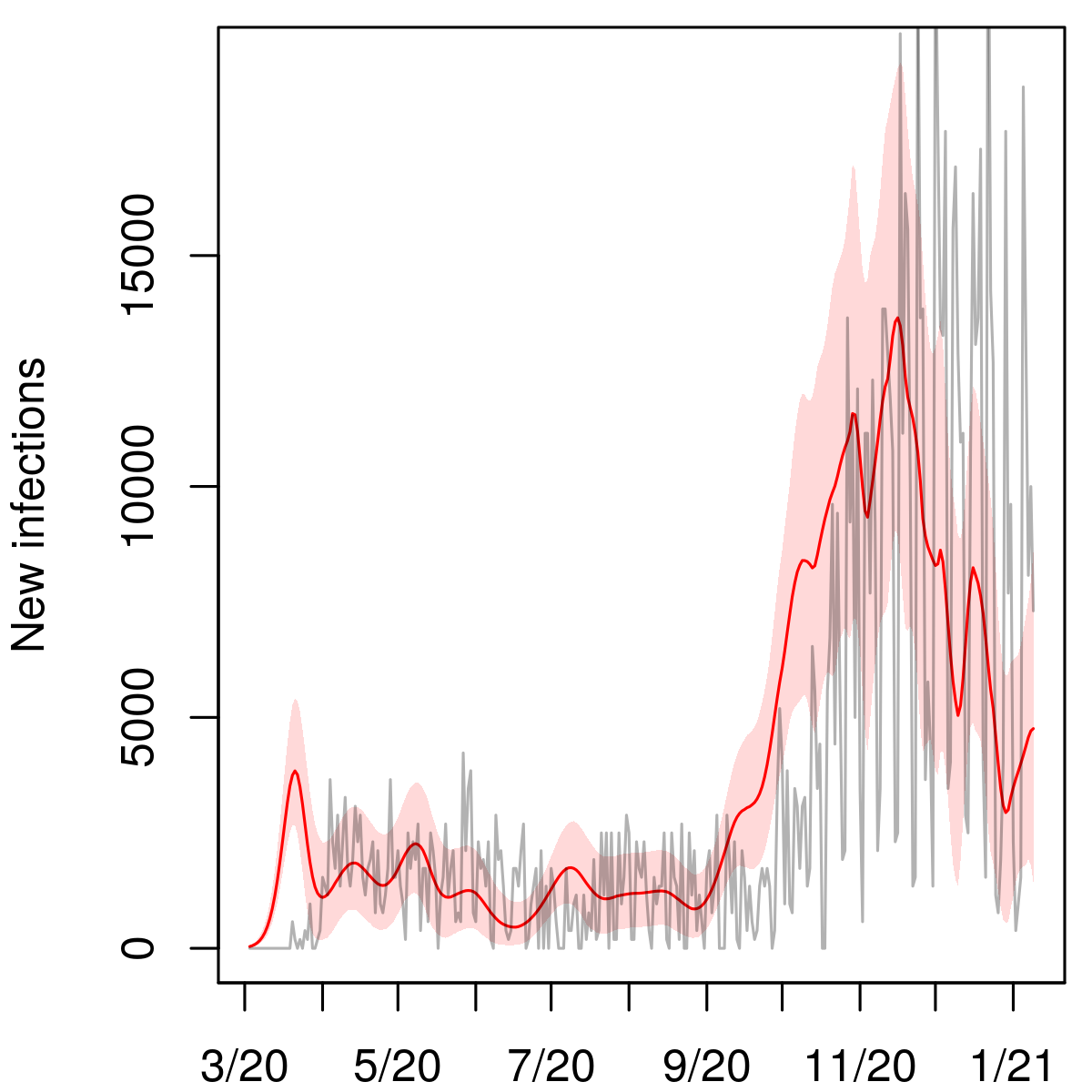}
&
\includegraphics[scale=0.77]{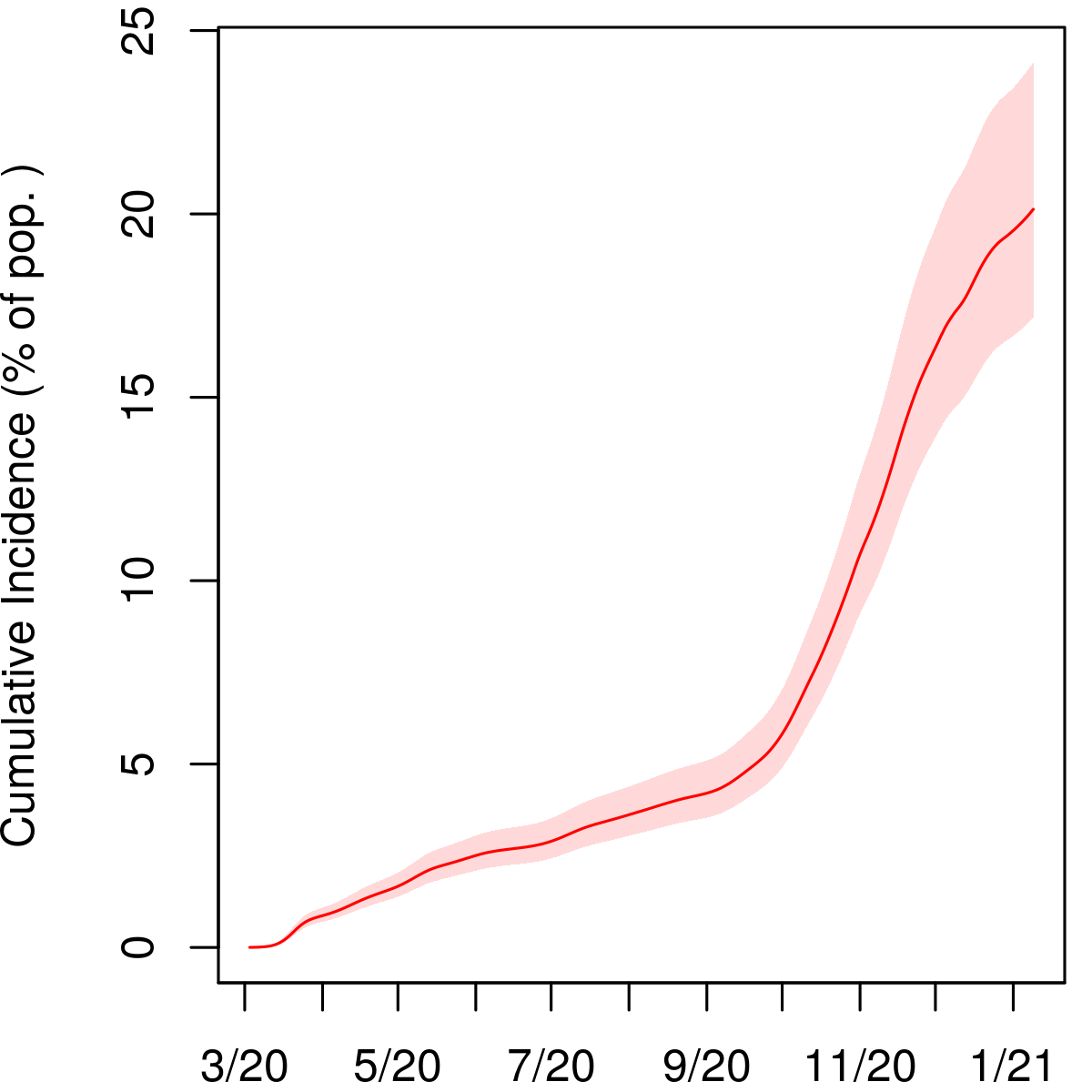} \\
\includegraphics[scale=0.77]{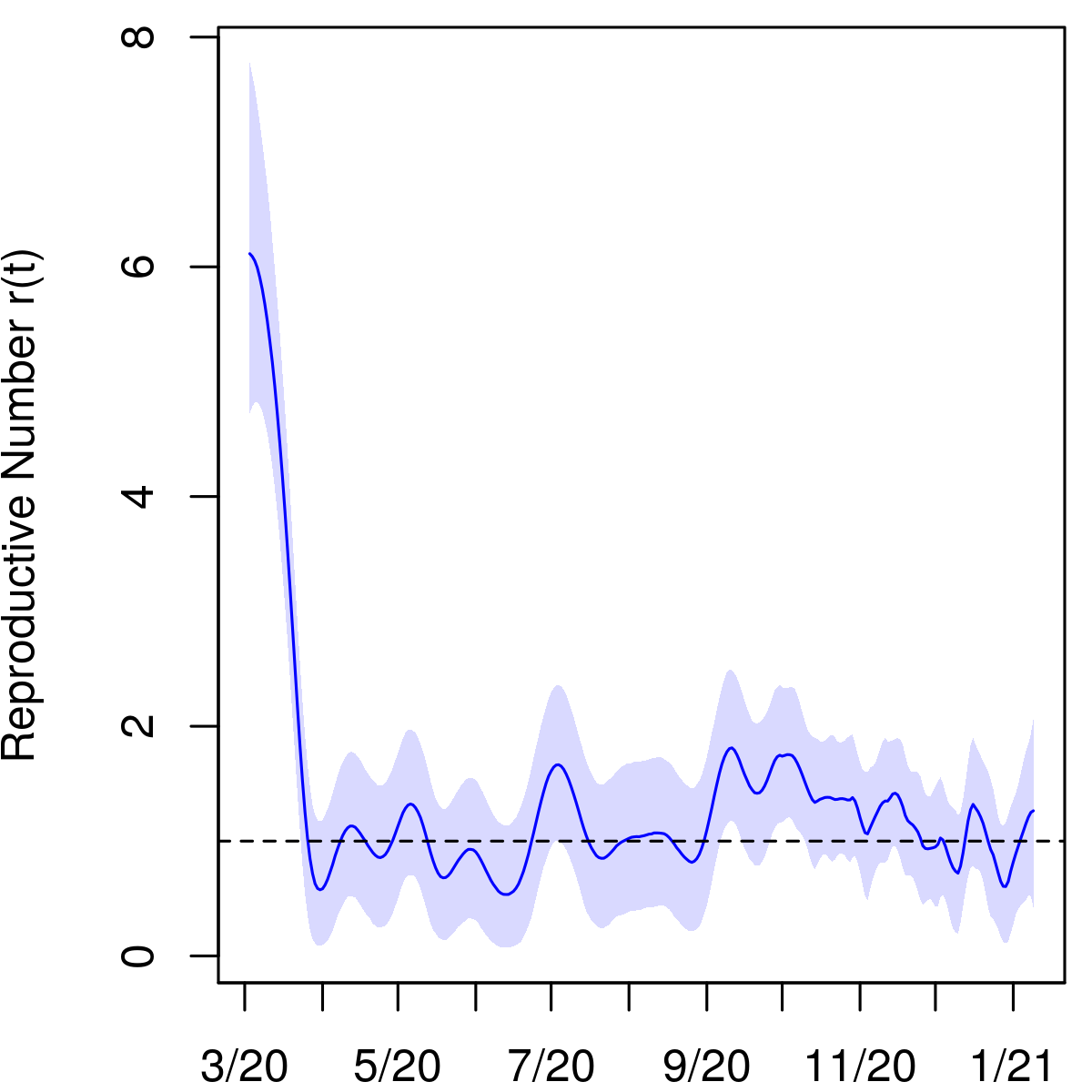}
&
\includegraphics[scale=0.77]{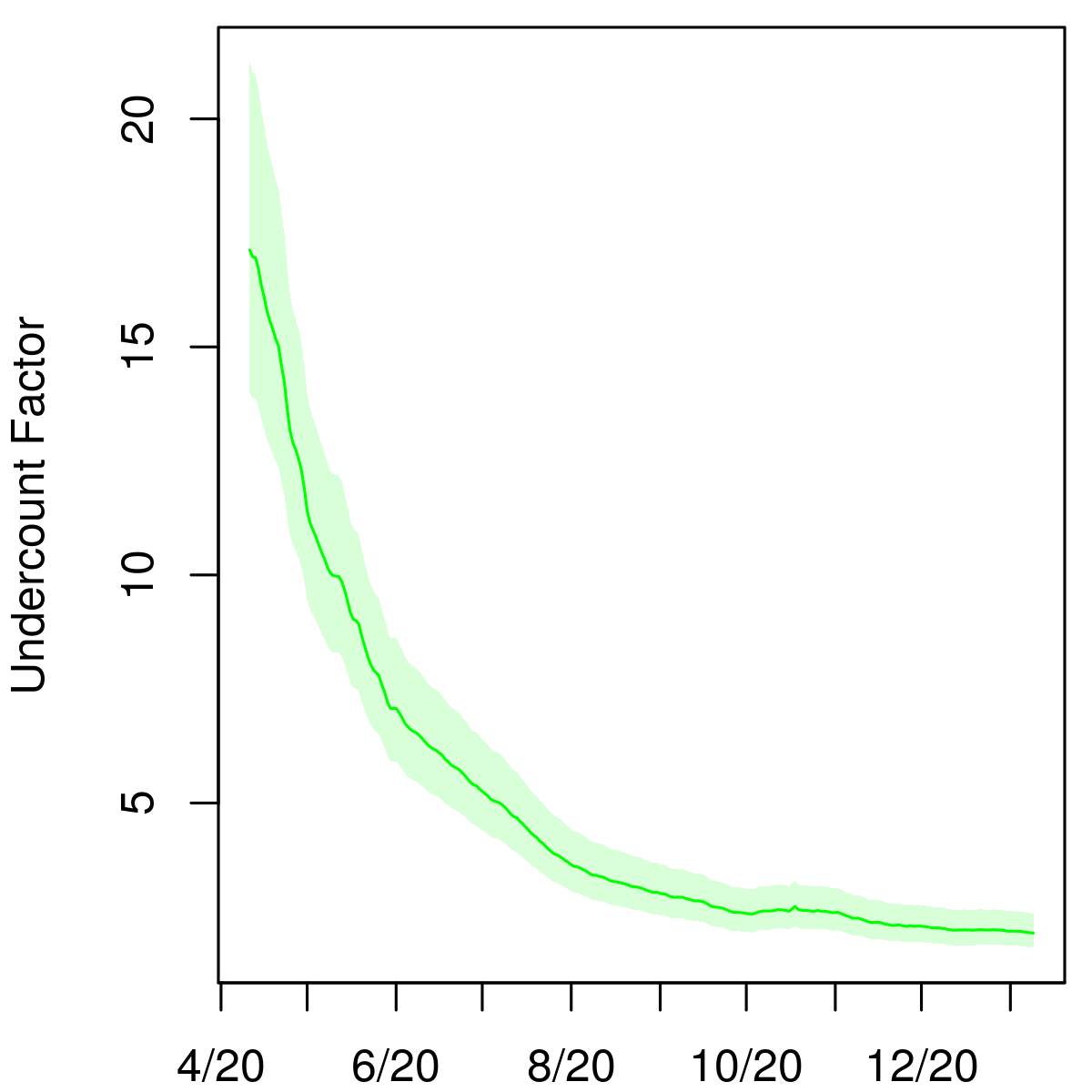} 
\end{tabular}
\caption{Posterior median and middle 95\% intervals for daily new infections, cumulative incidence, $r(t)$, and cumulative undercount from March 2020 to January 2021. In the top left panel, deaths divided by the posterior median IFR are plotted in grey for comparison.}
\end{figure}
\newpage
\begin{figure}[htbp!]
\textbf{West Virginia}
\centering
\begin{tabular}{ll}
\includegraphics[scale=0.77]{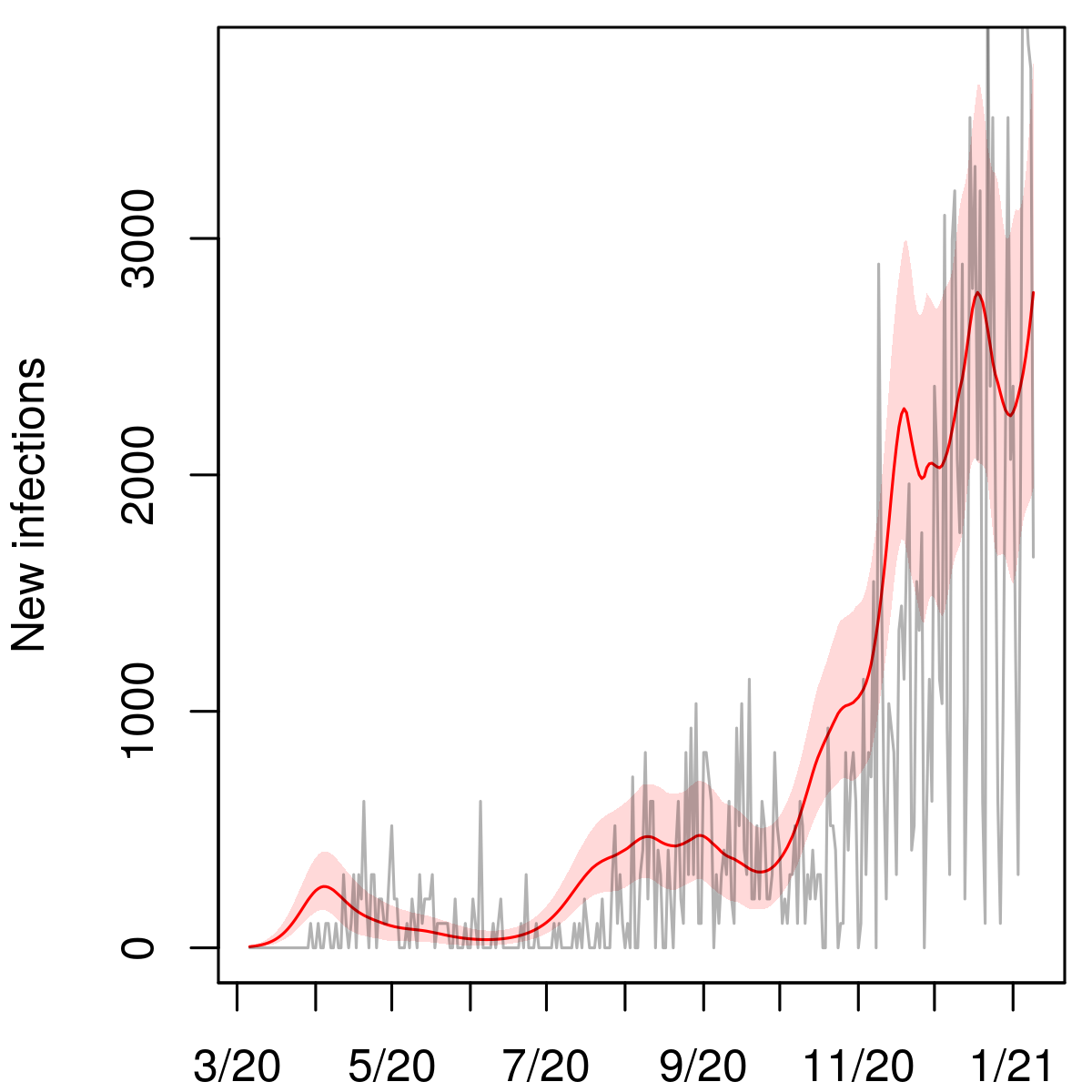}
&
\includegraphics[scale=0.77]{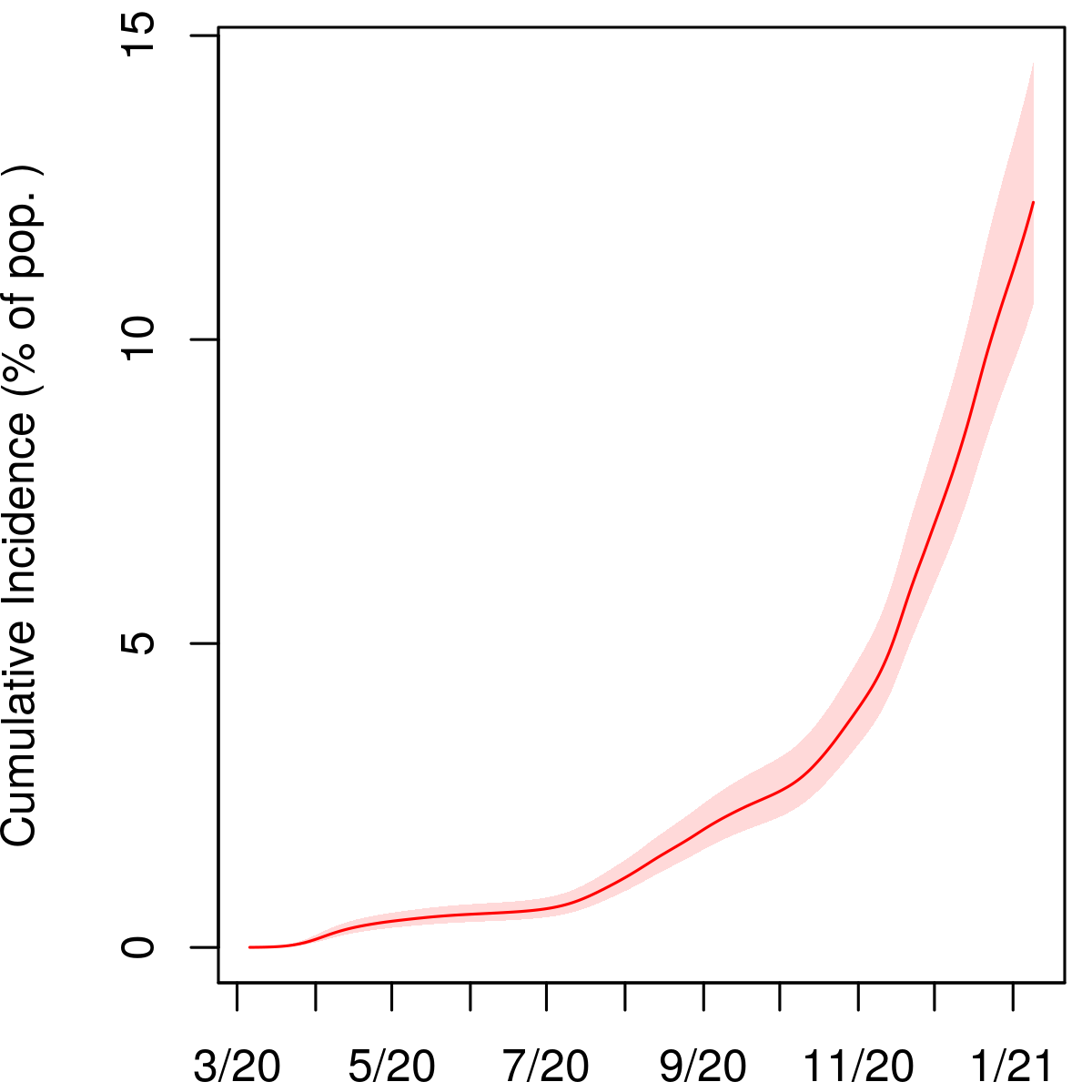} \\
\includegraphics[scale=0.77]{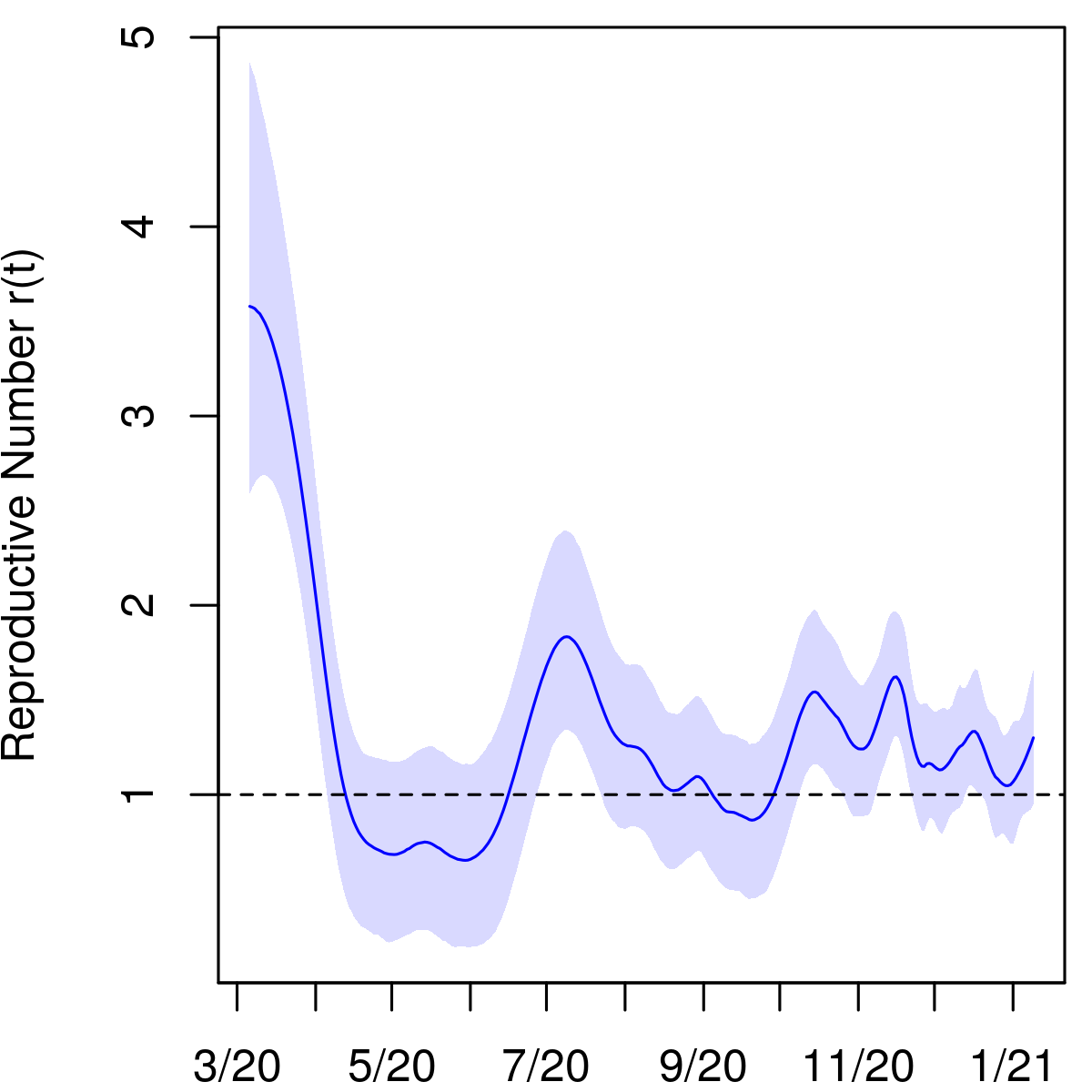}
&
\includegraphics[scale=0.77]{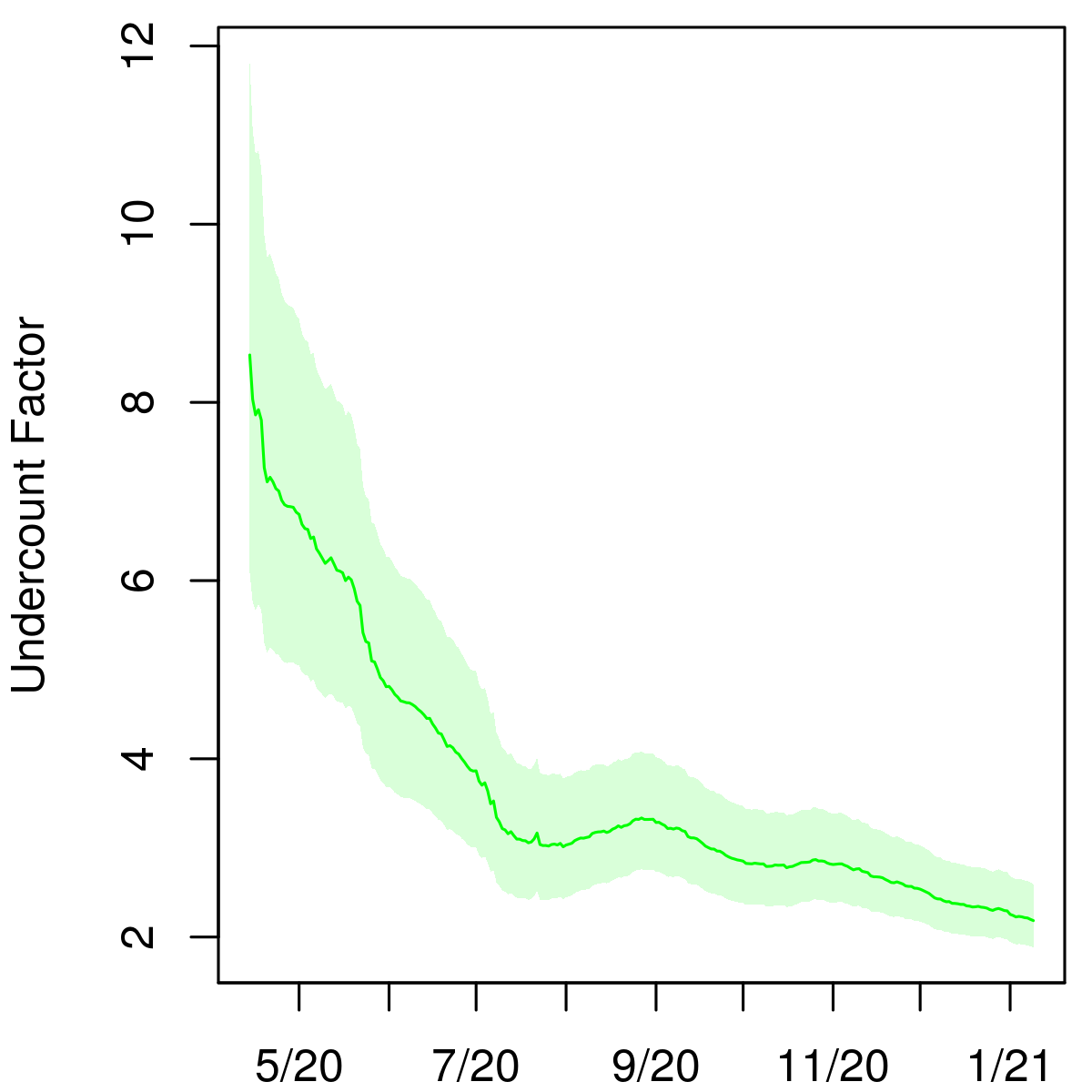} 
\end{tabular}
\caption{Posterior median and middle 95\% intervals for daily new infections, cumulative incidence, $r(t)$, and cumulative undercount from March 2020 to January 2021. In the top left panel, deaths divided by the posterior median IFR are plotted in grey for comparison.}
\end{figure}
\newpage
\begin{figure}[htbp!]
\textbf{Wyoming}
\centering
\begin{tabular}{ll}
\includegraphics[scale=0.77]{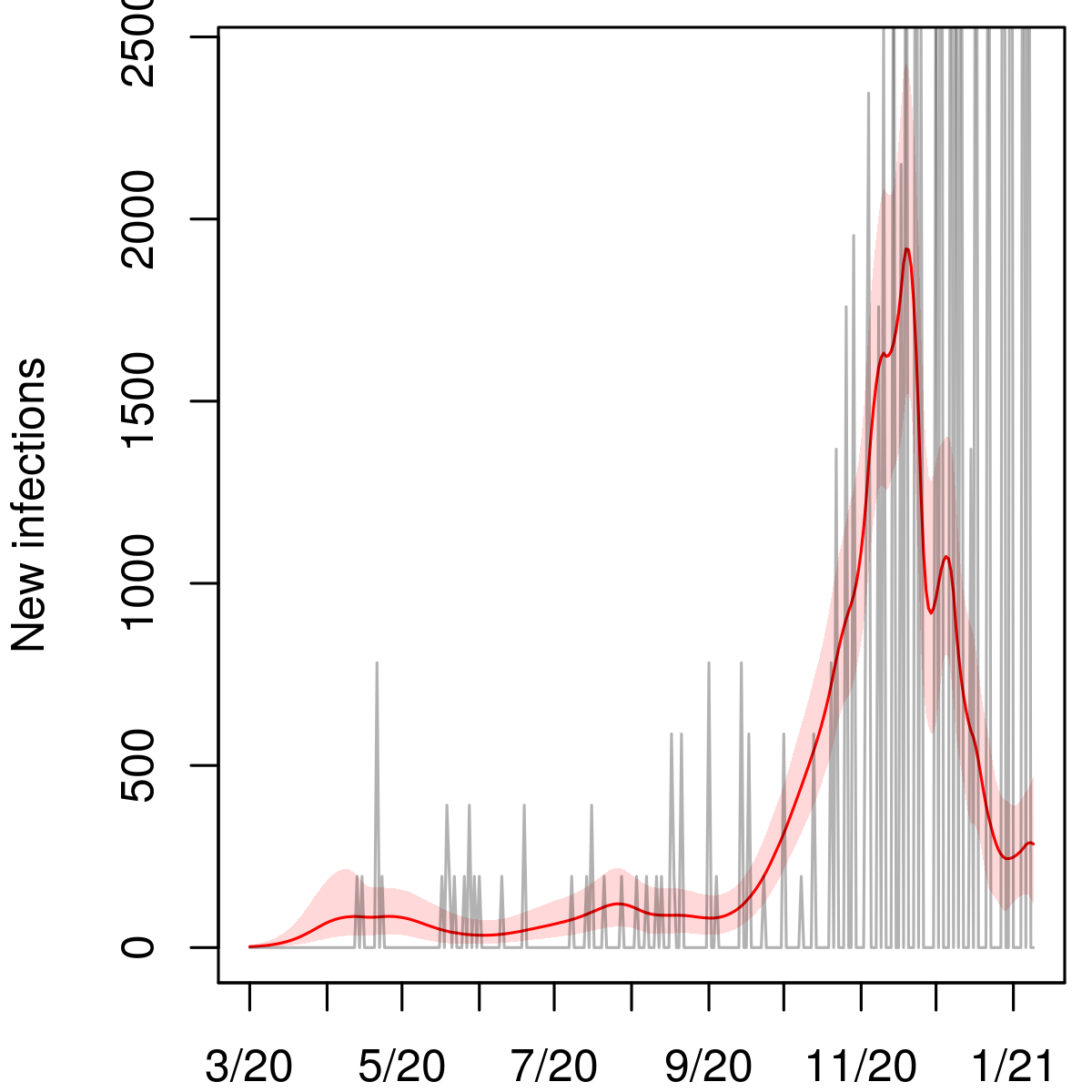}
&
\includegraphics[scale=0.77]{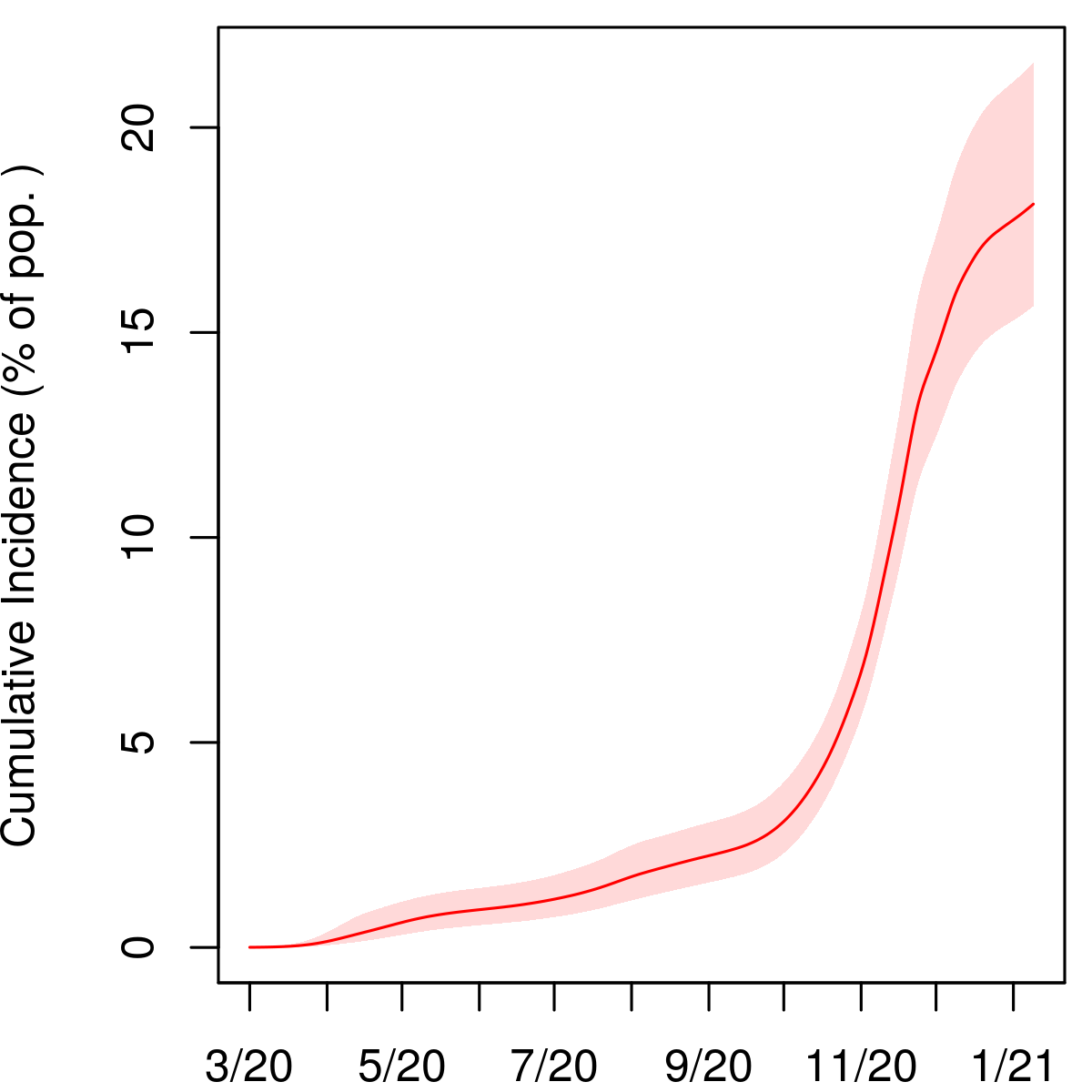} \\
\includegraphics[scale=0.77]{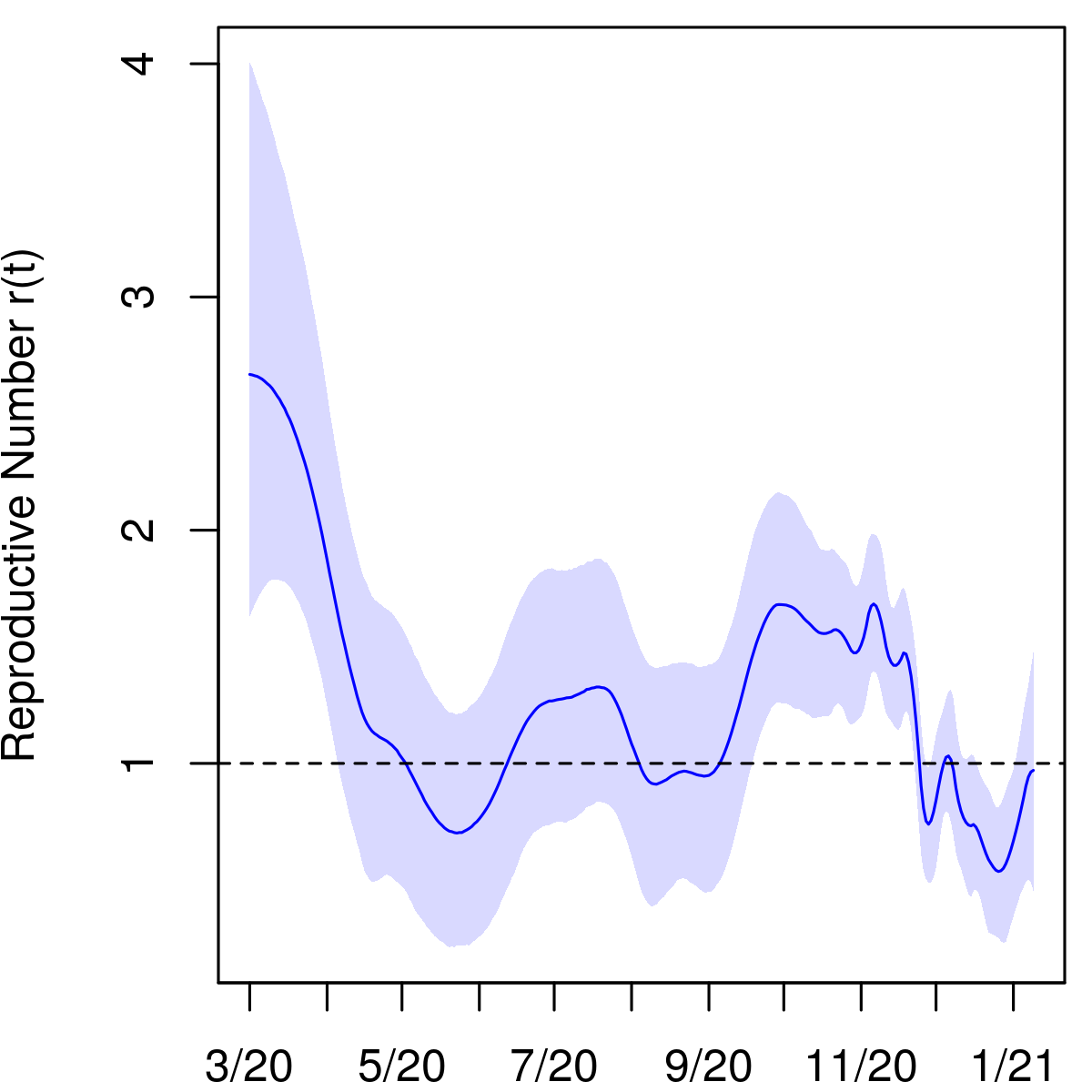}
&
\includegraphics[scale=0.77]{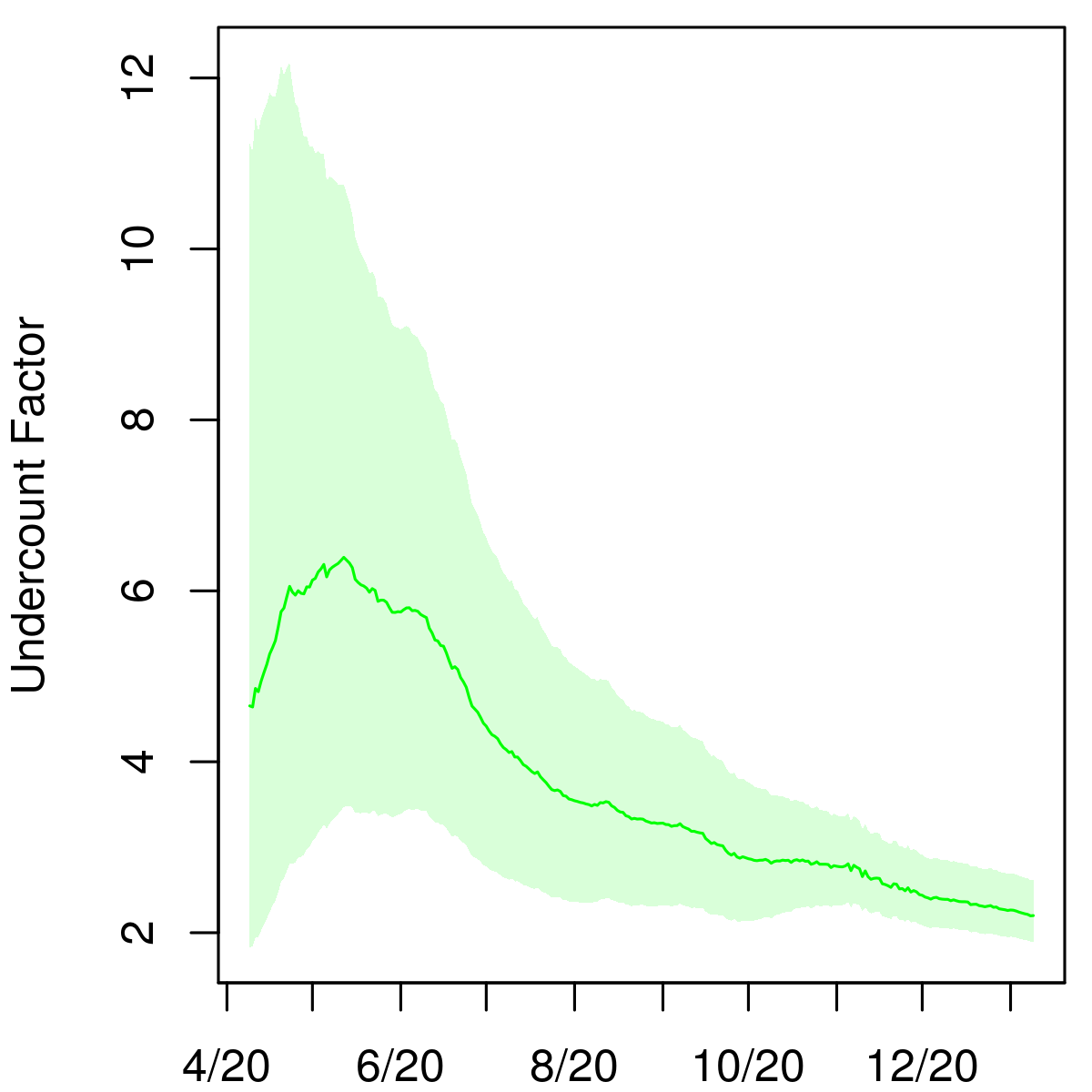} 
\end{tabular}
\caption{Posterior median and middle 95\% intervals for daily new infections, cumulative incidence, $r(t)$, and cumulative undercount from March 2020 to January 2021. In the top left panel, deaths divided by the posterior median IFR are plotted in grey for comparison.}
\end{figure}

\end{document}